\documentclass[11pt,a4paper]{article}
\usepackage{jstylemarg}
\usepackage[utf8]{inputenc}
\usepackage{amsmath}
\usepackage{mathtools}
\usepackage{tikz}
\usepackage{empheq}
\usepackage{enumitem}
\usepackage{graphics}
\usepackage{wrapfig}
\usepackage{caption}

\newcommand{\bi}{\begin{itemize}}
\newcommand{\ei}{\end{itemize}}
\newcommand{\bea}{\begin{align}}
\newcommand{\eea}{\end{align}}
\newcommand{\be}{\begin{equation}}
\newcommand{\ee}{\end{equation}}

\newcommand{\pl}{{\partial}}

\newcommand{\tcb}{\textcolor{blue}}

\makeatletter
\renewcommand*\env@matrix[1][\arraystretch]{%
  \edef\arraystretch{#1}%
  \hskip -\arraycolsep
  \let\@ifnextchar\new@ifnextchar
  \array{*\c@MaxMatrixCols c}}
\makeatother

\author[a,b]{Charlotte SLEIGHT}

\author[b]{\quad Massimo TARONNA}

\affiliation[a]{Universit\'e Libre de Bruxelles
and International Solvay Institutes\\
ULB-Campus Plaine CP231, 1050 Brussels, Belgium\\}

\affiliation[b]{Department of Physics, Princeton University, Princeton, NJ 08544}

\emailAdd{charlotte.sleight@gmail.com, mtaronna@princeton.edu}

\title{\centering
\huge{Spinning Mellin Bootstrap:\\Conformal Partial Waves, Crossing Kernels\\ and Applications}}

\abstract{We study conformal partial waves (CPWs) in Mellin space with totally symmetric external operators of arbitrary integer spin. The exchanged spin is arbitrary, and includes mixed symmetry and (partially)-conserved representations. In a basis of CPWs recently introduced in arXiv:1702.08619, we find a remarkable factorisation of the external spin dependence in their Mellin representation.
This property allows a relatively straightforward study of inversion formulae to extract OPE data from the Mellin representation of spinning 4pt correlators and in particular, to extract closed-form expressions for crossing kernels of spinning CPWs in terms of the hypergeometric function ${}_4F_3$. We consider numerous examples involving both arbitrary internal and external spins, and for both leading and sub-leading twist operators. As an application, working in general $d$ we extract new results for ${\cal O}\left(1/N\right)$ anomalous dimensions of double-trace operators induced by double-trace deformations constructed from single-trace operators of generic twist and integer spin. In particular, we extract the anomalous dimensions of double-trace operators $[\mathcal{O}_J\Phi]_{n,l}$ with ${\cal O}_J$ a single-trace operator of integer spin $J$.}

\begin{document}
\begin{flushright}    
  {\texttt{PUPT-2560}}
\end{flushright}
\maketitle

\section{Introduction}

The conformal bootstrap program has experienced a wave of successes in the past decade. A pivotal role has been played by the incredible progress of the available numerical methods to address the bootstrap quantitatively in $d>2$ (see e.g. \cite{Rattazzi:2008pe,ElShowk:2012ht,Beem:2013qxa, El-Showk:2014dwa,Kos:2016ysd,Dymarsky:2017yzx}). The development of suitable analytic methods has also been gaining traction, which have also clarified the successes of the numerical results from a theoretical standpoint. By now there are various complementary analytic techniques available, which include: Applications of slightly broken higher-spin symmetry \cite{Alday:2015ewa,Alday:2015ota,Alday:2016jfr}, large spin expansion \cite{Alday:2007mf,Fitzpatrick:2012yx,Komargodski:2012ek,Alday:2016njk,Simmons-Duffin:2016wlq}, Regge limit \cite{Costa:2012cb,Camanho:2014apa,Hartman:2015lfa,Kulaxizi:2017ixa,Li:2017lmh,Costa:2017twz}, inversion formulas \cite{Costa:2013zra,Hogervorst:2017sfd,Hogervorst:2017kbj,Gadde:2017sjg,Caron-Huot:2017vep,Simmons-Duffin:2017nub,Cardona:2018nnk} and also Mellin space techniques \cite{Mack:2009gy,Mack:2009mi,Penedones:2010ue,Paulos:2011ie,Costa:2012cb,Sen:2015doa,Gopakumar:2016cpb,Gopakumar:2016cpb,Nizami:2016jgt,Dey:2016mcs,Rastelli:2016nze,Cardona:2017tsw,Dey:2017fab,Faller:2017hyt,Dey:2017oim,Yuan:2018qva,Cardona:2018nnk} which conveniently encode complicated position space functions.

Correlators involving only scalar external operators have rightfully played a central role, providing an ideal testing ground owing to the number of explicit analytic results available. Among these are the explicit analytic expressions for conformal blocks in even dimensions \cite{Petkou:1994ad,Dolan:2000ut,Dolan:2003hv,Dolan:2011dv} and the simple form of the quadratic and quartic Casimir operators. On the other hand, limited progress has been made in the case where the external operators have non-trivial spin. This is mostly due to complications related to keeping track of the various tensor structures, which has somewhat hindered the development and application of analogous tools to those which worked so well for scalar correlators. So far, only a few numerical results are available for spinning correlators \cite{Iliesiu:2015qra,Dymarsky:2017yzx,Dymarsky:2017xzb}.\footnote{See however \cite{Li:2015itl,Hofman:2016awc,Elkhidir:2017iov} for progress on the analytic bootstrap for external spinning operators.} These results were made possible by the virtue of recursion relations for spinning conformal blocks \cite{Costa:2011dw,Costa:2014rya,Iliesiu:2015akf,Echeverri:2015rwa,Echeverri:2016dun,Costa:2016hju,Costa:2016xah,Karateev:2017jgd}, which are particularly well-suited for numerical implementation. The latter results however, being tuned to set up the numerical problem, did not yet allow for a detailed analysis of the actual structure of the spinning conformal blocks and applications thereof.

In this work we aim to lay down the groundwork for a detailed study of the analytic conformal bootstrap for arbitrary spinning external legs in Mellin space.\footnote{The Mellin space approach to the conformal bootstrap was initiated in \cite{Sen:2015doa,Gopakumar:2016wkt,Gopakumar:2016cpb} for external scalar operators.} We study in detail the explicit form of spinning conformal partial waves (CPWs) with the aim of acquiring a better handle on their structure, which in a particular basis \cite{Sleight:2017fpc} turns out to exhibit a rather attractive factorisation of the external spin dependence. This property of the basis \cite{Sleight:2017fpc} makes it particularly apt for the extension of bootstrap methods for scalar correlators to those with arbitrary spinning operators. We present various direct applications of our results, including the study of crossing kernels and $1/N$ corrections to OPE data induced by double-trace deformations. We postpone the application of these results to more advanced bootstrap problems to future works (see e.g. \cite{Sleight:2018ryu} and \cite{TTTT}), which include the $\epsilon$- and large-spin expansions.

The plan of the paper and a brief summary of results is as follows:
\begin{itemize}
    \item In Section \S\tcb{\ref{NewBasis}} we review the pertinent details of the basis of 3pt conformal structures recently introduced in \cite{Sleight:2017fpc}, which has simple transformation properties under global higher-spin symmetry transformations.\footnote{The basis itself was originally obtained in the context of simplifying the kinematic map between bulk cubic couplings and 3pt conformal structures, so our proposed framework also lends itself nicely to studies of bulk physics.} In \S \tcb{\ref{subsec::cpws}} we discuss the corresponding CPWs in position space, in particular their representation within the shadow formalism \cite{Ferrara:1972xe,Ferrara:1972uq} as an integrated product of the latter 3pt conformal structures. In \S\tcb{\ref{Shadow}} we evaluate in great detail the shadow transform for a large class of spinning 3pt correlators and obtain a simple closed formula. In this context we also discuss in detail the bulk counterpart of the shadow transform and how to use the corresponding basis of bulk cubic couplings to conveniently obtain closed form expressions for it (see \S\tcb{\ref{shadowbulk}}).
    
    \item In \S\tcb{\ref{MackPolinomials}} we move to Mellin space. We review how to obtain the Mellin representation of CPWs, which are expressed in terms of so-called Mack polynomials \cite{Mack:2009mi}. We refrain from presenting the generally complicated explicit expressions for Mack polynomials, which involve nested sums.\footnote{Some of such nested formulas recently appeared in the literature in a different CFT basis \cite{Chen:2017xdz}. For external scalars, such formulas were originally given in \cite{Mack:2009mi}.} Instead we focus on their orthogonality properties which arise when restricting to the leading pole in the Mellin representation of the CPW. These correspond to the contribution of the primary operator in the exchanged conformal multiplet. 
    
    In particular, the basis \cite{Sleight:2017fpc} of 3pt conformal structures gives rise to a remarkable factorisation of the dependence on the internal spin with respect to the external spins. This allows for a direct extension of the orthogonality observed for external scalars in \cite{Costa:2012cb} to spinning external operators, which appears through orthogonal polynomials known as continuous Hahn polynomials \cite{andrews_askey_roy_1999}. This analysis is carried out for various types of correlators with arbitrary spinning legs. 
    
    We furthermore derive inversion formulae to extract OPE data from the Mellin representation of a given 4pt spinning CFT correlator. In \S \tcb{\ref{OPEJ1J2J3J4}} we test the formalism by using it to extract the OPE coefficients from connected 4pt correlation functions in the free scalar $O\left(N\right)$ model in $d$-dimensions. In particular, we recover all known results for the OPE coefficients of single-trace conserved currents $J_l$ \cite{Sleight:2016dba}
    and double-trace operators $\left[J_0J_0\right]_{n,l}$ \cite{Dolan:2000ut,Bekaert:2015tva,Sleight:2016hyl}. We furthermore extract new results in general $d$ for leading twist $\left[J_{l_1}J_{l_2}\right]_{0,l}$ spinning double-trace operators built from single-trace conserved currents $J_{l_1}$ and $J_{l_2}$ of spins $l_1$ and $l_2$.
    
    \item In \S\tcb{\ref{Crossing Kernels}} we use the inversion formulas established in \S \tcb{\ref{MackPolinomials}} to study explicitly crossing kernels for CPWs with arbitrary spinning external operators. In general, upon restricting to a particular 4pt tensor structure, we can access a weighted average of the ${\sf s}$-channel CPW expansion coefficients for ${\sf u}$- and ${\sf t}$-channel CPWs. In \S \tcb{\ref{subsec::liftdeg}}, by considering all tensor structures we can go beyond the weighted average to obtain full crossing kernels for CPWs with two spinning external operators of spins $J_1$ and $J_2$, and an exchanged scalar. In \S \tcb{\ref{kernelarbitrary}} we also obtain all full crossing kernels for CPWs with external scalars and an exchanged operator of arbitrary integer spin.
    In \S \tcb{\ref{partiallyConserved}} we discuss crossing kernels for CPWs for the exchange of partially conserved currents, which are dual to (partially-)massless fields in the bulk. In \S \tcb{\ref{LargeL}} we discuss the large spin limit. 
    
    \item In \S\tcb{\ref{Application}} we give further concrete applications of our results. In particular, we use the crossing kernels of \S \tcb{\ref{Crossing Kernels}} to compute the change in the anomalous dimensions under double-trace flows in the large $N$ limit. We consider flows induced by the general double-trace perturbation
    \begin{equation}\label{dtflow0}
    S_{\lambda}=S_{\text{CFT}}+\lambda \int d^dy\, {\cal O}_{\mu_1...\mu_{l^\prime}}\left(y\right){\cal O}^{\mu_1...\mu_{l^\prime}}\left(y\right),
    \end{equation}
    where ${\cal O}_{\mu_1...\mu_{l^\prime}}$ is a single-trace operator of spin-$l^\prime$ and generic twist $\tau$. In \S \tcb{\ref{0000corr}} we begin by considering the case of external scalar operators $\Phi$ and extract anomalous dimensions of all double-trace operators $\left[\Phi_1\Phi_2\right]_{n,l}$ at ${\cal O}\left(1/N\right)$ for general double-trace flows \eqref{dtflow0}, including non-unitary spinning double-trace flows and flows induced by (partially-)conserved operators. In $d=4-\epsilon$ the $1/N$ corrections to double-trace anomalous dimensions are rather simple, which for the Wilson-Fisher fixed point read:
    \begin{equation}
    \gamma_{n,l}^{[\Phi\bar{\Phi}]}=\frac{4}{N}\,\frac{\epsilon\,(-1)^l}{(n+1)^2}\, _4F_3\left(\begin{matrix}1,1,-l,l +2n+3\\2,n+2,n+2\end{matrix};1\right)\,.
    \end{equation}
    
    In \S\tcb{\ref{1010corr}} we apply our formalism to spinning correlators in arbitrary space-time dimensions, which also requires to consider mixed symmetry CPWs and their crossing kernels -- which we derive. In particular, using the methods of \S \tcb{\ref{MackPolinomials}} we first extract mean-field theory OPE coefficients for leading twist double-trace operators $[{\cal O}_J\Phi]_{0,l}$ built from a spin-$J$ single-trace operator ${\cal O}_J$ and a scalar operator $\Phi$. Combining the latter with the results for crossing kernels, we then obtain the anomalous dimensions of the double-trace operators $[{\cal O}_J\Phi]_{0,l}$ induced by the double-trace flow \eqref{dtflow0} with $l^\prime=0$.
\end{itemize}

\newpage

\section{The basis}\label{NewBasis}

\subsection{3pt functions}

The OPE coefficients of primary operators with arbitrary spin are naturally encoded in the polynomial expansion of their 3pt functions in terms of the $6$ basic conformal structures ${\sf Y}_i$ and ${\sf H}_i$ \cite{Mack:1976pa,Sotkov:1976xe,Osborn:1993cr,Erdmenger:1996yc,Costa:2011mg}:\footnote{Here, $\left(y_{ij}\right)_{\mu}=\left(y_i-y_j\right)_{\mu}$ with $\mu=1,...,d$. The $\left(z_i\right)_\mu$ are null auxiliary vectors used to encode traceless indices. For example
\begin{equation}
    {\cal O}\left(y\right)=\frac{1}{J!}O_{\mu_1...\mu_J}\left(y\right)z^{\mu_1}...z^{\mu_{J}}, \qquad z^2=0,
\end{equation}
encodes the traceless spin-$J$ operator $O_{\mu_1...\mu_J}$.}
{\small \begin{subequations}
\begin{align}
& {\sf Y}_{1}= \frac{z_1\cdot y_{12}}{y_{12}^2}-\frac{z_1\cdot y_{13}}{y_{13}^2}, 
&& {\sf H}_1=\frac1{y_{23}^2}\left(z_2\cdot z_3+2\,\frac{z_2\cdot y_{23}\,z_3\cdot y_{32}}{y_{23}^2}\right),\\
&{\sf Y}_{2}= \frac{z_2\cdot y_{23}}{y_{23}^2}-\frac{z_2\cdot y_{21}}{y_{21}^2},
&& {\sf H}_2=\frac1{y_{31}^2}\left(z_3\cdot z_1+2\,\frac{z_3\cdot y_{31}\,z_1\cdot y_{13}}{y_{31}^2}\right),\\
& {\sf Y}_{3}= \frac{{z_3}\cdot y_{31}}{y_{31}^2}-\frac{{z_3}\cdot y_{32}}{y_{32}^2},
&& {\sf H}_3=\frac1{y_{12}^2}\left(z_1\cdot z_2+2\,\frac{z_1\cdot y_{12}\,z_2\cdot y_{21}}{y_{12}^2}\right).
\end{align}
\end{subequations}}

\noindent In particular, these define the following canonical expansion of spinning 3pt conformal correlators\footnote{For concision we define:
\begin{equation}
\sum_{{\bf{n}}}\equiv \sum_{n_1=0}^{\text{Min}(J_2-n_0,J_3-n_2)}\sum_{n_2=0}^{\text{Min}(J_3,J_1-n_0)}\sum_{n_0=0}^{\text{Min}(J_1,J_2)}. 
\end{equation}}
\begin{align}
\langle {\cal O}_{\Delta_1,J_1}(y_1){\cal O}_{\Delta_2,J_2}(y_2) {\cal O}_{\Delta_3,J_3}(y_3) \rangle &= \sum_{{\bf n}}C^{n_1,n_2,n_0}_{J_1,J_2,J_3}\langle \langle {\cal O}_{\Delta_1,J_1}(y_1){\cal O}_{\Delta_2,J_2}(y_2) {\cal O}_{\Delta_3,J_3}(y_3)  \rangle \rangle^{(\text{{\bf n}})}, 
\end{align}
in terms of the simple basis of 3pt conformal structures
\begin{subequations}\label{canonicalnbasis}
\begin{align}
\langle \langle {\cal O}_{\Delta_1,J_1}(y_1){\cal O}_{\Delta_2,J_2}(y_2) {\cal O}_{\Delta_3,J_3}(y_3)  \rangle \rangle^{(\text{{\bf n}})}&= \frac{\mathfrak{I}_{{\bf J}}^{{\bf n}}}{(y_{12}^2)^{\tfrac{\tau_1+\tau_2-\tau}{2}}(y_{23}^2)^{\tfrac{\tau_2+\tau-\tau_1}{2}}(y_{31}^2)^{\tfrac{\tau+\tau_1-\tau_2}{2}}},\\
\mathfrak{I}_{{\bf J}}^{{\bf n}}\left(y_1,y_2,y_3\right)&={\sf Y}_1^{J_1-n_2-n}{\sf Y}_2^{J_2-n-n_1}{\sf Y}_3^{J_3-n_1-n_2}{\sf H}_1^{n_1}{\sf H}_2^{n_2}{\sf H}_3^{n_0}\,,
\end{align}
\end{subequations}
where ${\bf n}=\left(n_1,n_2,n_0\right)$, ${\bf J}=(J_1,J_2,J_3)$ and each basis element is a monomial in the ${\sf Y}_i$ and ${\sf H}_i$. For unit normalisation of the 2pt functions,\footnote{In our conventions canonically normalised 2pt functions for totally symmetric operators read:
\begin{equation}
    \left\langle\mathcal{O}_{\Delta,J}(y_1)\mathcal{O}_{\Delta,J}(y_2)\right\rangle=\frac{{\sf H}_{3}^J}{(y_{12}^2)^{\Delta-J}}
\end{equation}} the $C^{n_1,n_2,n_0}_{J_1,J_2,J_3}$ are the OPE coefficients in the canonical basis \eqref{canonicalnbasis}.

Recently, with the aim of simplifying the kinematic 1:1 map between bulk cubic couplings and boundary 3pt conformal structures, in \cite{Sleight:2017fpc} the above canonical basis \eqref{canonicalnbasis} was repackaged into a different basis of 3pt conformal structures which is defined as:
\begin{align} 
&\left[\left[{\cal O}_{\Delta_1,J_1}(y_1){\cal O}_{\Delta_2,J_2}(y_2) {\cal O}_{\Delta_3,J_3}(y_3)\right]\right]^{(\text{{\bf n}})} \equiv\frac{\mathfrak{S}_{\bf J,\tau}^{\bf n}}{(y_{12})^{\delta_{12}}(y_{23})^{\delta_{23}}(y_{31})^{\delta_{31}}}, \label{nicebasis}
\end{align}
in terms of the following polynomials $\mathfrak{S}_{\bf J,\tau}^{\bf n}$ in ${\sf Y}_i$ and ${\sf H}_i$:
\begin{subequations}
\begin{align}
&\mathfrak{S}_{{\bf J},\mathbf{\tau}}^{0}\left(y_1,y_2,y_3\right)\equiv \left[\prod_{i=1}^32^{\tfrac{\delta_{(i+1)(i-1)} }{2}-1} \Gamma \left(\tfrac{\delta_{(i+1)(i-1)} }{2}\right)\right]\\ \nonumber
& \hspace*{3cm} \times 
\left[\prod_{i=1}^3 q_i^{\frac{1}{2}-\frac{\delta_{(i+1)(i-1)} }{4}} J_{(\delta_{(i+1)(i-1)} -2)/2}\left(\sqrt{q_i}\right)\right]
\,{\sf Y}_1^{J_1}{\sf Y}_2^{J_2}{\sf Y}_3^{J_3}\,,\\ 
&\mathfrak{S}_{\bf J,\tau}^{\bf n}\left(y_1,y_2,y_3\right)\equiv {\sf H}_1^{n_1}{\sf H}_2^{n_2}{\sf H}_3^{n_0}\,\mathfrak{S}_{J_i-n_{i+1}-n_{i-1},\tau_i+2(n_{i+1}+n_{i-1})}^{0}\,,
\end{align}
\end{subequations}
where {\small $q_i=2{\sf H}_i\pl_{{\sf Y}_{i+1}}\pl_{{\sf Y}_{i-1}}$} and $\delta_{ij}=\frac12\left(\tau_i+\tau_j-\tau_k\right)$, with ${i,j,k}$ cyclically ordered among the external legs in the correlator. In \cite{Sleight:2017fpc} it was argued that this alternative basis \eqref{nicebasis} is naturally selected for evaluating spinning Witten diagrams within the ambient space formalism.\footnote{See \cite{Joung:2011ww,Joung:2012hz} for the ambient space formalism for cubic couplings of arbitrary integer spin fields on AdS and \cite{Sleight:2016dba,Sleight:2016hyl,Sleight:2017fpc} for the evaluation of their corresponding tree level 3pt Witten diagrams.} Indeed, a way to derive the basis \eqref{nicebasis} is to consider an ansatz for a bulk cubic coupling at fixed spins of the type\footnote{For concision we define $\sum\limits_{{\bf m}} = \sum\limits^{\text{min}\left\{J_1,J_2,n_0\right\}}_{m_3=0}  \sum\limits^{\text{min}\left\{J_1-n_0,J_3,n_2\right\}}_{m_2=0}\sum\limits^{\text{min}\left\{J_2-n_0,J_3-n_2,n_1\right\}}_{m_1=0}.$}
\begin{equation}\label{Ical}
{\cal I}_{J_1,J_2,J_3}^{n_1,n_2,n_0}(\tau_i)=\sum_{{\bf m}} C_{J_1,J_2,J_3;m_1,m_2,m_3}^{n_1,n_2,n_0}(\tau_i)\,I^{m_1,m_2,m_3}_{J_1,J_2,J_3}\,,
\end{equation}
with a truncated summation (terminating at $m_i=n_i$) over the canonical basis of on-shell cubic vertices in the ambient space formalism between totally symmetric fields $\varphi_{J_i}$ of spins $J_i$ and mass $m^2_i R^2 = \Delta_i\left(\Delta_i-d\right)-J_i$, which is given by
\begin{align}\label{massivebasis}
& I^{n_1,n_2,n_0}_{J_1,J_2,J_3} = {\cal Y}^{J_1-n_2-n_0}_1{\cal Y}^{J_2-n_0-n_1}_2{\cal Y}^{J_3-n_1-n_2}_3 \\ \nonumber
& \hspace*{3cm} \times {\cal H}^{n_1}_1{\cal H}^{n_2}_2{\cal H}^{n_0}_3\, \varphi_{J_1}\left(X_1,U_1\right)\varphi_{J_2}\left(X_2,U_2\right)\varphi_{J_3}\left(X_3,U_3\right)\Big|_{X_i=X},
\end{align}
in terms of the six $SO (d + 1, 1)$-covariant
contractions
\begin{subequations}\label{6cont}
\begin{align}
    \mathcal{Y}_1&=\pl_{U_1}\cdot\pl_{X_2}\,,&\mathcal{Y}_2&=\pl_{U_2}\cdot\pl_{X_3}\,,&\mathcal{Y}_3&=\pl_{U_3}\cdot\pl_{X_1}\,,\\
    \mathcal{H}_1&=\pl_{U_2}\cdot\pl_{U_3}\,,&\mathcal{H}_2&=\pl_{U_3}\cdot\pl_{U_1}\,,&\mathcal{H}_3&=\pl_{U_1}\cdot\pl_{U_2}\,.
\end{align}
\end{subequations}

Tree level 3pt Witten diagrams for arbitrary spinning external legs were first computed in \cite{Sleight:2016dba} using the canonical basis \eqref{massivebasis} of bulk cubic couplings. The Witten diagram $\mathfrak{W}_{\tau}\left[\mathcal{I}_{\bf J}^{\bf n}(\tau)\right]$ generated by the cubic coupling \eqref{Ical} takes the following form in the canonical basis \eqref{canonicalnbasis} of 3pt conformal structures (dropping the overall dependence on $(y_{ij}^2)^{\delta_{ij}}$):
\begin{equation}\label{wittendinewbasis}
    \mathfrak{W}_{\tau}\left[\mathcal{I}_{\bf J}^{\bf n}(\tau)\right]=\sum_{{\bf p}}\mathfrak{C}_{{\bf J};{\bf p}}^{{\bf n}}\mathfrak{I}_{\bf J}^{\bf p}\,.
\end{equation}
The new basis \eqref{nicebasis} of 3pt conformal structures is selected by the condition that the Witten diagram \eqref{wittendinewbasis} satisfies the condition:
\begin{align}
    \mathfrak{C}_{J_1,J_2,J_3;p_1,p_2,p_3}^{n_1,n_2,n_0}=0\qquad \forall\quad p_1<n_1\,,\ \ p_2<n_2\,,\ \ p_3<n_0\,.
\end{align}
\begin{figure}
    \centering
    \includegraphics[width=0.9\textwidth]{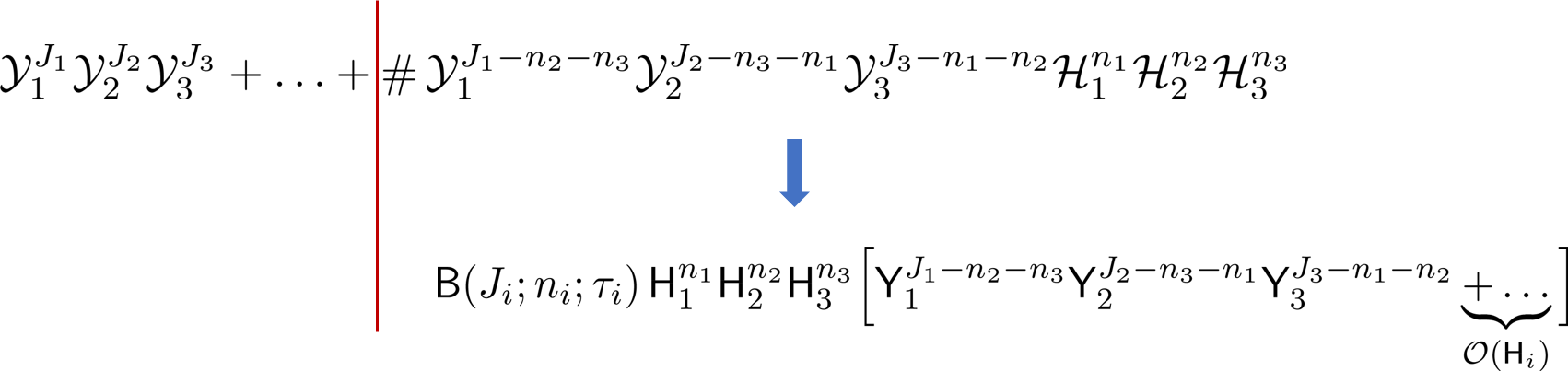}
    \caption{An illustration of the defining feature of the basis \eqref{IntegralBasisAdS} dual to the bulk cubic couplings \eqref{Ical}. The non-vanishing coefficients of the bulk couplings are exactly the same in number as the CFT structures that are set to zero. This condition has a unique solution which defines the basis proposed in \cite{Sleight:2017fpc} and which is employed in this work.}
    \label{fig:WitBasis}
\end{figure}
This is depicted schematically in fig.~\ref{fig:WitBasis}.
The above equations are exactly as many as the number of unknowns present in the ansatz \eqref{Ical} for the bulk couplings. Solving these conditions with the normalisation $C^{n_1,n_2,n_0}_{J_1,J_2,J_3;0,0,0}(\tau_i)=1$, the solution is given by
\begin{multline}
   C_{J_1,J_2,J_3;m_1,m_2,m_3}^{n_1,n_2,n_0}(\tau_i)= \left(\tfrac{d-2 (J_1+J_2+J_3-1)-\left(\tau_1+\tau_2+\tau_3\right)}{2} \right)_{m_1+m_2+m_3}\\\times\, \prod_{i=1}^3 \left[2^{m_i}\,\binom{n_i}{m_i} (n_i+\delta_{(i+1)(i-1)}-1)_{m_i}\right]\,.
\end{multline}
Furthermore the structure $\mathfrak{S}_{\bf J,\tau}^{\bf n}$ pops up automatically. In particular, we have:
\begin{equation}\label{IntegralBasisAdS}
\mathfrak{W}_{\tau_1,\tau_2,\tau_3}[\mathcal{I}^{n_1,n_2,n_0}_{J_1,J_2,J_3}]=\frac{{\sf B}(J_i;n_i;\tau_i)\mathfrak{S}_{\bf J,\tau}^{\bf n}}{(y_{12}^2)^{\tfrac{\tau_1+\tau_2-\tau_3}2}(y_{23}^2)^{\tfrac{\tau_2+\tau_3-\tau_1}2}(y_{31}^2)^{\tfrac{\tau_3+\tau_1-\tau_2}2}}\,,
\end{equation}
with the coefficient ${\sf B}(J_i;n_i;\tau_i)$ given by
{\begin{multline}\label{beeeeeee}
{\sf B}(J_i;n_i;\tau_i)=\pi ^{-d}  (-2)^{(J_1+J_2+J_3)-(n_1+n_2+n_0)-4}\,\Gamma \left(\tfrac{\tau_1+\tau_2+\tau_3-d+2 (J_1+J_2+J_3)}{2} \right)\\  \times\,\prod_{i=1}^3 \tfrac{\Gamma \left(J_i- n_{i+1}+ n_{i-1}+\frac{\tau_i+\tau_{i+1}-\tau_{i-1}}{2}\right)\Gamma \left(J_i+n_{i+1}- n_{i-1}+\frac{\tau_i+\tau_{i-1}-\tau_{i+1}}{2}\right)\Gamma (J_i+n_{i+1}+n_{i-1}+\tau_i-1)}{\Gamma \left(J_i+\tau_i-\tfrac{d}{2}+1\right)\Gamma \left(2 n_i+\frac{\tau_{i+1}+\tau_{i-1}-\tau_i}{2}\right)\Gamma (2 J_i+\tau_i-1)}\,,
\end{multline}}
or recursively as:
{\small\begin{multline}
    {\sf B}(J_i;n_i;\tau_i)=(-2)^{n_1+n_+n_0}\prod_i\left(1-\tfrac{d}{2}+J_i+\tau_i\right)_{n_{i-1}+n_{i+1}}\\\times\,{\sf B}(J_1-n_2-n_0,J_2-n_0-n_1,J_3-n_1-n_2;0;\tau_1+2(n_2+n_0),\tau_2+2(n_0+n_1),\tau_3+2(n_1+n_2)).
\end{multline}}
In order to employ the above basis for efficient Witten diagram evaluation, in \cite{Sleight:2017fpc} the inverse map to \eqref{Ical} was found. Starting from a coupling of the form
\begin{equation}
\mathcal{V}_{J_1,J_2,J_3}=\sum_{{\bf n}} g_{J_1,J_2,J_3}^{n_1,n_2,n_0}I_{J_1,J_2,J_3}^{n_1,n_2,n_0}\,,
\end{equation}
one can determine the coefficients ${\tilde g}_{J_1,J_2,J_3}^{n_1,n_2,n_0}$ in the basis:
\begin{equation}
\mathcal{V}_{J_1,J_2,J_3}=\sum_{{\bf n}}{\tilde g}_{J_1,J_2,J_3}^{n_1,n_2,n_0}{\cal I}_{J_1,J_2,J_3}^{n_1,n_2,n_0}\,.
\end{equation}
Inverting the infinite dimensional matrix we obtain\footnote{Here, for concision we define:
\begin{equation}
\sum_{{\bf m}}=\sum_{m_3=n_0}^{\text{Min}\{J_1,J_2\}}\sum_{m_2=n_2}^{\text{Min}\{J_3,J_1-m_3\}}\sum_{m_1=n_1}^{\text{Min}\{J_2-m_3,J_3-m_2\}}
\end{equation}
}
\begin{multline}
{\tilde g}_{J_1,J_2,J_3}^{n_1,n_2,n_0}=\sum_{{\bf m}}\Bigg[\frac{g^{m_1,m_2,m_3}_{J_1,J_2,J_3}}{ \left(\frac{d}{2}+1+\sum_{\alpha}(m_\alpha-J_\alpha-\tfrac{\tau_\alpha}2)\right)_{m_1+m_2+m_3}}\\\times\,\prod_{i=1}^3(-1)^{n_i+m_i}\frac{(2 n_i+\delta_{jk}-1)}{2^{m_i}\left( n_i+\delta_{jk}-1\right)_{m_i+1}}\binom{m_i}{n_i}\Bigg]\,.
\end{multline}

In the following section \S \tcb{\ref{subsec::cpws}} we consider the corresponding conformal partial waves in the basis \eqref{nicebasis} which, via the shadow formalism \cite{Ferrara:1972xe,Ferrara:1972ay,Ferrara:1972uq}, are an integrated product of 3pt conformal structures. In the bulk these encode exchanges of massive and (partially-)massless higher-spin fields. 

\subsubsection{Comments, higher-spin symmetry and relation to weight-shifting operators} 

Most of the simplifications which we will see using the new basis \eqref{nicebasis} can be regarded as a consequence of its transformation properties under higher-spin transformations. The complete classification of higher-spin transformations on totally symmetric tensors was obtained in the so-called metric-like formulation of higher-spin fields in \cite{Joung:2013nma}. On the CFT side, the recently introduced weight-shifting operators \cite{Karateev:2017jgd} appear to be particular examples of such higher-spin generators, whose action on cubic couplings or CFT correlators rotates them among each other, organising the bulk couplings and CFT correlators into infinite-dimensional multiplets. From the bulk perspective, such weight-shifting operators can be realised in terms of building blocks associated to the so called $\sigma_{\pm}$ operators (see e.g. \cite{Shaynkman:2000ts} and chapter 5.2 of \cite{Rahman:2015pzl} for a pedagogical review) describing the irreducible decomposition of the tensor product of a conformal module and a vector representation:
\begin{equation}\label{sigmaOp}
    P^a\,T^{a_1(l_1);\ldots;a_n(l_n)}=\sum_{i}(\sigma_i^+[T])^{a_1(l_1);\ldots;a_i(l_i+1);\ldots;a_n(l_n)}+ (\sigma_i^-[T])^{a_1(l_1);\ldots;a_i(l_i-1);\ldots;a_n(l_n)} \,,
\end{equation}
where on-shell the above tensors are all traceless. Depending on which row the $\sigma^\pm_i$ operators act upon, the tensors are shifted according to $l_i\to l_i\pm 1$. Up to analytic continuation we can identify the first row with the conformal dimension quantum number (of primary and descendants components) of the conformal module. Explicit expressions of \eqref{sigmaOp} for 2 row Young tableaux labelled by Lorentz tensors $T_{a(k),b(l)}$ read (see \cite{Rahman:2015pzl} for notation):
{\allowdisplaybreaks
\begin{subequations}
\begin{align}
&\sigma^-_1\left(T_{a(k),\,b(l)}\right)~=~T_{a(k-1)m,\,b(l)}\,P^m+\tfrac{1}{k-l+1}\,T_{a(k-1)b,\,b(l-1)m}\,P^m\,,\\[5pt]
&\sigma^+_1\left(T_{a(k),\,b(l)}\right)~=~P_a T_{a(k),\,b(l)}-\tfrac{2}{d+2k-2}\eta_{aa}T_{a(k-1)m,\,b(l)}\,P^m\nonumber\\[5pt]
&-\tfrac1{d+k+l-3}\eta_{ab}T_{a(k),\,b(l-1)m}\,P^m+\tfrac{2}{(d+2k-2)(d+k+l-3)}\eta_{aa}T_{a(k-1)b,\,b(l-1)m}\,P^m\,,\\[5pt]
&\sigma^-_2\left(T_{a(k),\,b(l)}\right)~=~T_{a(k),\,b(l-1)m}\,P^m\,,\\[5pt]
&\sigma^+_2\left(T_{a(k),\,b(l)}\right)~=~P_b T_{a(k),\,b(l)}-\tfrac{1}{k-l}\,P_a T_{a(k-1)b,\,b(l)}-\tfrac2{d+2l-4}\eta_{b(2)}T_{a(k),\,b(l-1)m}\,P^m\nonumber\\[5pt]
&~~~~~~~~~~~-\tfrac{k-l-1}{(k-l)(d+k+l-3)}\eta_{ab}T_{a(k-1)m,\,b(l)}P^m+\tfrac2{(k-l)(d+k+l-3)}\eta_{a(2)}T_{a(k-2)mb,\,b(l)}\,P^m\nonumber\\[5pt]
&~~~~~~~~~~~+\tfrac{d+2k-4}{(k-l)(d+2l-4)(d+k+l-3)}\eta_{ab}T_{a(k-1)b,\,b(l-1)m}\,P^m\nonumber\\[5pt]
&~~~~~~~~~~~-\tfrac{4}{(k-l)(d+k+l-3)(d+2l-4)} \eta_{a(2)}T_{a(k-1)b(2),\,b(l-1)m}\,P^m\,,
\end{align}
\end{subequations}}

\begin{figure}[t]
    \centering
    \includegraphics[width=0.7\textwidth]{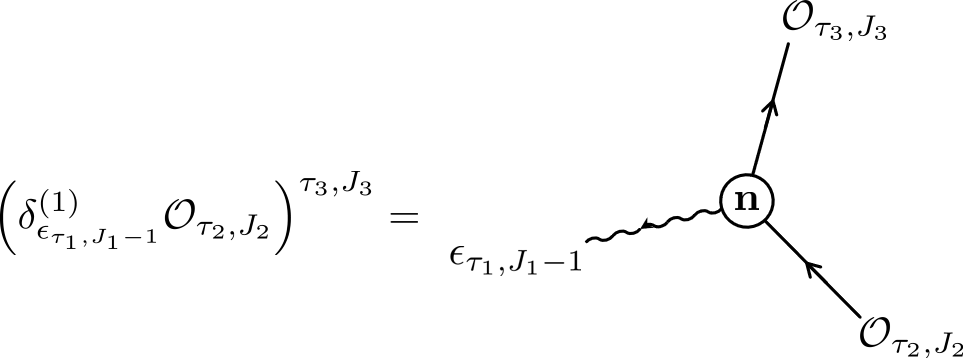}
    \caption{Pictorial representation of the action of higher-spin transformations on CFT operators/bulk fields via the bulk-to-boundary map. $\bf n$ lables the OPE structures. The basis used in this work should be thought of as an analytic continuation in dimensions of the higher-spin generators which one recovers for $\tau_1=d-2-k$ with $0\leq k< J_1$. In the bulk such analytic continuation is natural when going off-shell and considering the action of the deformed bulk gauge symmetries on off-shell couplings. The bulk to boundary map then gives the corresponding families of cubic structures on the boundary.}
    \label{fig:DiffOpDelta}
\end{figure}

Our basis has the virtue of making manifest the covariance properties under such transformations. This can be seen in various ways both on the boundary and on the bulk side \cite{Sleight:2016xqq,Sleight:2017pcz}. On the bulk side one considers the limit in which one of the external legs is a gauge field (either massless or partially-massless $\tau_1=d-2-k$ with $0\leq k<J_1$) and extracts the corresponding higher spin symmetry deformations $\delta^{(1)}$ from the terms proportional to the equations of motion:
\begin{equation}\label{delta1}
    \int_{\text{AdS}}\left[\delta^{(0)}_{\epsilon_{J_1}}\mathcal{V}_{J_1,J_2,J_3}+\delta^{(1)}\varphi_{J_2}\Box\varphi_{J_2}+\delta^{(1)}\varphi_{J_3}\Box\varphi_{J_2}\right]\,,
\end{equation}
where we assume $\epsilon_{J_1}$ to be a higher-spin killing tensor and $\delta^{(0)}_\epsilon$ is the the standard linear gauge symmetry transformation of (partially-)massless higher-spin field.\footnote{In ambient space formalism it is sufficient to restrict to the terms proportional to the ambient space Laplacian (see e.g. \cite{Joung:2013nma,Sleight:2016xqq,Sleight:2017cax}).} There is actually an explicit form for $\delta^{(1)}$ as a differential operator directly in terms of the cubic coupling $\mathcal{\mathcal{V}}_{J_1,J_2,J_3}$ in \eqref{Ical} in ambient space \cite{Joung:2013nma}. From this perspective, the crossing relations studied recently in \cite{Karateev:2017jgd} encode in general the transformation properties of infinite families of couplings with arbitrary spin external legs as infinite modules under some higher-spin transformations. Notice that from a bulk perspective one can go off-shell with respect to the bulk fields retaining the action of the higher-spin generators. This entails an analytic continuation in dimensions which are now expanded over the whole principal series. Our basis can be thought of as such an analytic continuation of the on-shell couplings fixed by higher-spin symmetry.\footnote{See \cite{Fradkin:1986ka,Mikhailov:2002bp,Eastwood:2002su,Vasiliev:2003ev,Joung:2014qya} and references therein for higher-spin algebras and their structure constants.}
In particular eq.~\eqref{delta1} with the choice \eqref{Ical} constructively identifies a convenient basis for the most general action of a conformal invariant differential operators on totally symmetric representations directly in the ambient space formalism. One can further distinguish the above differential operators into abelian or non-abelian ones as discussed in \cite{Boulanger:2008tg,Joung:2013nma}. The boundary action of the above operators is obtained by acting with the higher-spin generators on bulk-to-boundary propagator as done for instance in \cite{Chang:2012kt}. It is also interesting to note that our basis of cubic couplings depends analytically on the external dimensions implying that also conformal differential operators are nicely organised into such analytic families. It is tempting to think of this feature as related to the construction of massive multiple singleton tensor product representations of higher-spin algebras which would require such operator families to exist.

\subsection{Conformal Partial Waves}
\label{subsec::cpws}

The normalisable part of a 4pt correlation function\footnote{When referring to the non-normalisable part we mean any contribution of dimension $\Delta\leq\frac{d}{2}$, which in particular includes the identity.} in CFT admit an expansion in terms of an orthogonal basis of conformal partial waves (CPWs) \cite{Mack1974,Mack:1974sa,Dobrev:1975ru,Dobrev:1977qv}
\begin{equation}\label{cpwe}
    \langle {\cal O}_{\Delta_1,J_1}\left(y_1\right)...{\cal O}_{\Delta_4,J_4}\left(y_4\right) \rangle = \sum_{{\bf n}, {\bf \bar n},l} \int^{\frac{d}{2}+i\infty}_{\frac{d}{2}-i\infty} \frac{d\Delta}{2\pi i}\,a^{{\bf n},{\bf \bar n}}\left(\Delta,l\right)F^{{\bf n},{\bf \bar n}}_{\Delta,l}\left(y_i\right)+\text{non-normalisable},
\end{equation}
where the integral is over the principal series and the coefficient function $a^{{\bf n},{\bf \bar n}}\left(\Delta,l\right)$, which is meromorphic in $\Delta$, is the dynamical piece which contains the OPE data of the normalisable exchanged operators. The latter is encoded in poles of $\Delta$, which are the scaling dimensions of the physical exchanged spin-$l$ primary operators in the chosen expansion channel, and the residue gives their OPE coefficients. The superscripts ${\bf n}$ and ${\bf \bar n}$ label the 3pt conformal structures that enter in the integral representation of the CPW \cite{Ferrara:1972xe,Ferrara:1972uq}, which for an ${\sf s}$-channel\footnote{To avoid confusion with the Mellin variables $s$ and $t$, we use sans serif font to denote the ${\sf s}$-, ${\sf t}$- and ${\sf u}$-channels.} CPW in the basis \eqref{nicebasis} of 3pt conformal structures is given by
\begin{multline} \label{nbscpw}
{}^{({\sf s})}F^{{\bf n},{\bf \bar n}}_{\Delta,l}\left(y_i\right)=\frac{\kappa_{d-\Delta,l}}{\pi^{d/2}} \\\times\int d^dy_0\, \left[\left[ {\cal O}_{\Delta_1,J_1}(y_1){\cal O}_{\Delta_2,J_2}(y_2) {\cal O}_{\Delta,l}(y_0)  \right]\right]^{(\text{{\bf n}})} [[{\tilde {\cal O}}_{d-\Delta,l}(y_0)  {\cal O}_{\Delta_3,J_3}(y_3){\cal O}_{\Delta_4,J_4}(y_4)]]^{({\bf \bar n})},
\end{multline}
where ${\tilde {\cal O}}_{\Delta,l}$ is the shadow operator which has scaling dimension $d-\Delta$ and spin-$l$. It is related to ${\cal O}_{\Delta,l}$ via the shadow transform:\footnote{$I_{\mu \nu}\left(y\right)$ is the inversion tensor \begin{equation}\label{inversont}
    I_{\mu\nu}\left(y\right) = \delta_{\mu \nu} - \frac{2y_{\mu}y_{\nu}}{y^2} \: ; \qquad z_1 \cdot I\left(y\right) \cdot z_2 = z_1 \cdot z_2 - 2 \frac{z_1 \cdot y\, z_2 \cdot y }{y^2}.
\end{equation} The Thomas derivative \cite{10.2307/84634} (see also \cite{Dobrev:1975ru})
\begin{equation}
{\hat \partial}_{z^i} = \partial_{z^i} - \frac{1}{d-2+2 z \cdot \partial_z} z_i \partial^2_z,
\end{equation}
accounts for tracelessness, i.e. $z^2=0$.
}
\begin{equation}\label{shad}
{\tilde {\cal O}}_{d-\Delta,l}\left(y;z\right) = \kappa_{\Delta,l}\frac{1}{\pi^{d/2}}\int d^dy^\prime \frac{1}{\left[\left(y-y^\prime\right)^2\right]^{d-\Delta}} \left(z \cdot I(y-y^\prime) \cdot {\hat \partial}_{z^\prime}\right)^l  {\cal O}_{\Delta,l}\left(y^\prime;z^\prime\right).
\end{equation}
The normalisation
\begin{equation}\label{kds}
    \kappa_{\Delta,l} =  \frac{\Gamma\left(d-\Delta+l\right)}{\Gamma\left(\Delta-\frac{d}{2}\right)} \frac{1}{\left(\Delta-1\right)_l},
\end{equation}
ensures that applying \eqref{shad} twice gives the identity. Note that, applying the definition \eqref{shad}, we have
\begin{multline}\label{spinshad3pt}
    [[{\tilde {\cal O}}_{d-\Delta,l}(y_0;z_0)  {\cal O}_{\Delta_3,J_3}(y_3;z_3){\cal O}_{\Delta_4,J_4}(y_4;z_4)]]^{({\bf \bar n})}\\ = \kappa_{\Delta,l}\frac{1}{\pi^{d/2}}\int d^dy^\prime \frac{1}{\left[\left(y_0-y^\prime\right)^2\right]^{d-\Delta}} \left(z_0 \cdot I(y_0-y^\prime) \cdot {\hat \partial}_{z^\prime_0}\right)^l \\\times\,[[ {\cal O}_{\Delta,l}(y^\prime;z^\prime_0)  {\cal O}_{\Delta_3,J_3}(y_3;z_3){\cal O}_{\Delta_4,J_4}(y_4;z_4)]]^{({\bf \bar n})},
\end{multline}
which, upon evaluation of the conformal integral in $y^\prime$, is generally a linear combination of basis elements \eqref{nicebasis}. This is shown explicitly in \S \tcb{\ref{Shadow}}.\footnote{In particular, when there is more than one spinning operator in the 3pt correlator, it is crucial to keep in mind that the shadow 3pt conformal structure \eqref{spinshad3pt} cannot be obtained by simply replacing $\Delta \rightarrow d-\Delta$ since there is more than one 3pt conformal structure for such spinning correlators and the shadow is a linear combination of them. In this case one must apply the definition of the shadow transform as in the r.h.s of \eqref{spinshad3pt}.}

Mellin space uncovers some interesting orthogonality properties for CPWs, as has been observed in \cite{Costa:2012cb}. In particular, since the descendent contributions are entirely specified by the primary operator contribution via conformal symmetry, this allows to use orthogonality of primary operator contributions to uncover the CPW expansion of correlators. In the following we extend this idea to spinning CPWs \eqref{nbscpw} in the basis \eqref{nicebasis}. From the position space perspective this corresponds to a detailed analysis of the collinear limit of spinning CPWs, which we use to systematically work out the conformal partial wave expansion of 4pt correlators involving arbitrary totally symmetric operators. This analysis also serves to illustrate the strengths of the basis \eqref{nicebasis} for analysing crossing symmetry. In the following, for generality and ease of presentation we shall keep $\tau$ arbitrary. However, we stress here that we implicitly assume $\tau$ to be on the principal series for most of our discussions. In this way the crossing kernels we derive can be systematically used inside the spectral integral in the CPWE \eqref{cpwe} of 4pt correlators.

\section{Mack Polynomials and Inversion Formulae}\label{MackPolinomials}

The conformal partial wave expansion (CPWE) was first formulated in Mellin space by Mack \cite{Mack:2009mi} for external scalar operators. In this section we extend the formalism to include arbitrary spinning external operators in the totally symmetric representation. See \cite{Chen:2017xdz} for other recent work on the Mellin formalism with spinning external operators. We work primarily in the basis \eqref{nicebasis} which leads to a factorisation of the external spin dependence. This allows us to derive orthogonality relations for spinning Mellin amplitudes and study their inversion formulae.

\paragraph{External scalar operators} Let us first consider the ${\sf s}$-channel expansion of 4pt correlation function of scalar operators (i.e. $J_i=0$, ${\bf n}={\bf \bar n}=0$):
\begin{subequations}\label{00004pt}
\begin{align}\label{red4pt0000}
     \langle {\cal O}_{\tau_1,0}\left(y_1\right)...{\cal O}_{\tau_4,0}\left(y_4\right) \rangle &= \frac{1}{\left(y_{12}^2\right)^{\frac{1}{2}(\tau_1 + \tau_2)} \left(y_{34}^2\right)^{\frac{1}{2}(\tau_3 + \tau_4)}}\left(\frac{y_{24}^2}{y_{14}^2}\right)^{\tfrac{\tau_1-\tau_2}2}\left(\frac{y_{14}^2}{y_{13}^2}\right)^{\tfrac{\tau_3-\tau_4}2}  {}^{({\sf s})}\mathcal{A}\left(u,v\right),\\ \label{auv}
     {}^{({\sf s})}\mathcal{A}\left(u,v\right) &= \sum_{l} \int^{\frac{d}{2}+i\infty}_{\frac{d}{2}-i\infty} \frac{d\Delta}{2\pi i}\,a^{{\bf 0},{\bf 0}}\left(\Delta,l\right){}^{({\sf s})}{\cal F}^{{\bf 0},{\bf 0}}_{\Delta,l}\left(u,v\right),
\end{align}
\end{subequations}
where in \eqref{red4pt0000} we pulled out the appropriate overall factor for an ${\sf s}$-channel decomposition with cross ratios
\begin{align}
    u&=\frac{y_{12}^2y_{34}^2}{y_{13}^2y_{24}^2},& v&=\frac{y_{14}^2y_{23}^2}{y_{13}^2y_{24}^2}.
\end{align}
The Mellin representation of the 4pt correlator \eqref{red4pt0000} is defined by the (inverse) Mellin transform of the function of cross ratios \eqref{auv} where following the conventions of \cite{Costa:2012cb} we have:\footnote{Note that these conventions interchange the standard Mandelstam variables $t$ and $s$, so that poles in $t$ correspond to $\sf{s}$-channel physical exchanges.}
\begin{equation}
   {}^{({\sf s})}\mathcal{A}\left(u,v\right) = \int \frac{ds\, dt}{(4\pi i)^2}\,u^{t/2}v^{-(s+t)/2}\rho_{\left\{\tau_i\right\}}\left(s,t\right)\, {\cal M}\left(s,t\right).
\end{equation}
Here ${\cal M}\left(s,t\right)$ is the Mellin amplitude associated to the original position space amplitude \eqref{00004pt} and is defined with respect to the Mellin measure which, in our conventions, reads:
\begin{align}\label{rhoM}
\rho_{\left\{\tau_i\right\}}\left(s,t\right)&=\Gamma \left(\tfrac{-t+\tau_1+\tau_2}2\right) \Gamma \left(\tfrac{-t+\tau_3+\tau_4}2\right)\nonumber\\&\hspace{100pt}\times\Gamma \left(\tfrac{s+t}{2}\right) \Gamma \left(\tfrac{-s-\tau_1+\tau_2}2\right) \Gamma \left(\tfrac{-s+\tau_3-\tau_4}2\right) \Gamma \left(\tfrac{ s+t+\tau_1-\tau_2-\tau_3+\tau_4}2\right).
\end{align}
Similarly, CPWs admit the Mellin representation
\begin{subequations}\label{scmrep}
\begin{align}
     {}^{(\sf s)}\mathcal{F}^{\text{{\bf 0}},\text{{\bf 0}}}_{\tau,l}\left(y_i\right) &=\tfrac{1}{\left(y_{12}^2\right)^{\frac{1}{2}(\tau_1 + \tau_2)} \left(y_{34}^2\right)^{\frac{1}{2}(\tau_3 + \tau_4)}}\left(\tfrac{y_{24}^2}{y_{14}^2}\right)^{\tfrac{\tau_1-\tau_2}2}\left(\tfrac{y_{14}^2}{y_{13}^2}\right)^{\tfrac{\tau_3-\tau_4}2}  {}^{(\sf s)}\mathcal{F}^{\text{{\bf 0}},\text{{\bf 0}}}_{\tau,l}\left(u,v\right),
     \\ 
   {}^{(\sf s)}\mathcal{F}^{\text{{\bf 0}},\text{{\bf 0}}}_{\tau,l}\left(u,v\right) &= \int \frac{ds\, dt}{(4\pi i)^2}\,u^{t/2}v^{-(s+t)/2}\rho_{\left\{\tau_i\right\}}\left(s,t\right)\, {}^{(\sf s)}\mathcal{F}^{\text{{\bf 0}},\text{{\bf 0}}}_{\tau,l}\left(s,t\right).
   \end{align}
\end{subequations}
It is convenient to express the Mellin representation ${}^{(\sf s)}\mathcal{F}^{\text{{\bf 0}},\text{{\bf 0}}}_{\tau,l}\left(s,t\right)$ of the CPW \eqref{scmrep} in the form \cite{Mack:2009mi}
\begin{align}\label{mrepexsc}
    {}^{(\sf s)}\mathcal{F}^{\text{{\bf 0}},\text{{\bf 0}}}_{\tau,l}\left(s,t\right)&= \mathcal{C}_{l,\tau}(\tau_i)\,\Omega_l(t)\, {}^{(\sf s)}P^{\text{{\bf 0}},\text{{\bf 0}}}_{l,\tau}(s,t)\,,
\end{align}
where
\begin{subequations}
\begin{align}\label{cltau}
    \mathcal{C}_{l,\tau}(\tau_i)&=\underbrace{\frac{\Gamma (2 l+\tau )}{(d-l-\tau -1)_l\,\Gamma \left(\tfrac{d}2- l-\tau \right)}}_{\kappa_{d-(l+\tau),l}}\frac{\alpha_{0,0,l;\tau_3,\tau_4,\tau}}{\Gamma \left(\tfrac{\tau +\tau_1-\tau_2}{2}\right) \Gamma \left(\tfrac{\tau -\tau_1+\tau_2}{2}\right) \Gamma \left(\tfrac{d-\tau +\tau_3-\tau_4}{2}\right) \Gamma \left(\tfrac{d-\tau -\tau_3+\tau_4}{2}\right)}\\ \label{omegalt}
    \Omega_l(t)&=\frac{\Gamma \left(\tfrac{\tau -t}{2}\right) \Gamma \left(\tfrac{d-2 l-t-\tau}{2}\right)}{\Gamma \left(\tfrac{-t+\tau_1+\tau_2}{2}\right) \Gamma \left(\tfrac{-t+\tau_3+\tau_4}{2}\right)}\,,
\end{align}
\end{subequations} 
and the coefficient $\alpha_{0,0,l;\tau_3,\tau_4,\tau}$ is defined by eq.~\eqref{ShadowNorm} and arises from the shadow transform \eqref{shad}. Equation \eqref{mrepexsc} defines the so-called Mack polynomials ${}^{({\sf s})}P^{\text{{\bf 0}},\text{{\bf 0}}}_{l,\tau}(s,t)$, which are degree $l$ polynomials in both Mellin variables $s$ and $t$.\footnote{Our normalisation convention for the Mack polynomials pulls out the factor of $\pi^{d/2}$ coming from the conformal integral \eqref{0000conf} together with an overall product of $\Gamma$-functions. We furthermore pull out overall factors generated by the shadow transform \eqref{shad}, which includes $\tfrac{\kappa_{d-\Delta,l}}{\pi^{d/2}}$ and the shadow normalisation $\alpha_{0,0,l;\tau_3,\tau_4,\tau}$ defined in \eqref{ShadowNorm}. All of these overall factors are included in ${\cal C}_{l,\tau}\left(\tau_i\right)$, given by equation \eqref{cltau}. Although the coefficient ${\cal C}_{l,\tau}\left(\tau_i\right)$ is not symmetric, this convention will turn out convenient for our discussion below.} Explicit expressions for the Mack polynomials are complicated in general, and they are currently available in the form of nested sums (see \cite{Mack:2009mi} appendix 12) which are extracted from the integral form \eqref{nbscpw} of the CPWs by employing the Symanzik star formula \cite{Symanzik:1972wj}. For our purposes, we find it is useful\footnote{The virtue of the expression \eqref{MackPscalar} is that it explicitly disentangles the contributions of trace terms in the expansion of Gegenbauer polynomials \eqref{Pseries} labelled by $k$. Trace terms encodes the lower spin traces of the tensor structures which are fixed by conformal invariance and correspond to descendant contributions.} to express them in the following form
\begin{multline}\label{MackPscalar}
    P^{(\sf s)}_{\tau,l^\prime}(s,t|\tau_1,\tau_2,\tau_3,\tau_4)\\=\frac{4^{l^\prime} \left(\tfrac{\tau +\tau_1-\tau_2}{2}\right)_{l^\prime} \left(\tfrac{\tau -\tau_1+\tau_2}{2}\right)_{l^\prime}}{(d-l^\prime-\tau -1)_{l^\prime} (l^\prime+\tau -1)_{l^\prime}}\sum_{k=0}^{[l^\prime/2]}c_{l^\prime,k}\left[\sum_{\sum_ir_i=l^\prime-2k}p_{r_1,r_2,r_3,r_4}^{l^\prime,k}(s,t|\tau_1,\tau_2,\tau_3,\tau_4)\right]\,,
\end{multline}
where the normalisation ensures that $P_{\tau,l^\prime}(s,t)\sim s^{l^\prime}+\ldots$, the coefficients $c_{l^\prime,k}$ are the Gegenbauer expansion coefficients defined in \eqref{Gcoeff} and we introduced:
\begin{multline}
    p_{r_i}^{l^\prime,k}(s,t|\tau_1,\tau_2,\tau_3,\tau_4)=\left(\frac{\tau -t}{2}\right)_k \left(\frac{d-2 l^\prime-t-\tau}{2}\right)_k\\\times\,\frac{\left(\frac{-d+\tau +\tau_3-\tau_4+2}{2} \right)_{k+r_1+r_2} \left(\frac{-d+\tau -\tau_3+\tau_4+2}{2}\right)_{k+r_3+r_4}}{\left(\frac{\tau -\tau_1+\tau_2}{2}\right)_{k+r_1+r_3} \left(\frac{\tau +\tau_1-\tau_2}{2}\right)_{k+r_2+r_4}}\,\mathfrak{p}_{r_i}(s,t|\tau_1,\tau_2,\tau_3,\tau_4),
\end{multline}
with
\begin{multline}
    \mathfrak{p}_{r_i}(s,t|\tau_1,\tau_2,\tau_3,\tau_4)=\frac{(-1)^{r_1+r_4}}{2^{r_1+r_2+r_3+r_4}}(r_1,r_2,r_3,r_4)!\\\times\, \left(\frac{s+t}{2}\right)_{r_1} \left(\frac{-s+\tau_3-\tau_4}{2}\right)_{r_2} \left(\frac{-s-\tau_1+\tau_2}{2}\right)_{r_3} \left(\frac{s+t+\tau_1-\tau_2-\tau_3+\tau_4}{2}\right)_{r_4}\,.
\end{multline}
The function $\Omega_l(t)$ in the Mellin representation \eqref{mrepexsc} of CPWs exhibits poles at $t=\tau +2m$ $(m=0,1,2,...)$ and also at the shadow values $t=d-\tau-2l+2m$, where each string of poles arises from one of the two the Gamma function factors in the numerator. Here we label with $l^\prime$ and $\tau$ the spin and the twist of the CPW.\footnote{A CPW is a sum of two conformal blocks with twist $\tau$ and shadow twist $d-\tau-2l$. To obtain the Mellin representation of a single conformal block it is sufficient to project away all shadow poles in the variable $t$, which can be achieved by multiplying the CPW with the unique function having zeros at all shadow poles and which is normalised to one at non-shadow poles \cite{Fitzpatrick:2011hu}.} The Mellin representation of the CPWs \eqref{mrepexsc} factorises on these poles, whose residues are kinematical polynomials ${\cal Q}^{{\bf 0},{\bf 0}}_{l,\tau,m}\left(s\right)$ of degree-$l$ in the Mellin variable $s$ which, up to an overall coefficient, are given by the Mack polynomials \eqref{MackPscalar} at fixed $t=\tau+2m$. In particular, we have \cite{Mack:2009mi}:
\begin{subequations}
\begin{align}\label{expcpwme}
   \rho_{\{\tau_i\}}(s,t)\ {}^{({\sf s})}\mathcal{F}^{\text{{\bf 0}},\text{{\bf 0}}}_{\tau,l}\left(s,t\right)&=\tilde{\rho}_{\{\tau_i\}}(s,t)\sum_{m}\frac{{\cal Q}^{{\bf 0},{\bf 0}}_{l,\tau,m}\left(s\right)}{t-\tau-2m} + \text{shadow}\\ \label{kpolyq0}
 {\cal Q}^{{\bf 0},{\bf 0}}_{l,\tau,m}\left(s\right) &=\tfrac{(-1)^m}{m!} \,\mathcal{C}_{l,\tau}(\tau_i)\,\Gamma \left(\tfrac{d}{2}-l-m-\tau \right)\, {}^{({\sf s})}P^{\text{{\bf 0}},\text{{\bf 0}}}_{l,\tau}(s,\tau+2m),
\end{align}
\end{subequations}
where for future convenience we also defined the reduced Mellin measure:
\begin{align}\label{rhotilde0}
    \tilde{\rho}_{\{\tau_i\}}\left(s,t\right)=\Gamma\left(\tfrac{s+t}2\right)\Gamma\left(\tfrac{s+t+\tau_1-\tau_2-\tau_3+\tau_4}2\right)\Gamma\left(\tfrac{-s-\tau_1+\tau_2}2\right)\Gamma\left(\tfrac{-s+\tau_3-\tau_4}2\right).
\end{align}
The contributions from descendent poles $m>0$ are subleading in the limit $u\ll 1$ and furthermore they are entirely specified by symmetry.

In \cite{Costa:2012cb} it was reported that the kinematical polynomials \eqref{kpolyq0} generated by the contribution from the lowest weight (primary) operator ($m=0$) are orthogonal. In particular, they can be expressed in terms of so-called continuous Hahn polynomials \cite{andrews_askey_roy_1999}:\footnote{We normalise the continuous Hahn polynomials so that the leading power in $\sf{s}$ has unit normalisation: $Q_{l}^{(a,b,c,d)}(s)\sim s^l+\ldots$. See \S\tcb{\ref{Hahn}} for a review of various relevant properties of continuous Hahn polynomials.}
\begin{multline}\label{kpolyqhahn}
    \mathcal{Q}_{l,\tau,0}^{\bf{0},\bf{0}}(s)=\frac{2^{-2 l} (l+\tau -1)_l \Gamma (2 l+\tau )}{\Gamma \left(l+\tfrac{\tau +\tau_1-\tau_2}{2}\right) \Gamma \left(l+\tfrac{\tau -\tau_1+\tau_2}{2}\right) \Gamma \left(l+\tfrac{\tau +\tau_3-\tau_4}{2}\right) \Gamma \left(l+\tfrac{\tau -\tau_3+\tau_4}{2}\right)}\\\times\,Q_l^{(\tau,\tau+\tau_1-\tau_2-\tau_3+\tau_4,-\tau_1+\tau-2,\tau_3-\tau_4)}(s)\\=(-1)^l\,l!\,\left(\mathfrak{N}_l^{(\tau,\tau+\tau_1-\tau_2-\tau_3+\tau_4,-\tau_1+\tau-2,\tau_3-\tau_4)}\right)^{-1}\,Q_l^{(\tau,\tau+\tau_1-\tau_2-\tau_3+\tau_4,-\tau_1+\tau_2,\tau_3-\tau_4)}(s)\,,
\end{multline}
where in the last line we use the normalisation of the bilinear form for continuous Hahn polynomials \eqref{Qnorm} to express the overall normalisation. We review the most pertinent properties of the continuous Hahn polynomials $Q_{l}^{a,b,c,d}(s)$ for this work in appendix \S\tcb{\ref{Hahn}}.

The orthogonality of the kinematic polynomials can be used to seamlessly extract operator data from CFT 4pt functions in Mellin space. In particular, given a 4pt function \eqref{00004pt}, its conformal block decomposition in Mellin space reads (see e.g. \cite{Costa:2012cb}):
\begin{align}\label{cbemellin}
   \rho_{\left\{\tau_i\right\}}\left(s,t\right){\cal M}\left(s,t\right) &= \sum_{\tau,l} a^{{\bf 0},{\bf 0}}_{\tau,l} \left[\sum_{m}\frac{\tilde{\rho}_{\left\{\tau_i\right\}}\left(s,t\right){\cal Q}^{{\bf 0},{\bf 0}}_{l,\tau,m}\left(s\right)}{t-\tau-2m}+...\right],
\end{align}
where the $...$ is an entire function of the Mellin variables $\left(s,t\right)$ and
\begin{subequations}\label{c00taul}
\begin{align}
    a^{{\bf 0},{\bf 0}}_{\tau,l} &= -\text{Res}_{{\bar \tau}=\tau}\,a^{{\bf 0},{\bf 0}}\left({\bar \tau},l\right)\label{Acoeffs}\\
    &=c_{{\cal O}_1{\cal O}_2 {\cal O}_{\tau,l}}c^{{\cal O}_{\tau,l}}{}_{{\cal O}_3{\cal O}_4},
\end{align}
\end{subequations}
is the product of OPE coefficients $c_{{\cal O}_1{\cal O}_2 {\cal O}_{\tau,l}}$ and $c_{{\cal O}_{\tau,l}{\cal O}_3{\cal O}_4}$. Using the orthogonality \eqref{Qnorm} of the ${\cal Q}^{{\bf 0},{\bf 0}}_{l,\tau,0}\left(s\right)$ polynomials, the OPE coefficients can be extracted from the Mellin amplitude on the l.h.s. of \eqref{cbemellin}. Starting from the contributions from the lowest (leading) twist $\tau_{\text{min}}$ operators\footnote{I.e. $\tau_{\text{min}} \ne \tau +2m$ for some $\tau$ in the spectrum and some $m$. In practice there are two possible choices which we will consider in examples.} which are encoded in the pole $t=\tau_{\text{min}}$:
\begin{subequations}
\begin{align}
     \tilde{\rho}_{\{\tau_i\}}\left(s,\tau_{\text{min}}\right){\cal M}_{\tau_{\text{min}}}\left(s\right)&=-\text{Res}_{t=\tau_{\text{min}}}\left[\rho_{\{\tau_i\}}\left(s,t\right){\cal M}\left(s,t\right)\right]\\
     &= \tilde{\rho}_{\{\tau_i\}}\left(s,\tau_{\text{min}}\right) \sum_{l}\,a^{{\bf 0},{\bf 0}}_{\tau_{\text{min}},l}{\cal Q}^{{\bf 0},{\bf 0}}_{l,\tau_{\text{min}},0}\left(s\right),
\end{align}
\end{subequations}
using orthogonality of the ${\cal Q}^{{\bf 0},{\bf 0}}_{l,\tau,0}\left(s\right)$ polynomials \eqref{kpolyqhahn} the conformal block expansion can be readily inverted to recovering the coefficients \eqref{Acoeffs} for $\tau=\tau_\text{min}$:
\begin{multline}\label{OPE simple}
    a^{{\bf 0},{\bf 0}}_{\tau_{\text{min}},l} =\frac{(-1)^l}{l!} \int^{i\infty}_{-i\infty}\frac{ds}{4\pi i} \tilde{\rho}_{\{\tau_i\}}(s,\tau_{\text{min}})\mathcal{M}_{\tau_{\text{min}}}(s)\\\times\,Q_{l}^{(\tau_{\text{min}},\tau_{\text{min}}+\tau_1-\tau_2-\tau_3+\tau_4,-\tau_1+\tau_2,\tau_3-\tau_4)}(s).
\end{multline}
For contributions from generic twist operators $\tau$, the residue of the l.h.s of \eqref{cbemellin} at $t=\tau$ in general contains contributions from descendents of lower twist $\tau^\prime$ operators since $\tau=\tau^\prime+2m$ for positive integer $m$. To extract the coefficients \eqref{c00taul} for generic $\tau$, these lower twist contributions first have to be subtracted from the original Mellin amplitude, after which the coefficients can be extracted in the same way as for the lowest twist contributions outlined above. In \S \tcb{\ref{0000corr}} we present an efficient method to project away lower twist contributions which avoids having to directly subtract infinite sums.

The above manifestation of the orthogonality of CPWs in Mellin space has thus far proven instrumental in obtaining crossing kernels and inversion formulas for external scalar operators \cite{Gopakumar:2016wkt,Gopakumar:2016cpb}. Crossing kernels invert, say, a ${\sf t}$ or ${\sf u}$-channel CPW onto a particular ${\sf s}$-channel CPW along the lines of \eqref{OPE simple} where a given Mellin amplitude is expanded into the $\sf{s}$-channel. In the following we shall extend the above framework to include arbitrary totally symmetric spinning external operators, in the view of obtaining crossing kernels and inversion formulae for the case of external spinning operators. We shall make extensive use of the new CFT basis \eqref{nicebasis}, which will turn out to conveniently disentangle contributions from different spins in the $\sf{s}$-channel.

\paragraph{External spinning operators}

The definitions of Mack polynomials and their corresponding kinematic polynomials $\mathcal{Q}$ for totally symmetric spinning external operators is a straightforward extension of the definitions \eqref{mrepexsc} and \eqref{expcpwme} for external scalar operators:
\begin{subequations}
\begin{align}\label{mrepexscsp}
    {}^{({\sf s})}\mathcal{F}^{{\bf n},{\bf \bar n}}_{\tau,l}\left(s,t|W_{ij}\right)&= \mathcal{C}_{l,\tau}(\tau_i)\,\Omega_l(t)\,{}^{({\sf s})}P^{{\bf n},{\bf \bar n}}_{l,\tau}(s,t|W_{ij}), \\ \label{mrepexscsp2}
    \rho_{\{\tau_i\}}(s,t){}^{({\sf s})}\mathcal{F}^{{\bf n},{\bf \bar n}}_{\tau,l}\left(s,t|W_{ij}\right)&= \tilde{\rho}_{\{\tau_i\}}(s,t) \sum_{m}\frac{{\cal Q}^{{\bf n},{\bf \bar n}}_{l,\tau,m}\left(s|W_{ij}\right)}{t-\tau-2m} + \text{shadow},\\ \label{mrepexscsp3}
     {\cal Q}^{{\bf n},{\bf \bar n}}_{l,\tau,m}\left(s|W_{ij}\right) &=\frac{(-1)^m}{m!} \,\mathcal{C}_{l,\tau}(\tau_i)\,\Gamma \left(\tfrac{d}{2}-l-m-\tau \right)\,{}^{({\sf s})}P^{{\bf n},{\bf \bar n}}_{l,\tau}(s,\tau+2m|W_{ij}),
\end{align}
\end{subequations}
where the normalisation $\mathcal{C}_{l,\tau}(\tau_i)$ is the same normalisation we used for scalar external legs \eqref{cltau}, ${}^{({\sf s})}P^{{\bf n},{\bf \bar n}}_{l,\tau}(s,t|W_{ij})$ and ${\cal Q}^{{\bf n},{\bf \bar n}}_{l,\tau,m}\left(s|W_{ij}\right)$ are naturally the spinning extension of the Mack polynomial \eqref{MackPscalar} and kinematic polynomial \eqref{kpolyqhahn}. Like for the case of external scalars, they are obtained by using the Symanzik star formula to express the integral representation \eqref{nbscpw} of CPWs with external spinning operators in Mellin form. This is explained in appendix \S \tcb{\ref{4pt}}. In the above and throughout this work, for simplicity and without loss of generality we set to zero all tensor structures proportional to $z_i\cdot z_j$, which can be reconstructed via conformal symmetry. This means that we can focus on the tensorial structures $W_{ij}=z_i\cdot y_{ij}/y_{ij}^2$.\footnote{It is important to note that one can only drop all $z_i\cdot z_j$ \emph{after} performing the conformal integrals in the representation \eqref{nbscpw} of the CPWs.}

Of course, the explicit form of the spinning Mack polynomials ${}^{({\sf s})}P^{{\bf n},{\bf \bar n}}_{l,\tau}(s,t|W_{ij})$ depend on the choice of basis for the CPWs. As we shall see in the following sections, the basis \eqref{nbscpw} with 3pt conformal structures \eqref{nicebasis} allows for a remarkably simple extension of the above results on orthogonality and inversion formulae for external scalar operators. In particular, in the basis \eqref{nicebasis} the dependence on the external spins is completely factorised from that of the exchanged spin $l$, which can be regarded a consequence of the transformation properties of the basis \eqref{nicebasis} under global higher-spin transformations \cite{Sleight:2017fpc}. This property reduces the infinite dimensional inversion problem which involves infinitely many exchanged operators to a finite dimensional one, since the factorisation implies that also in the case of external spins we may restrict to a fixed internal spin by projecting with the continuous Hahn polynomials. This furthermore leads to a straightforward extension of the orthogonal polynomials \eqref{kpolyqhahn} to arbitrary spinning external legs.

The following sections are organised as follows. We evaluate explicitly the polynomials \eqref{mrepexscsp3} for $m=0$ in the basis \eqref{nicebasis} in various cases of increasing complexity, starting from the simplest case of $\sf{s}$-channel CPWs with external operators of spins $J$-$0$-$0$-$0$ and generic twist $\tau_i$ in \S \tcb{\ref{J000}}, before moving on to $J_1$-$J_2$-$0$-$0$ in \S \tcb{\ref{J1J200}} and finally to $J_1$-$J_2$-$J_3$-$J_4$ in \S \tcb{\ref{J1J2J3J4}}. This allows to study the extension the orthogonality relations for external scalars to spinning external legs. In each case we furthermore study the extension of the inversion formula \eqref{OPE simple} to spinning external legs and in \S \tcb{\ref{subsec::leadsub}} present an efficient method to disentangle descendent contributions from those of subleading twist operators. In \S \tcb{\ref{OPEJ1J2J3J4}} we give a simple application of the formalism to extract OPE coefficients from correlation functions of higher-spin currents in the free scalar $O\left(N\right)$.

In the following sections it will prove convenient to fix the normalisation asymmetrically as:
\begin{equation}\label{kpoly0}
    \mathcal{Q}^{{\bf n},{\bf \bar n}}_{l,\tau,0}(s|W_{ij})=\mathcal{A}_{l,\tau}(\tau_i)\,\mathfrak{q}^{{\bf n},{\bf \bar n}}_{l,\tau}(s|W_{ij})\,,
\end{equation}
with
\begin{equation}
    \mathcal{A}_{l,\tau}(\tau_i)=\frac{\Gamma (2 l+\tau )}{(d-l-\tau -1)_l}\tfrac1{\Gamma \left(\tfrac{\tau +\tau_1-\tau_2}{2}\right) \Gamma \left(\tfrac{\tau -\tau_1+\tau_2}{2}\right) \Gamma \left(\tfrac{2l+\tau +\tau_3-\tau_4}{2}\right) \Gamma \left(\tfrac{2l+\tau -\tau_3+\tau_4}{2}\right)},
\end{equation}
which is related to $\mathcal{C}_{l,\tau}$ given in equation \eqref{cltau}.

\subsection{$J$-$0$-$0$-$0$ correlators}\label{J000}

The simplest case to begin with is CPWs in the ${\sf s}$-channel expansion of 4pt correlators with operators of spins $J$-$0$-$0$-$0$ and generic twists $\tau_i$. In this case we have ${\bf n}=\left(n,0,0\right)$ and ${\bf \bar n}=0$ and with $ n\leq \text{min}(J,l)$, and we therefore can switch to the more streamlined notation: 
\begin{equation}
    {}^{({\sf s})}\mathcal{F}_{l,\tau}^{{\bf n},{\bf \bar n}}\rightarrow {}^{({\sf s})}\mathcal{F}_{l,\tau}^{(n)}.
\end{equation}
In the basis \eqref{nicebasis}, the kinematic polynomials \eqref{kpoly0} take the form: 
\begin{subequations}
\begin{align}\label{ql000}
\mathcal{Q}^{(n)}_{l,\tau,0}(s|W_{ij})&=\mathcal{A}_{l,\tau}(\tau_i)\,\mathfrak{q}^{(n)}_{l,\tau}(s|W_{ij})\\
    \mathfrak{q}_{l,\tau}^{(n)}(s|W_{ij})&=\frac{(d-l-\tau -1)_l (n+l+\tau -1)_{l-n}}{4^{l-n} \left(\frac{\tau +\tau_1-\tau_2}{2}\right)_{l+n} \left(\frac{\tau -\tau_1+\tau_2}{2}\right)_{l-n}}\Upsilon^{(n)}_J(s|W_{ij})\,\label{ql000bis}\\& \hspace{-1.5cm}\times\,\left(\frac{-s+\tau_3-\tau_4}{2}\right)_{n} \left(\frac{s+\tau +\tau_1-\tau_2-\tau_3+\tau_4}2\right)_{n} Q_{l-n}^{(\tau,\tau+\tau_1-\tau_2-\tau_3+\tau_4+2n,-\tau_1+\tau_2,\tau_3-\tau_4+2n)}(s)\,. \nonumber
\end{align}
\end{subequations}
Notice that the dependence on the external spin is completely factorised from the dependence on the exchanged spin $l$, where the latter furthermore appears through a continuous Hahn polynomial. In particular, the polynomial \eqref{ql000} for a single spin-$J$ external operator is a simple dressing of the result \eqref{kpolyqhahn} for $J=0$ by the $l$-independent tensor structure $\Upsilon^{(n)}_J(s|W_{ij})$ up to a shift of the arguments of the continuous Hahn polynomial. The structure $\Upsilon^{(n)}_J(s|W_{ij})$ is a polynomial in both the Mellin variable $s$ and $W_{ij}$. Its explicit form turns out to be rather simple: 
\begin{equation}\label{upsl000}
    \Upsilon^{(n)}_J(s|W_{ij})=(W_{13}-W_{14})^n\,\zeta_{J-n}(\tau_1+n,\tau_2-n,\tau_3+n,\tau_4-n,\tau+2n)\,,
\end{equation}
where we introduced the basic polynomial:
\begin{align}
    \zeta_J(\tau_i,\tau)&=\sum_{k_3+k_4\leq J}\binom{J}{k_3,k_4}\frac{\left(-\frac{s-\tau_3+\tau_4}{2} \right)_{k_3} \left(\frac{s+\tau +\tau_1-\tau_2-\tau_3+\tau_4}{2}\right)_{k_4}}{\left(\frac{\tau +\tau_1-\tau_2}{2}\right)_{k_3+k_4}}\,W_{12}^{J-k_3-k_4}(-W_{13})^{k_3}(-W_{14})^{k_4}\,.
\end{align}
An attractive feature of the factorised kinematic polynomial \eqref{ql000} is that the leading term in $W_{12}$ is orthogonal with respect to the Mellin-Barnes bi-linear product \eqref{Qnorm}.\footnote{Note that the dependence on the Mellin variable $s$ in the Pochammer factors on the second line of \eqref{ql000bis} can be absorbed into the Mellin measure in \eqref{mrepexscsp2}, so that the argument of the continuous Hahn polynomial in \eqref{ql000bis} is shifted in precisely the same way as the re-defined Mellin measure. One can then apply the orthogonality of the continuous Hahn polynomials to conclude orthogonality of \eqref{orthl0000}.} This orthogonal component is given explicitly by:
\begin{subequations}\label{orthl0000}
\begin{align}
\bar{\mathcal{Q}}^{(n)}_{l,\tau,0}(s|W_{ij})&=\mathcal{A}_{l,\tau}(\tau_i)\,\bar{\mathfrak{q}}^{(n)}_{l,\tau}(s|W_{ij})\\
\bar{\mathfrak{q}}_{l,\tau}^{(n)}(s|W_{ij})&=\frac{(d-l-\tau +1-2)_l (n+l+\tau -1)_{l-n}}{4^{l-n} \left(\frac{\tau +\tau_1-\tau_2}{2}\right)_{n+l} \left(\frac{\tau -\tau_1+\tau_2}{2}\right)_{l-n}}\bar{\Upsilon}^{(n)}_J(s|W_{ij})\,  \\  & \hspace*{-1.5cm} \times\,\left(\frac{-s+\tau_3-\tau_4}{2}\right)_{n} \left(\frac{s+\tau +\tau_1-\tau_2-\tau_3+\tau_4}2\right)_{n}Q_{l-n}^{(\tau,\tau+\tau_1-\tau_2-\tau_3+\tau_4+2n,-\tau_1+\tau_2,\tau_3-\tau_4+2n)}(s) \nonumber \\
    \bar{\Upsilon}^{(n)}_J&=(W_{13}-W_{14})^n\,W_{12}^{J-n}\,, 
\end{align}
\end{subequations}
and originates from the $k_3=k_4=0$ term in the tensor structure \eqref{upsl000}. The $k_3>0$ and $k_4>0$ terms in \eqref{upsl000} which involve lower powers of $W_{12}$ play an analogous role to the descendent ($m>0$) contributions in eq.~\eqref{mrepexscsp2} which are fixed by the primary ($m=0$) contributions due to conformal symmetry. 

This $J$-$0$-$0$-$0$ example illustrates two attractive properties associated to the basis \eqref{nicebasis} of conformal structures, which furthermore carry over to the general case of external operators with arbitrary spins $J_1$-$J_2$-$J_3$-$J_4$ which is presented in \S \tcb{\ref{J1J200}} and \S \tcb{\ref{J1J2J3J4}}:

\begin{enumerate}
    \item Factorisation of the dependence on the internal spin $l$ from the dependence on the external spin $J$.
    \item The manifest orthogonality of leading terms in the kinematic polynomials \eqref{upsl000}, which arises as a consequence of the latter factorisation. 
\end{enumerate}

Given the above orthogonality properties, like the external scalar case discussed earlier, we can study the conformal block decomposition of a given Mellin amplitude $\mathcal{M}(s,t|W_{ij})$ with one spinning leg in, say, the ${\sf s}$-channel: 
\begin{align}\label{cbeJ000}
   \rho_{\left\{\tau_i\right\}}\left(s,t\right){\cal M}\left(s,t|W_{ij}\right) &= \sum_{\tau,l,n} a^{(n)}_{\tau,l} \left[\sum_{m}\frac{\tilde{\rho}_{\left\{\tau_i\right\}}\left(s,t\right){\cal Q}^{(n)}_{l,\tau,m}\left(s|W_{ij}\right)}{t-\tau-2m}+...\right],
\end{align}
and in particular extract the expansion coefficients. To do so, we evaluating the residue at a certain physical pole $t=\bar{\tau}$:\footnote{The measure $\rho$ and $\tilde{\rho}$ are defined in \eqref{rhoM} and \eqref{rhotilde0}.}
\begin{subequations}
\begin{align}
   \tilde{\rho}_{\{\tau_i\}}(s,\bar{\tau}) \mathcal{M}_{\bar{\tau}}(s|W_{ij})&=-\,\text{Res}_{t=\bar{\tau}}\left[ {\rho}_{\{\tau_i\}}(s,t)\mathcal{M}(s,t|W_{ij})\right]\\
    &= \tilde{\rho}_{\{\tau_i\}}(s,\bar{\tau})\sum_{l,n}a^{(n)}_{\bar{\tau},l}\mathcal{Q}_{l,\bar{\tau},0}^{(n)}(s|W_{ij}).
\end{align}
\end{subequations}
For generic twists $\bar{\tau}$, this residue will in general include contributions from more than one primary operator of twist $\bar{t}$ and also contributions from descendants of primary operators of lower twist $\bar{t}-2m$ for some positive integer $m$ which we assume to have subtracted away (see \S \tcb{\ref{0000corr}} for precisely how). To extract the coefficients $a^{(n)}_{\bar{\tau},l}$, we proceed iteratively in $n$: First one identifies the tensor structure with highest power $J-\bar{n}$ of $W_{12}$. This structure can only be produced by a single CPW and we can thus restrict to the orthogonal component \eqref{orthl0000} thereof. One then extracts the corresponding coefficients of $\mathcal{Q}_{l,{\bar \tau},0}^{(\bar{n})}(s|W_{ij})$ using orthogonality:
\begin{equation}
    a_{\bar{\tau},l}^{(\bar{n})}=\frac{(-1)^{l-\bar{n}}}{(l-\bar{n})!}\int\frac{ds}{4\pi i}\,\tilde{\rho}_{\{\tau_i\}}\left(s,\bar{\tau}\right)\,\Big[\mathcal{M}_{\bar{\tau}}(s|W_{ij})\Big]_{\bar{\Upsilon}^{(\bar{n})}}Q_l^{(\bar{\tau},\bar{\tau}+\tau_1-\tau_2-\tau_3+\tau_4+2n,-\tau_1+\tau_2,\tau_3-\tau_4+2n)}(s),
\end{equation}
where $[\ldots]_{\bar{\Upsilon}^{(n)}}$ denotes the projection onto the coefficient of the structure $\bar{\Upsilon}^{(n)}$. Once all coefficients $a_{{\bar \tau},l}^{\bar{n}}$ have been determined, one can use the factorisation of the $l$-dependence from the tensorial structure to re-sum over $l$ and subtract it from the original amplitude:
\begin{equation}
    \widetilde{\mathcal{M}}_{\bar{\tau}}(s|W_{ij})=\mathcal{M}_{\bar{\tau}}(s|W_{ij})-\sum_l a_{{\bar \tau},l}^{(\bar{n})}\mathcal{Q}_{l,\bar{\tau},0}^{(\bar{n})}(s|W_{ij})\,.
\end{equation}
At this point we are left with a new Mellin amplitude $\widetilde{\mathcal{M}}_{\bar{\tau}}(s|W_{ij})$ whose leading power of $W_{12}$ is lower than $J-\bar{n}$ and we can iterate the same steps as before with the help of \eqref{QQintegral} until all the coefficients $a^{(n)}_{\tau,l}$ in the expansion \eqref{cbeJ000} have been determined.

\subsection{$J_1$-$J_2$-$0$-$0$ correlators}\label{J1J200}

In this subsection we consider $\sf{s}$-channel CPWs with external operators of spins $J_1$-$J_2$-$0$-$0$ and generic twist $\tau_i$. In this case, ${\bf \bar n}=0$ and we may adopt the more streamlined notation
\begin{equation}
    {}^{({\sf s})}\mathcal{F}_{l,\tau}^{{\bf n},{\bf \bar n}}\rightarrow {}^{({\sf s})}\mathcal{F}_{l,\tau}^{\bf n}\,.
\end{equation}
This case is the most general one in which mixed-symmetry exchanges do not contribute in generic dimensions.\footnote{For instance, one would need to consider mixed symmetry blocks already in $J_1$-$0$-$J_3$-$0$ correlators expanded in the $\sf{s}$-channel, as discussed in sections \S\tcb{\ref{1010corr}}, \S\tcb{\ref{2020corr}} and \S\tcb{\ref{J0J0corr}}.} The corresponding kinematic polynomials \eqref{kpoly0} in the basis \eqref{nicebasis} take the factorised form:
\begin{subequations}\label{j1j200Qn}
\begin{align}
\mathcal{Q}^{{\bf n}}_{l,\tau,0}(s|W_{ij})&=\mathcal{A}_{l,\tau}(\tau_i)\,\mathfrak{q}^{{\bf n}}_{l,\tau}(s|W_{ij}) \\\label{j1j200Qn2}
    \mathfrak{q}_{l,\tau}^{{\bf n}}(s|W_{ij})&=\frac{2^{n_0}(d-l-\tau -1)_l (n_1+n_2+l+\tau -1)_{l-n_1-n_2}}{4^{l-n_1-n_2} \left(\frac{\tau +\tau_1-\tau_2}{2}\right)_{l-n_1+n_2} \left(\frac{\tau -\tau_1+\tau_2}{2}\right)_{l-n_2+n_1}}\Upsilon^{{\bf n}}_{J_1,J_2}(s|W_{ij})\\& \times\,\left(\frac{-s-\tau_1+\tau_2}{2}\right)_{n_1} \left(\frac{s+\tau}2\right)_{n_1}\,\left(\frac{-s+\tau_3-\tau_4}{2}\right)_{n_2} \left(\frac{s+\tau +\tau_1-\tau_2-\tau_3+\tau_4}2\right)_{n_2} \nonumber \\ & \hspace*{4cm}\times Q_{l-n_1-n_2}^{(\tau+2n_1,\tau+\tau_1-\tau_2-\tau_3+\tau_4+2n_2,-\tau_1+\tau_2+2n_1,\tau_3-\tau_4+2n_2)}(s)\,.\nonumber
\end{align}
\end{subequations}
As anticipated, the dependence on the external spins $J_i$ is completely factorised into the tensor structure $\Upsilon^{{\bf n}}_{J_1,J_2}(s|W_{ij})$, which can be conveniently expressed in terms of the basic polynomials in $W_{ij}$ associated to the $J_1$-$0$-$0$-$0$ and $0$-$J_2$-$0$-$0$ CPWs:
\begin{subequations}\label{zetaCPW}
\begin{align}
    \zeta^{(1)}_J(\tau_i,\tau)&=\sum_{k_3+k_4\leq J}\binom{J}{k_3,k_4}\frac{\left(-\tfrac{s-\tau_3+\tau_4}{2} \right)_{k_3} \left(\tfrac{s+\tau +\tau_1-\tau_2-\tau_3+\tau_4}{2}\right)_{k_4}}{\left(\tfrac{\tau +\tau_1-\tau_2}{2}\right)_{k_3+k_4}}\,W_{12}^{J-k_3-k_4}(-W_{13})^{k_3}(-W_{14})^{k_4},\\
    \zeta^{(2)}_J(\tau_i,\tau)&=\sum_{k_3+k_4\leq J}\binom{J}{k_3,k_4}\frac{\left(-\tfrac{s+\tau_1-\tau_2}{2} \right)_{k_4} \left(\tfrac{s+\tau}{2}\right)_{k_3}}{\left(\tfrac{\tau -\tau_1+\tau_2}{2}\right)_{k_3+k_4}}\,(-W_{21})^{J-k_3-k_4}W_{23}^{k_3}W_{24}^{k_4}.
\end{align}
\end{subequations}
In terms of the above polynomials we have:\footnote{Here we use the short-hand notation
\begin{equation}
\widetilde{\sum}\equiv \sum_{k=0}^{\text{Min}(J_1-n_2-n_0,J_2-n_1-n_0)}
\end{equation}}
\begin{align}\label{Upsilon12}
    \Upsilon^{{\bf n}}_{J_1,J_2}(s)&=\widetilde{\sum}_k\frac{(J_1-n_2-n_0-k+1)_k(J_2-n_1-n_0-k+1)_k}{k!\left(2n_0-\tfrac{\tau -\tau_1-\tau_2}{2}\right)_k}\,(W_{12}W_{21})^{n_0+k}\\
    &\times\,(W_{13}-W_{14})^{n_2}\,\zeta^{(1)}_{J_1-n_2-n_0-k}(\tau_1+n_2,\tau_2-n_2,\tau_3+n_2,\tau_4-n_2,\tau+2n_2)\nonumber\\
    &\times\,(W_{23}-W_{24})^{n_1}\,\zeta^{(2)}_{J_2-n_1-n_0-k}(\tau_1-n_1,\tau_2+n_1,\tau_3+n_1,\tau_4-n_1,\tau+2n_1).\nonumber
\end{align}
In this case, the leading monomial in $W_{12}$ and $W_{21}$ is orthogonal and reads explicitly:
\begin{subequations}
\begin{align}
\bar{\mathcal{Q}}^{{\bf n}}_{l,\tau,0}(s|W_{ij})&=\mathcal{A}_{l,\tau}(\tau_i)\,\bar{\mathfrak{q}}^{{\bf n}}_{l,\tau}(s|W_{ij})\\
    \bar{\mathfrak{q}}_{l,\tau}^{{\bf n}}(s|W_{ij})&=\frac{2^{n_0}(d-l-\tau -1)_l (n_1+n_2+l+\tau -1)_{l-n_1-n_2}}{4^{l-n_1-n_2} \left(\frac{\tau +\tau_1-\tau_2}{2}\right)_{l-n_1+n_2} \left(\frac{\tau -\tau_1+\tau_2}{2}\right)_{l-n_2+n_1}}\\&\times\,(\frac{-s-\tau_1+\tau_2}{2})_{n_1} (\frac{s+\tau}2)_{n_1}\,(\frac{-s+\tau_3-\tau_4}{2})_{n_2} (\frac{s+\tau +\tau_1-\tau_2-\tau_3+\tau_4}2)_{n_2}\nonumber\\ \nonumber
    &\times\,\bar{\Upsilon}^{{\bf n}}_{J_1,J_2}(s|W_{ij})Q_{l-n}^{(\tau+2n_1,\tau+\tau_1-\tau_2-\tau_3+\tau_4+2n_2,-\tau_1+\tau_2+2n_1,\tau_3-\tau_4+2n_2)}(s)\,\\\label{barUpsilon12}
    \bar{\Upsilon}^{{\bf n}}_{J_1,J_2}(s)&=\, \underbrace{(-1)^{J_2-n_1-n_0}\frac{\left(-\tfrac{\tau-\tau_1-\tau_2}{2}+J_1-n_2+n_0\right)_{J_2-n_1-n_0}}{\left(-\tfrac{\tau-\tau_1-\tau_2}{2}+2n_0\right)_{J_2-n_1-n_0}}}_{w_{n_0}}\\& \nonumber \hspace*{3cm}\times\,(W_{13}-W_{14})^{n_2}(W_{23}-W_{24})^{n_1}W_{12}^{J_1-n_2}W_{21}^{J_2-n_1}\,.
\end{align}
\end{subequations}
Combining together the overall coefficients the above can be neatly expressed in terms of the normalisation of the bilinear product for continuous Hahn polynomials \eqref{Qnorm}:
\begin{multline}\label{Qfinalform}
    \bar{\mathcal{Q}}_{l,\tau}^{\bf{n}}(s|W_{ij})=(-1)^{l-n_1-n_2}(l-n_1-n_2)!\left(\mathfrak{N}_{l-n_1-n_2}^{(\tau+2n_1,\tau+\tau_1-\tau_2-\tau_3+\tau_4+2n_2,-\tau_1+\tau_2+2n_1,\tau_3-\tau_4+2n_2)}\right)^{-1}\\\times\,\left(\tfrac{-s-\tau_1+\tau_2}{2}\right)_{n_1} \left(\tfrac{s+\tau}2\right)_{n_1}\,\left(\tfrac{-s+\tau_3-\tau_4}{2}\right)_{n_2} \left(\tfrac{s+\tau +\tau_1-\tau_2-\tau_3+\tau_4}2\right)_{n_2}\\\times\,\bar{\Upsilon}^{{\bf n}}_{J_1,J_2}(s|W_{ij})Q_{l-n_1-n_2}^{(\tau+2n_1,\tau+\tau_1-\tau_2-\tau_3+\tau_4+2n_2,-\tau_1+\tau_2+2n_1,\tau_3-\tau_4+2n_2)}(s)\,,
\end{multline}

Unlike for the simpler $J$-$0$-$0$-$0$ CPWs considered in the previous section, in this case the orthogonal component \eqref{Qfinalform} is degenerate in $0\leq n_0\leq\text{min}(J_2-n_1,J_1-n_2)$. I.e. the tensor structure in \eqref{barUpsilon12} does not depend on $n_0$, but the contribution of a given CPW with $n_0$ to that tensor structure comes with weight $w_{n_0}$. Nonetheless, like for the $J$-$0$-$0$-$0$ case in the previous section, given a $J_1$-$J_2$-$0$-$0$ Mellin amplitude with ${\sf s}$-channel conformal block expansion 
\begin{align}\label{cbeJ1J2000}
   \rho_{\left\{\tau_i\right\}}\left(s,t\right){\cal M}\left(s,t|W_{ij}\right) &= \sum_{\tau,l} \sum_{{\bf n}} a^{({\bf n})}_{\tau,l} \left[\sum_{m}\frac{\tilde{\rho}_{\left\{\tau_i\right\}}\left(s,t\right){\cal Q}^{({\bf n})}_{l,\tau,m}\left(s|W_{ij}\right)}{t-\tau-2m}+...\right],
\end{align}
by considering the each physical residue of the Mellin amplitude:
\begin{subequations}
\begin{align}
    \tilde{\rho}_{\{\tau_i\}}(s,\bar{\tau})\mathcal{M}_{\bar{\tau}}(s|W_{ij})&=-\text{Res}_{t=\bar{\tau}}\left[\rho_{\{\tau_i\}}(s,t)\mathcal{M}(s,t|W_{ij})\right]\,\\
    &=\tilde{\rho}_{\{\tau_i\}}(s,\bar{\tau})\sum_{l}\sum_{{\bf n}}a_{\bar{\tau},l}^{(\bf{n})}\mathcal{Q}_{l,\bar{\tau},0}^{\bf{n}}(s|W_{ij})\,,
\end{align}
\end{subequations}
we can apply the orthogonality relations above to access a weighted average in $n_0$ of the conformal block expansion coefficients $a_{\bar{\tau},l}^{\bf{n}}=a_{\bar{\tau},l}^{\left(n_1,n_2,n_0\right)}$:
\begin{multline}\label{j1j2Inv}
    \underbrace{\sum_{n_0}a_{\bar{\tau},l}^{\left(n_1,n_2,n_0\right)}\,w_{n_0}}_{\left\langle a_{\bar{\tau},l}^{\left(n_1,n_2,n_0\right)}\right\rangle}=\frac{(-1)^{l-n_1-n_2}}{\left(l-n_1-n_2\right)!}\int\frac{ds}{4\pi i} \,\rho_{\left\{\tau_i\right\},\bar{\tau}}(s)\\\times\,\left[ {\cal M}_{\bar{\tau}}(s|W_{ij})\right]_{\bar{\Upsilon}}Q_{l-n_1-n_2}^{(\bar{\tau}+2n_1,\bar{\tau}+\tau_1-\tau_2-\tau_3+\tau_3+2n_2,-\tau_1+\tau_2+2n_1,\tau_3-\tau_4+2n_2)}(s)\,.
\end{multline}
In order to access more than the weighted average one would need to consider the full $\Upsilon$-polynomial and not just its leading term ${\bar \Upsilon}$. In general this is not a difficult task and is discussed in more detail towards the end of \S \tcb{\ref{J1J2J3J4}}, and is put into practice in the applications section \S \tcb{\ref{Application}}, \S \tcb{\ref{OPEJ1J2J3J4}} and \S \tcb{\ref{subsec::liftdeg}}

\subsection{$J_1$-$J_2$-$J_3$-$J_4$ correlators}\label{J1J2J3J4}

Here we consider CPWs with external operators of spins $J_1$-$J_2$-$J_3$-$J_4$ and generic twist $\tau_i$. For simplicity we restrict to CPWs with ${\bf n}=\left(0,0,n_0\right)$ and ${\bf \bar n}=\left(0,0,{\bar n}_0\right)$:\footnote{The general case $n_i\neq0$ can be also worked out in detail but will be presented elsewhere \cite{TTTT}.}
\begin{equation}
    {}^{({\sf s})}\mathcal{F}_{l,\tau}^{{\bf n},{\bf \bar n}}\rightarrow {}^{({\sf s})}\mathcal{F}_{l,\tau}^{n_0,{\bar n}_0}\,.
\end{equation}
This projection removes all contributions from mixed-symmetry operators in ${\sf s}$-channel.

In this case, the kinematic polynomials \eqref{kpoly0} in the basis \eqref{nicebasis} take the form:
\begin{subequations}\label{1234Q0}
\begin{align}
\mathcal{Q}^{n_0,{\bar n}_0}_{l,\tau,0}(s|W_{ij})&=\mathcal{A}_{l,\tau}(\tau_i)\,\mathfrak{q}^{n_0,{\bar n}_0}_{l,\tau}(s|W_{ij})\\
   \mathfrak{q}_{l,\tau}^{n_0,{\bar n}_0}(s|W_{ij})&=\frac{2^{n_0+{\bar n}_0}(d-l-\tau -1)_l (l+\tau -1)_{l}}{4^{l} \left(\tfrac{\tau +\tau_1-\tau_2}{2}\right)_{l} \left(\tfrac{\tau -\tau_1+\tau_2}{2}\right)_{l}}\\ \nonumber &\hspace*{1cm}\times\, (-1)^{J_3+J_4}\Upsilon^{n_0,{\bar n}_0}_{J_1,J_2,J_3,J_4}(s|W_{ij})Q_{l}^{(\tau,\tau+\tau_1-\tau_2-\tau_3+\tau_4,-\tau_1+\tau_2,\tau_3-\tau_4)}(s)\,,
\end{align}
\end{subequations}
where the dependence on the external spins $J_1$-$J_2$-$J_3$-$J_4$ is encoded in the tensor structure $\Upsilon^{n_0,{\bar n}_0}_{J_1,J_2,J_3,J_4}(s|W_{ij})$. In this case is the leading monomial in $W_{12}$, $W_{21}$, $W_{34}$ and $W_{43}$ is orthogonal, which is given explicitly by
\begin{subequations}
\begin{align}\label{Qbar12340}
\bar{\mathcal{Q}}^{n_0,{\bar n}_0}_{l,\tau,0}(s|W_{ij})&=\mathcal{A}_{l,\tau}(\tau_i)\,\bar{\mathfrak{q}}^{n_0,{\bar n}_0}_{l,\tau}(s|W_{ij})\\
 \bar{\mathfrak{q}}_{l,\tau}^{n_0,{\bar n}_0}(s)&=\frac{2^{n_0+{\bar n}_0}(d-l-\tau -1)_l (l+\tau -1)_{l}}{4^{l} \left(\tfrac{\tau +\tau_1-\tau_2}{2}\right)_{l} \left(\tfrac{\tau -\tau_1+\tau_2}{2}\right)_{l}}\\ \nonumber & \hspace*{3cm}\times\,(-1)^{J_3+J_4}\bar{\Upsilon}^{n_0,{\bar n}_0}_{J_1,J_2,J_3,J_4}(s|W_{ij})Q_{l}^{(\tau,\tau+\tau_1-\tau_2-\tau_3+\tau_4,-\tau_1+\tau_2,\tau_3-\tau_4)}(s)\,,\\ \label{Ybar1234}
    \bar{\Upsilon}^{n_0,{\bar n}_0}_{J_1,J_2,J_3,J_4}&=(-1)^{J_2+J_3-n_0-{\bar n}_0}\tfrac{\left(-\tfrac{\tau-\tau_1-\tau_2}{2}+J_1+n_0\right)_{J_2-n_0}\left(-\tfrac{\tau-\tau_3-\tau_4}{2}+J_3+{\bar n}_0\right)_{J_4-{\bar n}_0}}{\left(-\tfrac{\tau-\tau_1-\tau_2}{2}+2n_0\right)_{J_2-n_0}\left(-\tfrac{\tau-\tau_3-\tau_4}{2}+2{\bar n}_0\right)_{J_4-{\bar n}_0}}\\\nonumber &\hspace{250pt}\times\,W_{12}^{J_1}W_{21}^{J_2}W_{34}^{J_3}W_{43}^{J_4}\,.
\end{align}
\end{subequations}
In this case, in restricting to the structure \eqref{Ybar1234}, there is a degeneracy in both $n_0$ and ${\bar n}_0$. Combining the overall coefficients \eqref{Qbar12340} is given by
\begin{equation}\label{Qbar1234}
  \bar{\mathcal{Q}}_{\tau,l,0}^{n_0,{\bar n}_0}(s|W_{ij})=\mathfrak{K}_{\tau,l}^{n_0,{\bar n}_0}W_{12}^{J_1}W_{21}^{J_2}W_{34}^{J_3}W_{43}^{J_4}\, Q_{l}^{(\tau,\tau+\tau_1-\tau_2-\tau_3+\tau_4,-\tau_1+\tau_2,\tau_3-\tau_4)}(s),
\end{equation}
with
\begin{multline}\label{wdef}
    \mathfrak{K}_{\tau,l}^{n_0,{\bar n}_0}=\underbrace{\frac{4^{-l} (l+\tau -1)_l \Gamma (2 l+\tau )}{\Gamma \left(\frac{2 l+\tau +\tau_1-\tau_2}{2}\right) \Gamma \left(\frac{2 l+\tau -\tau_1+\tau_2}{2}\right) \Gamma \left(\frac{2 l+\tau +\tau_3-\tau_4}{2}\right) \Gamma \left(\frac{2 l+\tau -\tau_3+\tau_4}{2}\right)}}_{\mathcal{N}_{\tau,l}}\\\times\,\underbrace{2^{n_0+{\bar n}_0} (-1)^{J_2+J_4-n_0-{\bar n}_0}\,\frac{ \left(J_1+n_0+\tfrac{-\tau +\tau_1+\tau_2}{2}\right)_{J_2-n_0} \left(J_3+{\bar n}_0+\frac{-\tau +\tau_3+\tau_4}{2}\right)_{J_4-{\bar n}_0}}{\left(\tfrac{4 {\bar n}_0-\tau +\tau_1+\tau_2}{2}\right)_{J_2-n_0} \left(\tfrac{4 {\bar n}_0-\tau +\tau_3+\tau_4}{2}\right)_{J_4-{\bar n}_0}}}_{w_{n_0;{\bar n}_0}}\,,
\end{multline}
where we have defined the weights $w_{ n_0;{\bar n}_0}$ associated to partial wave coefficients labelled by $n_0$ and ${\bar n}_0$.

Given a Mellin amplitude ${\cal M}(s,t)$, using the orthogonal component \eqref{Qbar1234} of the kinematic polynomial \eqref{kpoly0} we can then access the weighted average of the coefficients in the conformal block expansion:
\begin{align}\label{cbeJ1J2J3J4}
   \rho_{\left\{\tau_i\right\}}\left(s,t\right){\cal M}\left(s,t|W_{ij}\right) &= \sum_{\tau,l}\sum_{{\bf n},{\bf \bar{n}}} a^{{\bf n},{\bf \bar{n}}}_{\tau,l} \left[\sum_{m}\frac{\tilde{\rho}_{\left\{\tau_i\right\}}\left(s,t\right){\cal Q}^{{\bf n},{\bf \bar{n}}}_{l,\tau,m}\left(s|W_{ij}\right)}{t-\tau-2m}+...\right],
\end{align}
where, taking the residue of the physical poles,
\begin{subequations}
\begin{align}
    {\cal M}_{\bar{\tau}}(s|W_{ij})&:= -\text{Res}_{t=\bar{\tau}} {\cal M}(s,t|W_{ij}) = \sum_{l,\bf{n},\bf{\bar n}}\,a^{\bf{n},\bf{\bar n}}_{\bar{\tau},l} \mathcal{Q}^{\bf{n},\bf{\bar n}}_{l,\bar{\tau},0}(s|W_{ij}),
\end{align}
\end{subequations}
we have
\begin{multline}\label{SpinningInv}
    \underbrace{\sum_{n_0,{\bar n}_0}a_{\bar{\tau},l}^{0,0,n_0;0,0,{\bar n}_0}\,w_{n_0,{\bar n}_0}}_{\left\langle a_{\bar{\tau},l}^{0,0,n_0;0,0,{\bar n}_0}\right\rangle}=\frac{(-1)^l}{l!}\int\frac{ds}{4\pi i} \,\rho_{\left\{\tau_i\right\},\bar{\tau}}(s)\\\times\,\left[ {\cal M}_{\bar{\tau}}(s|W_{ij})\right]_{W_{12}^{J_1}W_{21}^{J_2}W_{34}^{J_3}W_{43}^{J_4}}Q_{l}^{(\bar{\tau},\bar{\tau}+\tau_1-\tau_2-\tau_3+\tau_3,-\tau_1+\tau_2,\tau_3-\tau_4)}(s)\,,
\end{multline}
which is an average of the $a_{\bar{\tau},l}^{0,0,n_0;0,0,\bar{n}_0}$ with respect to the weights $w_{n_0;\bar{n}_0}$. We discuss how to lift this degeneracy in $n_0$ and $\bar{n}_0$ in the following subsection, and also put it into practice towards the end of \S \tcb{\ref{OPEJ1J2J3J4}}, \S \tcb{\ref{subsec::liftdeg}} and in the applications considered in \S \tcb{\ref{Application}}.

To conclude, it is useful to note that the orthogonal component \eqref{Qbar1234} does not receive any contributions from $\sf{s}$-channel CPWs with at least one of the $n_1$, $n_2$, ${\bar n}_1$, ${\bar n}_2$ non-zero, and therefore also from possible mixed-symmetry operators in the $\sf{s}$-channel which contribute in general to spinning 4pt correlators. This is because mixed symmetry operators can only be generated when an internal leg has indices contracted with at least two building blocks, say ${\sf H}$ and ${\sf  Y}$, in the integral representation \eqref{nbscpw}. Otherwise, over-symmetrisation would set the corresponding OPE structure to zero. As a consequence, in any generic dimensions $d$ part of the CFT data of solely totally symmetric operators can be read off by restricting to the tensorial structure:
\begin{equation}
    W_{12}^{J_1}W_{21}^{J_2}W_{34}^{J_3}W_{43}^{J_4}\,,
\end{equation}
within an otherwise generally complicated 4pt correlator, and applying the orthogonality properties of the continuous Hahn polynomials. Further weighted averages can be accessed by focusing on different tensor structures, as we will see in various examples. These give further equations to extract the OPE data completely, as is discussed in the following.

\subsection*{Beyond the weighted average}

In the preceding sections we observed that the orthogonality of the leading component $\bar{\Upsilon}$ of the $\Upsilon$-polynomials \eqref{barUpsilon12} generally gives access to only a weighted average of the OPE coefficients at fixed spins. This degeneracy can be lifted by looking at other, subleading, tensor structures in the $\Upsilon$-polynomials in order to obtain further constraints to fix all coefficients $a^{{\bf n},{\bf \bar{n}}}_{\tau,l}$, and is nothing but the tensorial analogue of disentangling the contributions from descendants when considering subleading twist operators.\footnote{It would be interesting to investigate the existence of differential operators which would project away specific conformal blocks and use them to lift the degeneracy in the same way as when disentangling lower and higher twist operators in \S \tcb{\ref{subsec::leadsub}}.} 

The basic idea is to focus on tensor structures in, say, \eqref{zetaCPW} with either $k_3$ or $k_4$ non zero and rewrite the $k$-dependent Pochhammer symbols as a sum of Pochhammer symbols with shifted arguments in such a way that they can be telescopically combined with the overall Pochhammer symbols in \eqref{j1j200Qn2}, e.g.:
\begin{equation}
    \left(\frac{s+\gamma}2\right)_k=\sum_{i=0}^{k}\binom{k}{i}\left(-n\right)_{k-i}\left(\frac{s+\gamma}2+n\right)_i\,.
\end{equation}
Combining the above decomposition with the prefactor in \eqref{j1j200Qn} simply gives a shifted $\tilde{\rho}_{\{\tau_i\}}$-measure.
In this way one can reduce the inversion formula to applications of eq.~\eqref{QQintegral} which gives a decomposition of continuous Hahn polynomials with arbitrary arguments in terms of those which are orthogonal with respect to a given $\tilde{\rho}$-measure:
\begin{equation}
    Q_l^{(\beta_1,\beta_2,\beta_3,\beta_4)}=Q_l^{(\alpha_1,\alpha_2,\alpha_3,\alpha_4)}+\sum_{l^\prime=0}^{l-1}\mathfrak{Z}_{\alpha_1,\alpha_2,\alpha_3,\alpha_4}^{\beta_1,\beta_2,\beta_3,\beta_4}Q_{l^\prime}^{(\alpha_1,\alpha_2,\alpha_3,\alpha_4)}\,,
\end{equation}
where the $\mathfrak{Z}_{\alpha_1,\alpha_2,\alpha_3,\alpha_4}^{\beta_1,\beta_2,\beta_3,\beta_4}$ are defined by \eqref{QQintegral}. The above shows that for a given fixed external spins the overlap between CPW with different exchanged spins in a given tensor structure is always finite dimensional, allowing to completely lift the degeneracy that comes with restricting to a single structure in finitely many steps. 
In particular, in the case that $\alpha_i-\beta_i=2n_i$, in terms of positive integer shifts $n_i$ we also have that the sum over $l^\prime$ truncates from below.

In some cases when there is an infinite twist degeneracy in spin the resummation over the exchanged spin $l$ typically produces distributions of the type $\delta(s+a)$ (see e.g. the end of \S\tcb{\ref{OPEJ1J2J3J4}}). Using factorisation of the $l$-dependence from the dependence on external spins in the basis \eqref{nicebasis}, the problem reduces again to decomposing the tensor structures in the correlator into a sum of $\Upsilon$-polynomials yielding at fixed spins a finite dimensional linear system for the tensor structures which can be solved for the OPE coefficients.

\subsection{Disentangling leading and subleading twists}
\label{subsec::leadsub}

As described in the preceding sections, contributions from sub-leading twist operators require extra work compared to those of leading twist owing to the mixing between the descendants of lower twist operators and the sub-leading primary operators themselves.

In order to study efficiently subleading twist operators it is convenient to recall that conformal blocks associated to totally symmetric representations of twist $\tau$ and spin $l$ propagating in the internal leg satisfy the following quadratic and quartic Casimir equations:\footnote{Our definition of the quadratic and quartic conformal Casimir operators is: 
\begin{align}
    \hat{\mathcal{C}}_2&=\frac12\,L^{AB}L_{BA}\,,&  \hat{\mathcal{C}}_4&=\frac12\,L^{AB}L_{BC}L^{CD}L_{DA}\,,
\end{align}
in terms of the generators $L_{AB}$ of the conformal group whose indices are raised and lowered with the $d+2$ dimensional metric $\eta_{AB}$.}
\begin{subequations}\label{c2c4}
\begin{align}
    \widehat{\mathcal{C}}_2\,u^{\Delta/2}g_{\tau,l}(u,v|W_{ij})&=[l (d+l-2)+\Delta(\Delta-d)]\,u^{\Delta/2}g_{\tau,l}(u,v|W_{ij})\,,\\
    \widehat{\mathcal{C}}_4\,u^{\Delta/2}g_{\tau,l}(u,v|W_{ij})&=[\Delta ^2 (\Delta -d)^2+\frac{1}{2} d (d-1) \Delta  (\Delta -d)\\\nonumber&\hspace{1cm}+l^2 (d+l-2)^2+\frac{1}{2} (d-4) (d-1) l (d+l-2)]\,u^{\Delta/2}g_{\tau,l}(u,v|W_{ij})\,.
\end{align}
\end{subequations}
Above $\widehat{\mathcal{C}}_2$ and $\widehat{\mathcal{C}}_4$ are differential operators representing the action of the Casimirs on 4pt functions.\footnote{Note that in the case of spinning correlators one needs to include derivatives with respect to the spinning structures.} 
Following Alday \cite{Alday:2016njk}, from \eqref{c2c4} a differential operator can be constructed whose kernel consists of any conformal block of a given twist. This differential operator can be fixed in general to be of the form
\begin{align}\label{tbop}
    \widehat{\mathcal{T}}_\tau=\widehat{\mathcal{C}}_4+a_1\,\left(\widehat{\mathcal{C}}_2\right)^2+a_2\,\widehat{\mathcal{C}}_2+a_3\,,
\end{align}
in terms of the corresponding representations for the Casimir operators as differential operators on a given 4pt correlator. The coefficients are given by:
\begin{subequations}
\begin{align}
    a_1&=-\frac12\,,\\
    a_2&=+\frac{1}{2} \left(d (4 \tau +7)-2 \tau  (\tau +2)-4-3 d^2\right)\,,\\
    a_3&=-\frac{1}{2} \tau  (d-\tau -2) (d-\tau ) (2 d-\tau -2)\,,
\end{align}
\end{subequations}
simply from the knowledge of Eigenvalues of the Casimirs. Similar but different choices of the coefficients $a_i$ would give the corresponding twist block operator for mixed-symmetry representations. In the totally symmetric case, the corresponding Eigenvalue equation is given by:
\begin{multline}\label{lambdatautaubar}
     \widehat{\mathcal{T}}_\tau\,u^{\bar{\tau}/2}g_{\bar{\tau}/2}(u,v|W_{ij})\\=\underbrace{\frac{(\bar{\tau}-\tau) (2 d-\tau -\bar{\tau}-2) (d-2 l-\tau -\bar{\tau}) (d+2 l-\tau +\bar{\tau}-2)}{2} }_{\lambda_{\tau,\bar{\tau}}}\,u^{\bar{\tau}/2}g_{\bar{\tau}/2}(u,vW_{ij}),
\end{multline}
where the r.h.s is zero for $\tau=\bar{\tau}$ and we can thus use $ \widehat{\mathcal{T}}_\tau$ to project away all conformal blocks with a certain twist.\footnote{Similar idea was used in \cite{Gliozzi:2017hni} to project away conformal blocks of a given dimension $\Delta$ and spin $l$, which requires only the quadratic Casimir $\mathcal{C}_2$. The latter was employed to extract the conformal block expansion of mean-field theory correlators.} Such a projection reduces the problem of extracting OPE data of subleading twist operators to the problem of extracting leading twist OPE data from projected CFT correlators.

In the following section we demonstrate how this idea works in practice by extracting the OPE coefficients of subleading twist double-trace operators from connected 4pt correlators in the free scalar $O\left(N\right)$ model.

\subsection{Consistency check: free ${\cal O}\left(N\right)$ model}\label{OPEJ1J2J3J4}

An immediate application of the methods presented in this section is to extract the OPE coefficients from known 4pt functions of totally symmetric spinning operators using the inversion formulae derived in the previous sections. A simple example in which we can test our framework is the free scalar $O\left(N\right)$ model in $d$-dimensions. Its spectrum includes a tower of conserved, single-trace primary operators $J_l$ of even spin $l$, whose connected 4pt correlation functions are known explicitly in general $d$ \cite{Sleight:2016dba} (see also \cite{Didenko:2012tv,Gelfond:2013xt,Bonezzi:2017vha}) together with their OPE coefficients (single-trace \cite{Sleight:2016dba}, $\left[J_0J_0\right]_{n,l}$ double-trace \cite{Dolan:2000ut,Bekaert:2015tva}). Below we demonstrate how we can seamlessly recover the latter from the former by applying the inversion formula \eqref{SpinningInv}. At the end of this section we present some new results for leading twist $\left[J_{l_1}J_{l_2}\right]_{n,l}$ double-trace operators.\footnote{The disconnected correlators and the extraction of the corresponding $\left[J_0J_0\right]_{n,l}$ mean field theory OPE coefficients \cite{Dolan:2000ut,Heemskerk:2009pn,Fitzpatrick:2011dm}, together with new results for spinning external legs, is considered in the context of other applications of this formalism in \S \tcb{\ref{Application}}.}

 To apply the inversion formula \eqref{SpinningInv}, following the prescription outlined in \S \tcb{\ref{J1J2J3J4}} we restrict to the tensor structure $W_{12}^{l_1}W_{21}^{l_2}W_{34}^{l_3}W_{43}^{l_4}$ in the connected 4pt correlators of higher-spin conserved currents $J_{l_i}$, which for canonically normalised 2pt functions reads \cite{Sleight:2016dba}:
\begin{multline}\label{JJJJ4pt}
    \left\langle J_{l_1}(y_1)J_{l_2}(y_2)J_{l_3}(y_3)J_{l_4}(y_4)\right\rangle_{\text{connected}}\sim (-1)^{l_2+l_4}\left(\prod_{i=1}^4c_i\right)\,\frac{\left(\tfrac{\Delta}2+l_2\right)_{l_1}\left(\tfrac{\Delta}2+l_3\right)_{l_4}}{\left(\tfrac{\Delta}2\right)_{l_2}\left(\tfrac{\Delta}2\right)_{l_4}}\,W_{12}^{l_1}W_{21}^{l_2}W_{34}^{l_3}W_{43}^{l_4}\\\times\frac{1}{(y_{12}^2)^{\Delta}(y_{34}^2)^{\Delta}}\,\left[u^{\Delta/2}+\left(\frac{u}{v}\right)^{\Delta/2}+u^{\Delta/2}\left(\frac{u}{v}\right)^{\Delta/2}\right]\,,
\end{multline}
where $\Delta=d-2$ is the twist of the conserved currents $J_l$ and we have defined
\begin{equation}
    c^2_i=\frac{2^{3-l_i-\Delta}\,\sqrt{\pi}\,l_i!\,\Gamma\left(\tfrac{\Delta}2+l_i\right)\Gamma(l_i+\Delta-1)}{N\,\Gamma\left(\tfrac{\Delta}2\right)^2\Gamma\left(l_i+\tfrac{\Delta-1}2\right)}\,.
\end{equation}
Recalling that in the $\sf{s}$-channel limit $u\rightarrow0$ an operator of twist $\tau$ contributes as $u^{\tau/2}$, there are two types of $\sf{s}$-channel contribution to the correlator \eqref{JJJJ4pt}: The two terms proportional to $u^{\Delta/2}$ encode the contributions from the tower of conserved currents $J_l$ of twist $\Delta$, while the term proportional to $u^\Delta$ encodes contributions from their double-trace operators. The OPE coefficients of these contributions can thus be extracted by projecting the corresponding terms in the correlator \eqref{JJJJ4pt} onto conformal blocks in the $\sf{s}$-channel using the orthogonality of continuous Hahn polynomials, which we carry out in the following.

\paragraph{Single-trace} Let us first extract the OPE coefficients of the single-trace currents $J_l$, which each have the same twist $\Delta=d-2$.

At fixed spins, the 3pt conformal structures for conserved currents in free scalar theories are those with ${\bf n}=0$ in the basis \eqref{nicebasis} -- see \cite{Stanev:2012nq,Zhiboedov:2012bm,Sleight:2017fpc}. We can therefore use the inversion formula \eqref{SpinningInv} to extract the full CPW expansion coefficients (i.e. not just a weighted average) of the conserved currents $J_l$ since the only CPWs which contribute are those with ${\bf n}={\bar {\bf n}}=\left(0,0,0\right)$.\footnote{This is most clearly seen from the integral representation \eqref{nbscpw} of CPWs. The vectors ${\bf n}$ and ${\bar {\bf n}}$ label the three-point conformal structures entering the integral representation, which in a free scalar theory can only be those with ${\bf n}={\bar {\bf n}}=\left(0,0,0\right)$ for single-trace operators.} We can then expand the ${\sf s}$-channel single-trace contributions to the correlator \eqref{JJJJ4pt} in the form
\begin{multline}\label{cbexpon1234}
    (-1)^{l_2+l_4}\left(\prod_{i=1}^4c_i\right)\frac{\left(\tfrac{\Delta}2+l_2\right)_{l_1}\left(\tfrac{\Delta}2+l_3\right)_{l_4}}{\left(\tfrac{\Delta}2\right)_{l_2}\left(\tfrac{\Delta}2\right)_{l_4}}\left[u^{\Delta/2}+\left(\frac{u}{v}\right)^{\Delta/2}\right]\\=u^{\Delta/2}\,\int_{-i\infty}^{i\infty}\frac{ds}{4\pi i}\,v^{-(s+\Delta)/2}\,\rho_{\left\{\Delta\right\}}\left(s,\Delta\right)\left[\sum_{l}a^{({\bf 0},{\bf 0})}_{l}\,\mathcal{Q}^{{\bf 0,0}}_{l,\Delta,0}(s)\right]_{W_{12}^{l_1}W_{21}^{l_2}W_{34}^{l_3}W_{43}^{l_4}}\,.
\end{multline}
Using the inversion formula \eqref{SpinningInv}, the conformal block expansion coefficients $a^{({\bf 0},{\bf 0})}_{l}$ are given by 
\begin{subequations}
\begin{align}
    a^{({\bf 0},{\bf 0})}_{l}&=\frac{(-1)^{l+l_2+l_4}}{2l!}(w_{0,0})^{-1}\left(\prod_{i=1}^4c_i\right)\frac{\left(\tfrac{\Delta}2+l_2\right)_{l_1}\left(\tfrac{\Delta}2+l_3\right)_{l_4}}{\left(\tfrac{\Delta}2\right)_{l_2}\left(\tfrac{\Delta}2\right)_{l_4}}\int_{-i\infty}^{i\infty} \frac{ds}{2\pi i}\,\left[\delta(s)+\delta(s+\tfrac{\Delta}2)\right]Q_l^{\left(\Delta/2,\Delta/2,0,0\right)}(s)\\  &=\frac{ (-1)^{l}}{l!}\left(\prod_{i=1}^4c_i\right)\left[Q_l^{\left(\Delta/2,\Delta/2,0,0\right)}(0)+Q_l^{\left(\Delta/2,\Delta/2,0,0\right)}(-\Delta/2)\right]\\ \label{3ptscalarOPE} & =
   \left[\frac{1+(-1)^l}2\right]\left(\prod_{i=1}^4c_i\right)\frac{\sqrt{\pi }2^{-\Delta -l+2} \Gamma \left(l+\frac{\Delta }{2}\right) \Gamma (l+\Delta -1)}{l!\,\Gamma \left(\frac{\Delta }{2}\right)^2 \Gamma \left(l+\frac{\Delta }{2}-\frac{1}{2}\right)}\,.
\end{align}
\end{subequations}
In the above we used that the Mellin transform of a free CFT correlator is a distribution.\footnote{\label{foo::freemellin} In particular, the formula \cite{Taronna:2016ats,Bekaert:2016ezc}
\begin{equation}
\delta(s+\Delta)\equiv\int_0^\infty dx \, x^{s-1}\,x^\Delta\,.
\end{equation}
The above distribution is defined by the property
\begin{equation}
    \int_{-i\infty}^{i\infty}\frac{ds}{2\pi i}\,\delta(s+\Delta)\,f(s)=f(-\Delta)\,.
\end{equation}}
Equation \eqref{3ptscalarOPE} precisely recovers the known expressions for the OPE coefficients of all single-trace higher-spin conserved currents in the $d$-dimensional free scalar $O\left(N\right)$ model, which were obtained in \cite{Sleight:2016dba} via Wick contractions. 

We can also verify the non-trivial cancellation of contributions from descendant operators in the conformal block expansion upon summing over spins -- i.e. all the $u^{\Delta/2+m}$ for $m>0$ in the conformal block expansion of \eqref{cbexpon1234} cancel by themselves. This can be seen by acting on \eqref{cbexpon1234} with the twist operator $\widehat{\mathcal{T}}_\tau$ defined in \eqref{tbop} for $\tau=\Delta$, which for instance in the case of external scalars gives:
\begin{equation}
    \widehat{\mathcal{T}}_\Delta\left[u^{\Delta/2}+\left(\frac{u}{v}\right)^{\Delta/2}\right]=0\,.
\end{equation}
The terms proportional to $u^{\Delta/2}$ in the correlator therefore comprise a twist block \cite{Alday:2016njk} with no higher-twist component since it lies in the kernel of $\widehat{\mathcal{T}}_\Delta$.\footnote{This is also in accordance with the explicit calculation of Diaz and Dorn \cite{Diaz:2006nm} for external scalars.}
We have checked the above also for spinning correlators using the explicit form of the twist operator \eqref{tbop} for when the Casimir operators are acting on spinning correlators, but do not present the derivation here as the corresponding Casimir operators are rather lengthy.

\subsubsection*{$\left[J_0J_0\right]_{n,l}$ Double-trace}

The double-trace contributions are encoded in the term proportional to $u^{\Delta}$ in the correlator \eqref{JJJJ4pt}. Unlike the term proportional to $u^{\frac{\Delta}{2}}$ in \eqref{JJJJ4pt} which encodes the single-trace contributions, this term is not a twist block but decomposes as an infinite sum of twist blocks owing to the different twists $\tau_n=2\Delta+2n$ of double-trace operators labelled by $n=0,1,2,3,...\,$.\footnote{The implication of this is that in the conformal block expansion of \eqref{aj0j0} the descendent contributions $u^{\tau_n/2+m}$ ($n=0,1,2,...$ and $m>0$) and higher-twist primary contributions $u^{\tau_n/2}$ with $n>0$ must cancel among each other. We thank Tassos Petkou for discussions on this point.} To extract the OPE coefficients for a given twist $\tau_n$ using the inversion formulae, we first have to project away all lower twist $\tau_{i}$ contributions and their descendants (where $i=0,...,n-1$) from the $u^{\Delta}$ term in the correlator \eqref{JJJJ4pt}, which can be done by acting successively with the operators $\widehat{\mathcal{T}}_{2\Delta+2n}$.

For simplicity, in the following we only consider the case of external scalar operators $J_0$ for which the OPE coefficients of the corresponding double-trace operators $\left[J_0J_0\right]_{n,l}$ in general $d$ are available in the literature \cite{Bekaert:2015tva,Sleight:2016hyl}. In the next section we  will study the case of external spinning operators and present new results for conformal block expansion coefficients of $\left[J_{l_1}J_{l_2}\right]_{0,l}$ double-trace operators. 

The term $u^{\Delta}$ encoding all double-trace contributions in the connected correlator \eqref{JJJJ4pt} for external scalars ($J_i=0$) reads
\begin{equation}\label{aj0j0}
    \mathcal{A}^{[J_0J_0]}_{\text{conn.}}(u,v)=\frac{4}{N}u^{\Delta/2}\left(\frac{u}{v}\right)^{\Delta/2}.
\end{equation}
To extract the OPE coefficients of the double-trace operator $[J_0J_0]_{n,l}$, we must first project away all
contributions from lower twist double trace operators $[J_0J_0]_{i,l}$ with $i=0,1,...,n-1$ (see \S \tcb{\ref{subsec::leadsub}}):
\begin{multline}
    \left[\prod_{i=0}^{n-1}\widehat{\mathcal{T}}_{2\Delta+2i} \mathcal{A}^{[J_0J_0]}(u,v)\right]_{\Delta=d-2}\\= \frac{4}{N}(-2)^{3 n}\,\left(\tfrac{d-2}{2}\right)_n^4\frac{ u^{d-2+n}}{v^{\frac{d-2}{2}+n}}\left[\sum_{i=0}^{n-1}\frac{\binom{n}{i} \left(\frac{d}{2}-1\right)_i}{\left(\frac{d}{2}+n-i-1\right)_i}\,v^i\right]+\mathcal{O}(u^{d-2+n+1})\,.
\end{multline}
For 4pt correlators with external scalar operators, the only CPWs which contribute are those with ${\bf n}=\bar{{\bf n}}=0$, and so we may expand:
\begin{multline}
\left[\prod_{i=0}^{n-1}\widehat{\mathcal{T}}_{2\Delta+2i} \mathcal{A}^{[J_0J_0]}(u,v)\right]_{\Delta=d-2}
\\=u^{\Delta+n}\int_{-i\infty}^{i\infty}\frac{ds}{4\pi i}\,v^{-(s+2\Delta+2n)/2}\,\rho_{\left\{\Delta\right\},\tau_n}\left(s\right)\sum_{l}a_{\tau_n,l}\,\mathcal{Q}^{{\bf 0,0}}_{l,\tau_n,0}(s)+\mathcal{O}(u^{\Delta+n+1})\,.
\end{multline}
 We can now use the orthogonality of continuous Hahn polynomials to extract all the double-trace conformal block expansion coefficients $a_{\tau_n,l}$, following the same steps as for the single-trace contributions above:
\begin{equation}\label{ataunl}
a_{\tau_n,l}=\frac{4}{N}\,\frac{(-2)^{3 n}}{\alpha_n}\,\left(\tfrac{d-2}{2}\right)_n^4\sum_{i=0}^{n}\frac{\binom{n}{i} \left(\tfrac{d-2}{2}\right)_i}{\left(\frac{d}{2}+n-i-1\right)_i}\,Q_l^{2(d-2)+2n,2(d-2)+2n,0,0}(2-d-2i)\,,
\end{equation}
where:
\begin{equation}\label{alphaCoeff}
    \alpha_n(\tau)=\prod_{i=0}^{n-1}\lambda_{\tau+2i,\tau+2n}=(-2)^{3 n}\, n! \left(\tfrac{d}{2}+l\right)_n (d-2 n-\tau )_n \left(-\tfrac{d}{2}+l+n+\tau \right)_n\,,
\end{equation}
is the product of Eigenvalues $\lambda_{\tau+2i,\tau+2n}$ of the operators $\widehat{\mathcal{T}}_{\tau+2i}$ defined by equation \eqref{lambdatautaubar}, where for \eqref{ataunl} we have $\tau=\tau_0=2\Delta$. The sum \eqref{ataunl} can be performed analytically by first expanding the hypergeometric function ${}_3F_2$ inside the continuous Hahn polynomial and then performing the sum over $i$ term by term. The latter sum gives a ${}_2F_1$ at argument\footnote{Upon expanding the ${}_3F_2$ in the definition of the continuous Hahn polynomial, the sum over $i$ can be performed using the identity:
\begin{align}
    \sum_{i=0}^n\binom{n}{i}\Gamma\left(\tfrac{d}{2}-i+k+n-1\right)\left(\tfrac{d-2}2\right)_i=\Gamma \left(\tfrac{d+2 k+2 n-2}{2}\right)\, _2F_1\left(\begin{matrix}-n,\frac{d}{2}-1\\-\frac{d}{2}-k-n+2\end{matrix};1\right)
\end{align}} $z=1$ which can be re-summed in a single term using Gauss' formula. After this step we arrive to:
\begin{multline}
a_{\tau_n,l}=\frac{4}{N}\,\frac{(-2)^{3 n}}{\alpha_n}\,\left(\frac{d-2}{2}\right)_n^4\sum_{k=0}^{l}\tfrac{\sqrt{\pi } (-1)^k \Gamma \left(\frac{d}{2}+k-1\right) 2^{-2 d-l-2 n+6} \Gamma (d+l+n-2) \Gamma (2 d+k+l+2 n-5)}{\Gamma (k+1) \Gamma (d+k-2) \Gamma \left(\frac{d}{2}+n-1\right) \Gamma (-k+l+1) \Gamma (d+k+n-2) \Gamma \left(d+l+n-\frac{5}{2}\right)}\\=\frac{(-2)^{3 n}}{\alpha_n}\,\left(\frac{d-2}{2}\right)_n^4\frac{\pi  2^{-3 d-l-2 n+9} \Gamma (d+l+n-2) \Gamma (2 d+l+2 n-5) }{\Gamma \left(\frac{d-1}{2}\right) \Gamma (l+1) \Gamma (d+n-2) \Gamma \left(\frac{d+2 n-2}{2}\right) \Gamma \left(\frac{2 d+2 l+2 n-5}{2}\right)}\\\times\, _3F_2\left(\begin{matrix}-l,\tfrac{d}{2}-1,2 d+l+2 n-5\\d-2,d+n-2\end{matrix};1\right)\,.
\end{multline}
The above can be further simplified to a simple term using the identity:
\begin{equation}
    _3F_2\left(\begin{matrix}-l,\tfrac{d}{2}-1,2 d+l+2 n-5\\d-2,d+n-2\end{matrix};1\right)=\frac{1+(-1)^l}2\frac{(l-1)\text{!!} \left(\tfrac{d+2 n-2}{2}\right)_{\frac{l}{2}}}{2^{l/2} \left(\frac{d-1}{2}\right)_{\frac{l}{2}} (d+n-2)_{\frac{l}{2}}}\,.
\end{equation}
Simplifying everything we arrive to the result:
\begin{equation}\label{dtopean}
    a_{\tau_n,l}=\frac{4}{N}\,\frac{1+(-1)^l}2\,\tfrac{\sqrt{\pi } 2^{-d-l+3} \left(\frac{d-2}{2}\right)_n^4 \Gamma \left(d+\frac{l-5}{2}+n\right) \Gamma (d+l+n-2) \Gamma \left(\frac{d+l-2}{2}+n\right)}{ n!\,\left(\frac{l}{2}\right)!\, \Gamma \left(\frac{d+l-1}{2}\right) (4-d-2 n)_n \Gamma \left(\frac{d}{2}+n-1\right)^2 \left(\frac{d}{2}+l\right)_n \left(\frac{3 d}{2}+l+n-4\right)_n \Gamma \left(d+l+n-\frac{5}{2}\right)}\,,
\end{equation}
which precisely agrees with the $d=4$ result \cite{Dolan:2000ut} and confirms the guess for the general $d$ result made in \cite{Bekaert:2016ezc,Sleight:2016hyl} based on the consideration of simple cases in general $d$ obtained via Wick contractions. 

\subsection*{$\left[J_{l_1}J_{l_2}\right]_{0,l}$ Double-trace}

For the double-trace contribution to the correlator with external spinning operators, restricting to the leading tensor structure $W_{12}^{l_1}W_{21}^{l_2}W_{34}^{l_3}W_{43}^{l_4}$ as in \eqref{JJJJ4pt} only gives access to a weighted average of the conformal block expansion coefficient since, unlike for the case of single-trace operators above, for a given exchanged spin $l$ it is not known which CPWs contribute and there could be more than one contributing in principle. To go beyond the weighted average, the full tensor structure of the correlator must therefore be considered. We demonstrate this in the following, where for simplicity we consider $l_1$-$l_2$-$0$-$0$ correlators and the contributions of double-trace operators $\left[J_{l_1}J_{l_2}\right]_{0,l}$ of leading twist $\tau_0=2\Delta$.

To this end, we express the result for the connected 4pt correlation function in terms of the $\Upsilon$ polynomials \eqref{Upsilon12}:
\begin{align}\label{corrJ1J2}
    \left\langle J_{l_1}(y_1,z_1)J_{l_2}(y_2,z_2)\mathcal{O}(y_3)\mathcal{O}(y_4)\right\rangle_{\text{conn.}}&=\frac{c_{l_1l_200}}{N(y_{12}^2)^{\Delta}(y_{34}^2)^{\Delta}}\left(\,u^{\Delta/2}\Upsilon^{0,0,0}_{l_1,l_2}(-\Delta|W_{ij})\right.\\
    &\left.+\left(\frac{u}{v}\right)^{\Delta/2}\Upsilon_{l_1,l_2}^{0,0,0}(0|W_{ij})+\,u^{\Delta } v^{-\Delta/2} \Upsilon_{l_1,l_2}^{l_2,l_1,0}(-\Delta|W_{ij})\right)\,,\nonumber
\end{align}
where
\begin{equation}
    c_{l_1l_200}^2=\frac{2^{8-2 \Delta -l_1-l_2}\pi \, \Gamma \left(l_1+\frac{\Delta }{2}\right) \Gamma (l_1+\Delta -1) \Gamma \left(l_2+\frac{\Delta }{2}\right) \Gamma (l_2+\Delta -1)}{l_1!l_2!\,\Gamma \left(\frac{\Delta }{2}\right)^4 \Gamma \left(l_1+\frac{\Delta }{2}-\frac{1}{2}\right) \Gamma \left(l_2+\frac{\Delta }{2}-\frac{1}{2}\right)}\,,
\end{equation}
and we recall that the contribution proportional to $u^\Delta$ encodes the double-trace contributions in the ${\sf s}$-channel. The latter is proportional to a single $\Upsilon$ polynomial \eqref{Upsilon12} with ${\bf n}=\left(l_2,l_1,0\right)$ and thus in this case only a single conformal partial wave contributes in the ${\sf s}$-channel for each exchanged spin $l$. We may now therefore restrict to the orthogonal component $\bar{\Upsilon}_{l_1,l_2}^{l_2,l_1,0}$ of the $\Upsilon_{l_1,l_2}^{l_2,l_1,0}$ polynomial defined in equation \eqref{barUpsilon12}. To extract the contributions from double-trace operators of leading twist $\tau_0=2\Delta$ we expand
\begin{multline}
   c_{l_1l_200}\,u^{\Delta}\bar{\Upsilon}_{l_1,l_2}^{l_2,l_1,0} \int_{-i\infty}^{i\infty} \frac{ds}{2\pi i}\,v^{-(s+2\Delta)/2}\delta\left(s+\Delta\right)\\=u^{\Delta} \int_{-i\infty}^{i\infty} \frac{ds}{4\pi i}\,\tilde{\rho}_{\{\Delta\}}(s,2\Delta)\,\sum_{l=0}^\infty a^{[J_{l_1}J_{l_2}][\Phi\Phi]}_{0,l}\,\bar{\mathcal{Q}}_{l,2\Delta,0}^{l_2,l_1,0}(s|W_{ij})\,,
\end{multline}
where the $a^{[J_{l_1}J_{l_2}][\Phi\Phi]}_{n=0,l}$ are the conformal block expansion coefficients in the ${\sf s}$-channel\footnote{Note that in this case both double trace operators $[J_{l_1}J_{l_2}]$ and $[\Phi\Phi]$ contribute. Since they are degenerate in spin and scaling dimension the coefficient $a^{[J_{l_1}J_{l_2}][\Phi\Phi]}_{0,l}$ gives an average of both contributions.} and
\begin{align}
    \tilde{\rho}_{\{\Delta\}}(s,2\Delta)=\Gamma\left(\tfrac{s+2\Delta}2\right)^2\Gamma\left(-\tfrac{s}2\right)^2.
\end{align}
Using the orthogonality of the $\bar{\cal Q}$ polynomials (see \S \tcb{\ref{J1J200}}), the above can be inverted as:\footnote{Note that, in the first arXiv version of this paper, the factor $c_{l_1l_200}$ was missing in the second equality of \eqref{OPEj1j2} due to a copy-paste typo from the first line to the second.}
\begin{multline}\label{OPEj1j2}
    a_{0,l}^{[J_{l_1}J_{l_2}][\Phi\Phi]}=\frac{(-1)^{l-l_1-l_2}\,c_{l_1l_200}}{(l-l_1-l_2)!}\int_{-i\infty}^{i\infty} \frac{ds}{2\pi i}\, \delta\left(s+\Delta\right)Q_{l-l_1-l_2}^{2\Delta+2l_2,2\Delta+2l_1,2l_2,2l_1}(s)\\=c_{l_1l_200}\,\frac{1+(-1)^{l-l_1-l_2}}2\frac{ \left(l_1+\frac{\Delta }{2}\right)_{\tfrac{l-l_1-l_2}{2}} \left(l_2+\frac{\Delta }{2}\right)_{\frac{l-l_1-l_2}{2}}}{ \left(\tfrac{l-l_1-l_2}{2}\right)! \left(\tfrac{l_1+l_2+l+2 \Delta -1}{2}\right)_{\frac{l-l_1-l_2}{2}}}.
\end{multline}

\section{Crossing Kernels}\label{Crossing Kernels}

In this section we employ our formalism to study the crossing kernels of spinning CPWs. We shall restrict to the orthogonal 4pt tensor structures \eqref{Qbar1234}, which in general allow us to access the weighted averages of the expansion coefficients of $\sf{t}$- and $\sf{u}$-channel CPWs into the $\sf{s}$-channel for totally symmetric spectra (i.e. the 6j symbols for generic spinning totally symmetric operators). In order to lift these degeneracies, in general the analysis for tensor structure \eqref{Qbar1234} needs to be supplemented with other tensor structures, which we demonstrate in \S\tcb{\ref{subsec::liftdeg}}, \S\tcb{\ref{1010corr}}, \S\tcb{\ref{2020corr}} and \S\tcb{\ref{J0J0corr}}.

For ease of presentation, we begin in \S \tcb{\ref{crossingScalar}} by considering crossing kernels of CPWs for an exchanged scalar primary operator and generic spinning external legs. In the subsequent section \S \tcb{\ref{kernelarbitrary}} we extend the analysis to also include exchanged primary operators of arbitrary spin, keeping generic the spinning external legs.\footnote{It might be useful to note here that, although Mack polynomials are derived from conformal partial waves defined for integer values of the spin, the final result for the crossing kernel after performing the Mellin-space integral obtained by applying the inversion formulas derived in this work is expressed in terms of ${}_4F_3$ -- whose dependence on the spin is analytic.}

For simplicity we focus on the crossing kernels of spinning OPE structures with ${\bf n}={\bf \bar n}={\bf 0}$, which are relevant for instance in applications to spinning correlators in the $O(N)$-model. For particular choices of the spins, this computation can be applied also to more general CFTs like conformal QED or similar examples.\footnote{For instance, in the case of only external scalars there is only one structure with ${\bf n}={\bf \bar n}={\bf 0}$ and therefore our result applies in full generality. Similar examples can be found in the case that $J_1$ and $J_2$ are arbitrary and $J_3=J_4=0$ with $l^\prime=0$, where also in this case the number of possible structures reduces to a single one for certain spinning correlators.} 

Finally, in this section we will keep the exchanged twist arbitrary and not distinguish between leading and sub-leading twists. Such contributions are straightforwardly disentangled by acting with the differential operators presented in \S\tcb{\ref{subsec::leadsub}}.

\subsection{Exchanged Scalar: $W_{12}^{J_1}W_{21}^{J_2}W_{34}^{J_3}W_{43}^{J_4}$}\label{crossingScalar}

In this section we study the $\sf{s}$-channel expansion of $\sf{t}$- and $\sf{u}$-channel spinning CPWs for an exchanged scalar primary operator. We focus on their projection onto $\sf{s}$-channel CPWs with ${\bf n}_{\sf{s}}=\left(0,0,n_{\sf{s},0}\right)$ and ${\bf \bar{n}}_{\sf{s}}=\left(0,0,m_{\sf{s},0}\right)$, for which the relevant orthogonal polynomials were derived in \S \tcb{\ref{MackPolinomials}}.\footnote{Here we introduced the label $\sf{s}$, $\sf{t}$ and $\sf{u}$ on the $\bf{n}$ and $\bf{m}$ to indicate that we are considering an ${\sf{s}}$, $\sf{t}$ or $\sf{u}$-channel CPW expansion respectively.} The restriction to such structures in the $\sf{s}$-channel and to the leading term $W_{12}^{J_1}W_{21}^{J_2}W_{34}^{J_3}W_{43}^{J_4}$ allows us to focus on the crossing kernels of $\sf{t}$- and $\sf{u}$-channel CPWs with ${\bf n}_{\sf{t}}={\bf \bar{n}}_{\sf{t}}=\left(0,0,0\right)$ and ${\bf n}_{\sf{u}}={\bf \bar{n}}_{\sf{u}}=\left(0,0,0\right)$. The latter are the only $\sf{t}$- and $\sf{u}$-channel CPWs with an exchanged scalar $(l^\prime=0)$ which generate contributions involving the tensor structure $W_{12}^{J_1}W_{21}^{J_2}W_{34}^{J_3}W_{43}^{J_4}$. Since spinning CPWs for an exchanged scalar operator in any channel are characterised by ${\bf n}=\left(0,0,n_{0}\right)$ and ${\bf \bar n}=\left(0,0,{\bar n}_{0}\right)$, this study is complete when restricting to contributions from the exchange of scalar primary operators in all channels. The external integer spins $J_1$-$J_2$-$J_3$-$J_4$ and twists $\tau_i$ are kept generic throughout.

We first determine the explicit Mellin representation of the $\sf{t}$- and $\sf{u}$-channel CPWs. Note that, since the goal is to decompose them into their $\sf{s}$-channel contributions, we establish their Mellin representations using the conventions for an $\sf{s}$-channel expansion (which are given at the beginning of \S \tcb{\ref{MackPolinomials}}). In particular, as in \eqref{red4pt0000} we pull out the overall factor appropriate for an $\sf{s}$-channel decomposition:
\begin{subequations}\label{schuandt}
\begin{align}
    {}^{(\sf t)}{\cal F}^{{\bf 0,0}}_{\tau,0}(y_i) & = \tfrac{1}{\left(y_{12}^2\right)^{\frac{1}{2}(\tau_1 + \tau_2)} \left(y_{34}^2\right)^{\frac{1}{2}(\tau_3 + \tau_4)}}\left(\tfrac{y_{24}^2}{y_{14}^2}\right)^{\tfrac{\tau_1-\tau_2}2}\left(\tfrac{y_{14}^2}{y_{13}^2}\right)^{\tfrac{\tau_3-\tau_4}2} 
    {}^{({\sf t})}{\cal F}^{{\bf 0,0}}_{\tau,0}(u,v|W_{ij})  \\
    {}^{(\sf u)}{\cal F}^{{\bf 0,0}}_{\tau,0}(y_i) & = \tfrac{1}{\left(y_{12}^2\right)^{\frac{1}{2}(\tau_1 + \tau_2)} \left(y_{34}^2\right)^{\frac{1}{2}(\tau_3 + \tau_4)}}\left(\tfrac{y_{24}^2}{y_{14}^2}\right)^{\tfrac{\tau_1-\tau_2}2}\left(\tfrac{y_{14}^2}{y_{13}^2}\right)^{\tfrac{\tau_3-\tau_4}2} 
    {}^{(\sf u)}{\cal F}^{{\bf 0,0}}_{\tau,0}(u,v|W_{ij})
\end{align}
\end{subequations}
and establish the Mellin representation
\begin{subequations}\label{tucpe0}
\begin{align}\label{tcpwem}
  {}^{(\sf t)}{\cal F}^{{\bf 0,0}}_{\tau,0}(u,v|W_{ij})&= \int \frac{dsdt}{\left(4\pi i\right)^2} u^{t/2}v^{-\left(s+t\right)/2}\,\rho_{\left\{\tau_i\right\},\tau}\left(s,t\right){}^{(\sf t)}{\cal F}^{{\bf 0,0}}_{\tau,0}(s,t|W_{ij})\\ \label{ucpwem}
  {}^{(\sf u)}{\cal F}^{{\bf 0,0}}_{\tau,0}(u,v|W_{ij})&= \int \frac{dsdt}{\left(4\pi i\right)^2} u^{t/2}v^{-\left(s+t\right)/2}\, \rho_{\left\{\tau_i\right\},\tau}\left(s,t\right){}^{(\sf u)}{\cal F}^{{\bf 0,0}}_{\tau,0}(s,t|W_{ij}).
\end{align}
\end{subequations}
The above Mellin representation of the $\sf{t}$-and-$\sf{u}$-channel CPWs are straightforwardly obtained from the $\sf{s}$-channel expression \eqref{mrepexscsp} by making the following replacements: to obtain the $\sf{t}$-channel \eqref{tcpwem}, in the $\sf{s}$-channel expression \eqref{mrepexscsp} we swap $2 \leftrightarrow 4$ for all indices associated to external legs and replace $\left(s,t\right) \rightarrow \left(s+\tau_4-\tau_2,-s-t+\tau_2+\tau_3\right)$. For the $\sf{u}$-channel \eqref{ucpwem}, we swap $2 \leftrightarrow 3$ for all indices associated to external legs and replace $\left(s,t\right) \rightarrow \left(t-\tau_1-\tau_4,s+\tau_1+\tau_4\right)$. The shifts in the Mellin variables are a consequence of using the conventions \eqref{schuandt} for an $\sf{s}$-channel expansion.

To expand the $\sf{t}$- and $\sf{u}$-channel CPWs \eqref{tucpe0} in terms of ${\sf s}$-channel contributions, following \S \tcb{\ref{J1J2J3J4}} we focus on the tensor structure $W_{12}^{J_1}W_{21}^{J_2}W_{34}^{J_3}W_{43}^{J_4}$:
\begin{equation}\label{LeadingM}
   \left[ {}^{(i)}{\cal F}^{{\bf 0,0}}_{\tau,0}(s,t|W_{ij})\right]_{W_{12}^{J_1}W_{21}^{J_2}W_{34}^{J_3}W_{43}^{J_4}}=\left(-\tfrac{t-\tau_1-\tau_2}2\right)_{J_1+J_2}\left(-\tfrac{t-\tau_3-\tau_4}2\right)_{J_3+J_4}\,\mathfrak{M}^{(i)}_{J_1,J_2,J_3,J_4}(s,t)\,.
\end{equation}
with $i$ labelling the channel: $i=\sf{t}$ or $\sf{u}$. In particular, we have: 
\begin{subequations}\label{ScalarTUmellinTau}
\begin{align}
    \mathfrak{M}^{(\sf t)}_{J_1,J_2,J_3,J_4|0}(s,t)&=\frac{(-1)^{J_1+J_3}\,\Gamma(\tau)}{\Gamma \left(\frac{d}{2}-\tau \right) \Gamma \left(J_1+\tfrac{\tau +\tau_1-\tau_4}{2}\right) \Gamma \left(J_2+\tfrac{\tau +\tau_2-\tau_3}{2}\right) \Gamma \left(J_3+\tfrac{\tau -\tau_2+\tau_3}{2}\right) \Gamma \left(J_4+\frac{\tau -\tau_1+\tau_4}{2}\right)}\nonumber \\&\hspace{5cm}\label{ScalarTUmellinTaut} \times\frac{\Gamma\left(\tfrac{s+t+d-\tau-\tau_2-\tau_3}2\right)\Gamma\left(\tfrac{s+t+\tau-\tau_2-\tau_3}2\right)}{\Gamma\left(\tfrac{s+t}2\right)\Gamma\left(\tfrac{s+t+\tau_1-\tau_2-\tau_3+\tau_4}2\right)}\\
    \mathfrak{M}^{(\sf u)}_{J_1,J_2,J_3,J_4|0}(s,t)&=\frac{(-1)^{J_1+J_2}\,\Gamma(\tau)}{\Gamma \left(\frac{d}{2}-\tau \right)\Gamma\left(J_1+\tfrac{\tau+\tau_1-\tau_3}2\right)\Gamma\left(J_3+\tfrac{\tau-\tau_1+\tau_3}2\right)\Gamma\left(J_2+\tfrac{\tau+\tau_2-\tau_4}2\right)\Gamma\left(J_4+\tfrac{\tau-\tau_1+\tau_4}2\right)}\nonumber\\&\hspace{5cm}\times\frac{\Gamma\left(\tfrac{d-s-\tau-\tau_1-\tau_4}2\right)\Gamma\left(\tfrac{-s+\tau-\tau_1-\tau_4}2\right)}{\Gamma\left(\tfrac{-s-\tau_1+\tau_2}2\right)\Gamma\left(\tfrac{-s+\tau_3-\tau_4}2\right)}\,,\label{ScalarTUmellinTauu}
\end{align}
\end{subequations}
which can be obtained using the techiques in Appendix \ref{4pt}. Note that \eqref{ScalarTUmellinTaut} and \eqref{ScalarTUmellinTauu} exhibit no poles for positive $t$. All poles for positive $t$ thus arise from the Mellin measure $\rho_{\{\tau_i\}}(s,t)$, in accordance with the fact that an $\sf{t}$ or $\sf{u}$-channel CPW/conformal blocks decompose in the $\sf{s}$-channel in terms of double-trace/double-twist operators.

Using the orthogonality relations established in the previous section we can access the weighted averages of the $\sf{s}$-channel expansion coefficients for the $\sf{t}$ and $\sf{u}$-channel CPWs above, which we carry out in the following.

\subsubsection*{Crossing kernels}
Given \eqref{LeadingM}, we can study the crossing kernels of the $\sf{t}$- and $\sf{u}$-channel CPWs \eqref{schuandt} into the $\sf{s}$-channel by expanding their expressions \eqref{LeadingM} in terms of the kinematic polynomials \eqref{Qbar1234} in the Mellin variable $s$ -- which corresponds to an $\sf{s}$-channel conformal block expansion in Mellin space -- and using the orthogonal projections derived in \S \tcb{\ref{MackPolinomials}}. 

\paragraph{Equal twists:} For simplicity let us first consider the case in which the external operators have equal twist $\tau_i=\Delta$, before giving the result for generic twist. We have:
\begin{subequations}\label{ScalarTUmellinDelta}
\begin{align}
    \mathfrak{M}^{(\sf t)}_{J_1,J_2,J_3,J_4|0}(s,t)&=\frac{(-1)^{J_1+J_3}\,\Gamma(\tau)}{\Gamma \left(\frac{d}{2}-\tau \right) \Gamma \left(J_1+\tfrac{\tau}{2}\right) \Gamma \left(J_2+\tfrac{\tau}{2}\right) \Gamma \left(J_3+\tfrac{\tau}{2}\right) \Gamma \left(J_4+\frac{\tau}{2}\right)}\\ \nonumber
    & \hspace*{6cm} \times  \frac{\Gamma\left(\tfrac{s+t+d-\tau-2\Delta}2\right)\Gamma\left(\tfrac{s+t+\tau-2\Delta}2\right)}{\Gamma\left(\tfrac{s+t}2\right)^2},\\
    \mathfrak{M}^{(\sf u)}_{J_1,J_2,J_3,J_4|0}(s,t)&=\frac{(-1)^{J_1+J_2}\,\Gamma(\tau)}{\Gamma \left(\frac{d}{2}-\tau \right)\Gamma\left(J_1+\tfrac{\tau}2\right)\Gamma\left(J_3+\tfrac{\tau}2\right)\Gamma\left(J_2+\tfrac{\tau}2\right)\Gamma\left(J_4+\tfrac{\tau}2\right)}\\ \nonumber
    & \hspace*{6cm} \times \frac{\Gamma\left(\tfrac{d-s-\tau-2\Delta}2\right)\Gamma\left(\tfrac{-s+\tau-2\Delta}2\right)}{\Gamma\left(-\tfrac{s}2\right)^2}.
\end{align}
\end{subequations}
Focusing only on the $s$ dependence (i.e. assuming to have performed the $t$ integral and that $t$ is thus fixed by the value at the corresponding double-trace residue), we consider the following expansion (see \S \tcb{\ref{J1J2J3J4}}):\footnote{Recall that, when restricting to the leading structure $W_{12}^{J_1}W_{21}^{J_2}W_{34}^{J_3}W_{43}^{J_4}$, $\sf{t}$- and $\sf{u}$-channel CPWs with ${\bf n}_{\sf{t}}={\bf n}_{\sf{u}}={\bar {\bf n}}_{\sf{t}}={\bar {\bf n}}_{\sf{u}}=\left(0,0,0\right)$ only contribute to $\sf{s}$-channel CPWs with ${\bf n}_{\sf{s}}=\left(0,0,n_{0}\right)$ and ${\bar {\bf n}_{\sf{s}}}=\left(0,0,{\bar n}_{0}\right)$.}\footnote{Note that, for poles corresponding to double-trace operators of subleading twist, this expansion also contributions from descendents of leading twist operators (with scaling dimension equal to that of the subleading twist operator) which should be disentangled. We explain and demonstrate how to disentangle such descendent contributions in \S\tcb{\ref{0000corr}}.} 
\begin{equation}
    \mathfrak{M}_{J_1,J_2,J_3,J_4|0}^{(i)}(s,t)=\sum_{l,n_0,\bar{n}_x} c^{(n_0,\bar{n}_0)}_{l}(t)\,\left[\mathcal{Q}^{(n_0,\bar{n}_0)}_{t,l,0}(s)\right]_{W_{12}^{J_1}W_{21}^{J_2}W_{34}^{J_3}W_{43}^{J_4}}
\end{equation}
where, using the orthogonality relations of the continuous Hahn polynomials, we can extract the weighted average: 
\begin{equation}\label{CrossingKernel}
   \left\langle c^{(i)}_{l}(t)\right\rangle=\underbrace{\frac{(-1)^{l}}{l!}
    \int_{-i\infty}^{i\infty}\frac{ds}{4\pi i}\,\rho_{\left\{\Delta\right\},t}(s)\,\mathfrak{M}^{(i)}_{J_1,J_2,J_3,J_4|0}(s,t)\,Q_{l}^{(t,t,0,0)}(s)}_{\mathfrak{I}_{J_1,J_2,J_3,J_4|l,0}^{(i)}\left(t\right)}\,.
\end{equation}
The coefficients $\mathfrak{I}^{(i)}_{J_1,J_2,J_3,J_4|l,0}$ are related to the 6j symbols (or Racah-Wigner coefficients) for the conformal group. Their explicit form can be obtained by evaluating the integral using methods similar to those employed in \cite{Gopakumar:2016cpb}, which we review in \S\tcb{\ref{KernelDetails}}. This gives
\begin{multline}\label{inv}
    \mathfrak{I}^{(i)}_{J_1,J_2,J_3,J_4|l,0}\left(t\right)= \mathcal{B}^{(i)}_{J_1,J_2,J_3,J_4}\,\frac{(-2)^{l}\Gamma\left(\tfrac{d+t-2\Delta-\tau}2\right)^2\Gamma\left(\tfrac{t-2\Delta+\tau}2\right)^2\left(\tfrac{t}{2}\right)_{l}^2}{l!\,\Gamma\left(\tfrac{d}2+t-2\Delta\right)(t+l-1)_{l}}\\\times\,{}_4 F_3\left(\begin{matrix}-l,t+l-1,\tfrac{d+t-\tau}2-\Delta,\tfrac{t+\tau}2-\Delta\\\tfrac{t}{2},\tfrac{t}2,\tfrac{d}{2}+t-2\Delta\end{matrix};1\right)
\end{multline}
where the dependence on the external spins $J_i$ is completely factorised into the coefficient $\mathcal{B}^{(i)}_{J_1,J_2,J_3,J_4}$\,, which in the $\sf{t}$-channel reads:
\begin{equation}
    \mathcal{B}^{(\sf t)}_{J_1,J_2,J_3,J_4}=\frac{(-1)^{J_1+J_3} \Gamma (\tau )}{\Gamma \left(\frac{d}{2}-\tau \right) \Gamma \left(J_1+\frac{\tau }{2}\right) \Gamma \left(J_2+\frac{\tau }{2}\right) \Gamma \left(J_3+\frac{\tau }{2}\right) \Gamma \left(J_4+\frac{\tau }{2}\right)},
\end{equation}
and, in the $\sf{u}$-channel:
\begin{equation}
    \mathcal{B}^{(\sf u)}_{J_1,J_2,J_3,J_4}=(-1)^l\,\frac{(-1)^{J_1+J_2} \Gamma (\tau )}{\Gamma \left(\frac{d}{2}-\tau \right) \Gamma \left(J_1+\frac{\tau }{2}\right) \Gamma \left(J_2+\frac{\tau }{2}\right) \Gamma \left(J_3+\frac{\tau }{2}\right) \Gamma \left(J_4+\frac{\tau }{2}\right)}.
\end{equation}

\paragraph{Generic twists} The above steps straightforwardly extend to external operators of different twists $\tau_i$, in which case one obtains the following crossing kernels:
\begin{subequations}
\begin{align}\label{invTau}
    \mathfrak{I}^{(\sf t)}_{J_1,J_2,J_3,J_4|l,0}\left(t\right)&= \mathcal{B}^{(\sf t)}_{J_1,J_2,J_3,J_4}\,\frac{(-2)^{l}\left(\frac{t+\tau_1-\tau_2}{2}\right)_{l}\left(\frac{t+\tau_3-\tau_4}{2}\right)_{l}}{2l!(t+l-1)_{l}}\\\nonumber
    &\times\frac{\Gamma\left(\frac{d+t-\tau-\tau_1-\tau_3}2\right)\Gamma\left(\frac{d+t-\tau-\tau_2-\tau_4}2\right)\Gamma\left(\frac{t+\tau-\tau_1-\tau_3}2\right)\Gamma\left(\frac{t+\tau-\tau_2-\tau_4}2\right)}{\Gamma\left(\frac{d-\tau_1-\tau_2-\tau_3-\tau_4}2+t\right)}\\\nonumber
    &\times\,{}_4 F_3\left(\begin{matrix}-l,t+l-1,\tfrac{d+t-\tau-\tau_2-\tau_4}2,\frac{t+\tau-\tau_2-\tau_4}2\\\frac{t+\tau_1-\tau_2}{2},\tfrac{t+\tau_3-\tau_4}2,\frac{d-\tau_1-\tau_2-\tau_3-\tau_4}{2}+t\end{matrix};1\right)\\
    \mathfrak{I}^{(\sf u)}_{J_1,J_2,J_3,J_4|l,0}\left(t\right)&= \mathcal{B}^{(\sf u)}_{J_1,J_2,J_3,J_4}\,\frac{(-2)^{l}\left(\tfrac{t-\tau_1+\tau_2}{2}\right)_{l}\left(\tfrac{t+\tau_3-\tau_4}{2}\right)_{l}}{2l!(t+l-1)_{l}}\\\nonumber
    &\times\frac{\Gamma\left(\tfrac{d+t-\tau-\tau_2-\tau_3}2\right)\Gamma\left(\tfrac{d+t-\tau-\tau_1-\tau_4}2\right)\Gamma\left(\tfrac{t+\tau-\tau_2-\tau_3}2\right)\Gamma\left(\tfrac{t+\tau-\tau_1-\tau_4}2\right)}{\Gamma\left(\tfrac{d-\tau_1-\tau_2-\tau_3-\tau_4}2+t\right)}\\\nonumber
    &\times\,{}_4 F_3\left(\begin{matrix}-l,t+l-1,\tfrac{d+t-\tau-\tau_1-\tau_4}2,\tfrac{t+\tau-\tau_1-\tau_4}2\\\tfrac{t-\tau_1+\tau_2}{2},\tfrac{t+\tau_3-\tau_4}2,\tfrac{d-\tau_1-\tau_2-\tau_3-\tau_4}{2}+t\end{matrix};1\right),
\end{align}
\end{subequations}
the coefficients read in this case:
\begin{subequations}\label{Bfunct}
\begin{align}
    \mathcal{B}^{(\sf t)}_{J_1,J_2,J_3,J_4}&=\frac{(-1)^{J_1+J_3}\,\Gamma(\tau)}{\Gamma \left(\frac{d}{2}-\tau \right) \Gamma \left(J_1+\tfrac{\tau +\tau_1-\tau_4}{2}\right) \Gamma \left(J_2+\tfrac{\tau +\tau_2-\tau_3}{2}\right) \Gamma \left(J_3+\tfrac{\tau -\tau_2+\tau_3}{2}\right) \Gamma \left(J_4+\frac{\tau -\tau_1+\tau_4}{2}\right)}\\
    \mathcal{B}^{(\sf u)}_{J_1,J_2,J_3,J_4}&=\frac{(-1)^{J_1+J_2+l}\,\Gamma(\tau)}{\Gamma \left(\frac{d}{2}-\tau \right)\Gamma\left(J_1+\tfrac{\tau+\tau_1-\tau_3}2\right)\Gamma\left(J_3+\tfrac{\tau-\tau_1+\tau_3}2\right)\Gamma\left(J_2+\tfrac{\tau+\tau_2-\tau_4}2\right)\Gamma\left(J_4+\tfrac{\tau-\tau_1+\tau_4}2\right)}\,.
\end{align}
\end{subequations}
For $t=2\Delta$ -- i.e. for leading twist double-trace operators -- upon expanding the ${}_4F_3$ Hypergeometric function, the above simple result matches the analogous result of \cite{Gopakumar:2016cpb} and thus provides a simple re-summation of it.

\subsection{Arbitrary exchanged spin $l^\prime$: $W_{12}^{J_1}W_{21}^{J_2}W_{34}^{J_3}W_{43}^{J_4}$}\label{kernelarbitrary}

Our proposed framework straightforwardly extends the results for exchanged scalar operators ($l^\prime=0$) in the previous section to generic exchanged spin $l^\prime$. We restrict to studying the crossing kernels of $\sf{t}$- and $\sf{u}$-channel CPWs with ${\bf n}_t={\bar {\bf n}}_t={\bf n}_u={\bar {\bf n}}_u=\left(0,0,0\right)$. Note that no structure ${\bf n}=\left(n_1,n_2,n_0\right)$ with $n_0>0$ in the ${\sf t}$ or ${\sf u}$-channel contributes to the tensor structure $W_{12}^{J_1}W_{21}^{J_2}W_{34}^{J_3}W_{43}^{J_4}$. This list of structures is thus complete for scalar partial waves. Instead structures with non-vanishing $n_1$ and $n_2$ exist for higher-spin partial waves and require a separate study which will be considered in detail elsewhere \cite{TTTT}.

Remarkably, also in this case we obtain a simple factorised form for the crossing kernel, which in turn can be expressed in terms of the crossing kernel for CPWs with external scalars dressed with a spin dependent coefficient. The complete expressions for the full Mack polynomials \eqref{mrepexscsp} with generic $l^\prime$ and arbitrary external spins is rather involved. However, restricting to a single tensor structure provides some simplifications: In restricting to the component $W_{12}^{J_1}W_{21}^{J_2}W_{34}^{J_3}W_{43}^{J_4}$ in the $\sf{t}$- and $\sf{u}$-channel CPWs in the \eqref{nicebasis}, the dependence on the external spins $J_i$ is completely factorised from the dependence on the exchanged spin-$l^\prime$. In particular, in this case performing the conformal integrals (see e.g. appendix \S\tcb{\ref{4pt}}) we have have:
\begin{multline}\label{LeadingMspinning}
   \left[ {}^{(i)}{\cal F}^{{\bf 0,0}}_{\tau,l^\prime}(s,t|W_{ij})\right]_{W_{12}^{J_1}W_{21}^{J_2}W_{34}^{J_3}W_{43}^{J_4}}\\=\left(-\frac{t-\tau_1-\tau_2}2\right)_{J_1+J_2}\left(-\frac{t-\tau_3-\tau_4}2\right)_{J_3+J_4}\,\mathfrak{M}^{(i)}_{J_1,J_2,J_3,J_4|l^\prime}(s,t)\,,
\end{multline}
where now
\begin{multline}\label{tchCPW}
    \mathfrak{M}^{(\sf t)}_{J_1,J_2,J_3,J_4|l^\prime}(s,t)=\frac{2^{-2l^\prime}\,(l^\prime+\tau-1)_{l^\prime}\,\Gamma(2l^\prime+\tau)\,\mathfrak{D}^{(\sf t)}_{J_1J_2J_3J_4|l^\prime}}{\Gamma \left(\frac{d}{2}-\tau-l^\prime \right) \Gamma \left(l^\prime+\tfrac{\tau +\tau_1-\tau_4}{2}\right) \Gamma \left(l^\prime+\tfrac{\tau +\tau_2-\tau_3}{2}\right) \Gamma \left(l^\prime+\tfrac{\tau -\tau_2+\tau_3}{2}\right) \Gamma \left(l^\prime+\frac{\tau -\tau_1+\tau_4}{2}\right)}\\\times\frac{\Gamma\left(\tfrac{s+t+d-\tau-2l^\prime-\tau_2-\tau_3}2\right)\Gamma\left(\tfrac{s+t+\tau-\tau_2-\tau_3}2\right)}{\Gamma\left(\tfrac{s+t}2\right)\Gamma\left(\tfrac{s+t+\tau_1-\tau_2-\tau_3+\tau_4}2\right)}\,P^{(\sf t)}_{\tau,l^\prime}(s,t|\tau_1,\tau_2,\tau_3,\tau_4),
\end{multline}
and it proves more convenient for $l^\prime>0$ to define the following coefficient 
\begin{align}\label{d1234t}
    \mathfrak{D}^{(\sf t)}_{J_1J_2J_3J_4}=\frac{(-1)^{J_1+J_3}}{ \left(\tfrac{\tau +\tau_1-\tau_4}{2}\right)_{J_1} \left(\tfrac{\tau +\tau_2-\tau_3}{2}\right)_{J_2} \left(\tfrac{\tau -\tau_2+\tau_3}{2}\right)_{J_3} \left(\frac{\tau -\tau_1+\tau_4}{2}\right)_{J_4}},
\end{align}
which carries dependence on the external spins $J_i$ and differs from ${\cal B}$ \eqref{Bfunct} by a $\Gamma$-function factor which we have included in the Mellin amplitude $\mathfrak{M}_{J_1J_2J_3J_4|l^\prime}^{(i)}$. Note that, as anticipated from the factorisation of the external spin dependence for spinning CPWs in the basis \eqref{nicebasis} in \S \tcb{\ref{MackPolinomials}}, this result is given in terms of the Mack polynomials \eqref{MackPscalar} for external scalars (i.e. $J_i=0$).  Similarly, for the $\sf{u}$-channel we have 
\begin{multline}\label{uchCPW}
    \mathfrak{M}^{(\sf u)}_{J_1,J_2,l_3,J_4|l^\prime}(s,t)=\frac{2^{-2l^\prime}\,(l^\prime+\tau-1)_{l^\prime}\,\Gamma(2l^\prime+\tau)\,\mathfrak{D}^{(\sf u)}_{J_1J_2J_3J_4|l^\prime}}{\Gamma \left(\frac{d}{2}-\tau-l^\prime \right) \Gamma \left(l^\prime+\tfrac{\tau +\tau_1-\tau_3}{2}\right) \Gamma \left(l^\prime+\tfrac{\tau +\tau_2-\tau_4}{2}\right) \Gamma \left(l^\prime+\tfrac{\tau -\tau_1+\tau_3}{2}\right) \Gamma \left(l^\prime+\frac{\tau -\tau_2+\tau_4}{2}\right)}\\\times\frac{\Gamma\left(\tfrac{s+d-\tau-2l^\prime-\tau_1-\tau_4}2\right)\Gamma\left(\tfrac{-s+\tau-\tau_1-\tau_4}2\right)}{\Gamma\left(\tfrac{-s-\tau_1+\tau_2}2\right)\Gamma\left(\tfrac{-s+\tau_3-\tau_4}2\right)}\,P^{(\sf u)}_{\tau,l^\prime}(s,t|\tau_1,\tau_2,\tau_3,\tau_4),
\end{multline}
with
\begin{align}\label{d1234u}
    \mathfrak{D}^{(\sf u)}_{J_1J_2J_3J_4}=\tfrac{(-1)^{J_1+J_2}}{ \left(\tfrac{\tau +\tau_1-\tau_3}{2}\right)_{J_1} \left(\tfrac{\tau +\tau_2-\tau_4}{2}\right)_{J_2}  \left(\tfrac{\tau -\tau_1+\tau_3}{2}\right)_{J_3} \left(\frac{\tau -\tau_2+\tau_4}{2}\right)_{J_4}}.
\end{align}

The $\sf{t}$- and $\sf{u}$-channel expressions of \eqref{tchCPW} and \eqref{uchCPW}, which we employ here, are obtained using the simple replacements given below equation \eqref{schuandt}.

Note that for non-zero external spins $J_i$ the Pochhammer factor in \eqref{LeadingMspinning} has the effect of removing some of the poles appearing in the Mellin measure $\rho_{\{\tau_i\}}(s,t)$. This means that the leading double-trace contributions (which would be proportional to $u^\Delta$) cancel when restricting to the conformal structure $W_{12}^{J_1}W_{21}^{J_2}W_{34}^{J_3}W_{43}^{J_4}$, so that the first non-vanishing contribution to this tensor structure is given by the subleading double-trace contribution $u^{\Delta+2(J_1+J_2)}$. 

\subsubsection*{Crossing Kernel}
The expressions \eqref{tchCPW} and \eqref{uchCPW} for the $\sf{t}$- and $\sf{u}$-channel spinning CPWs projected on the structure $W_{12}^{J_1}W_{21}^{J_2}W_{34}^{J_3}W_{43}^{J_4}$ make clear how the crossing kernel for spinning legs is a simple dressing of the crossing kernel for external scalar operators. As for the crossing kernels of spinning CPWs with $l^\prime=0$ in \S \tcb{\ref{crossingScalar}}, also in this case one can project the $\sf{t}$ and $\sf{u}$-channel Mellin amplitudes in the $\sf{s}$-channel by evaluating the Mellin-Barnes integral
\begin{equation}\label{ck1234}
 \mathfrak{I}_{J_1,J_2,J_3,J_4|l,l^\prime}^{(i)}\left(t\right)=\frac{(-1)^{l}}{l!}\int_{-i\infty}^{i\infty}\frac{ds}{4\pi i}\,\rho_{\left\{\Delta\right\},t}(s)\,\mathfrak{M}^{(i)}_{J_1,J_2,J_3,J_4|l^\prime}(s,t)\,Q_{l}^{(t,t,0,0)}(s)\,.
\end{equation}
We discuss the evaluation of the above integral in full generality in \S \tcb{\ref{KernelDetails}}. For simplicity, in the following we give the results for correlators with operators of equal twist, with the derivation mostly relegated to \S \tcb{\ref{KernelDetails}} where we also consider generic twists $\tau_i$. We focus on the crossing kernel of $\sf{t}$-channel CPWs, keeping in mind that for equal external twists we have:
\begin{equation}
    \mathfrak{J}^{(\sf t)}_{J_1,J_2,J_3,J_4|l,l^\prime}=(-1)^{l+l^\prime}\ \mathfrak{J}^{(\sf u)}_{J_1,J_2,J_3,J_4|l,l^\prime}.
\end{equation}

\paragraph{Leading twist $t=2\Delta$} For simplicity we first consider $t=2\Delta$, which for external scalars is the leading twist pole. We then give the result for generic $t$, which would be more relevant for spinning legs owing to the cancellation of leading twist contributions in that case -- which was noted at the end of the previous section.

Let us first consider the projection of spinning CPWs for exchanged spin $l^\prime$ primary operators in the $\sf{t}$-channel onto spinning CPWs for exchanged scalar primary operators ($l=0$) of twist $\tau$ in the $\sf{s}$-channel. In this case we obtain:
\begin{subequations}\label{Jlprime0}
\begin{align}
    \frac{\mathfrak{J}^{({\sf t})}_{J_1,J_2,J_3,J_4|l^\prime,0}\left(2\Delta\right)}{\mathfrak{J}^{({\sf t})}_{J_1,J_2,J_3,J_4|0,0}\left(2\Delta\right)}&=f_{l^\prime}\,\frac{2^{l^\prime} \left(\frac{\tau +1}{2}\right)_{l^\prime} \left(-\frac{d}{2}+\tau +1\right)_{l^\prime}}{\left(\frac{\tau }{2}\right)_{l^\prime} (d-l^\prime-\tau -1)_{l^\prime}}\,,\\
    \mathfrak{J}^{({\sf t})}_{J_1,J_2,J_3,J_4|0,0}\left(2\Delta\right)&=\mathcal{D}^{(t)}_{J_1,J_2,J_3,J_4}\frac{\Gamma (\tau ) \Gamma \left(\frac{d-\tau }{2}\right)^2}{\Gamma \left(\frac{d}{2}\right) \Gamma \left(\frac{\tau }{2}\right)^2 \Gamma \left(\frac{d}{2}-\tau \right)}
\end{align}
\end{subequations}
where $f_{l^\prime}$ is a numerical coefficient depending only on the space time dimension:
\begin{equation}
    f_{l^\prime}=\frac{\sqrt{\pi }\, 2^{3-d} (d+l^\prime-3)!}{(\tfrac{d-3}{2})! \left(\tfrac{d+2 l^\prime-4}{2}\right)!}\,.
\end{equation}
Remarkably the above result is independent of $\tau_i=\Delta$ for any $l^\prime$.

The result obtained above can be extended to $l>0$ by dressing the $l=0$ case \eqref{Jlprime0} with an appropriate combination of hypergeometric functions which we give below for the equal twists, relegating the result for generic twists to appendix \S\tcb{\ref{KernelDetails}} where further details on the computation are also provided. We obtain the following general expression valid for $t=2\Delta$ and $\tau_i=\Delta$:
\begin{align}\label{KernelEqualTau}
    \frac{\mathfrak{J}_{J_1,J_2,J_3,J_4|l^\prime,l}^{({\sf t})}\left(2\Delta\right)}{ \mathfrak{J}^{({\sf t})}_{J_1,J_2,J_3,J_4|0,0}\left(2\Delta\right)}=\mathcal{Z}_{{l^\prime}}\,\sum_{p=0}^{
{l^\prime}}\sum_{i=0}^{{l^\prime}-p}a^{{l^\prime}}_{p,i}\ {}_4F_3\left(\begin{matrix}p-l,l+p+2 \Delta-1,\tfrac{d-\tau }{2}-i,\tfrac{\tau}{2}+{l^\prime}-i\\\tfrac{d}{2},\Delta+p ,\Delta+p \end{matrix};1\right)\,,
\end{align}
with
\begin{subequations}
\begin{align}
    \mathcal{Z}_{{l^\prime}}&=\frac{2^{l^\prime}\left(\frac{\tau +1}{2}\right)_{l^\prime} \left(\tau +1-\tfrac{d}{2}\right)_{l^\prime}}{\left(\frac{\tau }{2}\right)_{l^\prime} (d-{l^\prime}-\tau -1)_{l^\prime}},\\
    a^{{l^\prime}}_{p,i}(t)&=\binom{{l^\prime}}{p}\frac{\sqrt{\pi }\,2^{2-l-2\Delta}\,\Gamma \left(l+\Delta\right) \Gamma (l+p+2\Delta-1)}{(l-p)! \Gamma \left(l+\Delta-\frac{1}{2}\right) \Gamma \left(p+\Delta\right)^2}\,
    \alpha^{l^\prime}_{p,i}(t)\,,\\
    \alpha^{l^\prime}_{p,i}(t)&=\binom{{l^\prime}-p}{i}\,\frac{\left(\tfrac{d+2 p-2}{2}\right)_i}{\left(\tfrac{d+2 {l^\prime}-2}{2}-i\right)_i}, 
\end{align}
\end{subequations}
where $\alpha^{l^\prime}_{p,i}$ plays the role of a generalised binomial coefficient. 

It is interesting to note that for $l^\prime=0$ the ${}_4F_3$ hypergeometric functions entering the kernel above can be expressed in terms of a Wilson polynomial \cite{doi:10.1137/0511064}:
\begin{multline}\label{WilsonP}
    P^{(a_i)}_n\left(-z^2\right)=\left(a_1+a_2\right)_n\left(a_1+a_3\right)_n\left(a_1+a_4\right)_n\\\times\,{}_4F_3\left(\begin{matrix}
    -n,n+a_1+a_2+a_3+a_4-1,a_1+z,a_1-z\\a_1+a_2,a_1+a_3,a_1+a_4
    \end{matrix};1\right)
\end{multline}
with
\begin{align}
    a_1&= \frac{d}{4},& a_2&= \frac{d}{4},& a_3&= \Delta -\frac{d}{4},& a_4&= \Delta -\frac{d}{4},& z&= \frac{1}{4} (d-2 \tau ).&
\end{align}
This property also persists in the case of generic twists (given in \S \tcb{\ref{KernelDetails}}) in which case:
\begin{subequations}
\begin{align}
    a_1&= \frac{1}{4} (d+2 \tau_1-2 \tau_4),& a_2&= \frac{1}{4} (d+2 \tau_2-2 \tau_3),& a_3&= \frac{1}{4} (-d+2 \tau_1+2 \tau_4),\\ a_4&= \frac{1}{4} (-d+2 \tau_2+2 \tau_3),& z&= \frac{1}{4} (d-2 \tau ),
\end{align}
\end{subequations}
and is consistent with the appearance of Wilson functions as $6-j$ symbols for representations of the $\mathfrak{sl}\left(2,\mathbb{R}\right)$ conformal algebra \cite{Groenevelt2005WilsonFT,Hogervorst:2017sfd} as relevant for CFTs in $d=1$.\footnote{A different manifestation of this property was also observed for external scalar operators with pairwise equal twists in \cite{Giombi:2018vtc}, as the anomalous dimensions of leading twist double-trace operators $\left[\Phi_1\Phi_2\right]_{0,l}$ induced by the perturbation \eqref{dtflow} with $l^\prime=0$. As explained in \S \tcb{\ref{Application}}, such anomalous dimensions induced by double-trace flows are given in terms of crossing kernels ($6-j$ symbols).} For $l^\prime>0$ however the above property does not hold in general.

\paragraph{Generic $t$}
For applications to the case of spinning external legs and to subleading poles in $t$, it is useful to also present the result for the crossing kernel \eqref{ck1234} for arbitrary $t$. The result reads: 
\begin{subequations}\label{KernelArbitraryTtau}
\begin{empheq}[box=\fbox]{align}
\frac{\mathfrak{J}^{(\sf t)}_{J_1J_2J_3J_4|{l^\prime},l}(t)}{\mathfrak{J}^{(\sf t)}_{J_1J_2J_3J_4|0,0}(t)}&=\mathcal{Z}_{{l^\prime}}(t)\,\sum_{p=0}^{
{l^\prime}}d_{p}^{({l^\prime},l)}\sum_{i,j=0}^{{l^\prime}-p}a^{({l^\prime},l)}_{p,i,j}\\
&\times\, {}_4F_3\left(\begin{matrix}-j,1-i-j-p-\frac{\tau }{2},1+i+p-\frac{d-\tau}{2},\Delta -\frac{t}{2}\\\nonumber \frac{t-2\Delta+2}{2}-j,i+p-{l^\prime}+\frac{d-\tau }{2},{l^\prime}-i-j-p+\frac{\tau }{2} \end{matrix};1\right)\\\nonumber
&\times {}_4F_3\left(\begin{matrix}p-l,l+p+t-1,\frac{d-\tau}{2} -i+\frac{t-2\Delta}{2}, {l^\prime}-i+\frac{t-2\Delta}{2}+\frac{\tau }{2}\\p+\frac{t}{2},p+\frac{t}{2},\frac{d}{2}-2 \Delta +t \end{matrix};1\right)\,,\\
\mathfrak{J}^{(\sf u)}_{J_1J_2J_3J_4|{l^\prime},l}(t)&=(-1)^{{l^\prime}+l}\,\mathfrak{J}^{(\sf u)}_{J_1J_2J_3J_4|{l^\prime},l}(t)
\end{empheq}
\end{subequations}
with
{\allowdisplaybreaks
\begin{subequations}
\begin{align}
    \mathfrak{J}_{J_1J_2J_3J_4|0,0}(t)&= \mathfrak{D}^{(\sf t)}_{J_1J_2J_3J_4}\frac{\Gamma (\tau ) \Gamma \left(\tfrac{t-2 \Delta +\tau }{2} \right)^2 \Gamma \left(\tfrac{d+t-2 \Delta -\tau}{2}\right)^2}{\Gamma \left(\frac{\tau }{2}\right)^4 \Gamma \left(\frac{d}{2}-\tau \right) \Gamma \left(\frac{d}{2}+t-2 \Delta \right)},\\
    \mathcal{Z}_{{l^\prime}}(t)&=\frac{(\tau )_{2 {l^\prime}} \left(\frac{d}{2}-{l^\prime}-\tau \right)_{l^\prime} \left(\tfrac{t-2 \Delta +\tau }{2}\right)_{l^\prime}}{2^{{l^\prime}} \left(\frac{\tau }{2}\right)_{l^\prime}^4 (d-{l^\prime}-\tau -1)_{l^\prime}}\,,\\
    d_p^{({l^\prime},l)}(t)&=\frac{(-2)^l p! }{l!}\binom{{l^\prime}}{p} \binom{l}{p}
    \\\nonumber&\times\frac{ \left(\frac{\tau +2-d}{2}\right)_p \left(\frac{\tau +2-d}{2} \right)_{{l^\prime}-p} \left({l^\prime}-p+\frac{\tau }{2}\right)_p \left(p+\frac{t}{2}\right)_{l-p}^2  \left(\frac{t-2 \Delta +\tau}{2}\right)_p \left(\frac{d+t-2 \Delta -\tau}{2}\right)_p}{ \left({l^\prime}+\frac{\tau }{2}\right)_{p-{l^\prime}} (l+p+t-1)_{l-p} \left(\frac{d-2 {l^\prime}+2 p+t-2 \Delta -\tau }{2}\right)_{l^\prime}\left(\tfrac{d-2 p+t-2 \Delta -\tau}{2}\right)_p \left(\tfrac{2 {l^\prime}-2 p+t-2 \Delta +\tau }{2}\right)_p},\\
    a^{({l^\prime},l)}_{p,i,j}(t)&=\frac{(-1)^j ({l^\prime}-p)!}{i! j! (-i-j+{l^\prime}-p)!}\left(\Delta -\tfrac{t}{2}\right)_j \left(\tfrac{d}{2}-j+t-2 \Delta \right)_j\\\nonumber
    &\times\frac{\left(p+\frac{t}{2}-\Delta +\frac{\tau }{2}\right)_i \left(\tfrac{d-2 {l^\prime}+2 p+t-2 \Delta -\tau }{2}\right)_i}{\left(\tfrac{d-2 (i+ j+ p)+t-2 \Delta -\tau }{2}\right)_{i+j} \left(\tfrac{2 {l^\prime}-2 (p+i+j)+t-2 \Delta +\tau }{2}\right)_{i+j}}\\\nonumber
    &\times\,\frac{\left(p+\frac{\tau -d}{2}+1\right)_i \left({l^\prime}-p-i-j+\frac{\tau }{2}\right)_{i+j}}{\left(p+\frac{\tau }{2}\right)_{i+j} \left({l^\prime}-p-i+\frac{\tau-d }{2}+1\right)_i}\,\tfrac{\left(\tfrac{d}{2}+ j+p-1\right)_i}{\left(\tfrac{d}{2}+{l^\prime}-i-1\right)_i}.
\end{align}
\end{subequations}}

\noindent
Although the overall coefficients appear quite cumbersome, the structure of this result is rather remarkable: The product of hypergeometric function ${}_4F_3$ is reminiscent of a product of 6j symbols, which in this case represent the action of a generic higher-spin generator on the scalar CPW. It is also interesting to notice that the sum has a tetrahedral structure with the sum over $i$ and $j$ concentrated for $i+j\leq {l^\prime}-p$, so that at fixed $p$ the sum covers a triangle of decreasing area as $p$ grows. 

\subsection{(Partially-)conserved currents}\label{partiallyConserved}

Notice that the result \eqref{KernelEqualTau} exhibits poles at values of the twist $\tau$ for which there is shortening of the conformal representation owing to the emergence of (partially-)conserved currents. In the bulk these is dual to partially-massless gauge fields either on dS or AdS \cite{Deser:2001us,Dolan:2001ih}.\footnote{See also \cite{Joung:2012rv,Joung:2012hz} for detailed analysis of bulk couplings involving partially-massless fields.} For spin $l^\prime$, such values of $\tau$ are given by
\begin{equation}
    \tau=d-2-k\qquad k=0,\ldots, l^\prime-1\,,
\end{equation}
at which a tower of descendant operators of the original long multiplet decouples the $m$-th order descendant decouples for $m\geq k$. Expanding around the pole, we then obtain:
\begin{align}
    \mathfrak{J}_{J_1,J_2,J_3,J_4|l^\prime,0}^{({\sf t})}\Big|_{\tau=d-2-k}&=\mathcal{D}^{(t)}_{J_1,J_2,J_3,J_4}f_{l^\prime}\frac{(-1)^{{l^\prime}+1}2^{d+2 {l^\prime}-2 k-2} \Gamma \left(\tfrac{k}{2}+1\right)\Gamma \left(\tfrac{d}{2}+{l^\prime}-k-1\right) \Gamma \left(\tfrac{d+2 {l^\prime}-k-1}{2}\right)}{\pi\,  \Gamma \left(\tfrac{d}{2}\right) \Gamma \left(\tfrac{k+1}{2}\right) \Gamma ({l^\prime}-k) \Gamma \left(\tfrac{d+2 {l^\prime}-k-2}{2}\right)}\nonumber\\\nonumber
    &\times\Big[\frac{\sin \left(\tfrac{\pi  d}{2}\right)}{\tau +2-d+k}+ \tfrac{1}{2}\sin \left(\tfrac{\pi  d}{2}\right) \Big(-H_{\tfrac{d}{2}+J-\tfrac{k}{2}-2}+2 H_{\tfrac{d}{2}+J-k-2}+H_{\tfrac{d+2 J-k-3}{2}}\\&\hspace{50pt}-2 H_{J-k-1}-H_{\tfrac{k}{2}}+H_{\tfrac{k-1}{2}}+\log (16)\Big)+\pi  \cos \left(\tfrac{\pi  d}{2}\right)\Big],
\end{align}
in terms of harmonic numbers $H_n=\sum_{i=1}^n\tfrac1{i}$. The above confirms the appearance of poles only for $0\leq k< {l^\prime}$. 

It is instructive to consider the above case from the perspective of a conformal partial wave expansion \eqref{cpwe} of a CFT correlator in which the CPWs are integrated over the principal series\footnote{Note that although we have kept $\tau$ arbitrary for ease of presentation, the whole discussion of this paper is implicitly assumed to be at the level of a CPW expansion of the CFT correlators for the principal series.} $\tau=\tfrac{d}2-i\nu-l^\prime$ in, say, the $\sf{t}$-channel with a weight function $a(\nu,l^\prime)$. In this case the full crossing kernel onto the ${\sf s}$-channel
\begin{align}
\left\langle u^{\bar{\tau}}g_{\bar{\tau},l}(u,v)|\mathcal{A}(u,v)\right\rangle
\end{align}
receives contribution from all principal series, which is given by
\begin{equation}
   \left\langle u^{\bar{\tau}}g_{\bar{\tau},l}(u,v)|\mathcal{A}(u,v)\right\rangle= \frac1{2\pi i}\int_{-\infty}^{+\infty}d\nu\sum_{l^\prime} a(\nu,l^\prime)\ \mathfrak{J}_{J_1,J_2,J_3,J_4|l^\prime,l}^{({\sf t})}\,\Big|_{\tau=\tfrac{d}{2}-i\nu-l^\prime}\,.
\end{equation}
Above we have included the factor $a(\nu,l^\prime)$ encoding the OPE coefficients. The virtue of the above expression is that there is no singularity of the CPW on the principal series. On the other hand, we can still evaluate what would be the contribution of a single CPW in the $\sf{t}$-channel by picking the residue of the integrand at a certain physical pole:
\begin{equation}
    \mathfrak{J}^{(\sf{t})}_{J_1,J_2,J_3,J_4|l^\prime,0}=-i\left(d-2\tau-2l^\prime\right)\,\text{Res}_{\nu=i(\tau+l^\prime-\tfrac{d}{2})}\left[\frac{1}{\left[\nu^2+(\Delta-\tfrac{d}{2})^2\right]}\ \mathfrak{J}_{J_1,J_2,J_3,J_4|l^\prime,0}^{({\sf t})}\,\Big|_{\tau=\tfrac{d}{2}-i\nu-l^\prime}\right]\,.
\end{equation}
The above formula reproduces exactly $\mathfrak{J}_{J_1,J_2,J_3,J_4|l^\prime,0}^{\sf(t)}$ for any $\tau\neq d-2-k$ with $0\leq k<l^\prime$. However, when $\tau= d-2-k$ one of the $\nu$ poles of the conformal block coincide with the physical pole we want to take the residue on. The effect is that the integrand now develops a double pole at $\nu=i(\Delta-\tfrac{d}2)$. 

From the spectral representation perspective one can still evaluate the latter residue obtaining a finite result.\footnote{This correspond to the fact that the unphysical polarisation have decoupled and do not contribute to the 4pt correlator \cite{Costa:2012cb,Bekaert:2014cea,Sleight:2017fpc,Sleight:2017cax} as it would be expected from a bulk computation where the unphysical components of the massive spin-1 field decouples automatically for current exchanges involving conserved currents. The individual blocks are still singular but the principal series expansion is still well defined.} This perspective makes clear why the limit $\tau\to d-2-k$ with $0\leq k\leq J-1$ integer cannot be approached after taking the residue. A shortcut to obtain the result for the $\tau=d-2-k$ case, which is analogous to the shortcut considered in \cite{Giombi:2013yva} to evaluate the same difference for the free energy $\delta F$, is then to evaluate the above residue after setting $\tau=d-2-k$. In this case we obtain the finite result:
\begin{multline}
    \mathfrak{J}_{J_1,J_2,J_3,J_4|l^\prime,0}^{(\sf{t})}\Big|_{\tau=d-2-k}=\mathcal{D}^{(t)}_{J_1,J_2,J_3,J_4}f_{l^\prime}\frac{(-1)^{J-k} 2^{d+J-2 k-2}\Gamma \left(\frac{k}{2}+1\right)\Gamma \left(\frac{d}{2}+J-k-1\right) \Gamma \left(\frac{d+2 J-k-1}{2}\right)}{ \Gamma (J-k) \Gamma \left(\frac{d}{2}\right) \Gamma \left(\frac{k+1}{2}\right) \Gamma \left(\frac{d+2 J-k-2}{2}\right)}\\\Big[ \cos \left(\tfrac{\pi  d}{2}\right)+\frac{1}{2\pi} \sin \left(\tfrac{\pi  d}{2}\right) \left(-H_{\frac{d}{2}+J-\frac{k}{2}-2}+2 H_{\frac{d}{2}+J-k-2}+H_{\frac{1}{2} (d+2 J-k-3)}\right.\\\left.-2 H_{J-k-1}-H_{\frac{k}{2}}+H_{\frac{k-1}{2}}+\log (16)-\frac{2}{d+2 J-2 k-4}\right)\Big]\,.
\end{multline}
The above simplifies in even or odd dimensions, for which we obtain the following results:
\begin{multline}
    \mathfrak{J}_{J_1,J_2,J_3,J_4|l^\prime,0}^{(\sf{t})}\Big|_{\tau=d-2-k}=\mathcal{D}^{(t)}_{J_1,J_2,J_3,J_4}f_{l^\prime}\frac{(-1)^{J+1} 2^{d+J-2 k-2}\Gamma \left(\frac{k}{2}+1\right)\Gamma \left(\frac{d}{2}+J-k-1\right) \Gamma \left(\frac{d+2 J-k-1}{2}\right)}{ \Gamma (J-k) \Gamma \left(\frac{d}{2}\right) \Gamma \left(\frac{k+1}{2}\right) \Gamma \left(\frac{d+2 J-k-2}{2}\right)}\\\times\left\{\begin{matrix}
      (-1)^{h} & \text{if} \quad d=2h,\quad h\in\mathbb{N}\\
      & \\
    \frac{(-1)^{h}}{2\pi} \Big(\frac{2}{3-2 h-2 J+2 k}+H_{h+J-\frac{k}{2}-1}-H_{h+J-\frac{k}{2}-\frac{3}{2}}& \text{if} \quad d=2h+1\quad, h\in\mathbb{N}
    \\\qquad 2 H_{h+J-k-\frac{3}{2}}-H_{\frac{k}{2}}+H_{\frac{k-1}{2}}-2 H_{J-k-1}+\log (16)\Big)& 
    \end{matrix}\right.\,.
\end{multline}
In general the role of singularities of conformal blocks and the corresponding cancellation thereof when summing over the spin $l^\prime$ is expected to play a key role in the context of slightly broken higher-spin symmetry (see e.g. \cite{Gliozzi:2017hni} for some related discussions).

\subsection{Full example: $J_1$-$J_2$-$0$-$0$}
\label{subsec::liftdeg}

In the previous sections we studied the crossing kernels of spinning ${\sf t}$- and ${\sf u}$-channel CPWs, restricting to the leading tensor structure $W_{12}^{J_1}W_{21}^{J_2}W_{34}^{J_3}W_{43}^{J_4}$. Using the results of \S \tcb{\ref{J1J2J3J4}} this allowed to access a weighted average of their ${\sf s}$-channel expansion coefficients. By instead considering the full tensor structure of the CPWs, each ${\sf s}$-channel coefficient can be extracted. This is illustrated in the following, where we revisit \S \tcb{\ref{crossingScalar}} for CPWs with an exchanged scalar and consider the full tensor structure of $J_1$-$J_2$-$0$-$0$ CPWs to extract their ${\sf s}$-channel expansion coefficients onto operators of leading twist.

In this case, the full tensor structure of ${\sf t}$- and ${\sf u}$-channel CPWs \eqref{LeadingM} in Mellin space read:
\begin{multline}\label{12tcpw}
    {}^{(\sf t)}{\cal F}^{\bf{0,0}}_{\tau,0}\left(s,t=\tau_1+\tau_2|W_{ij}\right)=\mathcal{C}^{\sf{t}}\frac{\Gamma \left(\tfrac{s +\tau +\tau_1-\tau_3}{2}\right) \Gamma \left(\tfrac{d+s -\tau +\tau_1-\tau_3}{2}\right)}{\Gamma \left(\tfrac{s+2\tau_1-\tau_3+\tau_4}{2}\right) \Gamma \left(\tfrac{s+\tau_1+\tau_2}{2}\right)}\\ \times\,\frac{(-1)^{J_1+J_2} \left(\tfrac{d-\tau-\tau_3+\tau_4 }{2}\right)_{J_2}\left(\tfrac{-s+\tau_3-\tau_4}{2}\right)_{J_1} \left(\tfrac{-s-\tau_1+\tau_2}{2}\right)_{J_2}}{\left(\tfrac{\tau-\tau_3+\tau_4 }{2}\right)_{J_2} \left(\tfrac{\tau +\tau_1-\tau_4}{2}\right)_{J_1} \left(\tfrac{d -\tau +\tau_2-\tau_3}{2}\right)_{J_2}}\,(W_{13}-W_{14})^{J_1} (W_{23}-W_{24})^{J_2}
\end{multline}
with
\begin{equation}
     \mathcal{C}^{\sf{t}}=\frac{\Gamma (\tau ) \Gamma \left(\frac{d-\tau+\tau_3-\tau_4 }{2}\right)\Gamma \left(\frac{d-\tau -\tau_3+\tau_4}{2}\right)}{\Gamma \left(\frac{d}{2}-\tau \right) \Gamma \left(\frac{\tau+\tau_3-\tau_4 }{2}\right)\Gamma \left(\frac{\tau-\tau_3+\tau_4 }{2}\right) \Gamma \left(\tfrac{\tau -\tau_1+\tau_4}{2}\right) \Gamma \left(\tfrac{ \tau +\tau_1-\tau_4}{2}\right) \Gamma \left(\tfrac{d -\tau -\tau_2+\tau_3}{2}\right) \Gamma \left(\tfrac{d -\tau +\tau_2-\tau_3}{2}\right)}
\end{equation}
and 
\begin{multline}\label{12ucpw}
    {}^{(\sf u)}{\cal F}^{\bf{0,0}}_{\tau,0}\left(s,t=\tau_1+\tau_2|W_{ij}\right)=\mathcal{C}^{\sf{u}}\frac{\Gamma \left(\tfrac{-s +\tau -\tau_1-\tau_4}{2}\right) \Gamma \left(\tfrac{d-s -\tau -\tau_1-\tau_4}{2}\right)}{\Gamma \left(\tfrac{-s+\tau_3-\tau_4}{2}\right)\Gamma \left(\tfrac{-s-\tau_1+\tau_2}{2}\right)}\\\times\,\frac{(-1)^{J_2} \left(\tfrac{d-\tau+\tau_3-\tau_4 }{2}\right)_{J_2}\left( \tfrac{s+2\tau_1-\tau_3+\tau_4}{2}\right)_{J_1}  \left(\tfrac{s+\tau_1+\tau_2}{2}\right)_{J_2}}{\left(\tfrac{\tau+\tau_3-\tau_4 }{2}\right)_{J_2} \left(\tfrac{\tau +\tau_1-\tau_3}{2}\right)_{J_1} \left(\tfrac{d -\tau +\tau_2-\tau_4}{2}\right)_{J_2}}\,(W_{13}-W_{14})^{J_1} (W_{23}-W_{24})^{J_2}
\end{multline}
with
\begin{equation}
     \mathcal{C}^{\sf{u}}=\frac{\Gamma (\tau ) \Gamma \left(\frac{d-\tau+\tau_3-\tau_4 }{2}\right)\Gamma \left(\frac{d-\tau -\tau_3+\tau_4}{2}\right)}{\Gamma \left(\frac{d}{2}-\tau \right) \Gamma \left(\frac{\tau+\tau_3-\tau_4 }{2}\right)\Gamma \left(\frac{\tau-\tau_3+\tau_4 }{2}\right) \Gamma \left(\tfrac{\tau -\tau_1+\tau_4}{2}\right) \Gamma \left(\tfrac{ \tau +\tau_1-\tau_4}{2}\right) \Gamma \left(\tfrac{d -\tau -\tau_2+\tau_3}{2}\right) \Gamma \left(\tfrac{d -\tau +\tau_2-\tau_3}{2}\right)}
\end{equation}
and we assumed without loss of generality that $J_1\geq J_2$. We remind the reader that in this case we are restricting to the leading twist contributions, which have $t=\tau_1+\tau_2$.

The crossing kernels of the above CPWs in the ${\sf s}$-channel are the coefficients in the expansions
\begin{subequations}\label{12utchexpQ}
\begin{align}
    {}^{(\sf t)}{\cal F}^{\bf{0,0}}\left(s,\tau_1+\tau_2|W_{ij}\right)&=\sum_{l,\bf{n}}{}^{(\sf t)}\mathfrak{J}_{J_1,J_2,0,0|0,l}^{\bf{n}}\,\mathcal{Q}_{l,\tau_1+\tau_2,0}^{\bf{n}}(s|W_{ij}) \\
    {}^{(\sf u)}{\cal F}^{\bf{0,0}}\left(s,\tau_1+\tau_2|W_{ij}\right)&=\sum_{l,\bf{n}}{}^{(\sf u)}\mathfrak{J}_{J_1,J_2,0,0|0,l}^{\bf{n}}\,\mathcal{Q}_{l,\tau_1+\tau_2,0}^{\bf{n}}(s|W_{ij}),
\end{align}
\end{subequations}
in terms of the kinematic polynomials \eqref{j1j200Qn}. By inspecting the tensor structure of the CPWs \eqref{12tcpw} and \eqref{12ucpw} and comparing with the tensor structure of the kinematic polynomials \eqref{j1j200Qn}, we see that the only contribution to the expansions \eqref{12utchexpQ} comes from those polynomials with ${\bf n}=\left(J_2,J_1,0\right)$. Upon this observation, following \S \tcb{\ref{J1J200}} we can restrict to the leading term of the polynomial $\mathcal{Q}_{l,\tau_1+\tau_2,0}^{\left(J_2,J_1,0\right)}(s|W_{ij})$, which is given by \eqref{Qfinalform} with $n_0=0$, $n_1=J_2$ and $n_2=J_1$, and use its orthogonality to extract the full crossing kernels. To wit,
   \begin{align}
    {}^{(\sf{t})}\mathfrak{J}_{J_1,J_2,0,0|0,l}^{{\bf n}}& = \frac{(-1)^{l-J_1-J_2}}{(l-J_1-J_2)!}\,\mathcal{C}^{\sf{t}}\int_{-i\infty}^{i\infty}\frac{ds}{4\pi i}\,\tilde{\rho}_{\{\tau_i\}}(s,\tau_1+\tau_2)\,\frac{\Gamma \left(\tfrac{s +\tau +\tau_1-\tau_3}{2}\right) \Gamma \left(\tfrac{d+s -\tau +\tau_1-\tau_3}{2}\right)}{\Gamma \left(\tfrac{s+2\tau_1-\tau_3+\tau_4}{2}\right) \Gamma \left(\tfrac{s+\tau_1+\tau_2}{2}\right)}\nonumber\\&  \hspace*{-2cm}\times\,\frac{(-1)^{J_1+J_2} \left(\tfrac{d-\tau-\tau_3+\tau_4 }{2}\right)_{J_2}\left(\tfrac{-s+\tau_3-\tau_4}{2}\right)_{J_1} \left(\tfrac{-s-\tau_1+\tau_2}{2}\right)_{J_2}}{\left(\tfrac{\tau-\tau_3+\tau_4 }{2}\right)_{J_2} \left(\tfrac{\tau +\tau_1-\tau_4}{2}\right)_{J_1} \left(\tfrac{d -\tau +\tau_2-\tau_3}{2}\right)_{J_2}}\,Q_{l-J_1-J_2}^{\tau_1+\tau_2+2J_2,2\tau_1+2J_1,-\tau_1+\tau_2+2J_2,2J_1}(s)\,
\end{align} 
and 
   \begin{align}\nonumber
    {}^{(\sf{u})}\mathfrak{J}_{J_1,J_2,0,0|0,l}^{{\bf n}}&=\frac{(-1)^{l-J_1-J_2}}{(l-J_1-J_2)!}\,\mathcal{C}^{\sf{u}}\int_{-i\infty}^{i\infty}\frac{ds}{4\pi i}\,\tilde{\rho}_{\{\tau_i\}}(s,\tau_1+\tau_2)\,
    \frac{\Gamma \left(\tfrac{-s +\tau -\tau_1-\tau_4}{2}\right) \Gamma \left(\tfrac{d-s -\tau -\tau_1-\tau_4}{2}\right)}{\Gamma \left(\tfrac{-s+\tau_3-\tau_4}{2}\right)\Gamma \left(\tfrac{-s-\tau_1+\tau_2}{2}\right)}\\&\hspace*{-2cm}\times\,\frac{(-1)^{J_2} \left(\tfrac{d-\tau+\tau_3-\tau_4 }{2}\right)_{J_2}\left( \tfrac{s+2\tau_1-\tau_3+\tau_4}{2}\right)_{J_1}  \left(\tfrac{s+\tau_1+\tau_2}{2}\right)_{J_2}}{\left(\tfrac{\tau+\tau_3-\tau_4 }{2}\right)_{J_2} \left(\tfrac{\tau +\tau_1-\tau_3}{2}\right)_{J_1} \left(\tfrac{d -\tau +\tau_2-\tau_4}{2}\right)_{J_2}}
    \,Q_{l-J_1-J_2}^{\tau_1+\tau_2+2J_2,2\tau_1+2J_1,-\tau_1+\tau_2+2J_2,2J_1}(s)\,
\end{align} 

The above integrals can be performed in a similar manner as those which have appeared previously, with the following results:
\begin{align}\label{CKj1j2t}
     {}^{(\sf{t})}\mathfrak{J}_{J_1,J_2,0,0|0,l}^{{\bf n}}&=\mathcal{C}^{\sf{t}}\frac{(-2)^{-J_1-J_2+l} \Gamma (J_1-J_2+l+\tau_1)  \Gamma (J_1+J_2+l+\tau_1+\tau_2-1)}{ (l-J_1-J_2)!\,\Gamma (2 J_1+\tau_1) \Gamma (2 l+\tau_1+\tau_2-1) }\\
    &\times\tfrac{\Gamma \left(\frac{2 J_1+\tau +\tau_1-\tau_4}{2}\right) \Gamma \left(\frac{2 J_2+\tau +\tau_2-\tau_3}{2}\right) \Gamma \left(\frac{d+2 J_1-\tau +\tau_1-\tau_4}{2}\right) \Gamma \left(\frac{d+2 J_2-\tau +\tau_2-\tau_3}{2}\right)\Gamma \left(\frac{2 l+\tau_1+\tau_2+\tau_3-\tau_4}{2}\right)}{\Gamma \left(\frac{2 J_1+2 J_2+\tau_1+\tau_2+\tau_3-\tau_4}{2}\right) \Gamma \left(\frac{d+2 J_1+2 J_2+\tau_1+\tau_2-\tau_3-\tau_4}{2}\right)}\nonumber\\\nonumber
    &\times \, _4F_3\left(\begin{matrix}J_1+J_2-l,J_1+J_2+l+\tau_1+\tau_2-1,\frac{d}{2}+J_1-\frac{\tau-\tau_1+\tau_4 }{2},J_1+\frac{\tau+\tau_1-\tau_4 }{2}\\2 J_1+\tau_1,\frac{d+\tau_1+\tau_2-\tau_3-\tau_4}{2}+J_1+J_2,J_1+J_2+\frac{\tau_1+\tau_2+\tau_3-\tau_4}{2}\end{matrix};1\right),
    \end{align}
    and
    \begin{align}\label{CKj1j2u}
     {}^{(\sf{u})}\mathfrak{J}_{J_1,J_2,0,0|0,l}^{{\bf n}}&=\mathcal{C}^{\sf{u}}\frac{(-2)^{-J_1-J_2+l} (-1)^{-J_1-J_2+l} \Gamma (-J_1+J_2+l+\tau_2) \Gamma (J_1+J_2+l+\tau_1+\tau_2-1)}{ (l-J_1-J_2)!\,\Gamma (2 J_2+\tau_2) \Gamma (2 l+\tau_1+\tau_2-1)}\\
    &\times \tfrac{\Gamma \left(\frac{2 J_1+\tau +\tau_1-\tau_3}{2}\right) \Gamma \left(\frac{2 J_2+\tau +\tau_2-\tau_4}{2} \right) \Gamma \left(\frac{d+2 J_1-\tau +\tau_1-\tau_3}{2}\right) \Gamma \left(\frac{d+2 J_2-\tau +\tau_2-\tau_4}{2}\right) \Gamma \left(\frac{2 l+\tau_1+\tau_2+\tau_3-\tau_4}{2}\right)}{\Gamma \left(\frac{2 J_1+2 J_2+\tau_1+\tau_2+\tau_3-\tau_4}{2}\right) \Gamma \left(\frac{d+2 J_1+2 J_2+\tau_1+\tau_2-\tau_3-\tau_4}{2}\right)}\nonumber\\
    &\times \, _4F_3\left(\begin{matrix}J_1+J_2-l,J_1+J_2+l+\tau_1+\tau_2-1,\frac{d-\tau+\tau_2-\tau_4}{2}+J_2,J_2+\frac{\tau+\tau_2-\tau_4 }{2}\\2 J_2+\tau_2,\frac{d+\tau_1+\tau_2-\tau_3-\tau_4}{2}+J_1+J_2,J_1+J_2+\frac{\tau_1+\tau_2+\tau_3-\tau_4}{2}\end{matrix};1\right)\,.\nonumber
\end{align}
As a consistency check, the above reduce to the result \eqref{inv} for external scalars and are proportional to a Wilson polynomial \eqref{WilsonP} with
\begin{subequations}
\begin{align}
    a_1&= \frac{1}{4} (d+4 J_1+2 \tau_1-2 \tau_4),& a_2&= \frac{1}{4} (-d+4 J_1+2 \tau_1+2 \tau_4),\\ 
    a_3&= \frac{1}{4} (d+4 J_2+2 \tau_2-2 \tau_3),& a_4&= \frac{1}{4} (-d+4 J_2+2 \tau_2+2 \tau_3),\\
    z&= \frac{1}{4} (d-2 \tau ),
\end{align}
\end{subequations}
for the $\sf{t}$-channel and
\begin{subequations}
\begin{align}
    a_1&= \frac{1}{4} (d+4 J_2+2 \tau_2-2 \tau_4),& a_2&= \frac{1}{4} (-d+4 J_2+2 \tau_2+2 \tau_4),\\
    a_3&= \frac{1}{4} (d+4 J_1+2 \tau_1-2 \tau_3),& a_4&= \frac{1}{4} (-d+4 J_1+2 \tau_1+2 \tau_3),\\
    z&= \frac{1}{4} (d-2 \tau ),
\end{align}
\end{subequations}
for the $\sf{u}$-channel. For $l^\prime>0$ this rewriting in terms of Wilson polynomials does not appear to be possible.

\subsection{Large spin limit}\label{LargeL}

The explicit results for the crossing kernels obtained in \eqref{KernelEqualTau}, \eqref{CKj1j2t} and \eqref{CKj1j2u} allow to study the large $l$ limit in full generality and for arbitrary $J$. The basic ingredient is the large $l$ behaviour of the ${}_4 F_3$. This can be systematically obtained as an expansion in $\tfrac1{l}$ with a trick: Focusing on the scalar kernels \eqref{KernelEqualTau}, one can first consider the series expansion of the hypergeometric function:
\begin{multline}
    {}_4F_3\left(\begin{matrix}p-l,l+p+2 \Delta-1,\tfrac{d-\tau }{2}-i,\tfrac{\tau}{2}+J-i\\\tfrac{d}{2},\Delta+p ,\Delta+p \end{matrix};z\right)\\=\frac{\left(\frac{d}{2}-1\right)!(\Delta +p-1)!^2}{(-l+p-1)!\left(\frac{d}{2}-i-\frac{\tau }{2}-1\right)! \left(-i+J+\frac{\tau }{2}-1\right)!(2 \Delta +l+p-2)!}\\ \times \sum_{k=0}^\infty\frac{ (k-l+p-1)! \left(\frac{d}{2}-i+k-\frac{\tau }{2}-1\right)! \left(-i+J+k+\frac{\tau }{2}-1\right)! (2 \Delta +k+l+p-2)!}{k! \left(\frac{d}{2}+k-1\right)! (\Delta +k+p-1)!^2}\,z^k\,.
\end{multline}
Being cavalier about the analytic continuation of the sum in $z$ (which can be performed more rigorously in Mellin-Barnes form)\footnote{Such expansions can be systematically performed (see e.g. \cite{MellinBook}) via a Mellin-Barnes representation of the Hypergeometric function:
\begin{multline}
    \, _4F_3\left(\begin{matrix}a_1,a_2,a_3,a_4\\b_1,b_2,b_3\end{matrix};z\right)=\frac{\Gamma \left(b_1\right) \Gamma \left(b_2\right) \Gamma \left(b_3\right) }{\Gamma \left(a_1\right) \Gamma \left(a_2\right) \Gamma \left(a_3\right) \Gamma \left(a_4\right)}\\\times\oint\frac{ds}{2\pi i}\frac{\Gamma (s) \Gamma \left(a_1-s\right) \Gamma \left(a_2-s\right) \Gamma \left(a_3-s\right) \Gamma \left(a_4-s\right)}{\Gamma \left(b_1-s\right) \Gamma \left(b_2-s\right) \Gamma \left(b_3-s\right)}\,(-z)^{-s} 
\end{multline}
employing series expansion of the type:
\begin{equation}
    \Gamma(p-l-s)\Gamma(l+p+2\Delta-1)\approx\pi  \csc (\pi  (l-p+s))\sum_{n}\,b_n(s)\,\left(\tfrac1{l}\right)^{-2 (\Delta +p-1)+n}l^{-2s}\,,
\end{equation} in terms of polynomials in the variable $s$ which we denote by $b_n(s)$. Similar expansions can be performed for the overall pre-factor and combined.} one can perform the large $l$ expansion term by term for each power of $z$. We then arrive to
\begin{multline}
    {}_4F_3\left(\begin{matrix}p-l,l+p+2 \Delta-1,\tfrac{d-\tau }{2}-i,\tfrac{\tau}{2}+J-i\\\tfrac{d}{2},\Delta+p ,\Delta+p \end{matrix};z\right)\\=\sum_{k=0}^\infty\left[\frac{(-1)^k \Gamma \left(\frac{d}{2}\right) \Gamma (p+\Delta )^2 \Gamma \left(\frac{d}{2}-i+k-\frac{\tau }{2}\right) \Gamma \left(-i+J+k+\frac{\tau }{2}\right)}{\Gamma (k+1) \Gamma \left(\frac{d}{2}+k\right) \Gamma \left(\frac{d}{2}-i-\frac{\tau }{2}\right) \Gamma \left(-i+J+\frac{\tau }{2}\right) \Gamma (k+p+\Delta )^2}+\mathcal{O}\left(1/l\right)\right]\,(l^2z)^k\,.
\end{multline}
Dropping the $\mathcal{O}(1/l)$ terms the series can be resummed into:
\begin{multline}
    {}_4F_3\left(\begin{matrix}p-l,l+p+2 \Delta-1,\tfrac{d-\tau }{2}-i,\tfrac{\tau}{2}+J-i\\\tfrac{d}{2},\Delta+p ,\Delta+p \end{matrix};z\right)\Big|_{l\to\infty}\sim \, _2F_3\left(\begin{matrix}\tfrac{d}{2}-i-\frac{\tau }{2},-i+J+\frac{\tau }{2}\\\tfrac{d}{2},\Delta +p,\Delta +p\end{matrix};-l^2 z\right).
\end{multline}
Finally, we arrive to the following expansion:
\begin{multline}\label{largeLhyper1}
    {}_4F_3\left(\begin{matrix}p-l,l+p+2 \Delta-1,\tfrac{d-\tau }{2}-i,\tfrac{\tau}{2}+J-i\\\tfrac{d}{2},\Delta+p ,\Delta+p \end{matrix};1\right)\sim\\
    \left(\frac{1}{l}\right)^{2 (J- i)+\tau }\frac{\Gamma \left(\tfrac{d}{2}\right) \Gamma (p+\Delta )^2 \Gamma \left(\tfrac{d-2 (J+\tau )}{2}\right)}{\Gamma \left(\frac{d-2 i-\tau}{2} \right) \Gamma \left(\frac{d}{2}+i-J-\frac{\tau }{2}\right) \Gamma \left(i-J+p+\Delta -\frac{\tau }{2}\right)^2}
    \\+\left(\frac{1}{l}\right)^{d-\tau-2i} \frac{\Gamma \left(\tfrac{d}{2}\right) \Gamma (p+\Delta )^2  \Gamma \left(-\frac{d}{2}+J+\tau \right)}{\Gamma \left(i+\frac{\tau }{2}\right) \Gamma \left(-i+J+\frac{\tau }{2}\right) \Gamma \left(-\frac{d}{2}+i+p+\Delta +\frac{\tau }{2}\right)^2}+\ldots
\end{multline}
from which it is also clear how the $i=0$ term gives the leading contribution. Furthermore, it is also straightforward to separate the contribution of the shadow operator from that of the physical one. The above also matches the standard asymptotic behaviour of Wilson polynomials \cite{doi:10.1137/0511064} and allows a systematic large spin expansion\footnote{See also \cite{Dey:2017fab} where the large spin expansion was performed at the level of continuous Hahn polynomials. We find it more convenient to consider the large spin expansion at the level of the crossing kernel directly, which should lead to the same conclusions.}. Taking the above leading terms for $i=0$ and summing over $p$ one gets the asymptotic behaviour of the full kernel, which matches the expectations from large spin bootstrap \cite{Fitzpatrick:2012yx,Komargodski:2012ek}. With similar techniques we can also get the asymptotic behaviour of crossing kernels with external spinning legs like \eqref{CKj1j2t} and \eqref{CKj1j2u} using:
\begin{multline}
    {}_4F_3\left(\begin{matrix}p-l+m,l+p+2 \Delta-m-1,\tfrac{d-\tau+m }{2}-i,\tfrac{\tau+m}{2}+J-i\\\tfrac{d}{2}+m,\Delta+p ,\Delta+p \end{matrix};1\right)\sim\\
    \left(\frac{1}{l}\right)^{2 (J- i)+\tau+m }\frac{\Gamma \left(\tfrac{d}{2}+m\right) \Gamma (p+\Delta )^2 \Gamma \left(\tfrac{d-2 (J+\tau )}{2}\right)}{\Gamma \left(\frac{d-2 i-\tau+m}{2} \right) \Gamma \left(\frac{d-\tau+m}{2}+i-J\right) \Gamma \left(i-J+p+\Delta -\frac{\tau +m}{2}\right)^2}
    \\+\left(\frac{1}{l}\right)^{d-\tau-2i+m} \frac{\Gamma \left(\tfrac{d}{2}+m\right) \Gamma (p+\Delta )^2  \Gamma \left(-\frac{d}{2}+J+\tau \right)}{\Gamma \left(i+\frac{\tau+m }{2}\right) \Gamma \left(-i+J+\frac{\tau }{2}\right) \Gamma \left(i+p+\Delta +\frac{\tau-d-m }{2}\right)^2}+\ldots
\end{multline}
which again allows easily to distinguish the shadow contribution from the physical one. For instance, normalising by the mean-field theory OPE coefficients the external scalar result for $l^\prime=0$ gives:
\begin{align}
    \mathfrak{J}^{(\sf{t})}_{0,0,0,0|0,l}\Big|_{l\to\infty}\sim\left(\frac1{l}\right)^\tau \frac{\Gamma (\Delta )^2 \Gamma (\tau )}{\Gamma \left(\frac{\tau }{2}\right)^2 \Gamma \left(\Delta -\frac{\tau }{2}\right)^2}-\left(\frac1{l}\right)^{d-\tau}\frac{\Gamma (\Delta )^2 \Gamma (d-\tau )}{\Gamma \left(\frac{d-\tau }{2}\right)^2 \Gamma \left(\Delta-\tfrac{d-\tau }{2}\right)^2}\,a^{\text{sh.}}_{0,0,0}+\ldots\,,
\end{align}
where the factor
\begin{equation}
    a^{\text{sh.}}_{0,0,0}=\left(\frac{\alpha_{0,0,0;\Delta,\Delta,\tau}}{\kappa_{\tau,0}/\pi^{d/2}}\right)^2\left(-\frac{\kappa_{\tau,0}\kappa_{d-\tau,0}}{\pi^d}\right)\,,
\end{equation}
nicely reproduces the shadow OPE coefficients taking into account the definition of $\alpha$ in \eqref{shad}, \eqref{ShadowNorm} and \eqref{shadowgeneralN3} and the 2pt function normalisation of the shadow field which is:
\begin{equation}
    C_{l^\prime=0}^{\text{sh.}}=-\frac{\kappa_{\tau,0}\kappa_{d-\tau,0}}{\pi^d}\,,
\end{equation}
directly in terms of the shadow transform normalisation. Similar results follow straightforwardly for general $l^\prime$. Using \eqref{KernelEqualTau} and \eqref{largeLhyper1} we obtain the general asymptotic behaviour:
\begin{multline}
    \mathfrak{J}^{(\sf{t})}_{0,0,0,0|l^\prime,l}\Big|_{l\to\infty}\sim\left(\frac1{l}\right)^\tau \frac{\Gamma (\Delta )^2 \Gamma (\tau +2l^\prime)}{2^{l^\prime}\Gamma \left(\frac{\tau }{2}+l^\prime\right)^2 \Gamma \left(\Delta -\frac{\tau }{2}\right)^2}\\-\left(\frac1{l}\right)^{d-\tau-2l^\prime}\frac{\Gamma (\Delta )^2 \Gamma (d-\tau)}{2^{l^\prime}\Gamma \left(\frac{d-\tau }{2}\right)^2 \Gamma \left(\Delta -\frac{d-\tau-2l^\prime }{2}\right)^2}\,a^{\text{sh.}}_{0,0,l^\prime}+\ldots\,,
\end{multline}
where now the shadow OPE coefficient squared read:
\begin{equation}
    a^{\text{sh.}}_{0,0,l^\prime}=\left(\frac{\alpha_{0,0,l^\prime;\Delta,\Delta,\tau}}{\kappa_{\tau+l^\prime,0}/\pi^{d/2}}\right)^2\left(-\frac{\kappa_{\tau+l^\prime,l^\prime}\kappa_{d-\tau-l^\prime,l^\prime}}{\pi^d}\right)\,,
\end{equation}
again expressed in terms of $\alpha_{\bf{s},\bf{\tau}}$ in \eqref{ShadowNorm} with shadow 2pt shadow normalisation given by:
\begin{equation}
    C_{l^\prime}^{\text{sh.}}=-\frac{\kappa_{\tau+l^\prime,l^\prime}\kappa_{d-\tau-l^\prime,l^\prime}}{\pi^d}\,.
\end{equation}
The above is again in agreement with the large spin bootstrap result \cite{Komargodski:2012ek,Fitzpatrick:2012yx} showing both contribution from physical and shadow operator contributing to the $\sf{t}$-channel CPW.

We can also check the large spin behaviour of the general result \eqref{CKj1j2t} and \eqref{CKj1j2u}. Normalising the crossing kernel by the $\sf{s}$-channel OPE coefficients \eqref{OPEj1j2} and taking the large $l$ limit we get for $\tau_i=\Delta$:
\begin{multline}
    {}^{(\sf{t})}\mathfrak{J}_{J_1,J_2,0,0|0,l}^{{\bf n}}\Big|_{l\to\infty}=\left(\frac1{l}\right)^{\tau-\Delta}\frac{ 2^{J_1+J_2-1} \Gamma (\tau ) \Gamma \left(J_1+\frac{\Delta }{2}\right) \Gamma \left(J_2+\frac{\Delta }{2}\right)}{\Gamma \left(\frac{\tau }{2}\right)^2 \Gamma \left(J_1+\Delta -\frac{\tau }{2}\right) \Gamma \left(J_2+\Delta -\frac{\tau }{2}\right)}\\-\left(\frac1{l}\right)^{d-\tau-\Delta}\frac{2^{J_1+J_2-1} \Gamma (d-\tau ) \Gamma \left(J_1+\frac{\Delta }{2}\right) \Gamma \left(J_2+\frac{\Delta }{2}\right)}{\Gamma \left(\frac{d-\tau }{2}\right)^2 \Gamma \left(J_1+\Delta +\frac{\tau -d}{2}\right) \Gamma \left(J_2+\Delta +\frac{\tau -d}{2}\right)}\,a^{\text{sh.}}_{J_1,0,0}a^{\text{sh.}}_{J_2,0,0}\,,
\end{multline}
where we recall that ${\bf n}=\left(J_2,J_1,0\right)$ and now the shadow OPE coefficients read:
\begin{equation}
    a^{\text{sh.}}_{J_1,0,0}a^{\text{sh.}}_{J_2,0,0}=\frac{\alpha_{J_1,0,0;\Delta,\Delta,\tau}\alpha_{J_2,0,0;\Delta,\Delta,\tau}}{\kappa_{\tau,0}^2/\pi^d}\,\left(-\frac{\kappa_{\tau,0}\kappa_{d-\tau,0}}{\pi^d}\right)\,.
\end{equation}
Above the overall factor $l^\Delta$ originates from our normalisation of the crossing kernel with the $\sf{s}$-channel OPE coefficients \eqref{OPEj1j2}.\footnote{Note that for these correlators we dont have a mean-field theory OPE when $J_i>0$, since the first non-vanishing OPE arise at order $1/N$ \eqref{corrJ1J2}. Hence a diagonalisation of the corresponding mixing matrix is required.} 

\section{Application: Double-trace flows}
\label{Application}

An interesting direct application of the framework developed in \S \tcb{\ref{MackPolinomials}} and \S \tcb{\ref{Crossing Kernels}} is to extract corrections to CFT data under double-trace flows, which are neatly encoded in the crossing kernels of CPWs. For a large $N$ CFT whose spectrum includes a single trace operator ${\cal O}_{\mu_1...\mu_{l^\prime}}$ of scaling dimension $\Delta$ and spin $l^\prime$, the double-trace deformation 
\begin{equation}\label{dtflow}
    S_{\lambda}=S_{\text{CFT}}+\lambda \int d^dy\, {\cal O}_{\mu_1...\mu_{l^\prime}}{\cal O}^{\mu_1...\mu_{l^\prime}}
\end{equation}
triggers a flow to a new CFT in which the single-trace operator ${\cal O}_{\mu_1...\mu_{l^\prime}}$ has scaling dimension $d-\Delta + {\cal O}\left(1/N\right)$ \cite{Witten:2001ua,Gubser:2002vv,Giombi:2013yva}. At leading order in $1/N$, the spectrum of the new CFT is identical to that of the unperturbed CFT save for ${\cal O}_{\mu_1...\mu_{l^\prime}}$ and double(multi)-trace operators comprised of it. Beyond leading order, all operators in the spectrum receive corrections to their OPE coefficients and anomalous dimensions.

Under the double-trace flow \eqref{dtflow}, the change in the four-point function of generic single-trace operators ${\cal O}_i$ is elegantly encoded in terms of CPWs involving the exchanged operator ${\cal O}_{\mu_1...\mu_{l^\prime}}$. In particular, in large $N$ conformal perturbation theory we have:\footnote{Equation \eqref{cpt} can be obtained by introducing appropriate spinning $\sigma$-fields and using the Hubbard-Stratonovich transformation to re-write the conformal integrals in terms of the shadow transform, where the 2-pt function normalisation of the shadow $\sigma$-field is conveniently expressed in terms of $\kappa_{\Delta,d}$ as $C_\sigma=-\frac{\kappa_{\Delta_+,l^\prime}\kappa_{\Delta_-,l^\prime}}{\pi^d}$. The latter normalisation precisely matches also the normalisation that would be obtained from a bulk computation as discussed in appendix \S\tcb{\ref{shadowbulk}}. In this case we indeed have the identity:
\begin{equation}
    (\Delta_+-\Delta_-)^2\mathcal{C}_{\Delta_+,l^\prime}\mathcal{C}_{\Delta_-,l^\prime}=-\frac{\kappa_{\Delta_+,l^\prime}\kappa_{\Delta_-,l^\prime}}{\pi^d}\,,
\end{equation}
in terms of the bulk-to-boundary normalisation $\mathcal{C}_{\Delta,l^\prime}$.}
\begin{multline}\label{cpt}
    \delta \langle {\cal O}_1\left(y_1\right){\cal O}_2\left(y_2\right){\cal O}_1\left(y_3\right){\cal O}_2\left(y_4\right) \rangle \\
    =  - \frac{\kappa_{d-\Delta,l}}{\pi^{d/2}}\int d^dy_0\ \langle {\cal O}_1\left(y_1\right){\cal O}_2\left(y_2\right) {\cal O}_{\Delta,l}\left(y_0\right) \rangle  \langle  {\tilde {\cal O}}_{d-\Delta,l}\left(y_0\right) {\cal O}_1\left(y_3\right){\cal O}_2\left(y_4\right) \rangle \\  + {\sf t}\text{-channel} + {\sf u}\text{-channel} + {\cal O}\left(1/N^2\right),
\end{multline}
where the difference
\begin{multline}
    \delta\langle {\cal O}_1\left(y_1\right){\cal O}_2\left(y_2\right){\cal O}_1\left(y_3\right){\cal O}_2\left(y_4\right) \rangle \\=\langle {\cal O}_1\left(y_1\right){\cal O}_2\left(y_2\right){\cal O}_1\left(y_3\right){\cal O}_2\left(y_4\right) \rangle_{d-\Delta,l}-\langle {\cal O}_1\left(y_1\right){\cal O}_2\left(y_2\right){\cal O}_1\left(y_3\right){\cal O}_2\left(y_4\right) \rangle_{\Delta,l}\,,
\end{multline}
is defined to be minus the difference between the two correlators exchanging the operator $\mathcal{O}_{\Delta,l}$ and its shadow $\widetilde{\mathcal{O}}_{d-\Delta,l}$. This means that, depending on whether $\Delta>\tfrac{d}{2}$ or $\Delta<\tfrac{d}{2}$,  this is difference between UV and IR or IR and UV fixed points respectively.
From the integral representation \eqref{nbscpw} of CPWs, the conformal integral in \eqref{cpt} is a finite sum of ${\sf s}$-channel CPWs involving the exchange of ${\cal O}_{\mu_1...\mu_{l^\prime}}$:
\begin{multline}\label{schexpincpt}
    - \frac{\kappa_{d-\Delta,l}}{\pi^{d/2}}\int d^dy_0\ \langle {\cal O}_1\left(y_1\right){\cal O}_2\left(y_2\right) {\cal O}_{\Delta,l}\left(y_0\right) \rangle  \langle  {\tilde {\cal O}}_{d-\Delta,l}\left(y_0\right) {\cal O}_1\left(y_3\right){\cal O}_2\left(y_4\right) \rangle \\ = -\sum_{{\bf n}_s,{\bar {\bf n}}_s} a^{{\bf n}_s,{\bar {\bf n}}_s}_{\tau,l^\prime}{}^{({\sf s})}F^{{\bf n}_s,{\bar {\bf n}}_s}_{\Delta,l^\prime}\left(y_i\right).
\end{multline}
Similarly, the corresponding ${\sf t}$- and ${\sf u}$-channel expressions on the bottom line of \eqref{cpt} are, respectively, a finite sum of ${\sf t}$- and ${\sf u}$-channel CPWs in which the single-trace operator ${\cal O}_{\mu_1...\mu_{l^\prime}}$ is exchanged. While the $\sf{s}$-channel term \eqref{schexpincpt} above just encodes the change in dimension for the single-trace operator ${\cal O}_{\mu_1...\mu_{l^\prime}}$ inducing the flow, the contribution from $\sf{t}$ and $\sf{u}$-channels - when expanded in the $\sf{s}$-channel - give $1/N$ corrections to the scaling dimensions and OPE coefficients of double-trace operators $\left[{\cal O}_1{\cal O}_2\right]_{l,n}$, which are encoded in terms proportional to $u^{\tau_1+\tau_2+n}\log u$ and $u^{\tau_1+\tau_2+n}$ respectively. These corrections to double-trace data can thus be extracted using the framework introduced in \S \tcb{\ref{Crossing Kernels}} for obtaining the crossing kernels of spinning $\sf{t}$ and $\sf{u}$-channel CPWs in the ${\sf s}$-channel. 

Before we proceed, it is useful to review how $1/N$ corrections to double-trace operator data is encoded in the ${\sf s}$-channel expansion of 4pt correlators. A generic 4pt correlator of spinning operators ${\cal O}_i$
\begin{multline}
    \langle {\cal O}_1\left(y_1,z_1\right){\cal O}_2\left(y_2,z_2\right){\cal O}_1\left(y_3,z_3\right){\cal O}_2\left(y_4,z_4\right) \rangle \\= \left(\frac{y_{24}^2}{y_{14}^2}\right)^{\frac{\tau_1-\tau_2}2}\left(\frac{y_{14}^2}{y_{13}^2}\right)^{\tfrac{\tau_3-\tau_4}2} \frac{{\cal A}\left(u,v|W_{ij}\right)}{\left(y_{12}^2\right)^{\frac{1}{2}(\tau_1 + \tau_2)} \left(y_{34}^2\right)^{\frac{1}{2}(\tau_3 + \tau_4)}} \,,
\end{multline}
admits an ${\sf s}$-channel expansion in terms of conformal blocks
\begin{equation}
    \mathcal{A}(u,v|W_{ij})= \sum_{\delta,l}\left(\sum_{\bf{n},\bf{\bar{n}}}a^{\bf{n},\bf{\bar{n}}}_{\delta,l} \,u^{\frac{\delta}{2}}g^{\bf{n},\bf{\bar{n}}}_{\delta,l}(u,v|W_{ij})\right)\,,
\end{equation}
where $a^{\bf{n},\bf{\bar{n}}}_{\delta,l}$ is the product of OPE coefficients and $u^{\delta/2}g^{\bf{n},\bf{\bar{n}}}_{\delta,l}(u,v|W_{ij})$ is the corresponding spinning conformal block of scaling dimension $\delta$ and spin $l$. Notice that here the sum over spins runs in principle also over mixed-symmetry representations which can contribute to the conformal block expansion with spinning external legs. We now consider a $1/N$ expansion of the twists and OPE coefficients of the double-trace primaries $\left[{\cal O}_1{\cal O}_2\right]_{q,l}$:\footnote{We remind the reader here that there will be in general degeneracies in $q$ at leading order in $1/N$, corresponding to the mean-field theory result. This happens in particular when there exists more than one possible Young projection for the indices of the double-trace operator, which allows both totally symmetric and mixed symmetry operators with the same dimension. In the following we do not write down explicitly the additional index parameterising such degeneracies for ease of notation.}
\begin{subequations}
\begin{align}\label{deltaQ}
    \delta_q&=\underbrace{\tau_1+\tau_2}_{ 2\Delta}+2q+\frac1{N}\,\gamma_{q,l}+...\,,\\
    a^{\bf{n},\bf{\bar{n}}}_{q,l}& ={}^{(0)}a^{\bf{n},\bf{\bar{n}}}_{q,l}+\frac{1}{N}{}^{(1)}a^{\bf{n},\bf{\bar{n}}}_{q,l}+...
\end{align}
\end{subequations}
with $\tau_i$ the twists of the external legs and where $\Delta=\tfrac{\tau_1+\tau_2}{2}$ should not to be confused here with a dimension. For the double-trace contributions to the ${\sf s}$-channel conformal block expansion, this translates into
\begin{align}\label{Nexpbt}
    \mathcal{A}^{\left[{\cal O}_1{\cal O}_1\right]}(u,v|W_{ij})&=u^{\Delta}\sum_{q,l^\prime}\left[\sum_{\bf{n},\bf{\bar{n}}}{}^{(0)}a^{\bf{n},\bf{\bar{n}}}_{q,l^\prime} \,u^{q}g^{\bf{n},\bf{\bar{n}}}_{2\Delta+2q,l}(u,v|W_{ij})\right]
    \\
    &+\frac{1}{N}u^{\Delta}\sum_{q,l}\left[\sum_{\bf{n},\bf{\bar{n}}}\left(\frac{\gamma_{q,l}}{2}\,{}^{(0)}a^{\bf{n},\bf{\bar{n}}}_{q,l}\,\partial_q+{}^{(1)}a^{\bf{n},\bf{\bar{n}}}_{q,l}\right) \,u^{q}g^{\bf{n},\bf{\bar{n}}}_{2\Delta+2q,l}(u,v|W_{ij})\right]\nonumber\\
    &+...\,.\nonumber
\end{align}
where the $...$ denote terms of ${\cal O}\left(1/N^2\right)$ and higher. Under a double-trace flow, the $1/N$ expansion of the change in the ${\sf s}$-channel double-trace contributions to the 4pt function \eqref{cpt} is given by
\begin{equation}\label{Nexpbtdiff}
    \delta \mathcal{A}^{\left[{\cal O}_1{\cal O}_1\right]}(u,v|W_{ij})=\frac{1}{N}u^{\Delta}\sum_{q,l}\left[\sum_{\bf{n},\bf{\bar{n}}}\left(\frac{\delta \gamma_{q,l}}{2}\,{}^{(0)}a^{\bf{n},\bf{\bar{n}}}_{q,l}\,\partial_q+{}^{(1)}\delta a^{\bf{n},\bf{\bar{n}}}_{q,l}\right) \,u^{q}g^{\bf{n},\bf{\bar{n}}}_{2\Delta+2q,l}(u,v|W_{ij})\right]+...,
\end{equation}
where the mean field theory contributions (the first line of \eqref{Nexpbt}) cancel. To extract the anomalous dimensions $\delta\gamma_{q,l}$ induced by the double-trace flow \eqref{dtflow}, we thus need to: 
\begin{enumerate}
    \item Determine the mean-field theory coefficients ${}^{(0)}a^{\bf{n},\bf{\bar{n}}}_{q,l}$, which can be extracted along the same lines as in \S \tcb{\ref{OPEJ1J2J3J4}} from the mean-field theory part of the correlator using the orthogonality relations derived in \S \tcb{\ref{MackPolinomials}}. The mean-field theory part of the correlator itself is given simply by Wick contractions.
    \item Given ${}^{(0)}a^{\bf{n},\bf{\bar{n}}}_{q,l}$, we can then determine the anomalous dimensions $\delta\gamma_{l,q}$ from the ${\sf t}$- and ${\sf u}$-channel crossing kernels, which schematically reads:
    \begin{equation}\label{gammacrosslog}
        \frac{\delta\gamma_{q,l}}{2}\,{}^{(0)}a^{\bf{n},\bf{\bar{n}}}_{q,l} =-\left\langle u^{\Delta+q}g_{2\Delta+2q,l}^{\bf{n,\bar{n}}}|a^{\sf{t}}{F}^{(\sf{t})}+a^{\sf{u}}{F}^{(\sf{u})}\right\rangle\Big|_{\log u},
    \end{equation}
    where $a^{\sf{t}}$ and $a^{\sf{u}}$ schematically denote the expansion coefficients and ${F}^{(\sf{t})}$ and ${F}^{(\sf{u})}$ the CPWs.
    The anomalous dimensions can be read off from the $\log$ term $u^{\tau_1+\tau_2+q}\log u$ noting that in \eqref{Nexpbt}
    \begin{equation}
        u^{\Delta}\partial_q\left(u^{q}\right) = u^{\Delta+q}\log u.
    \end{equation}
\end{enumerate}

\subsection{Scalar correlators}\label{0000corr}

The simplest case, which has also been the subject of intense study in the bootstrap context, is the case of 4pt correlators of scalar operators. In the following we study the anomalous dimensions induced at ${\cal O}\left(1/N\right)$ by the general double-trace perturbation \eqref{dtflow} on double-trace operators $\left[\Phi_1\Phi_2\right]_{n,l}$ comprised of scalar single-trace operators $\Phi_i$. Later on in \S \tcb{\ref{subsec::spinningcorrelators}} we consider the effect of double-trace flows on 4pt correlators involving spinning operators, extracting the corresponding anomalous dimensions of double-trace operators comprised of spinning single-trace operators.

After using the orthogonality relations of \S \tcb{\ref{MackPolinomials}} to extract the mean field theory OPE coefficients in \S \tcb{\ref{subsub::MFT}}, in \S \tcb{\ref{0000corr}} we first consider the simplest case of the flow induced by \eqref{dtflow} with the double-trace operator comprised of scalar single-trace operators $l^\prime=0$. This case was recently considered in \cite{Giombi:2018vtc} using position space techniques, in which the anomalous dimensions of leading double-trace operators $\left[\Phi_1\Phi_2\right]_{0,l}$ (i.e. with $n=0$) were obtained.\footnote{We thank Simone Giombi for discussions and for sharing a draft of \cite{Giombi:2018vtc} prior to its publication, which allowed us to compare our independent derivations of this result.} We recover the latter results using the present Mellin space methods, and furthermore extend them to include the anomalous dimensions of all subleading double trace operators $\left[\Phi_1\Phi_2\right]_{n,l}$ with $n$ and $l$ arbitrary. In \S \tcb{\ref{subsub::sdt}} we further extend these results to the anomalous dimensions induced by the double-trace perturbation \eqref{dtflow} built from single-trace operators of arbitrary spin $l^\prime$. 

\subsubsection{Mean field theory}
\label{subsub::MFT}

Here we extract the mean field theory OPE coefficients. We first consider 4pt functions of identical scalars $\Phi_1=\Phi_2=\Phi$. The mean-field theory correlator is obtained simply by Wick contractions \cite{Dolan:2000ut}:
\begin{align}\label{0000mft}
    \left\langle \Phi(y_1)\Phi(y_2)\Phi(y_3)\Phi(y_4)\right\rangle\equiv \frac{\mathcal{A}^{(0)}_{0000}\left(u,v\right)}{(y_{12}^2)^{\Delta}(y_{34}^2)^{\Delta}}\,,
\end{align}
with
\begin{align}
    \mathcal{A}^{(0)}_{0000}&=\left[1+u^\Delta+\left(\frac{u}v\right)^\Delta\right]\,,
\end{align}
where we normalise the scalar 2pt function as
\begin{equation}
    \left\langle \Phi(y_1)\Phi(y_2)\right\rangle=\frac{1}{(y_{12}^2)^{\Delta}}\,.
\end{equation}
Its conformal block expansion contains contributions from the identity operator and all double-trace operators $\left[\Phi \Phi\right]_{q,l}$, which in the ${\sf s}$-channel reads
\begin{equation}\label{mft0000cbe}
    \mathcal{A}^{(0)}_{0000}\left(u,v\right) = 1 + \sum^{\infty}_{l,q=0}{}^{(0)}a^{\left[\Phi \Phi\right]}_{q,l} u^{\Delta+q}g_{2\Delta+2q,l}\left(u,v\right).
\end{equation}

We can straightforwardly extract the OPE coefficients of the leading twist ($q=0$) double-trace operators using the inversion formula \eqref{OPE simple}:
\begin{equation}\label{leadingmft}
   {}^{(0)}a_{0,l}^{[\Phi\Phi]}=\frac{(-1)^l}{l!}\int_{-i\infty}^{i\infty} \frac{ds}{2\pi i}\,\Big[\delta(s+2\Delta)+\delta(0)\Big]Q_l^{2\Delta,2\Delta,0,0}(s)=\frac{1+(-1)^l}2\,\frac{2^{l+1} (\Delta )_l^2}{l!\, (l+2 \Delta -1)_l}\,,
\end{equation}
where we remind the reader that in the second equality we used that the Mellin representation of free theory correlators is given as a linear sum of Dirac delta functions (see footnote \ref{foo::freemellin}).

To extract contributions from double-trace operators of subleading twist in \eqref{0000mft}, the leading twist contributions \eqref{leadingmft} must first be subtracted. As explained in \S \tcb{\ref{subsec::leadsub}}, to avoid cumbersome infinite summations one acts on \eqref{0000mft} with the operators \eqref{tbop}. These project away contributions from from conformal multiplets of a given twist. In this way, the subleading twist operator becomes leading and we can apply the same approach as we did to extract leading twist operator data to extract operator data for the (former) subleading operator. 
 
In particular, to extract the OPE coefficients of double-trace operators of twist $\tau_n=2\Delta+2n$ with $n>0$, we must first act with the operators $\mathcal{T}_{2\Delta+2i}$ defined in equation \eqref{tbop} for $i=0,...,n-1$, which projects away contributions from all lower twist $\tau_i$ double-trace conformal multiplets. To wit, 
\begin{align}
    \prod_{i=0}^{n-1}\widehat{\mathcal{T}}_{2\Delta+2i}\left[u^{\Delta}\left(1+v^{-\Delta}\right)\right]=8^n (\Delta )_n^2 \left(-\tfrac{d}{2}+\Delta +1\right)_n^2\left[u^{\Delta+n}\left(1+v^{-\Delta-n}\right)\right]\,.
\end{align}
which up to an overall coefficient $\alpha_n(2\Delta)$ in \eqref{alphaCoeff} encodes the contribution of double-trace operators of twist $\tau\geq 2\Delta+2n$. Using the orthogonality relations we straightforwardly obtain the OPE coefficients as for the leading twist contributions above:
\begin{multline}
  {}^{(0)}a_{n,l}^{[\Phi\Phi]}=\frac{(-1)^l}{l!}\frac{8^n (\Delta )_n^2 \left(-\tfrac{d}{2}+\Delta +1\right)_n^2}{\alpha_n(2\Delta)}\int_{-i\infty}^{i\infty} \frac{ds}{2\pi i}\Big[\delta(s+2\Delta+2n)+\delta(s)\Big]Q_{l}^{2\Delta+2n,2\Delta+2n,0,0}(s)\\=\frac{1+(-1)^l}2\,\frac{2^{l+1} (-1)^n (\Delta )_n^2 \left(-\frac{d}{2}+\Delta +1\right)_n^2 (\Delta )_{n+l}^2}{l!\, n! \left(\frac{d}{2}+l\right)_n (d-2 (n+\Delta ))_n (l+2 n+2 \Delta -1)_l \left(-\frac{d}{2}+l+n+2 \Delta \right)_n}\,,
\end{multline}
which perfectly matches the results of \cite{Dolan:2000ut,Heemskerk:2009pn,Fitzpatrick:2011dm}. 

In the same way, we can also consider the mixed correlator which has the mean-field theory contribution:\footnote{Note that, obviously, this is the most general case for which a mean field theory contribution exists.}
\begin{subequations}
\begin{align}
     \left\langle\Phi_1\Phi_2\Phi_1\Phi_2\right\rangle&=\frac{\mathcal{A}^{[\Phi_1\Phi_2]}}{(y_{12})^{\tau_1}(y_{34})^{\tau_2}}\\
    \mathcal{A}^{[\Phi_1\Phi_2]}&=u^{(\tau_1+\tau_2)/2}.
\end{align}
\end{subequations}
From the action of the operators \eqref{tbop}
\begin{equation}
    \left[\prod_{i=0}^{n-1}\widehat{\mathcal{T}}_{\tau_1+\tau_2+2i}\right]\mathcal{A}^{[\Phi_1\Phi_2]}=2^{3 n} (\tau_1)_n (\tau_2)_n \left(-\tfrac{d-2 (\tau_1+1)}{2}\right)_n \left(-\tfrac{d-2 (\tau_2+1)}{2}\right)_n\,u^{(\tau_1+\tau_2+2n)/2}\,,
\end{equation}
we quickly recover the corresponding double-trace operator $[\Phi_1\Phi_2]_{n,l}$ OPE coefficients for all $n$ and $l$ as:
\begin{multline}\label{OPEnl}
    {}^{(0)}a_{n,l}^{[\Phi_1\Phi_2]}=\tfrac{2^{3 n} (\tau_1)_n (\tau_2)_n \left(-\tfrac{d-2 (\tau_1+1)}{2}\right)_n \left(-\tfrac{d-2 (\tau_2+1)}{2}\right)_n}{\alpha_n\left(\tfrac{\tau_1+\tau_2}2\right)}\\\times\,\frac{(-1)^l}{l!}\int_{-i\infty}^{i\infty} \frac{ds}{2\pi i}\,\delta(s+\tau_1+\tau_2+2n)\,Q_{l}^{\tau_1+\tau_2+2n,\tau_1+\tau_2+2n,-\tau_1+\tau_2,\tau_1-\tau_2}(s)\\=\frac{2^l (-1)^{-n} (\tau_1)_n (\tau_2)_n \left(-\frac{d}{2}+\tau_1+1\right)_n \left(-\frac{d}{2}+\tau_2+1\right)_n (n+\tau_1)_l (n+\tau_2)_l}{l!\, n! \left(\frac{d}{2}+l\right)_n (d-2 n-\tau_1-\tau_2)_n (l+2 n+\tau_1+\tau_2-1)_l \left(-\frac{d}{2}+l+n+\tau_1+\tau_2\right)_n}\,,
\end{multline}
which matches the result of \cite{Fitzpatrick:2011dm}.\footnote{Note that in our conventions OPE coefficients are defined with respect to the conformal structures \eqref{nicebasis}. To make contact with the normalisation of \cite{Fitzpatrick:2011dm} we compare the corresponding normalisation of the conformal blocks, which in the $u\to 0$ and $v\to 1$ limit in our conventions reads:
\begin{equation}
    u^{\tau/2}\int \frac{ds}{4\pi i}\,v^{-(s+\tau)/2}\,\tilde{\rho}_{\{\tau_i\}}(s,\tau)\,\mathcal{Q}_{l,\tau}(s)\sim\left(-\tfrac12\right)^l\,u^{\tau/2}(1-v)^l\,.
\end{equation}
Accounting for the different normalisation $\sim u^{\tau/2}(1-v)^l$ in \cite{Fitzpatrick:2011dm} we obtain perfect agreement for the OPE coefficients \eqref{OPEnl}.}

\subsubsection{Scalar double-trace deformations: Anomalous dimensions}

\subsubsection*{Leading twist anomalous dimensions} 

As was outlined at the beginning of this section, once the mean field theory OPE coefficients are known, we can extract the anomalous dimensions induced by double-trace flows \eqref{dtflow} from equation \eqref{cpt} by extracting the $\log u$ terms arising in the $\sf{s}$-channel expansion of $\sf{t}$- and $\sf{u}$-channel CPWs. This computation coincides with the crossing kernel computation we performed in section \S\ref{Crossing Kernels}. Focusing first on the leading twist double-trace operators, which contribute a term proportional to $ u^\Delta \log u$, from \eqref{cpt} we obtain:
\begin{equation}\label{leadingtwistgamma}
   \frac{1}{2}  {}^{(0)}a_{0,l}^{[\Phi\Phi]}\delta\gamma^{[\Phi\Phi]}_{l^\prime|0,l}=\left(\frac{1+(-1)^{l+l^\prime}}2\right)2\,\mathfrak{J}^{(\sf t)}_{0,0,0,0|l,l^\prime}(t=2\Delta)\,c_{\Phi\Phi\mathcal{O}}\,,
\end{equation}
where we have also inserted the squared OPE coefficient $c_{\Phi\Phi\mathcal{O}}$ and where the $l^\prime$ labels the spin of the CPW and therefore the spin of the double-trace flow \eqref{dtflow}, while we used that for identical external legs $\mathfrak{J}_{0,0,0,0|l^\prime,l}^{(\sf{u})}=(-1)^{l+l^\prime}\mathfrak{J}_{0,0,0,0|l^\prime,l}^{(\sf{t})}$.

The simplest double-trace flow \eqref{dtflow} is for $l^\prime=0$ (for which we define $\delta\gamma_{0|n,l}\equiv\delta\gamma_{n,l}$), which corresponds to the case in which we deform the CFT with
\begin{equation}
    \lambda\int d^dy\,\mathcal{O}^2\,,
\end{equation}
where $\mathcal{O}$ is a scalar single-trace primary operator of twist $\tau$. If $\tau<\tfrac{d}2$ the CPW gives the difference between the IR and UV fixed points. Plugging into \eqref{leadingtwistgamma} the crossing kernal \eqref{inv} for $l^\prime=0$, we obtain: 
\begin{subequations}\label{dtanoml}
\begin{align}
    \delta\gamma^{[\Phi\Phi]}_{0,0}&=\frac{2\, \Gamma (\tau ) \Gamma \left(\frac{d-\tau }{2}\right)^2}{\Gamma \left(\frac{d}{2}\right) \Gamma \left(\frac{\tau }{2}\right)^2 \Gamma \left(\frac{d}{2}-\tau \right)}\,c_{\Phi\Phi\mathcal{O}}\,,\\
    \delta\gamma^{[\Phi\Phi]}_{0,l}&=\left(\frac{1+(-1)^{l}}2\right)\,\delta\gamma_{0,0}\, _4F_3\left(\begin{matrix}-l,2 \Delta +l-1,\frac{d-\tau}{2},\frac{\tau }{2}\\\frac{d}{2},\Delta ,\Delta\end{matrix} ;1\right)\,,
\end{align}
\end{subequations}
which matches the result recently obtained in \cite{Giombi:2018vtc} using position space methods.\footnote{Note that the results \eqref{dtanoml} and \eqref{mixdtanoml} for external scalars were first obtained in \cite{Gopakumar:2016cpb} in Appendix D, although they were not re-summed in terms of ${}_4F_3$ hypergeometric functions.}

\paragraph{Mixed correlators} The above result can be extended with little effort to the case of mixed correlators $\left\langle\Phi_1\Phi_2\Phi_1\Phi_2\right\rangle$. The mean-field theory OPE coefficients were computed in \S \tcb{\ref{subsub::MFT}}, which for leading twist double-trace operators read
\begin{equation}
   {}^{(0)}a^{[\Phi_1\Phi_2]}_{0,l}=\frac{(-1)^l}{l!}\,Q_l^{\tau_1+\tau_2,\tau_1+\tau_2,-\tau_1+\tau_2,\tau_2-\tau_1}(-\tau_1-\tau_2)=\frac{2^l (\tau_1)_l (\tau_2)_l}{(l+\tau_1+\tau_2-1)_l}\,.
\end{equation}
In this case, assuming that the only non-vanishing OPE coefficients involving the scalar operators are $c_{\Phi_1\Phi_2\mathcal{O}}$\footnote{This assumption fits naturally the case in which $\mathcal{O}$ is uncharged and by charge conservation $\Phi_1$ and $\Phi_2$ must have opposite charge. This also requires $c_{\Phi_i\Phi_i\mathcal{O}}=0$. If we consider instead uncharged operators such that $c_{\Phi_i\Phi_i\mathcal{O}}\neq0$ we get the same result with an overall factor $(-1)^l$.} , we only have one contribution to the $\log u$ term which is given by the $\sf{t}$-channel CPW:
\begin{equation}\label{leadingtwistgamma2}
   \frac{1}{2} {}^{(0)}a^{[\Phi_1\Phi_2]}_{0,l}\delta\gamma^{[\Phi_1\Phi_2]}_{0,l}=\mathfrak{J}^{(\sf t)}_{0,0,0,0|l,l^\prime}(t=\tau_1+\tau_2)\,c_{\Phi_1\Phi_2\mathcal{O}}\,,
\end{equation}
from which we obtain
\begin{subequations}\label{mixdtanoml}
\begin{align}
    \delta\gamma_{0,0}^{[\Phi_1\Phi_2]}&=\frac{2\Gamma (\tau ) \Gamma \left(\tfrac{d-\tau +\tau_1-\tau_2}{2}\right) \Gamma \left(\tfrac{d-\tau -\tau_1+\tau_2}{2}\right)}{\Gamma \left(\tfrac{d}{2}\right) \Gamma \left(\tfrac{d-2 \tau}{2}\right) \Gamma \left(\tfrac{\tau +\tau_1-\tau_2}{2}\right) \Gamma \left(\tfrac{\tau -\tau_1+\tau_2}{2}\right)}\,c_{\Phi_1\Phi_2\mathcal{O}}\,,\\
    \delta\gamma_{0,l}^{[\Phi_1\Phi_2]}&=(-1)^l\,\delta\gamma_{0,0}^{[\Phi_1\Phi_2]}\,\frac{(\tau_1)_l}{(\tau_2)_l}\, {}_4F_3\left(\begin{matrix}-l,\frac{d-\tau+\tau_1-\tau_2}{2},\frac{\tau +\tau_1-\tau_2}{2},l+\tau_1+\tau_2-1\\\frac{d}{2},\tau_1,\tau_1\end{matrix};1\right)
\end{align}
\end{subequations} 
which again is in agreement with the result of \cite{Giombi:2018vtc} obtained using position space methods.

\subsubsection*{Subleading double-trace operators}

To extract the anomalous dimensions of subleading twist double-trace operators $[\Phi\Phi]_{n,l}$ (i.e. with $n\neq0$), we have to act on the ${\sf t}$- and ${\sf u}$-channel CPWs in \eqref{cpt} with the operators \eqref{tbop} to project out the ${\sf s}$-channel contributions from conformal multiplets of all lower-twist double-trace operators $[\Phi\Phi]_{i,l}$, $i=0,...,n-1$.

\begin{figure}[t]
\centering
\includegraphics[width=\textwidth]{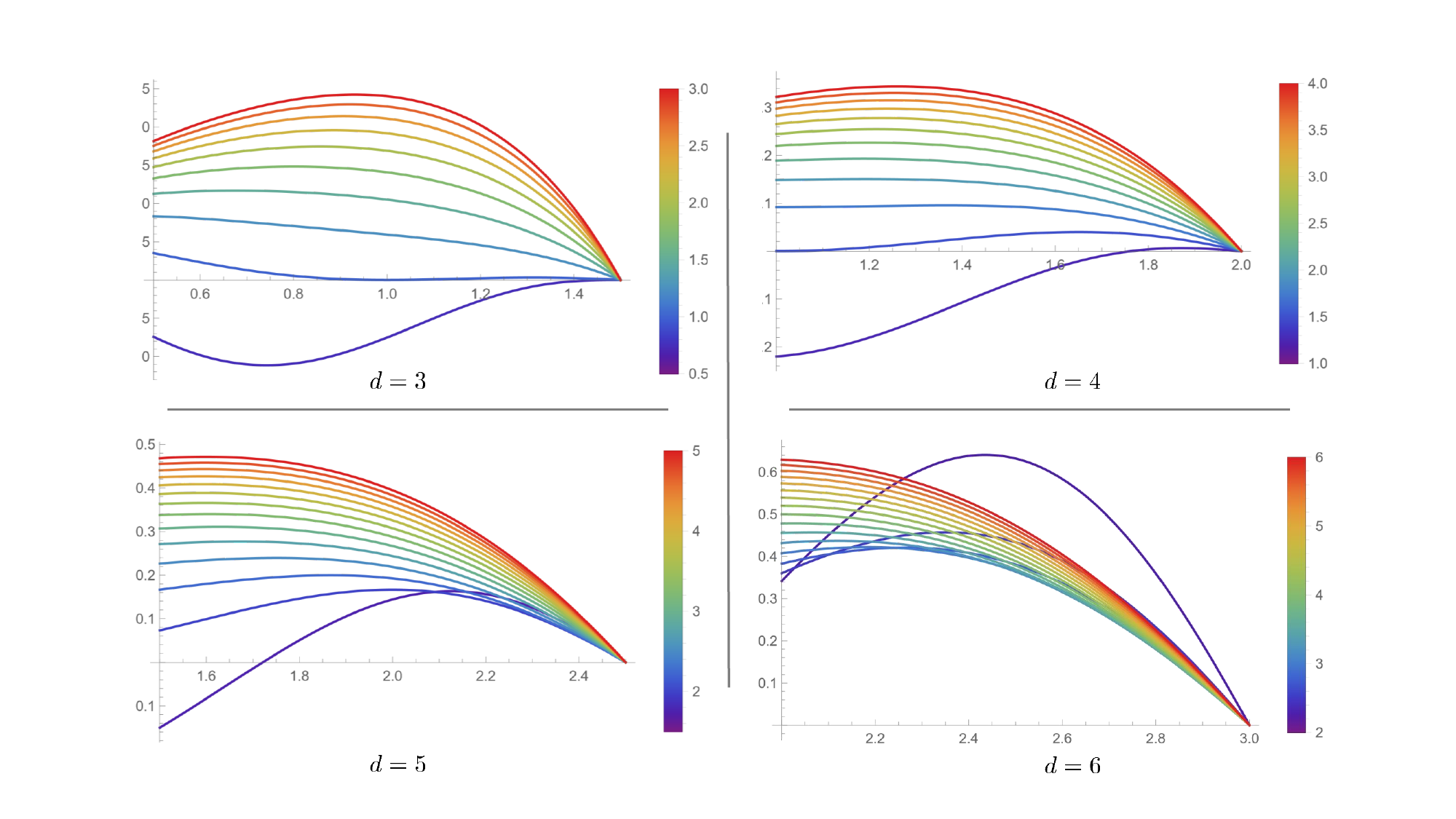}
\caption{Plot of $\delta\gamma_{1,0}$ (vertical axis) in various dimensions in the interval $\tau\in[\tfrac{d-2}2,\tfrac{d}2]$ (horizontal axis) for increasing values of $\Delta>\tfrac{d-2}2$ (color-bar in each graph).}
\label{g10}
\end{figure}

\begin{figure}[t]
\centering
\includegraphics[width=\textwidth]{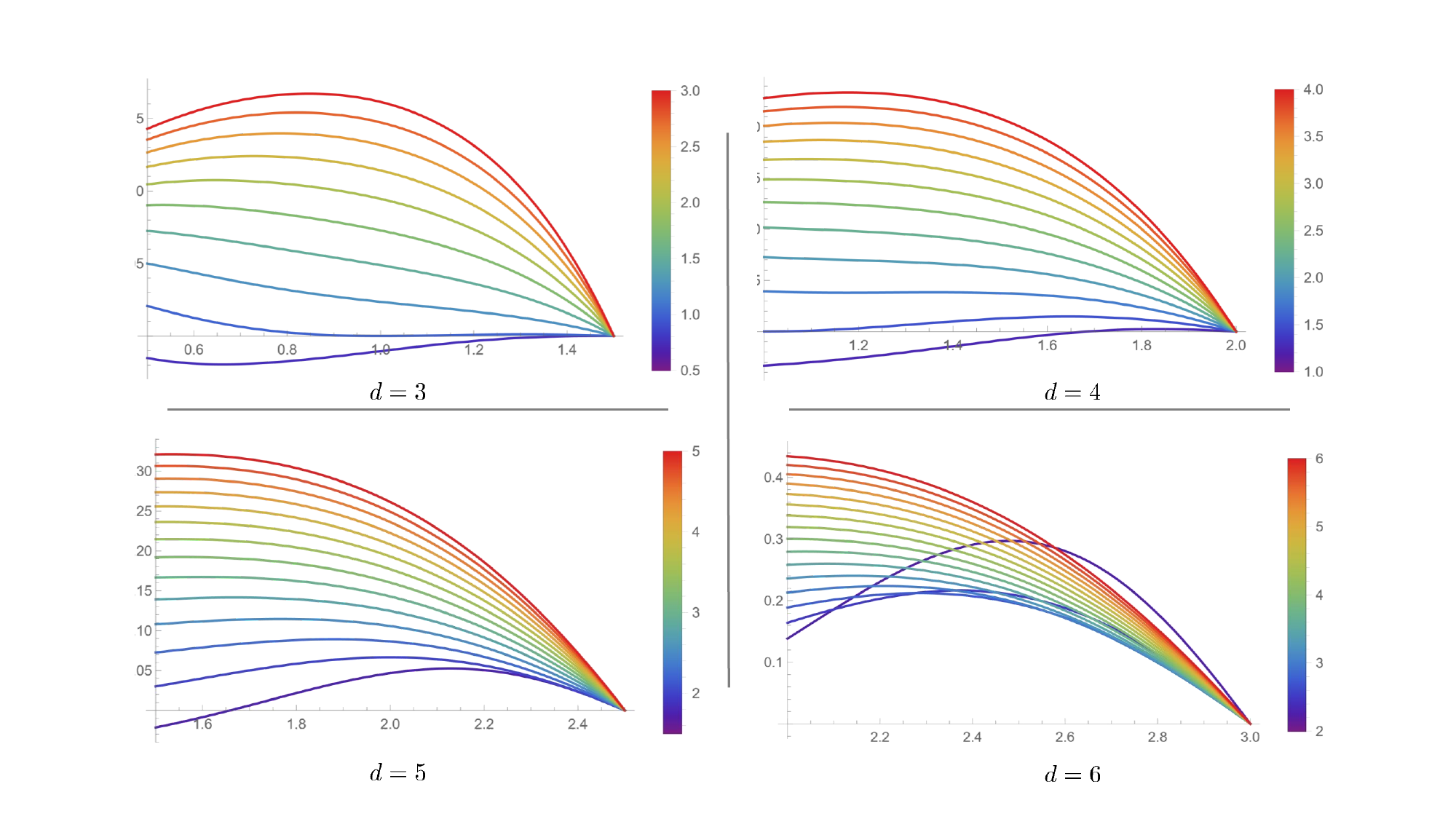}
\caption{Plot of $\delta\gamma_{2,0}$ (vertical axis) in various dimensions in the interval $\tau\in[\tfrac{d-2}2,\tfrac{d}2]$ (horizontal axis) for increasing values of $\Delta>\tfrac{d-2}2$ (color bar).}
\label{g20}
\end{figure}

The action of the operator $\widehat{\mathcal{T}}_\tau$ (given in \eqref{tbop}) on a generic Mellin amplitude ${\cal M}\left(s,t\right)$ with equal external twists $\tau_1=\tau_2=\Delta$ satisfies the following difference relation
\begin{align}\label{diffrelmellin}
    \widehat{\mathcal{T}}_{\tau}{\cal M}(s,t)&=\frac{(t-\tau ) (-2 d+t+\tau +2)}2\Bigg[\left(d^2-2 d (\tau +1)+2 (s (s+t)+t+\tau )+\tau ^2\right){\cal M}(s,t) \\\nonumber
    &\hspace{140pt}+(s+t)^2 {\cal M}(s+2,t)+ s^2 {\cal M}(s-2,t)\Bigg]\\\nonumber
    &+\frac{(t-2 \Delta )^2}{2} \Bigg[\left(d^2-2 d (s+\tau +2)+2 (s+1) t+\tau  (\tau +2)\right){\cal M}(s,t-2) \\\nonumber
    &\hspace{70pt}+\left(d^2+2 d (s+t-\tau -2)-2 t (s+t-1)+\tau  (\tau +2)\right) {\cal M}(s+2,t-2)\Bigg]\\\nonumber
    &+\frac{1}{2} (t-2 \Delta )^2 (t-2 (\Delta +1))^2 {\cal M}(s+2,t-4)\,.
\end{align}
In appendix \S\tcb{\ref{DiffRel}} we give the explicit form of the difference relation for Mellin amplitudes with arbitrary twist external legs. It is straightforward to obtain analogous difference relations for arbitrary spinning legs but the corresponding result rather involved and we will not present it explicitly here. Starting from the Mellin representation \eqref{ScalarTUmellinDelta} of the $\sf{t}$- and $\sf{u}$-channel CPWs:
\begin{subequations}\label{MellinScalar}
\begin{align}
    \mathfrak{M}^{(\sf{t})}(s,\tau_1+\tau_2)&=\frac{\Gamma (\tau )}{\Gamma \left(\frac{d}{2}-\tau \right) \Gamma \left(\tfrac{\tau +\tau_1-\tau_2}{2}\right)^2 \Gamma \left(\tfrac{\tau -\tau_1+\tau_2}{2}\right)^2}\frac{\Gamma \left(\tfrac{s+\tau }{2}\right) \Gamma \left(\tfrac{d+s-\tau}{2}\right)}{\Gamma \left(\tfrac{s+\tau_1+\tau_2}{2}\right)^2}\,,\\
    \mathfrak{M}^{(\sf{u})}(s,\tau_1+\tau_2)&=\frac{\Gamma \left(\tfrac{d-\tau +\tau_1-\tau_2}{2} \right) \Gamma \left(\tfrac{d-\tau -\tau_1+\tau_2}{2}\right)}{\Gamma \left(\tfrac{\tau }{2}\right)^2 \Gamma \left(\tfrac{d-\tau }{2}\right)^2 \Gamma \left(\tfrac{\tau +\tau_1-\tau_2}{2}\right) \Gamma \left(\tfrac{\tau -\tau_1+\tau_2}{2}\right)}\\\nonumber&\hspace{100pt}\times\,\frac{\Gamma \left(\tfrac{-s+\tau -\tau_1-\tau_2}{2}\right) \Gamma \left(\tfrac{d-s-\tau -\tau_1-\tau_2}{2}\right)}{\Gamma \left(\tfrac{-s+\tau_1-\tau_2}{2}\right) \Gamma \left(\tfrac{-s-\tau_1+\tau_2}{2}\right)}
\end{align}
\end{subequations}
we can act with the operators $\widehat{\mathcal{T}}_\tau$ using the difference relation \eqref{diffrelmellin} to project out all contributions from lower twist double-trace operators. We can then proceed as for the leading twist case, which leads to the following result for the anomalous dimensions of double-trace operators $[\Phi\Phi]_{n,l}$ under double-trace flows:
\begin{align}\label{gamma0000nl}
    \frac12\,\alpha_n( 2\Delta+2n){}^{(0)}a^{[\Phi\Phi]}_{n,l}\,\delta\gamma^{[\Phi\Phi]}_{n,l}=\frac{1+(-1)^l}2\tfrac{4\, \Gamma (\tau )}{\Gamma \left(\frac{\tau }{2}\right)^4 \Gamma \left(\frac{d}{2}-\tau \right)}\sum_{j=0}^n\,D_{j}T^n_{n-j,j}\,,
\end{align}
where
\begin{multline}
    T^n_{ij}=\,\int_{-i\infty}^{i\infty} \frac{ds}{4\pi i}\,\Gamma(-\tfrac{s}{2})^2\Gamma(\tfrac{d+s-\tau}{2}+i)\Gamma(\tfrac{s+\tau}2+j)\,Q_{l}^{2\Delta+2n,2\Delta+2n,0,0}(s)\\=\,\frac{2^l \Gamma \left(\tfrac{2 j+\tau}{2}\right)^2 \Gamma \left(\tfrac{d+2 i-\tau }{2}\right)^2\left(\tfrac{2 \Delta +2n}{2}\right)_l^2}{(l+2 \Delta+2n -1)_l \Gamma \left(\tfrac{d+2 i+2 j}{2} \right)}\, _4F_3\left(\begin{matrix}-l,2 \Delta +2n +l+1,\frac{d}{2}+i-\frac{\tau }{2},j+\frac{\tau }{2}\\\frac{d}{2}+i+j,\Delta +n,\Delta +n\end{matrix};1\right)\,.
\end{multline}
 The mean field theory OPE coefficients ${}^{(0)}a^{[\Phi\Phi]}_{n,l}$ are given by equation \eqref{OPEnl} and $D_j$ are coefficients depending on $d$, $\Delta$ and $\tau$:
\begin{subequations}
\begin{align}
    D^{(0)}_0=&\tfrac12\\
    D^{(1)}_0=&\frac{(d-2 (\Delta +1)) (\tau -2 \Delta )^2}{2 (d-2 \tau )}\left[d-2 \Delta -(\tau -2) \tau -2\right]\,,\\
    D^{(1)}_1=&\frac{(d-2 (\Delta +1)) (-d+2 \Delta +\tau )^2}{2 (d-2 \tau )}\,\left[ (d-3) d+2 (\Delta +\tau +1)+\tau(\tau-2 d)\right]\,,
\end{align}
\end{subequations}
Where the suffix indicates the value of $n$. This recovers the $n=0$ and $n=1$ result discussed previously. For $n>1$ the general results are a bit cumbersome and we avoid to write down the explicit form of the coefficients $D_i$ here. We give explicit general $n$ expressions for selected dimensions in the next section. It is useful to look at the $l=0$ case where the hypergeometric function collapses to a single term. In this case, for $n=1$ we get:
\begin{align}\label{deltagamma10}
    \delta\gamma^{[\Phi\Phi]}_{1,0}&=\frac{1}{2 d \Delta ^2 (d-2 (\Delta +1))^2}\left[D_0 (d-\tau )^2+D_1 \tau ^2\right]\delta\gamma_{0,0}^{[\Phi\Phi]}.
\end{align}
The above result, together with the $n=2$ result which we do not write down for brevity, is plotted for various values of $\Delta$ and $d$ in fig.~\ref{g10} and fig.~\ref{g20}. 
This exhibits positivity for $\frac{d-2}2<\tau<\frac{d}2$ and $\Delta>\Delta_0$ for some $\Delta_0$.

As has been demonstrated, the current framework allows to straightforwardly obtain any $\gamma_{n,l}$ by simply iterating with the operator $\widehat{\mathcal{T}}_\tau$ and evaluating the orthogonal projection with continuous Hahn polynomials, exactly mimicking the extraction of OPE coefficients discussed in preceding sections. In general the result for $\gamma_{n,l}$ is a linear combination of $T_{ij}^n$ with $i+j=n$ and coefficients $D_{ij}$ fixed by the action of the operators $\widehat{\mathcal{T}}_{2\Delta+2n}$.

\paragraph{Vector models in $d=4-\epsilon$ and $d=6-\epsilon$} 

The explicit form of the anomalous dimensions \eqref{gamma0000nl} for general $n$ is quite involved for arbitrary values of $\Delta$, $d$ and $\tau$, but simplifications arise for particular values. For instance, an interesting example is given by $d=4-\epsilon$ with $\tau=d-2=2-\epsilon$. In this case we can obtain the general $n$ result, which reads:\footnote{This result can be trivially extended to odd $l$ by considering the result for mixed correlators in \eqref{mixdtanoml}.}
\begin{equation}
    \delta\gamma_{n,l}=\epsilon\,c_{\Phi\Phi\mathcal{O}}\,\frac{1+(-1)^l}2\,\frac{(\Delta -1)^2}{(\Delta +n-1)^2}\, _4F_3\left(\begin{matrix}1,1,-l,l+2 \Delta +2n-1\\2,\Delta +n,\Delta +n\end{matrix};1\right)\,,
\end{equation}
and is a remarkably simple extension of \eqref{dtanoml}, to which it reduces for $n=0$. A slightly more involved result can be obtained for $d=6-\epsilon$ and $\tau=d-2$. In this case we have:
{\allowdisplaybreaks
\begin{subequations}
\begin{align}
    \delta\gamma_{0,l}&=\epsilon\,c_{\Phi\Phi\mathcal{O}}\frac{1+(-1)^l}2\frac{6 (\Delta -1)^2}{(l+1) (2 \Delta +l-2)}\,\Bigg[ {}_4F_3\left(\begin{matrix}1,1,-l-1,l+2 \Delta -2\\2,\Delta -1,\Delta -1\end{matrix};1\right)-1\Bigg]\,,\\
    \delta\gamma_{1,l}&=\epsilon\,c_{\Phi\Phi\mathcal{O}}\frac{1+(-1)^l}2\Bigg[\frac{3 \left(\Delta ^2-4\right)}{(l+1) (2 \Delta +l)}{}_4F_3\left(\begin{matrix}1,1,-l-1,l+2 \Delta \\3,\Delta ,\Delta \end{matrix};1\right)\\\nonumber
    &\hspace{80pt}-\frac{6 [(\Delta -4) \Delta +6] (\Delta -1)^2}{(l+1) (l+2) (2 \Delta +l-1) (2 \Delta +l)}{}_4F_3\left(\begin{matrix}1,1,-l-2,l+2 \Delta -1\\2,\Delta -1,\Delta -1\end{matrix};1\right)\\\nonumber
    &\hspace{80pt}-\frac{6 (\Delta -1)^2\left[l^2+2 \Delta  l+l-(\Delta -8) \Delta -8\right]}{(l+1) (l+2) (2 \Delta +l-1) (2 \Delta +l)}\Bigg]\\
     \delta\gamma_{n,l}&=\epsilon\,c_{\Phi\Phi\mathcal{O}}\frac{1+(-1)^l}2\,\frac{6 (\Delta -1)^2}{(l+1) (l+2) (2 \Delta +l+2 n-3) (2 \Delta +l+2 n-2)}\\\nonumber&\times\Bigg[\frac{1}{2} \left[\Delta  (\Delta +4 n-4)-8\right] \, {}_4F_3\left(\begin{matrix}1,1,-l-1,l+2 n+2 \Delta -2\\3,n+\Delta -1,n+\Delta -1\end{matrix};1\right)\\
     &\hspace{37pt}-\left[(\Delta -2)^2+2 n\right] \, {}_4F_3\left(\begin{matrix}1,1,-l-2,l+2 n+2 \Delta -3\\2,n+\Delta -2,n+\Delta -2\end{matrix};1\right)\nonumber\\\nonumber
     &\hspace{150pt}+\Delta ^2-l^2-2 \Delta  (l+4)-2 l n+l-2 n+10\Bigg]\,.
\end{align}
\end{subequations}
}
The above result in $d=4-\epsilon$ can be neatly applied to the Wilson-Fisher fixed point for a vector model with additional flavour symmetries $\phi^{a,i}$, where the index $a$ is rotated by $O(N)$ and the index $i$ is rotated by $O(M)$. In this case\footnote{We remind the reader that CPWs are directly related to anomalous dimensions only for correlators which do not include on the the operator exchanged in the internal leg also on the external legs of the CPW. Otherwise further contributions would need to be taken into account.} we can study singlet double-trace deformations for the operator $\mathcal{O}=\phi^{a,i}\phi^{a,i}$ in correlators where the external legs are the non-singlet operators with respect to $O(M)$ $\Phi^{ij}=\phi^{a,(i}\phi^{a,|j)}$. See e.g. \cite{Giombi:2018vtc} for similar discussions for leading twist $n=0$ double-trace operators and \cite{Lang:1992zw} on another approach to anomalous dimensions at the Wilson Fisher fixed point.

We consider the simplest case $M=2$ for which $O(2)\sim U(1)$ and the operators can be chosen to be $\Phi_1\equiv \Phi=\phi^a\phi^a$, $\Phi_2\equiv \bar{\Phi}=\bar{\phi}^a\bar{\phi}^a$ and $\mathcal{O}={\phi}^a\bar{\phi}^a$. In this case also $\Phi$ and $\bar{\Phi}$ have dimension $\Delta=d-2$ and Wick contraction gives: $c_{\Phi\bar{\Phi}\mathcal{O}}=\tfrac4{N}$. We thus obtain:
\begin{equation}
    \gamma_{n,l}=\frac{4}{N}\,\frac{\epsilon(-1)^l}{(n+1)^2}\, _4F_3\left(\begin{matrix}1,1,-l,l+2 n+3\\2,n+2,n+2\end{matrix};1\right)\,.
\end{equation}
The large $l$-limit following the discussion in \S\tcb{\ref{LargeL}} gives the same asympotic behaviour independently of $n$:\footnote{A similar logarithmic behaviour was observed in \cite{Giombi:2018vtc} for the $n=0$ case.}
\begin{equation}
    \gamma_{n,l\to\infty}^{[\Phi\Phi]}\sim\frac{8}{N}\,\frac{\log l}{l^2}\,\epsilon\,.
\end{equation}
In general, for arbitrary $\Delta$ the following result holds:
\begin{equation}
    \gamma_{n,l\to\infty}^{[\Phi\Phi]}\sim 2\, c_{\Phi\Phi\mathcal{O}}\,(\Delta-1)^2\,\frac{\log l}{l^2}\,\epsilon\,.
\end{equation}
Similar results can be obtained for mixed correlators $\left\langle\Phi_1\Phi_2\Phi_1\Phi_2\right\rangle$ with different values of $\tau_1$ and $\tau_2$ but they are slightly more involved when $\tau_1\neq\tau_2$.

\subsubsection{Higher-spin double-trace deformations: Anomalous dimensions}
\label{subsub::sdt}

In the subsequent sections we focus on the double-trace anomalous dimensions induced by a general double-trace deformation \eqref{dtflow} with $l^\prime$ arbitrary, which is generically non-unitary.

\begin{figure}[t]
\centering
\includegraphics[width=\textwidth]{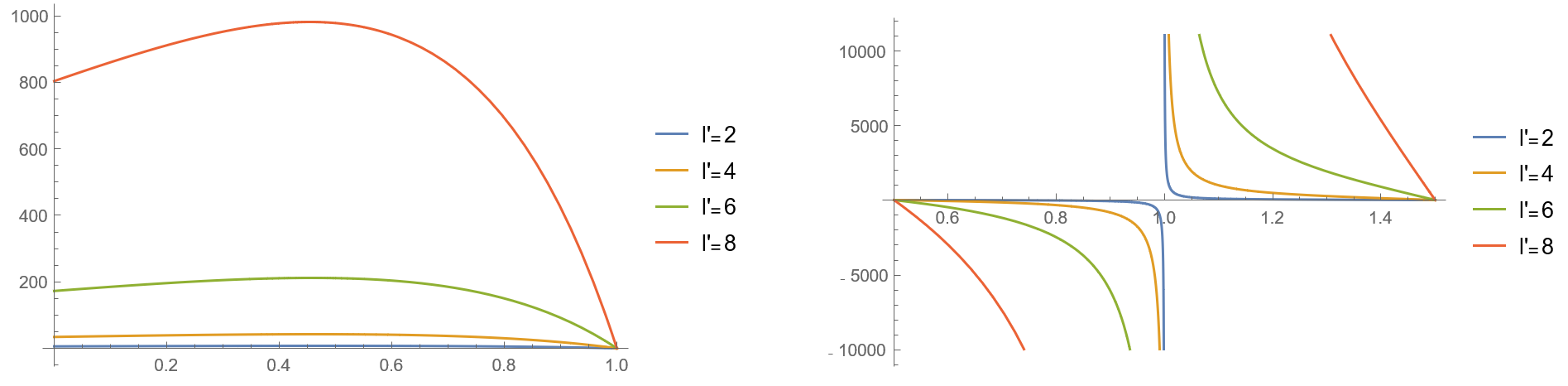}
\caption{Plot of $\delta\gamma^{[\Phi\Phi]}_{l^\prime|0,0}$ in $d=2$ on the left and $d=3$ on the right, in the interval $\tau\in[\tfrac{d-2}2,\tfrac{d}2]$ for even values of $l^\prime$.}
\label{gammaJ}
\end{figure}

In this case, using \eqref{leadingtwistgamma}, for $l=0$ we get:
\begin{equation}
    \delta\gamma_{l^\prime|0,0}^{[\Phi\Phi]}=\frac{1+(-1)^{l^\prime}}2\,\frac{\left(\frac{1}{2}\right)^{l^\prime-1} (d-2)_{l^\prime} (\tau )_{2 {l^\prime}} \left(\tfrac{\tau +2-d}{2}\right)_{l^\prime} \left(\tfrac{d-2 ({l^\prime}+\tau )}{2}\right)_{l^\prime}}{\left(\frac{d}{2}-1\right)_{l^\prime} \left(\frac{\tau }{2}\right)_{l^\prime}^2 \left(\tfrac{d-2 {l^\prime}-\tau }{2}\right)_{l^\prime} (d-{l^\prime}-\tau -1)_{l^\prime}}\,\delta\gamma_{0|0,0}^{[\Phi\Phi]}\,.
\end{equation}
It is interesting to notice that, for instance in $d=3$, the above result is positive for $\tau>1$ for any even $l^\prime$, which is shown in fig. \ref{gammaJ}.\footnote{We note that a similar positivity result was obtained in \cite{Giombi:2013yva} for the change in free energy $\delta F$ for the same range of dimensions.}
For $l^\prime=0$ the above positivity holds in the full range $\tau\in[\tfrac{d-2}2,\tfrac{d}2]$. It is also interesting to consider the generic $d=2$ result for some values of $l^\prime$ which is displayed in fig.~\ref{gammaJ}, from which one can see the positivity of the RG flow for all unitary values of $\tau$ for arbitrary even spin deformations.\footnote{Odd spin deformation have opposite sign in accordance with the intuition that they carry repulsive forces while even spin fields only carry attractive forces.}

The above results can be extended to mixed correlators $\left\langle\Phi_1\Phi_2\Phi_1\Phi_2\right\rangle$. We can read off the anomalous dimensions from the $\sf{t}$ channel crossing kernel only since $\left\langle\Phi_1\Phi_2 J_{\mu(2l+1)}\right\rangle\neq 0$ but $\left\langle\Phi_i\Phi_i J_{\mu(2l+1)}\right\rangle=0$ for both $i=1,2$. In this case we obtain:
\begin{equation}
    \delta\gamma^{[\Phi_1\Phi_2]}_{l^\prime|0,0}=\frac{(-2)^{-{l^\prime}} (d-2)_{l^\prime} (\tau )_{2 {l^\prime}} \left(\tfrac{d-2 ({l^\prime}+\tau )}{2}\right)_{l^\prime} \left(\tfrac{-\tau_1+\tau_2+\tau +2-d}{2}\right)_{l^\prime}}{\left(\frac{d}{2}-1\right)_{l^\prime} (d-{l^\prime}-\tau -1)_{l^\prime} \left(\tfrac{\tau_1-\tau_2+\tau }{2}\right)_{l^\prime} \left(\tfrac{-\tau_1+\tau_2+\tau}{2}\right)_{l^\prime} \left(\tfrac{d-2 {l^\prime}+\tau_1-\tau_2-\tau }{2}\right)_{l^\prime}}\,\delta\gamma^{[\Phi_1\Phi_2]}_{0|0,0}\,.
\end{equation}

\begin{figure}[t]
    \centering
    \includegraphics[width=0.9\textwidth]{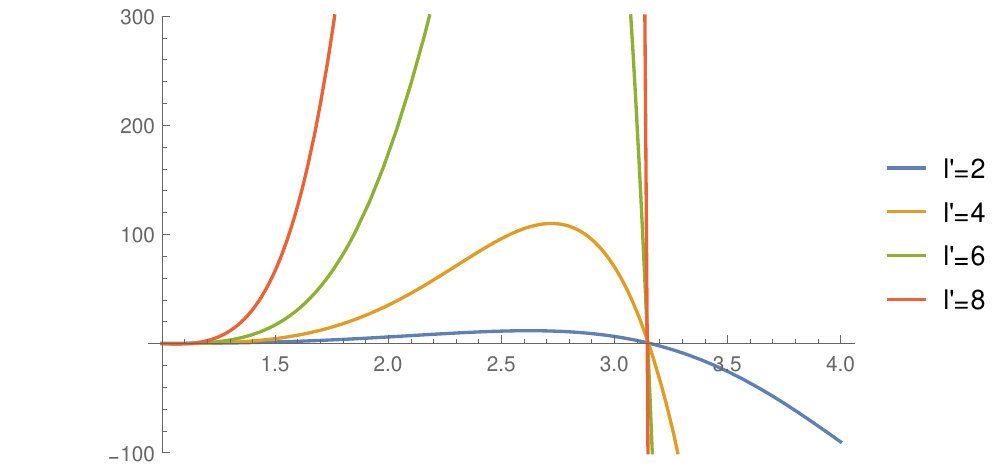}
    \caption{Plot of $\delta\gamma^{[\Phi\Phi]}_{l^\prime,k=0|0,0}$ (vertical axis) for $l^\prime=2,4,6$ varying the space time dimension from in $d\in [1,4]$ (horizontal axis). The finite part is positive up to $d\approx 3$. }
    \label{fig:Gammak0l246}
\end{figure}

The above computation is valid at generic values of $\tau$. However, as discussed in section \S\tcb{\ref{kernelarbitrary}}, the above diverges when $\tau=d-2-k$, with $k=0,\ldots,l^\prime-1$ since the conformal multiplet is shortened. These cases correspond to (partially-)conserved current deformations and the bulk dual would include partially-massless gauge fields \cite{Deser:2001us,Dolan:2001ih}.
Proceeding as discussed in \S\tcb{\ref{partiallyConserved}} in order to properly treat the divergences, the general result for $\delta\gamma_{l^\prime|0,0}^{[\Phi\Phi]}$ reads:
\begin{multline}
    \delta\gamma_{l^\prime,k|0,0}^{[\Phi\Phi]}=-\tfrac{1+(-1)^{l^\prime}}2\tfrac{2^{d+l^\prime-2 k-1}\Gamma \left(\frac{k}{2}+1\right) \Gamma (d+l^\prime-2) \Gamma \left(\frac{d}{2}+l^\prime-k-1\right) \Gamma \left(\tfrac{d+2 l^\prime-k-1}{2}\right)}{\Gamma (d-1) \Gamma \left(\frac{k+1}{2}\right) \Gamma \left(\frac{d}{2}+l^\prime-1\right) \Gamma (l^\prime-k) \Gamma \left(\frac{d}{2}+l^\prime-\frac{k}{2}-1\right)}\Big[ \cos \left(\tfrac{\pi  d}{2}\right)\\+\frac{1}{2\pi} \sin \left(\tfrac{\pi  d}{2}\right) \left(-H_{\frac{d}{2}+l^\prime-\frac{k}{2}-2}+2 H_{\frac{d}{2}+l^\prime-k-2}+H_{\frac{1}{2} (d+2 l^\prime-k-3)}\right.\\\left.-2 H_{l^\prime-k-1}-H_{\frac{k}{2}}+H_{\frac{k-1}{2}}+\log (16)-\tfrac{2}{d+2 l^\prime-2 k-4}\right)\Big]\,c_{\Phi\Phi\mathcal{O}_{l^\prime}}.
\end{multline}
For example, for $k=0$ (conserved currents) we obtain:
\begin{multline}
    \delta\gamma^{[\Phi\Phi]}_{l^\prime,0|0,0}=-\frac{1+(-1)^{l^\prime}}2\frac{2^{d+J-1} \Gamma \left(\frac{d-1}{2}+J\right) \Gamma (d+J-2)}{\sqrt{\pi } \Gamma (d-1) \Gamma (J) \Gamma \left(\frac{d}{2}+J-1\right)}\,\\ \times \left[\frac{1}{\pi}\sin \left(\frac{\pi  d}{2}\right)\left(H_{d+2 l^\prime-3}-H_{l^\prime-1}-\frac{1}{d+2 J-4}\right)+\cos \left(\frac{\pi  d}{2}\right)\right]\,c_{\Phi\Phi\mathcal{O}_{l^\prime}}\,.
\end{multline}
Figure \ref{fig:Gammak0l246} exhibits nice positivity properties in low dimensions which hold also for $k>0$.
Note that in this case $\Delta=d-2+s-k>\tfrac{d}2$ for all $0\leq k\leq s-1$ and therefore for all partially massless points when $d> 2$. Therefore, in our conventions, $\delta\gamma\equiv\gamma_{UV}-\gamma_{IR}$. In this case for $l=0$ and arbitrary $l^\prime$ we obtain:
\begin{subequations}
\begin{align}
    \delta\gamma^{[\Phi\Phi]}_{l^\prime,k=0|0,0}\Big|_{d=2}&=\frac{1+(-1)^{l^\prime}}{2}\,\frac{2^{l^\prime+1} \Gamma \left(l^\prime+\frac{1}{2}\right)}{\sqrt{\pi } \Gamma (l^\prime)}\, c_{\Phi\Phi \mathcal{J}_{l^\prime}}\,,\\
    \delta\gamma^{[\Phi\Phi]}_{l^\prime,k=0|0,0}\Big|_{d=4}&=-\frac{1+(-1)^{l^\prime}}{2}\,\frac{2^{l^\prime+2} l^\prime (l^\prime+1) \Gamma \left(l^\prime+\frac{3}{2}\right)}{\sqrt{\pi }\, l^\prime!}\, c_{\Phi\Phi \mathcal{J}_{l^\prime}}\,.
\end{align}
\end{subequations}
Similar results can be straightforwardly obtained for arbitrary $l$ by employing the explicit form of the crossing kernel in terms of the ${}_4F_3$.
It would be interesting to study the implications of this result in $d=2$ to the case of the $T\bar{T}$ deformation, which was introduced recently in \cite{Smirnov:2016lqw}.

\subsection{Spinning correlators}
\label{subsec::spinningcorrelators}

In this section we consider 4pt correlators of spinning operators under a double trace flow, and the corresponding anomalous dimensions of totally symmetric double-trace operators induced at ${\cal O}\left(1/N\right)$. We shall focus on examples of spinning correlators of the type $J$-$0$-$J$-$0$, looking at CPWs with exchanged scalars only -- i.e. double-trace flows induced by the perturbation \eqref{dtflow} with $l^\prime=0$. We also do not discuss explicitly subleading twist operators. None-the-less, the extension of these results to include these more general cases does not pose conceptual obstacle following the same steps as in the previous section \S \tcb{\ref{0000corr}} for the external scalar case, in which these more general cases where considered.

Let us note here that the CPWEs of spinning 4pt correlators generally receive contributions from exchanges of mixed-symmetry representations. In the $J$-$0$-$J$-$0$ cases considered in the following, under the double-trace flow such contributions are from the exchange of two-row mixed-symmetry double-trace operators. We label mixed symmetry representations in the manifestly symmetric representation by the notation:
\begin{equation}
    (l_1,l_2)=\begin{matrix}\includegraphics{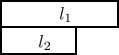}\end{matrix}\,,
\end{equation}
and refer to Appendix \ref{Mixed} for some details about the OPE structures contributing in this case.
The CPW of a spin $\left(l_1,l_2\right)$ double-trace operator $\left[J_1J_2\right]$ is then represented by ${\cal F}^{\left[J_1J_2\right]}_{\left(l_1,l_2\right)}$. See \cite{Costa:2014rya,Rejon-Barrera:2015bpa,Costa:2016hju,Costa:2016xah,Karateev:2017jgd} on conformal partial waves for the exchange of mixed symmetry representations in position space. 

\subsubsection{1010 correlator}\label{1010corr}

In this section we consider the correlator
\begin{equation}
    \left\langle J(y_1)\Phi(y_2) J(y_3)\Phi(y_4)\right\rangle, \label{JphiJphi}
\end{equation}
involving a spin-$1$ operator $J$ of twist $\tau_1$ and a scalar $\Phi$ operator of twist $\tau_2$. After determining the mean field theory ${\sf s}$-channel OPE coefficients in the following section, we extract the anomalous dimensions of the totally symmetric $\left[J\Phi\right]_{0,l}$ double-trace operators induced by the double-trace flow.\footnote{The anomalous dimensions of $\left[JJ\right]$ and $\left[\Phi\Phi\right]$ double-trace operators cannot be extracted from the above correlator alone at ${\cal O}\left(1/N\right)$, since they do not contribute to the mean field theory part.}

\paragraph{Mean field theory OPE coefficients}

\begin{figure}
    \centering
    \includegraphics[width=0.70\textwidth]{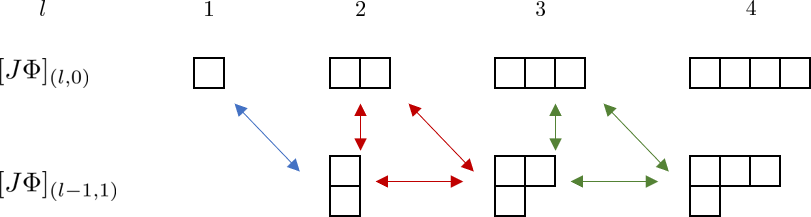}
    \caption{List of representations contributing in the ${\sf s}$-channel for the correlator $\left\langle J\Phi J\Phi\right\rangle$.}
    \label{fig:Jphitableaux}
\end{figure}

The mean-field theory part of the correlator \eqref{JphiJphi} is very simple and reads:
\begin{equation}\label{mean1010}
    \mathcal{A}^{(0)}_{1010}=2\,W_{13}W_{31}\,u^{\tau_1+\tau_2}\,.
\end{equation}
In this case, there are two possible representations which may contribute in the ${\sf s}$-channel for each spin $l$: totally symmetric $\left(l,0\right)$ and $\left(l-1,1\right)$ hook. To extract the OPE coefficients, we find that it is convenient to work in the following basis of CPWs:
\begin{subequations}\label{CPWtilde}
\begin{align}
    {}_1\widetilde{\mathcal{F}}^{[J\Phi]}_l(s,t|W_{ij})&=\mathcal{F}^{[J\Phi]}_{(l,0)}(s,t|W_{ij})+\alpha_l\,\mathcal{F}^{[J\Phi]}_{(l,1)}(s,t|W_{ij})+\beta_l\,\mathcal{F}^{[J\Phi]}_{(l-1,1)}(s,t|W_{ij})\,,\\
    {}_2\widetilde{\mathcal{F}}^{[J\Phi]}_l(s,t|W_{ij})&=\mathcal{F}^{[J\Phi]}_{(l-1,1)}(s,t|W_{ij})\,,
\end{align}
\end{subequations}
which includes a linear combination CPWs for the exchange of totally symmetric and mixed symmetry operators of different spin. I.e. this basis is not diagonal in spin. We pick a basis of this type to automatically cancel all tensor structures which do not appear in the mean-field theory correlator. In particular, we fix the coefficients $\alpha_l$ and $\beta_l$ above requiring that the dependence of ${}_1\widetilde{\mathcal{F}}^{[J\Phi]}_l(s,t|W_{ij})$ on $W_{ij}$ is the same as that in the mean-field theory correlator \eqref{mean1010}. This gives the solution:
\begin{subequations}
\begin{align}
    \alpha_l&=\frac{l+1}{l}\frac{2 (l+\tau_2-1)^2 (l+\tau_1+\tau_2) (2 l+\tau_1+\tau_2+1)}{(l+\tau_1) (l+\tau_2) (l+\tau_1+\tau_2-1) (2 l+\tau_1+\tau_2-1) (-d+l+\tau_1+\tau_2+2)}\,,\\
    \beta_l&=1\,.
\end{align}
\end{subequations}
Going to Mellin space and focusing on the leading twist contributions, the corresponding kinematic polynomials for this basis
\begin{equation}
  {}_q\widetilde{\mathcal{Q}}_{l,\tau_1+\tau_2,0}(s|W_{ij})=-\tfrac12\text{Res}_{t=\tau_1+\tau_2}\left[\rho_{\{\tau_i\}}(s,t)\,{}_q\widetilde{\mathcal{F}}_l(s,t|W_{ij})\right]\,,
\end{equation}
are given by:
\begin{subequations}
\begin{align}
    {}_1\widetilde{{\cal Q}}_{l,\tau_1+\tau_2,0}(s|W_{ij})&=[2W_{13}W_{31}]\,\mathcal{A}_{l}\,\left(\tfrac{-s+\tau_1-\tau_2}2\right)_2\,Q_{l-1}^{(\tau_1+\tau_2,\tau_1+\tau_2,-\tau_1+\tau_2,\tau_1-\tau_2+4)}(s)\,\\ \label{2qtilde}
    {}_2\widetilde{{\cal Q}}_{l,\tau_1+\tau_2,0}(s|W_{ij})&=W_{13}W_{31}\,\mathcal{B}^{(1)}_{l}\left(\tfrac{-s-\tau_1+\tau_2}{2}\right)_1\left(\tfrac{-s+\tau_1-\tau_2}{2}\right)_2Q_{l-2}^{(\tau_1+\tau_2,\tau_1+\tau_2,-\tau_1+\tau_2+2,\tau_1-\tau_2+4)}(s)\\\nonumber
    &+\left(W_{13} W_{32} +W_{14} W_{31}-W_{14} W_{32}\right)\\\nonumber
    &\hspace{50pt}\times\,\mathcal{B}^{(2)}_{l}\,\left(\tfrac{-s+\tau_1-\tau_2}{2}\right)_1\,\left(\tfrac{s+\tau_1+\tau_2}{2}\right)_1^2\,Q_{l-2}^{(\tau_1+\tau_2+2,\tau_1+\tau_2+2,-\tau_1+\tau_2,\tau_1-\tau_2+2)}(s)\,,
\end{align}
\end{subequations}
in terms of the continuous Hahn polynomials and the overall coefficients are:
\begin{subequations}
\begin{align}
  \mathcal{A}_{l}&=(-1)^{l-1}(l-1)!\,\left(\mathfrak{N}_{l-1}^{(\tau_1+\tau_2,\tau_1+\tau_2,-\tau_1+\tau_2,\tau_1-\tau_2+4)}\right)^{-1}\,\\
    \mathcal{B}^{(1)}_{l}&=\tfrac{(-1)^{l-1}}{l}\,(l-1)!\,\left(\mathfrak{N}_{l-2}^{(\tau_1+\tau_2,\tau_1+\tau_2,-\tau_1+\tau_2+2,\tau_1-\tau_2+4)}\right)^{-1}\\
    \mathcal{B}^{(2)}_{l}&=\tfrac{(-1)^{l}}l\,(l-1)!\,\left(\mathfrak{N}_{l-2}^{(\tau_1+\tau_2+2,\tau_1+\tau_2+2,-\tau_1+\tau_2,\tau_1-\tau_2+2)}\right)^{-1}\,.
\end{align}
\end{subequations}
The above kinematic polynomials allow to straightforwardly extract the OPE coefficients of the leading twist double-trace operators $[J\Phi]$ along the same lines as in \S \tcb{\ref{OPEJ1J2J3J4}} using the orthogonality of the continuous Hahn polynomials. In particular, the mean field theory part \eqref{mean1010} just receives contributions from ${}_1\widetilde{{\cal Q}}_{l,\tau_1+\tau_2,0}(s|W_{ij})$, so we may expand:
\begin{align}
    \mathcal{A}_{1010}^{(0)}&=(2W_{13}W_{31})u^{\tau_1+\tau_2}\int_{-i\infty}^{i\infty}\frac{ds}{2\pi i}\,v^{-(s+\tau_1+\tau_2)/2}\,\delta(s+\tau_1+\tau_2)\\ \nonumber &=u^{\tau_1+\tau_2}\int_{-i\infty}^{i\infty}\frac{ds}{2\pi i}\,v^{-(s+\tau_1+\tau_2)/2}\,\tilde{\rho}_{\{\tau_i\}}(s,\tau_1+\tau_2)\sum_l\,{}_1 a_l\,{}_1\widetilde{\mathcal{Q}}_{l,\tau_1+\tau_2}(s|W_{ij})\,.
\end{align}
The above expressions can then be inverted using the orthogonality of the continuous Hahn polynomials, which leads to the following expressions for the coefficients ${}_1 a_l$:
\begin{multline}
    {}_1 a_l=\frac{(-1)^{l-1}}{(l-1)!}\int_{-i\infty}^{i\infty}\frac{ds}{2\pi i}\,\delta(s+\tau_1+\tau_2)\,Q_{l-1}^{(\tau_1+\tau_2,\tau_1+\tau_2,-\tau_1+\tau_2,\tau_1-\tau_2+4)}(s)\\=\frac{(-1)^{l-1}}{(l-1)!}\,Q_{l-1}^{(\tau_1+\tau_2,\tau_1+\tau_2,-\tau_1+\tau_2,\tau_1-\tau_2+4)}(-\tau_1-\tau_2)=\frac{2^{l-1} (\tau_1+2)_{l-1} (\tau_2)_{l-1}}{(l-1)! (l+\tau_1+\tau_2)_{l-1}}
\end{multline}
From above OPE coefficients for the non-diagonal basis \eqref{CPWtilde} one can automatically read off the individual OPE coefficients of the totally symmetric and hook representations from the r.h.s of \eqref{CPWtilde} as
\begin{align}
    a_{(l,0)}^{[J\Phi]}&={}_1 a_l\,,& a_{(l-1,1)}^{[J\Phi]}\Big|_{l\geq2}&=\beta_l\,({}_1 a_l)+\alpha_{l-1}\,({}_1 a_{l-1})\,,
\end{align}
which gives:
\begin{subequations}\label{opejphi}
\begin{align}
    a_{(l,0)}^{[J\Phi]}&=\frac{2^{l-1} (\tau_1+2)_{l-1} (\tau_2)_{l-1}}{(l-1)! (l+\tau_1+\tau_2)_{l-1}}\,,\\
    a_{(l-1,1)}^{[J\Phi]}&=a_{(l,0)}^{[J\Phi]}\frac{2 l+\tau_1+\tau_2-2}{(l+\tau_1-1) (l+\tau_1)}\\\nonumber&\hspace{30pt}\times\left(\frac{l (l+\tau_2-2) (2 l+\tau_1+\tau_2-1)}{(l+\tau_2-1) (l+\tau_1+\tau_2-2) (l+\tau_1+\tau_2+1-d)}+\frac{(l+\tau_1-1) (l+\tau_1)}{2 l+\tau_1+\tau_2-2}\right)\,.
\end{align}
\end{subequations}

\paragraph{Anomalous dimensions}

With the mean field theory OPE coefficients \eqref{opejphi} of the leading twist $\left[J\Phi\right]$ double-trace operators, we can now extract their anomalous dimensions from the crossing kernels of ${\sf t}$- and ${\sf u}$-channel CPWs for the scalar single-trace operator ${\cal O}$ of twist $\tau$ in the double-trace perturbation \eqref{dtflow} (see e.g. \eqref{gammacrosslog}). Only the ${\sf u}$-channel CPWs contribute if we assume that $\mathcal{O}$ is not charged under $U(1)$. In this case, the contribution from the $\sf{t}$-channel in the difference \eqref{cpt} is proportional to $c_{J\Phi\mathcal{O}}=0$. 

For $J$-$0$-$J$-$0$ correlators, the ${\sf u}$-channel CPW with a scalar internal leg is unique. In equation \eqref{cpt} it yields the following $\log$ contribution of leading twist double-trace operators in the ${\sf s}$-channel:
\begin{equation}\label{opeuchu1010}
   c_{JJ\mathcal{O}}\,c_{\Phi\Phi\mathcal{O}} {}^{(\sf{u})}\mathcal{F}^{\bf{0,0}}_{\tau,0}(u,v|W_{ij})= u^{\tau_1+\tau_2}\log u\,f(v|W_{ij})\,+{\cal O}\left(u^{\tau_1+\tau_2+1}\right),
\end{equation}
with OPE coefficients $c_{JJ\mathcal{O}}\,$ and $c_{\Phi\Phi\mathcal{O}}$, and $f(v|W_{ij})$ in Mellin space reads:
\begin{equation}
    f(v|W_{ij})=c_{JJ\mathcal{O}}\,c_{\Phi\Phi\mathcal{O}}\,\int_{-i\infty}^{i\infty}\frac{ds}{4\pi i}\,v^{-(s+\tau_1+\tau_2)/2}\,\rho_{\{\tau_i\}}(s,\tau_1+\tau_2)\mathfrak{M}^{(\sf{u})}_{1010}(s,\tau_1+\tau_2|W_{ij})\,,
\end{equation}
where in this case we have:
\begin{align}\label{mu1010}
    \mathfrak{M}&_{1010}^{(\sf{u})}(s,\tau_1+\tau_2)=C_1(2W_{13}W_{31})+C_2\left[W_{14}W_{31}+W_{13}W_{32}-W_{14}W_{32}\right]\,,
\end{align}
with 
\begin{subequations}
\begin{align}
    C_i&=\tfrac{\Gamma \left(\frac{-s+\tau -\tau_1-\tau_2}{2}\right) \Gamma \left(\frac{d-s-\tau -\tau_1-\tau_2}{2}\right)}{\Gamma \left(\frac{-s+\tau_1-\tau_2}{2}\right) \Gamma \left(\frac{-s-\tau_1+\tau_2}{2}\right)}\mathcal{Z}_\tau\,c_i,\\
     \mathcal{Z}_\tau&=\frac{\Gamma (\tau ) \Gamma \left(\tfrac{d-\tau +\tau_1-\tau_2}{2}\right) \Gamma \left(\tfrac{d-\tau -\tau_1+\tau_2}{2}\right)}{\Gamma \left(\tfrac{\tau }{2}\right)^2 \Gamma \left(\tfrac{d}{2}-\tau \right) \Gamma \left(\tfrac{d-\tau }{2}\right)^2 \Gamma \left(\tfrac{\tau +\tau_1-\tau_2}{2}\right) \Gamma \left(\tfrac{\tau -\tau_1+\tau_2}{2}\right)},\\
      c_1&=-\frac{s^2+2 s (\tau_1+\tau_2-1)+2 \tau +(\tau_1+\tau_2-2) (\tau_1+\tau_2)}{\tau ^2}-\frac{2}{-\tau +\tau_1+\tau_2},\\
    c_2&=\frac{(s+\tau_1+\tau_2)^2}{\tau ^2}.
\end{align}
\end{subequations}

Like in the previous examples, the anomalous dimensions are the coefficients in the expansion of \eqref{mu1010} in terms of the kinematic polynomials \eqref{kpolyq0}:\footnote{Here we fixed the signs to reproduce $\mathcal{F}_{IR}-\mathcal{F}_{UV}$ with $\tau<\tfrac{d}2$.}
\begin{multline}\label{cmuqexp1010}
   c_{JJ\mathcal{O}}\,c_{\Phi\Phi\mathcal{O}}\, \mathfrak{M}^{(\sf{u})}(s,\tau_1+\tau_2|W_{ij})=\frac12\sum_{l=1}^\infty\left[a_{0,l}^{[J\Phi]}\left(\delta\gamma^{[J\Phi]}_{(l,0)}\mathcal{Q}_{(l,0),\tau_1+\tau_2,0}(s|W_{ij})\right.\right.\\\left.\left.+\alpha_l\,\delta\gamma^{[J\Phi]}_{(l,1)}\mathcal{Q}_{(l,1),\tau_1+\tau_2}(s|W_{ij})+\beta_l\delta\gamma^{[J\Phi]}_{(l-1,1)}\mathcal{Q}_{(l-1,1),\tau_1+\tau_2,0}(s|W_{ij})\right)\right]\,.
\end{multline}
I.e. they are given by the crossing kernels of the ${\sf u}$-channel CPW \eqref{opeuchu1010} onto the leading twist double-trace CPWs in the ${\sf s}$-channel. This in particular involves the projection onto CPWs of the mixed-symmetry type, which was not considered previously in \S \tcb{\ref{Crossing Kernels}}. To extract the coefficients we employ the non-diagonal basis \eqref{CPWtilde}, in terms of which \eqref{cmuqexp1010} reads:
\begin{multline}\label{10101basisenw}
    c_{JJ\mathcal{O}}\,c_{\Phi\Phi\mathcal{O}}\,\mathfrak{M}^{(\sf{u})}(s,\tau_1+\tau_2|W_{ij})=\sum_{l=1}^\infty\Bigg\{\underbrace{\frac12a_{0,l}^{[J\Phi]}\delta\gamma^{[J\Phi]}_{(l,0)}}_{p^{(1)}_l}{}_1\widetilde{\mathcal{Q}}_{l,\tau_1+\tau_2}(s|W_{ij})\\+\underbrace{\frac12a_{0,l}^{[J\Phi]}\left[\alpha_l(\delta\gamma_{l,1}^{[J\Phi]}-\delta\gamma_{l,0}^{[J\Phi]})+\beta_{l+1}(\delta\gamma_{l,1}^{[J\Phi]}-\delta\gamma_{l+1,0}^{[J\Phi]})\right]}_{p^{(2)}_l}{}_2\widetilde{\mathcal{Q}}_{l,\tau_1+\tau_2}(s|W_{ij})\Bigg\}.
\end{multline}
Diagram \ref{fig:Jphitableaux} shows the set of operators which mix at a given $l$. It is now straightforward to use the orthogonality of continuous Hahn polynomials to extract $p_l^{(1)}$ and $p_l^{(2)}$. We first extract $p^{(2)}_l$ by restricting to the tensor structure $W_{14}W_{31}+W_{13}W_{32}-W_{14}W_{32}$ which appears only in the kinematic polynomial ${}_2\widetilde{\mathcal{Q}}_{l,\tau_1+\tau_2}(s|W_{ij})$ (see \eqref{2qtilde}):
\begin{multline}\label{pl2int}
    p_l^{(2)}=\frac{(-1)^l\,l}{(l-1)!}\int\frac{ds}{4\pi i}\,\rho_{\{\tau_i\}}(s,\tau_1+\tau_2)\left[\mathfrak{M}^{(\sf{u})}(s,\tau_1+\tau_2|W_{ij})\right]_{W_{14}W_{31}+W_{13}W_{32}-W_{14}W_{32}}\\\times\,Q_{l-2}^{(\tau_1+\tau_2+2,\tau_1+\tau_2+2,-\tau_1+\tau_2,\tau_1-\tau_2+2)}(s)\,.
\end{multline}
From this we can then extract $p_l^{(1)}$ by focusing on the remaining structure $W_{13}W_{31}$, which is present in both ${}_2\widetilde{\mathcal{Q}}_{l,\tau_1+\tau_2,0}$ and ${}_1\widetilde{\mathcal{Q}}_{l,\tau_1+\tau_2,0}$:
\begin{multline}\label{p1inver}
    p_l^{(1)}=\frac{2(-1)^{l-1}l}{(l-1)!}\int\frac{ds}{4\pi i}\,\rho_{\{\tau_i\}}(s,\tau_1+\tau_2)\left[\mathfrak{M}^{(\sf{u})}(s,\tau_1+\tau_2|W_{ij})-\sum_{l^\prime=1}^\infty p_{l^\prime}^{(2)}{}_2\widetilde{{\cal Q}}_{l^\prime,\tau_1+\tau_2,0}(s|W_{ij})\right]_{2W_{13}W_{31}}\\\times\,Q_{l-2}^{(\tau_1+\tau_2,\tau_1+\tau_2,-\tau_1+\tau_2+2,\tau_1-\tau_2+4)}(s)\,,
\end{multline}
where we subtracted the contributions from the ${}_2\widetilde{\mathcal{Q}}_{l,\tau_1+\tau_2,0}$. Carrying out the algebra and performing the integral in \eqref{pl2int} using \eqref{u-channel4F3}, for $p_l^{(2)}$ we obtain:
\begin{multline}
    p_l^{(2)}=p_2^{(2)}\,a_{(l,0)}^{[J\Phi]}\,\frac{l (\tau_1+2) (2 l+\tau_1+\tau_2-2)}{(l+\tau_1) (\tau_1+\tau_2+2)} {}_4F_3\left(\begin{matrix}2-l,\frac{d-\tau}{2}+1,\frac{\tau }{2}+1,l+\tau_1+\tau_2\\\frac{d}{2}+2,\tau_1+2,\tau_2+1\end{matrix};1\right)\,,
\end{multline}
where
\begin{equation}
    p_2^{(2)}=\frac{\Gamma \left(\tfrac{\tau}{2}\right)^2 \Gamma \left(\tfrac{d-\tau +2}{2}\right)^2}{\Gamma \left(\tfrac{d+4}{2}\right)}\,\mathcal{Z}_\tau\,.
\end{equation}

\begin{figure}[t]
    \centering
    \includegraphics[width=\textwidth]{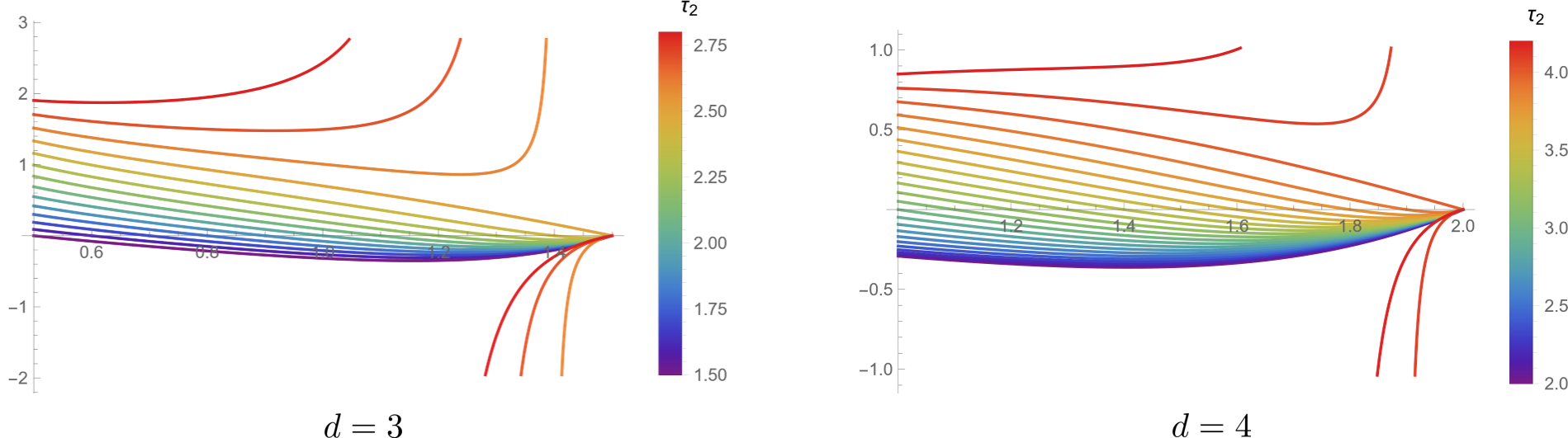}
    \caption{Plot of $\delta \gamma_{(1,0)}^{[J\Phi]}$ (vertical axis) for various values of $\tau_2$ and $\tau_1=d-2$ in $d=3,4$ for $\tau\in [\tfrac{d-2}2,\tfrac{d}2]$ (horizontal axis).}
    \label{GammaJ10d3}
\end{figure}

We can now focus on the tensor structure $2W_{13}W_{31}$ and use the corresponding inversion formula for ${}_1\widetilde{{\cal Q}}_{l,\tau_1+\tau_2}$ (equation \eqref{p1inver}) to extract $p_l^{(1)}$. To evaluate \eqref{p1inver}, for the second term in the bracket we employ the identity:\footnote{This confirms that the degeneracy of operators at fixed $l$ is among operators as shown in figure \ref{fig:Jphitableaux}.}
\begin{multline}
    \int_{-i\infty}^{i\infty}\frac{ds}{4\pi i}\,\rho_{\{\tau_i\}}(s,\tau_1+\tau_2)\left(\tfrac{-s-\tau_1+\tau_2}2\right)_1\left(\tfrac{-s+\tau_1-\tau_2}2\right)_2\\\times\,Q_{l^\prime-1}^{(\tau_1+\tau_2,\tau_1+\tau_2,-\tau_1+\tau_2+2,\tau_1-\tau_2+4)}(s)\,Q_{l-1}^{(\tau_1+\tau_2,\tau_1+\tau_2,-\tau_1+\tau_2,\tau_1-\tau_2+4)}(s)\\ =\mathfrak{N}^{(\tau_1+\tau_2,\tau_1+\tau_2,-\tau_1+\tau_2+2,\tau_1-\tau_2+4)}_{l^\prime-1}\left[\delta_{l^\prime,l}+\frac{2 (l-1) (l+\tau_1)^2}{(2 l+\tau_1+\tau_2-2) (2 l+\tau_1+\tau_2-1)}\delta_{l^\prime,l-1}\right]\,.
\end{multline}
For the first term in the bracket, we bring it into the form of the integral \eqref{int4F3} by decomposing it as: 
\begin{multline}
    \rho_{\{\tau_i\}}(s,\tau_1+\tau_2)\mathfrak{M}^{(\sf{u})}(s,\tau_1+\tau_2|W_{ij})\Big|_{2W_{13}W_{31}}\\=\mathcal{Z}_\tau\sum_{i=0}^2d_i\left(\tfrac{s+\tau_1+\tau_2}{2}\right)_i\,\Gamma \left(\tfrac{s+\tau_1+\tau_2}{2}\right)^2 \Gamma \left(\tfrac{-s+\tau -\tau_1-\tau_2}{2}\right) \Gamma \left(\tfrac{d-s-\tau -\tau_1-\tau_2}{2}\right)\,,
\end{multline}
where the coefficients $d_i$ are $s$-independent:
\begin{equation}
    d_0=\frac{\tau  (\tau_1+\tau_2)}{4 (\tau -\tau_1-\tau_2)}\,, \qquad  d_1=1\,, \qquad d_2=-\frac12\,.
\end{equation}
Combining all terms we find the following explicit expression for the anomalous dimensions of the totally symmetric leading twist double-trace operators of spin-$l$:
\begin{multline}
 \tfrac12 \delta \gamma_{(l,0)}^{[J,\Phi]}= \frac{p_l^{(1)}}{a_{\left(l,0\right)}^{[J\Phi]}}=
    c_{JJ\mathcal{O}}\,c_{\Phi\Phi\mathcal{O}}\Bigg\{\mathcal{Z}_\tau\sum_{i=0}^2d_i\,T_i\\-\left[\frac{2 (l-1) (l+\tau_1)^2}{(2 l+\tau_1+\tau_2-2) (2 l+\tau_1+\tau_2-1)}p_{l}^{(2)}-\frac{l}{l+1}p_{l+1}^{(2)}\right]\Bigg\}\,,
\end{multline}
where the mean field theory OPE coefficient $a_{\left(l,0\right)}$ is given by \eqref{opejphi} and
\begin{multline}
    T_i=\frac{(-1)^{l+1} \Gamma (l) \Gamma \left(\frac{\tau }{2}\right) \Gamma \left(\frac{d}{2}-\frac{\tau }{2}\right) \Gamma \left(i+\frac{\tau }{2}\right) \Gamma \left(\frac{d}{2}+i-\frac{\tau }{2}\right)}{\Gamma \left(\frac{d}{2}+i\right)}\\\times\, _4F_3\left(\begin{matrix}1-l,\frac{d}{2}-\frac{\tau }{2},\frac{\tau }{2},l+\tau_1+\tau_2\\\frac{d}{2}+i,\tau_1+2,\tau_2\end{matrix};1\right)\,.
\end{multline}
The simplest form is given by $l=1$:
\begin{equation}\label{gamma1Jphi}
    \delta \gamma_{(1,0)}^{[J\Phi]}=c_{JJ\mathcal{O}}\,c_{\Phi\Phi\mathcal{O}}\frac{(d-\tau +\tau_1+\tau_2)}{2(\tau -\tau_1-\tau_2)}\frac{2\Gamma (\tau ) \Gamma \left(\tfrac{d-\tau +\tau_1-\tau_2}{2}\right) \Gamma \left(\tfrac{d-\tau -\tau_1+\tau_2}{2}\right)}{\Gamma \left(\tfrac{d}{2}+1\right) \Gamma \left(\tfrac{d}{2}-\tau \right) \Gamma \left(\tfrac{\tau +\tau_1-\tau_2}{2}\right) \Gamma \left(\tfrac{\tau -\tau_1+\tau_2}{2}\right)}\,,
\end{equation}
which we represent explicitly in figure \ref{GammaJ10d3} for $d=3$ and for $d=4$ and various increasing choices of $\tau$ and $\tau_2$ while we fix $\tau_1=d-2$ which corresponding to a conserved current.

More simply the general $l$ result can be written in this case as:

\subsubsection{2020 correlators} \label{2020corr}

In this section we consider the correlator
\begin{equation}
    \langle T\left(y_1\right)\Phi\left(y_2\right)T\left(y_3\right) \Phi\left(y_4\right) \rangle 
\end{equation}
involving a spin $2$ operator $T$ of twist $\tau_1$ and a scalar operator $\Phi$ of twist $\tau_2$. 

\paragraph{Mean field theory OPE coefficients}

\begin{figure}[t]
    \centering
    \includegraphics[width=0.95\textwidth]{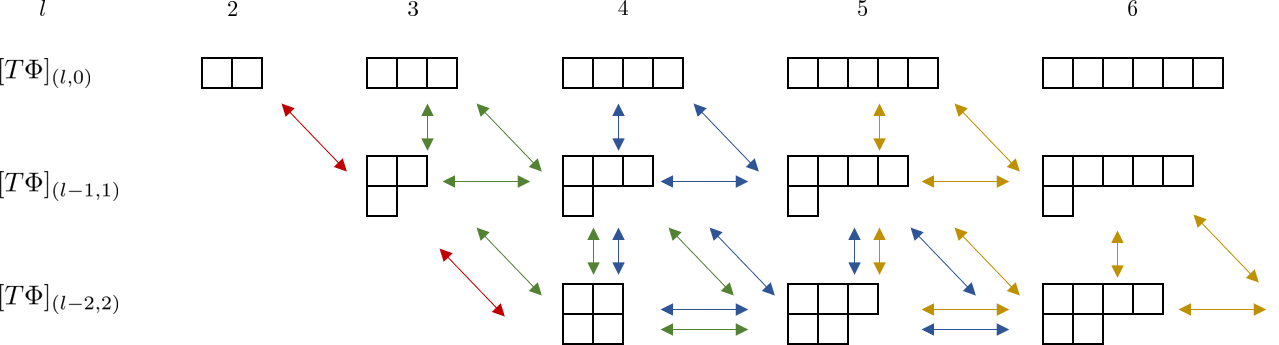}
    \caption{Double-trace operators contributing to the $\left\langle T\Phi T\Phi\right\rangle$. The colored triangles represent the operator which mix with a given totally symmetric operator.}
    \label{fig:Tphi}
\end{figure}

We start again from the mean-field theory contribution, which reads:
\begin{equation}
    \mathcal{A}_{2020}=(2W_{13}W_{31})^2\,u^{\frac{\tau_1+\tau_2}2}\,,
\end{equation}
where we work with arbitrary $\tau_1$ which can then be set to $d-2$ for the case of the stress tensor. In this case the list of conformal blocks exchanged in the ${\sf s}$-channel includes the representations $(l,0)$, $(l-1,1)$ and $(l-2,2)$ CPWs, see fig.~\ref{fig:Tphi}.

Generalising the $1010$ case considered in the previous section, we pick the following basis of CPWs (see figure \ref{fig:Tphi}):
\begin{align}\label{lowspin2020basis}
    {}_q\widetilde{\mathcal{F}}^{[T\Phi]}_l(s,t|W_{ij})&=\sum_{p=0}^{2-q}\sum_{k=0}^p\alpha^{q}_{p,k}\mathcal{F}^{[T\Phi]}_{(l-p-q+k+1,p+q-1)}(s,t|W_{ij})\,,
\end{align}
which is non-diagonal in spin and $q=1,2,3$. We fix $\alpha_{0,0}^q=1$, which is the coefficient of the CPW for the exchange of totally symmetric representations $\left(l,0\right)$. As before, we fix all remaining coefficients above by requiring that for each $q$ the terms not proportional to $(2W_{13}W_{31})^{3-q}$ are cancelled. Since we focus on extracting the OPE data of totally symmetric double-trace operators, the precise explicit result for the other coefficients $\alpha_{p,k}^q$ is not important and we will not discuss them in the following. 

The corresponding kinematic polynomials for leading twist double-trace operators are given by
\begin{equation}
 {}_q\widetilde{\mathcal{Q}}_{l,\tau_1+\tau_2,0}(s|W_{ij})= -\tfrac12\text{Res}_{t=\tau_1+\tau_2}\left[\rho_{\{\tau_i\}}(s,t)\,{}_q\widetilde{\mathcal{F}}_l(s,t|W_{ij})\right]\,,
\end{equation}
with 
\begin{subequations}
\begin{align}\label{l0polynomial2020}
    {}_1\widetilde{{\cal Q}}_{l,\tau_1+\tau_2,0}(s|W_{ij})&=[2W_{13}W_{31}]^2\,\mathcal{A}_{l}\,\left(\tfrac{-s+\tau_1-\tau_2}2\right)_{4}\,Q_{l-2}^{(\tau_1+\tau_2,\tau_1+\tau_2,-\tau_1+\tau_2,\tau_1-\tau_2+8)}(s)\,,\\
    {}_2\widetilde{\mathcal{Q}}_{l,\tau_1+\tau_2,0}(s|W_{ij})&=\mathcal{B}_l^{(1)}(W_{13}W_{31})^2
    \\\nonumber
    &\hspace{20pt}\times\,\left(\tfrac{-s-\tau_1+\tau_2}{2}\right)_1 \left(\tfrac{-s+\tau_1-\tau_2}{2}\right)_4Q_{l-3}^{(\tau_1+\tau_2,\tau_1+\tau_2,-\tau_1+\tau_2+2,\tau_1-\tau_2+8)}(s)
    \\\nonumber
    &+\mathcal{B}_l^{(2)}W_{13}W_{31}(W_{14}W_{31}+W_{13}W_{32}-W_{14}W_{32})
    \\\nonumber
    &\hspace{20pt}\times\,\left(\tfrac{s+\tau_1+\tau_2}{2}\right)_1^2 \left(\tfrac{-s+\tau_1-\tau_2}{2}\right)_3Q_{l-3}^{(\tau_1+\tau_2+2,\tau_1+\tau_2+2,-\tau_1+\tau_2,\tau_1-\tau_2+6)}(s)\,,
    \\ \label{Q3expre}
    {}_3\widetilde{\mathcal{Q}}_{l,\tau_1+\tau_2,0}(s|W_{ij})&=\mathcal{C}_l^{(1)}(W_{13}W_{31})^2\,\times
    \\\nonumber
    &\hspace{20pt}\times\,\left(\tfrac{-s-\tau_1+\tau_2}{2}\right)_2 \left(\tfrac{-s+\tau_1-\tau_2}{2}\right)_4Q_{l-4}^{(\tau_1+\tau_2,\tau_1+\tau_2,-\tau_1+\tau_2+4,\tau_1-\tau_2+8)}(s)
    \\\nonumber
    &+\mathcal{C}_l^{(2)}W_{13}W_{31}(W_{14}W_{31}+W_{13}W_{32})
    \\\nonumber
    &\hspace{20pt}\times\,\left(\tfrac{s+\tau_1+\tau_2}{2}\right)_1^2\left(\tfrac{-s-\tau_1+\tau_2}{2}\right)_1 \left(\tfrac{-s+\tau_1-\tau_2}{2}\right)_3Q_{l-4}^{(\tau_1+\tau_2+2,\tau_1+\tau_2+2,-\tau_1+\tau_2+2,\tau_1-\tau_2+6)}(s)
    \\\nonumber
    &+\mathcal{C}_l^{(3)}(W_{14}^2 W_{32}^2-2 W_{13} W_{14} W_{32}^2+W_{13}^2 W_{32}^2-2 W_{14}^2 W_{31} W_{32}+W_{14}^2 W_{31}^2)
    \\\nonumber
    &\hspace{20pt}\times\,\left(\tfrac{s+\tau_1+\tau_2}{2}\right)_2^2\left(\tfrac{-s+\tau_1-\tau_2}{2}\right)_2Q_{l-4}^{(\tau_1+\tau_2+4,\tau_1+\tau_2+4,-\tau_1+\tau_2,\tau_1-\tau_2+4)}(s)
    \\\nonumber
    &+\ldots
\end{align}
\end{subequations}
where the $\ldots$ in \eqref{Q3expre} represents the tensor structure $W_{13}W_{31}W_{14}W_{32}$ which is fixed by conformal invariance and does not play a role in the following discussion. For this reason we do not need to write it down explicitly. The coefficients are given by:
\begin{subequations}
\begin{align}
    \mathcal{A}_{l}&=(-1)^{l-2}(l-2)!\left(\mathfrak{N}_{l-2}^{(\tau_1+\tau_2,\tau_1+\tau_2,-\tau_1+\tau_2,\tau_1-\tau_2+8)}\right)^{-1}\,,\\
    \mathcal{B}_l^{(1)}&=(-1)^{l-2} (l-3)!\frac{2 (l-2)^2}{l (l-1)}\,\left(\mathfrak{N}_{l-3}^{(\tau_1+\tau_2,\tau_1+\tau_2,-\tau_1+\tau_2+2,\tau_1-\tau_2+8)}\right)^{-1},\\
    \mathcal{B}_l^{(2)}&=(-1)^{l-3} (l-3)!\frac{2 (l-2)^2}{l (l-1)}\,\left(\mathfrak{N}_{l-3}^{(\tau_1+\tau_2+2,\tau_1+\tau_2+2,-\tau_1+\tau_2,\tau_1-\tau_2+6)}\right)^{-1},\\
    \mathcal{C}_l^{(1)}&=(-1)^{l-4} \frac{ (l-3)!}{ l-1}\left(\mathfrak{N}_{l-4}^{(\tau_1+\tau_2,\tau_1+\tau_2,-\tau_1+\tau_2+4,\tau_1-\tau_2+8)}\right)^{-1},\\
    \mathcal{C}_l^{(2)}&=2(-1)^{l-3} \frac{ (l-3)!}{ l-1}\left(\mathfrak{N}_{l-4}^{(\tau_1+\tau_2+2,\tau_1+\tau_2+2,-\tau_1+\tau_2+2,\tau_1-\tau_2+6)}\right)^{-1},\\
    \mathcal{C}_l^{(3)}&=(-1)^{l-4} \frac{(l-3)!}{ l-1}\left(\mathfrak{N}_{l-4}^{(\tau_1+\tau_2+4,\tau_1+\tau_2+4,-\tau_1+\tau_2,\tau_1-\tau_2+4)}\right)^{-1}.
\end{align}
\end{subequations}
In the ${\sf s}$-channel expansion of the mean field theory correlator there are only contributions from ${}_1\widetilde{\mathcal{F}}_l$. Like in all previous examples, we can extract the mean field theory OPE coefficients of the totally symmetric leading twist double-trace operators $[T\Phi]$ using the orthogonality of the continuous Hahn polynomials. We obtain:
\begin{equation}\label{2020mftope}
    a_{(l,0)}^{[T\Phi]}=\frac{2^{l-2} (\tau_1+4)_{l-2} (\tau_2)_{l-2}}{(l-2)! (l+2+\tau_1+\tau_2-1)_{l-2}}\,,
\end{equation}
whose simple form is remarkable. For brevity we will not attempt to obtain here the OPE coefficients of the mixed symmetry double-trace operators, though the approach follows the same lines as for the totally symmetric double-trace operators above.

\paragraph{Anomalous dimensions}

With the result \eqref{2020mftope} for the mean field theory OPE coefficients, we can now extract the anomalous dimensions induced at ${\cal O}\left(1/N\right)$ by the perturbation \eqref{dtflow} with $l^\prime=0$. Like from the $1$-$0$-$1$-$0$ case considered previously, this comes from the crossing kernel of the $\sf{u}$-channel CPW in \eqref{cpt} for the exchange of the scalar single trace operator ${\cal O}$ of twist $\tau$. In this case we have:
\begin{align}\label{2020MellinA}
    \mathfrak{M}&_{2020}^{(\sf{u})}(s,\tau_1+\tau_2)=C_1(2W_{13}W_{31})^2+C_2(2W_{13}W_{31})\left[W_{14}W_{31}+W_{13}W_{32}\right]\\&+C_3\,(2W_{13}W_{31})\,W_{14}W_{32}+C_4\left[W_{14}^2W_{31}^2+W_{13}^2W_{32}^2-2W_{14}^2W_{31}W_{32}-2W_{13}W_{14}W_{32}^2\right]\,,\nonumber
\end{align}
where 
\begin{subequations}
\begin{align}
C_i&=\tfrac{\Gamma \left(\frac{-s+\tau -\tau_1-\tau_2}{2}\right) \Gamma \left(\frac{d-s-\tau -\tau_1-\tau_2}{2}\right)}{\Gamma \left(\frac{-s+\tau_1-\tau_2}{2}\right) \Gamma \left(\frac{-s-\tau_1+\tau_2}{2}\right)}\mathcal{Z}_\tau\,c_i\\
\mathcal{Z}_\tau&=\frac{\Gamma (\tau ) \Gamma \left(\tfrac{d-\tau +\tau_1-\tau_2}{2}\right) \Gamma \left(\tfrac{d-\tau -\tau_1+\tau_2}{2}\right)}{\Gamma \left(\tfrac{\tau }{2}\right)^2 \Gamma \left(\tfrac{d}{2}-\tau \right) \Gamma \left(\tfrac{d-\tau }{2}\right)^2 \Gamma \left(\tfrac{\tau +\tau_1-\tau_2}{2}\right) \Gamma \left(\tfrac{\tau -\tau_1+\tau_2}{2}\right)}\\
    c_1&=\frac{(s-\tau +\tau_1+\tau_2-6) (s-\tau +\tau_1+\tau_2-4) (s-\tau +\tau_1+\tau_2-2) (s-\tau +\tau_1+\tau_2)}{4\tau ^2 (\tau +2)^2}\\\nonumber
    &+\frac{ (s-\tau +\tau_1+\tau_2-4) (s-\tau +\tau_1+\tau_2-2) (s-\tau +\tau_1+\tau_2)}{\tau ^2 (\tau +2)}\\ \nonumber
    &+\frac{ (s-\tau +\tau_1+\tau_2-2) (s-\tau +\tau_1+\tau_2)}{\tau ^2}-\frac{2 (s-\tau +\tau_1+\tau_2-2) (s-\tau +\tau_1+\tau_2)}{\tau ^2 (\tau -\tau_1-\tau_2)}\\ \nonumber
    &+\frac{ (s-\tau +\tau_1+\tau_2-2) (s-\tau +\tau_1+\tau_2)}{2\tau  (\tau +2)}
    +\frac{ (s-\tau +\tau_1+\tau_2)}{\tau }
    +\frac{4 (s-\tau+\tau_1+\tau_2)}{\tau(-\tau +\tau_1+\tau_2)}\\\nonumber
    &+\frac{2}{-\tau +\tau_1+\tau_2}-\frac{2}{(\tau -\tau_1-\tau_2) (-\tau +\tau_1+\tau_2+2)}+\frac14\\
    c_2&=\frac{(s+\tau_1+\tau_2)^2 \left(-\tau  (s+\tau_1+\tau_2)^2+(\tau_1+\tau_2) (s+\tau_1+\tau_2)^2+4 \tau  (\tau_1+\tau_2+3)+4 (\tau_1+\tau_2+4)\right)}{\tau ^2 (\tau +2)^2 (\tau -\tau_1-\tau_2)}\\
    c_3&=\frac{2(s+\tau_1+\tau_2)^2}{\tau ^2} \left(\frac{(s-\tau +\tau_1+\tau_2) (s+\tau +\tau_1+\tau_2+2)}{(\tau +2)^2}+\frac{2}{\tau_1+\tau_2-\tau}+1\right)\\
    c_4&=\frac{(s+\tau_1+\tau_2)^2 (s+\tau_1+\tau_2+2)^2}{\tau ^2 (\tau +2)^2}
\end{align}
\end{subequations}

As before, to extract the anomalous dimensions we expand in the kinematic polynomials of the basis \eqref{lowspin2020basis}:
\begin{equation}\label{ndb20202mexp}
   c_{TT\mathcal{O}}\,c_{\Phi\Phi\mathcal{O}}\, \mathfrak{M}_{2020}^{(\sf{u})}(s,t)=\sum_{l=2}^\infty\left(p_l^{(1)}{}_1\widetilde{{\cal Q}}_{l,\tau_1+\tau_2,0}(s|W_{ij})+p_l^{(2)}{}_2\widetilde{{\cal Q}}_{l,\tau_1+\tau_2,0}(s|W_{ij})+p_l^{(3)}{}_3\widetilde{{\cal Q}}_{l,\tau_1+\tau_2,0}(s|W_{ij})\right)\,,
\end{equation}
where for the totally symmetric double-trace operators we have
\begin{equation}
    \frac12\,a_{(l,0)}^{[T\Phi]}\,\delta\gamma_{(l,0)}^{[T\Phi]}=c_{TT\mathcal{O}}c_{\Phi\Phi\mathcal{O}}\,p_l^{(1)}.
\end{equation}

\begin{figure}[t]
    \centering
    \includegraphics[width=0.9\textwidth]{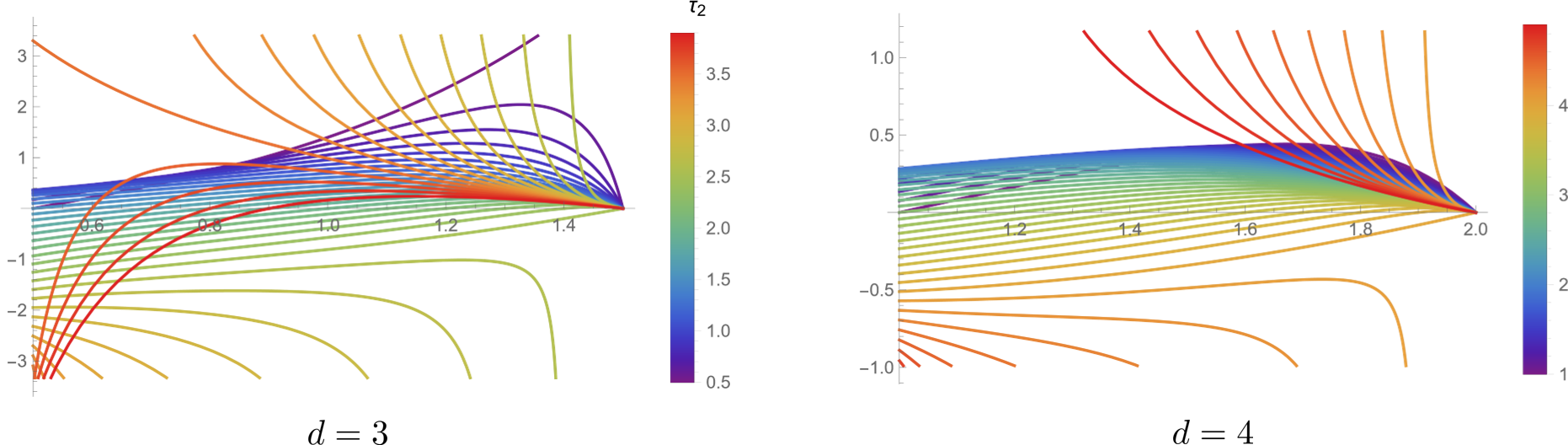}
    \caption{Plot of $\delta\gamma_{(2,0)}^{[T\Phi]}$ (vertical axis) against $\tau$ for $\tau_1=d-2$ and various increasing values of $\tau_2$. The dimension is set to $d=3,4$ respectively.} 
    \label{fig:T20}
\end{figure}

For ease of presentation we just focus here on the simplest case of the anomalous dimension of $[T\Phi]_{(2,0)}$. In this case we just need to disentangle the contributions $[T\Phi]_{(2,0)}$, $[T\Phi]_{(2,1)}$ and $[T\Phi]_{(2,2)}$. The mixing in general is only among terms belonging to the same triangle in figure \ref{fig:Tphi} and in the lowest spin case it corresponds to just the red arrows. The structure of the basis allows to iteratively solve first for the coefficient of ${}_3\widetilde{{\cal Q}}_{4,\tau_1+\tau_2,0}$ by focusing, say, on the tensor structure $W_{14}^2W_{32}^2$ which is not present in ${}_2\widetilde{{\cal Q}}_{3,\tau_1+\tau_2,0}$ or ${}_1\widetilde{{\cal Q}}_{2,\tau_1+\tau_2,0}$. In this way we obtain:
\begin{multline}
    p^{(3)}_4=3\,\int_{-i\infty}^{i\infty}\frac{ds}{4\pi i}\,\tilde{\rho}_{\{\tau_i\}}(s,\tau_1+\tau_2)\,C_4 \,Q_0^{(\tau_1+\tau_2+4,\tau_1+\tau_2+4,-\tau_1+\tau_2,\tau_1-\tau_2+4)}(s)\\=\frac{3\, \Gamma \left(\frac{\tau }{2}\right)^2 \Gamma \left(\frac{d-\tau +4}{2}\right)^2}{\Gamma \left(\frac{d}{2}+4\right)}\,\mathcal{Z}_\tau\,.
\end{multline}
Then one can subtract the contribution of ${}_3\widetilde{{\cal Q}}_{4,\tau_1+\tau_2,0}$ in \eqref{ndb20202mexp} and extract the coefficient of ${}_2\widetilde{{\cal Q}}_3$ from, say, the tensor structure $W_{13}W_{31}W_{14}W_{31}$ which is not present in ${}_1\widetilde{{\cal Q}}_{2,\tau_1+\tau_2,0}$:\footnote{Note that the tensor structure which we have not written explicitly in \eqref{Q3expre} indeed are not needed for this analysis. The consistency of the decomposition of such tensor structures follows by conformal invariance and does not impose new consistency conditions.}
\begin{multline}\label{p3T0}
    p_3^{(2)}=3\int_{-i\infty}^{i\infty}\frac{ds}{4\pi i}\tilde{\rho}_{\{\tau_i\}}(s,\tau_1+\tau_2)\,\left[C_2-p_4^{(3)}{}_3\widetilde{{\cal Q}}_{4,\tau_1+\tau_2,0}\right]_{W_{13}W_{31}W_{14}W_{31}}\\\times Q_3^{(\tau_1+\tau_2+2,\tau_1+\tau_2+2,-\tau_1+\tau_2,\tau_1-\tau_2+6)}(s)=\frac{12\, \Gamma \left(\tfrac{\tau }{2}\right)^2 \Gamma \left(\tfrac{d-\tau +2}{2}\right)^2 (d-\tau +\tau_1+\tau_2+4)}{\Gamma \left(\tfrac{d}{2}+3\right) (\tau -\tau_1-\tau_2)}\,\mathcal{Z}_\tau\,.
\end{multline}
Finally the coefficient of ${}_1\widetilde{{\cal Q}}_{2,\tau_1+\tau_2,0}$ can be extracted from the tensor structure proportional to $W_{13}^2W_{31}^2$, which gives:
\begin{multline}\label{p2T0}
    p_2^{(1)}=\frac14\int_{-i\infty}^{i\infty}\frac{ds}{4\pi i}\,\rho_{\{\tau_i\}}(s,\tau_1+\tau_2)\,\left[C_1-p_3^{(2)}{}_2\widetilde{{\cal Q}}_{3,\tau_1+\tau_2,0}(s|W_{ij})-p_4^{(3)}{}_3\widetilde{{\cal Q}}_{4,\tau_1+\tau_2,0}(s|W_{ij})\right]_{(W_{13}W_{31})^2}\\\times\,Q_{2}^{(\tau_1+\tau_2,\tau_1+\tau_2,-\tau_1+\tau_2,\tau_1-\tau_2+8)}(s)=\frac{(d-\tau +\tau_1+\tau_2+2) (d-\tau +\tau_1+\tau_2+4)}{2(\tau -\tau_1-\tau_2) (\tau -\tau_1-\tau_2-2)}\\\times\,\frac{\Gamma \left(\frac{\tau }{2}\right)^2 \Gamma \left(\frac{d-\tau }{2}\right)^2}{2 \Gamma \left(\frac{d}{2}+2\right)}\,\mathcal{Z}_\tau\,,
\end{multline}
from which we obtain the anomalous dimensions:
\begin{multline}\label{gammaT20}
    \delta\gamma_{(2,0)}^{[T\Phi]}=\frac{(d-\tau +\tau_1+\tau_2+2) (d-\tau +\tau_1+\tau_2+4)}{2(\tau -\tau_1-\tau_2) (\tau -\tau_1-\tau_2-2)}\\\times\,\frac{2\Gamma (\tau ) \Gamma \left(\tfrac{d-\tau +\tau_1-\tau_2}{2}\right) \Gamma \left(\tfrac{d-\tau -\tau_1+\tau_2}{2}\right)}{\Gamma \left(\tfrac{d}{2}+1\right) \Gamma \left(\tfrac{d}{2}-\tau \right) \Gamma \left(\tfrac{\tau +\tau_1-\tau_2}{2}\right) \Gamma \left(\tfrac{\tau -\tau_1+\tau_2}{2}\right)}\,c_{TT\mathcal{O}}c_{\Phi\Phi\mathcal{O}}\,.
\end{multline}
The above with $\tau_1=d-2$ and $\tau_2=\Delta$ gives the anomalous dimensions in the case of double-trace operator built from the stress tensor and a scalar operator of scaling dimension $\Delta$. We plot the result in $d=3$ for various values of $\Delta$ in figure \ref{fig:T20} starting at $\Delta=\tfrac12$.

The extension of the result \eqref{gammaT20} from $l=2$ to arbitrary $l$ can be straightforwardly obtained by following exactly the same steps as above using \eqref{int4F3} and \eqref{QQintegral} but keeping all contributions from the polynomials ${}_i\widetilde{\mathcal{Q}}_{l,\tau_1+\tau_2,0}$ inside a certain triangle in fig.~\ref{fig:Tphi} when considering the corresponding generalisation of \eqref{p3T0} and \eqref{p2T0} to arbitrary $l$. First one determines $p_l^{(3)}$ ($l\geq4$) using \eqref{t-channel4F3}:
\begin{multline}
    p_l^{(3)}=\frac{(-1)^{l-4}(l-1)}{(l-3)!}\int_{-i\infty}^{i\infty}\frac{ds}{4\pi i}\,\tilde{\rho}_{\{\tau_i\}}(s,\tau_1+\tau_2)\,C_4\,Q_{l-4}^{(\tau_1+\tau_2+4,\tau_1+\tau_2+4,-\tau_1+\tau_2,\tau_1-\tau_2+4}(s)\\=\frac{16 (-1)^l (l-1)}{\tau ^2 (\tau +2)^2 (l-3)!}\,\mathcal{Z}_\tau\,\mathfrak{I}_{\tau_1+\tau_2+4,\tau_1+\tau_2+4,d-\tau -\tau_1-\tau_2,\tau -\tau_1-\tau_2}^{\tau_1+\tau_2+4,\tau_1+\tau_2+4,\tau_2-\tau_1,\tau_1-\tau_2+4}(l-4)\,,
\end{multline}
where in this case the sum in \eqref{t-channel4F3} reduces to a single term. We can now extract $p_l^{(2)}$ as ($l\geq3$):
\begin{multline}
    p_l^{(2)}=\frac{(-1)^{l-3}}{(l-2)!}\frac{l(l-1)}{2(l-2)}\,\int_{-i\infty}^{i\infty}\frac{ds}{4\pi i}\,\tilde{\rho}_{\{\tau_i\}}(s,\tau_1+\tau_2)\,\left[\mathfrak{M}_{2020}^{(\sf{u})}-\sum_{l^\prime}p_{l^\prime}^{(3)}\,{}_3\widetilde{\mathcal{Q}}_{l^\prime,\tau_1+\tau_2,0}\right]_{W_{13}W_{31}^2W_{14}}\\\times\,Q_{l-3}^{(\tau_1+\tau_2+2,\tau_1+\tau_2+2,-\tau_1+\tau_2,\tau_1-\tau_2+6)}(s)\,,
\end{multline}
which can be evaluated using \eqref{u-channel4F3} and \eqref{QQintegral} as:
\begin{align}
    p_l^{(2)}&=
    \frac{(-1)^{l-3}}{(l-2)!}\frac{l(l-1)}{2(l-2)}\mathcal{Z}_\tau\sum_{i=0}^2d_i\,\mathfrak{I}_{\tau_1+\tau_2+2,2 i+\tau_1+\tau_2+2,d-\tau -\tau_1-\tau_2,\tau -\tau_1-\tau_2}^{\tau_1+\tau_2+2,\tau_1+\tau_2+2,\tau_2-\tau_1,\tau_1-\tau_2+6}(l-3)\\\nonumber
    &-\frac{l}{(l-2)^2}\,p_l^{(3)}\underbrace{\mathfrak{Z}_{\tau_1+\tau_2+2,\tau_1+\tau_2+2,-\tau_1+\tau_2+2,\tau_1-\tau_2+6}^{\tau_1+\tau_2+2,\tau_1+\tau_2+2,\tau_2-\tau_1,\tau_1-\tau_2+6}(l-4,l-3)}_{\frac{2 (l-3) (l+\tau_1)^2}{(2 l+\tau_1+\tau_2-3) (2 l+\tau_1+\tau_2-2)}}+\frac{1-l}{l-2}\,p_{l+1}^{(3)}\,,\nonumber
\end{align}
with
\begin{align}
    d_0&=-\frac{32 (\tau_1+\tau_2+2)}{\tau ^2 (\tau +2) (-\tau +\tau_1+\tau_2)}\,,& d_1&=\frac{96}{\tau ^2 (\tau +2)^2}\,,& d_2&=-\frac{32}{\tau ^2 (\tau +2)^2}\,.
\end{align}
Finally we can now compute $p_l^{(1)}$ as:
\begin{multline}
    p_l^{(1)}=\frac14\frac{(-1)^{l-2}}{(l-2)!}\int_{-i\infty}^{i\infty}\frac{ds}{4\pi i}\,\tilde{\rho}_{\{\tau_i\}}(s,\tau_1+\tau_2)\,\Bigg[\mathfrak{M}_{2020}^{(\sf{u})}-\sum_{l^\prime}\Big(p_{l^\prime}^{(2)}\,{}_2\widetilde{\mathcal{Q}}_{l^\prime,\tau_1+\tau_2,0}\\+p_{l^\prime}^{(3)}\,{}_3\widetilde{\mathcal{Q}}_{l^\prime,\tau_1+\tau_2,0}\Big)\Bigg]_{W_{13}^2W_{31}^2}\,Q_{l-2}^{(\tau_1+\tau_2,\tau_1+\tau_2,-\tau_1+\tau_2,\tau_1-\tau_2+8)}(s)\,,
\end{multline}
and carrying all integrals we obtain ($l\geq 2$):
\begin{align}
    p_l^{(1)}&=\frac14\frac{(-1)^{l-2}}{(l-2)!}\mathcal{Z}_\tau\sum_{i=0}^4\,e_i\,\mathfrak{I}_{\tau_1+\tau_2,2 i+\tau_1+\tau_2,d-\tau -\tau_1-\tau_2,\tau -\tau_1-\tau_2}^{\tau_1+\tau_2,\tau_1+\tau_2,\tau_2-\tau_1,\tau_1-\tau_2+8}(l-2)\\\nonumber
    &-\frac{l-2}{2 l(l-1) }\,p_l^{(2)}\underbrace{\mathfrak{Z}_{\tau_1+\tau_2,\tau_1+\tau_2,-\tau_1+\tau_2+2,\tau_1-\tau_2+8}^{\tau_1+\tau_2,\tau_1+\tau_2,\tau_2-\tau_1,\tau_1-\tau_2+8}(l-3,l-2)}_{\frac{2 (l-2) (l+\tau_1+1)^2}{(2 l+\tau_1+\tau_2-2) (2 l+\tau_1+\tau_2-1)}}+\frac{(l-1)^2}{2 l (l+1)}\,p_{l+1}^{(2)}\\\nonumber
    &-\frac{1}{4 (l-2) (l-1)}\,p_l^{(3)}\underbrace{\mathfrak{Z}_{\tau_1+\tau_2,\tau_1+\tau_2,-\tau_1+\tau_2+4,\tau_1-\tau_2+8}^{\tau_1+\tau_2,\tau_1+\tau_2,\tau_2-\tau_1,\tau_1-\tau_2+8}(l-4,l-2)}_{\frac{4 (l-2) (l-3) (l+\tau_1)^2 (l+\tau_1+1)^2}{(2 l+\tau_1+\tau_2-3) (2 l+\tau_1+\tau_2-2)^2 (2 l+\tau_1+\tau_2-1)}}\\\nonumber
    &+\frac{1}{4 l}\,p_{l+1}^{(3)}\,\underbrace{\mathfrak{Z}_{\tau_1+\tau_2,\tau_1+\tau_2,-\tau_1+\tau_2+2,\tau_1-\tau_2+8}^{\tau_1+\tau_2,\tau_1+\tau_2,\tau_2-\tau_1,\tau_1-\tau_2+8}(l-3,l-2)}_{\frac{4 (l-2) (l+\tau_1+1)^2}{(2 l+\tau_1+\tau_2-2) (2 l+\tau_1+\tau_2)}}-\frac{l-1}{4 (l+1)}\,p_{l+2}^{(3)}\,,\nonumber
\end{align}
where the coefficients $e_i$ are defined as:
\begin{subequations}
\begin{align}
    e_0&=\frac{4 (\tau +4)}{\tau  (-\tau +\tau_1+\tau_2)}-\frac{4}{-\tau +\tau_1+\tau_2+2}+\frac{8}{\tau  (\tau +2)}\,,\\
    e_1&=-\frac{64 (\tau_1+\tau_2+2)}{\tau ^2 (\tau +2) (-\tau +\tau_1+\tau_2)}\,,\\
    e_2&=\frac{32}{\tau ^2} \left(\frac{1}{-\tau +\tau_1+\tau_2}+\frac{\tau +8}{(\tau +2)^2}\right)\,,\\
    e_3&=-\frac{128}{\tau ^2 (\tau +2)^2}\,,\\
    e_4&=\frac{16}{\tau ^2 (\tau +2)^2}\,.
\end{align}
\end{subequations}
Finally, the anomalous dimensions of totally symmetric double-trace operators $[T\Phi]$ are given by:
\begin{equation}
    \gamma_{(l,0)}^{[T\Phi]}=\frac{2\,p_l^{(1)}}{a_{(l,0)}^{[T\Phi]}}\,,
\end{equation}
which extends \eqref{gammaT20} to any $l>2$ and is a sum of fourteen ${}_4 F_3$ hypergeometric functions. 

\subsubsection{$J0J0$ correlators}\label{J0J0corr}

In this section we briefly consider correlators involving generic spinning operators of the type 
\begin{equation}\label{genojpojp}
    \left\langle {\cal O}_J\left(y_1\right)\Phi\left(y_2\right) {\cal O}_J\left(y_3\right)\Phi\left(y_4\right)\right\rangle,
\end{equation}
involving a spin-$J$ operator ${\cal O}_J$ of twist $\tau_1$ and a scalar operator $\Phi$ of twist $\tau_2$. 

\begin{figure}[t]
    \centering
    \includegraphics[width=\textwidth]{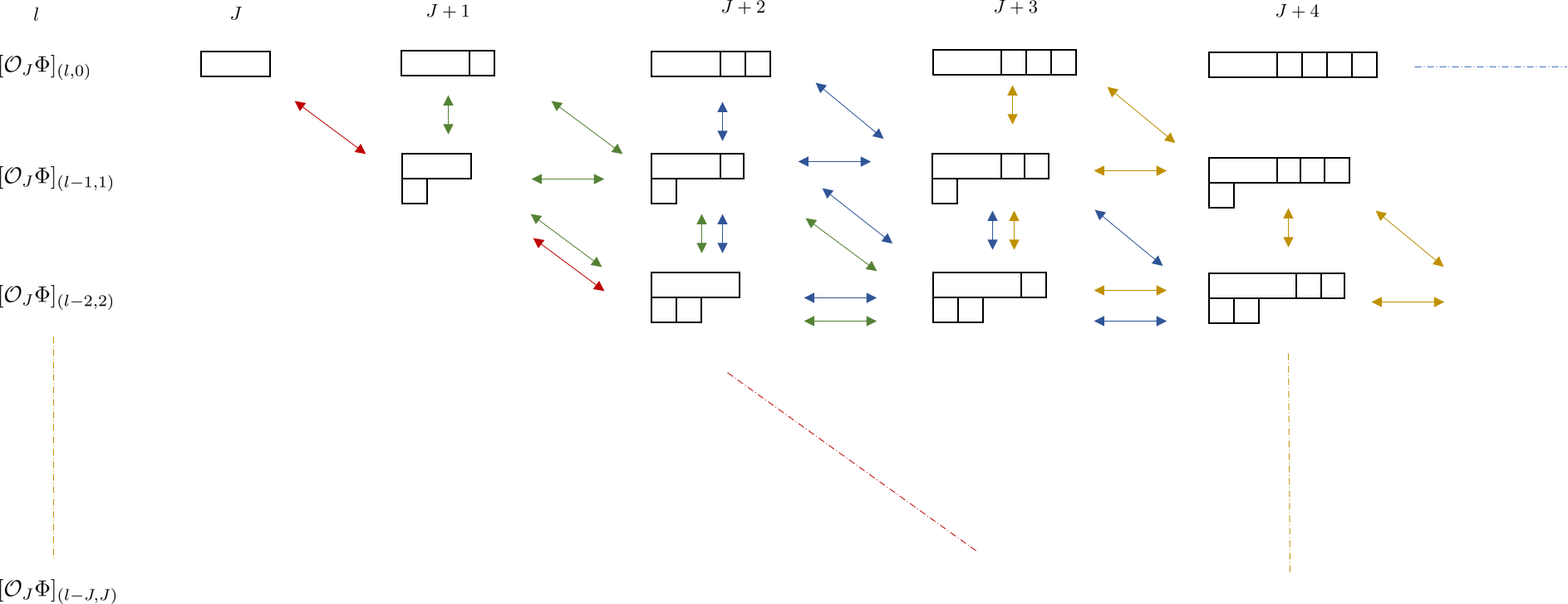}
    \caption{Double-trace operators contributing to the $\left\langle {\cal O}_J\Phi {\cal O}_J\Phi\right\rangle$. The colored triangles represent the operator which mix with a given totally symmetric operator.}
    \label{fig:Jphi}
\end{figure}

The mean field theory contribution to the correlator \eqref{genojpojp} is given by
\begin{equation}\label{mftj0j0}
    \mathcal{A}_{J0J0}=(2W_{13}W_{31})^J\,u^{\frac{\tau_1+\tau_2}2}\,.
\end{equation}
In this case the possible representations which may contribute in the ${\sf s}$-channel for each spin $l$ consist of: $(l-j,j)$ for $0\leq j\leq J$ (see e.g. fig.~\ref{fig:Jphi}). 

The generalisation of the non-diagonal bases \eqref{CPWtilde} and \eqref{lowspin2020basis} in this case is (see figure \ref{fig:Jphi}):
\begin{align}\label{lowspinJ0J0basis}
    {}_q\widetilde{\mathcal{F}}^{[T\Phi]}_l(s,t|W_{ij})&=\sum_{i=0}^{J-q}\sum_{k=0}^i\alpha^{q}_{i,k}\mathcal{F}^{[J\Phi]}_{(l-i-q+k+1,i+q-1)}(s,t|W_{ij})\,,
\end{align}
where we fix $\alpha_{0,0}^q=1$, which is the coefficient of the CPW for the exchange of the totally symmetric representation $\left(l,0\right)$. We can fix all other coefficients above by requiring that for each $q$ the terms not proportional to $(2W_{13}W_{31})^{J-q}$ are cancelled. Like in the previous examples, since we focusing on the OPE data of totally symmetric double-trace operators, the precise expressions for the other coefficients $\alpha_{i,k}^q$ is not important and we will not discuss it in the following.

Like for the $1$-$0$-$1$-$0$ and $2$-$0$-$2$-$0$ cases considered previously, only the CPWs \eqref{lowspinJ0J0basis} with $q=1$ contribute to the mean field theory correlator \eqref{mftj0j0} in the ${\sf s}$-channel. To extract the mean field theory OPE coefficients of the leading twist double-trace operators $\left[{\cal O}_J\Phi\right]$ we therefore only need to expand the Mellin representation of \eqref{mftj0j0} in terms of the corresponding kinematic polynomials, which read
\begin{align}\label{l0polynomialJ0J0}
    {}_1\widetilde{{\cal Q}}^{(J)}_{l,\tau_1+\tau_2}(s|W_{ij})=[2W_{13}W_{31}]^J\,\mathcal{A}^{(J)}_{l}\,\left(\tfrac{-s+\tau_1-\tau_2}2\right)_{2J}\,Q_{l-J}^{(\tau_1+\tau_2,\tau_1+\tau_2,-\tau_1+\tau_2,\tau_1-\tau_2+4J)}(s)\,,
\end{align}
with
\begin{equation}
    \mathcal{A}^{(J)}_{l}=(-1)^{l-J}(l-J)!\left(\mathfrak{N}_{l-J}^{(\tau_1+\tau_2,\tau_1+\tau_2,-\tau_1+\tau_2,\tau_1-\tau_2+4J)}\right)^{-1}\,.
\end{equation}
Following the same steps as in the previous cases we obtain the following expressions for the OPE coefficients of totally symmetric double-trace operators $\left[{\cal O}_J\Phi\right]$ of leading twist:
\begin{equation}
    a_{(l,0)}^{[\mathcal{O}_J\Phi]}=\frac{2^{l-J} (2 J+\tau_1)_{l-J} (\tau_2)_{l-J}}{(l-J)! (l+J+\tau_1+\tau_2-1)_{l-J}}\,,
\end{equation}
whose simple form is remarkable. On the other hand a little more work needs to be done in order to extract mixed-symmetry OPE coefficients, for which the explicit value of $\alpha^q_{i,k}$ is needed.

The above result leads to the following expression for the anomalous dimension of the $l=J$ lowest spin double-trace totally symmetric operator:
\begin{multline}\label{gammaOJ0}
    \delta\gamma_{(J,0)}^{[\mathcal{O}_J\Phi]}=\frac{J!}{(-2)^J}\frac{\left(\tfrac{d-\tau +\tau_1+\tau_2}2+J-1\right)_J}{ \left(\tfrac{\tau_1+\tau_2-\tau}2\right)_J}\\\times\,\frac{2\Gamma (\tau ) \Gamma \left(\tfrac{d-\tau +\tau_1-\tau_2}{2}\right) \Gamma \left(\tfrac{d-\tau -\tau_1+\tau_2}{2}\right)}{\Gamma \left(\tfrac{d}{2}+1\right) \Gamma \left(\tfrac{d}{2}-\tau \right) \Gamma \left(\tfrac{\tau +\tau_1-\tau_2}{2}\right) \Gamma \left(\tfrac{\tau -\tau_1+\tau_2}{2}\right)}\,c_{{\cal O}_J{\cal O}_J{\cal O}}c_{{\cal O}\Phi\Phi}\,.
\end{multline}
This result can be applied, for instance, to correlators of composite operators in the $O(N)$ model. In this case we set $\tau_i=d-2$ and Wick contraction gives the OPE coefficients \cite{Sleight:2016dba}
\begin{equation}
    c_{{\cal O}_J{\cal O}_J{\cal O}}c_{{\cal O}\Phi\Phi}=\frac{\sqrt{\pi } 2^{-d-J+7} \Gamma \left(\frac{d}{2}+J-1\right) \Gamma (d+J-3)}{\Gamma \left(\frac{d-2}{2}\right)^2 \Gamma (J+1) \Gamma \left(\frac{d-3}{2}+J\right)}\,.
\end{equation}
Putting everything together we get the remarkably simple result:
\begin{equation}\label{OjPhi}
    \gamma_{J}^{[\mathcal{O}_J\Phi]}=\frac{(-1)^{J+1}(d+2 J-3)}{(d+J-3) (d+2 J-2)}\frac{2^{d+2} \sin \left(\frac{\pi  d}{2}\right) \Gamma \left(\frac{d-1}{2}\right)}{N\,\pi ^{3/2} \Gamma \left(\frac{d-2}{2}\right)}\,,
\end{equation}
which we evaluate in various dimension in $d-\epsilon$ in table~\ref{tabgammaJ}.
\begin{center}
\begin{tabular}{|c|c|c|c|c|c|}
\hline
$d$ & $2$ & $3$ & $4$ & $5$ & $6$\\
 \hline
$\gamma_J^{[\mathcal{O}_J\Phi]}$ & $\frac{(-1)^{J} (4 J-2) \epsilon ^2}{N\,(J-1) J}$ & $\frac{64 (-1)^{J}}{N\,\pi ^2 (2 J+1)}$ & $\frac{8 (-1)^{J} (2 J+1) \epsilon }{N\,(J+1)^2}$ & $-\frac{512 (-1)^{J} (J+1)}{N\,\pi ^2 (J+2) (2 J+3)}$ & $-\frac{48 (-1)^{J} (2 J+3) \epsilon }{N\,(J+2) (J+3)}$\\
 \hline
\end{tabular}
\captionof{table}{Explicit values of \eqref{OjPhi} in $d-\epsilon$ dimensions for integer $d$.}\label{tabgammaJ}
\end{center}
In $d=1$ the result is singular in $\epsilon$ for $J\neq 1$:
\begin{multline}
    \gamma_J^{[\mathcal{O}_J\Phi]}\Big|_{d=1-\epsilon}=\frac{(-1)^{J} (32 (J-2) (J-1) (2 J-1) \log (2)-8 J (2 J (2 J-5)+7))}{N\,\pi ^2 \left(2 J^2-5 J+2\right)^2}\\-\frac{16 (-1)^{J} (J-1)}{N\,\pi ^2 (J-2) (2 J-1) \epsilon }\,.
\end{multline}
The $J=1$ case gives $\gamma_1^{[\mathcal{O}_1\Phi]}\Big|_{d=1-\epsilon}=\frac{8}{N\,\pi^2}$. In this case the spin $1$ gauge boson has vanishing dimension $\Delta=0$ (see e.g. \cite{Mezei:2017kmw}).

\subsection*{Acknowledgements}

We thank Simone Giombi, Igor Klebanov, Tassos Petkou, Silviu Pufu, Leonardo Rastelli, Herman Verlinde and Yifan Wang for discussions. C.S. holds a Marina Solvay Fellowship. The research of M.T. is partially supported by the European Union’s Horizon 2020 research and innovation programme under the Marie Skłodowska-Curie grant agreement No 747228 and by the Russian Science Foundation grant 14-42-00047 in association with Lebedev Physical Institute. M.T. also acknowledges support from the Fund for Scientific Research-FNRS Belgium, grant FC 6369 from which he is now on leave.

\begin{appendix}

\section{Spinning the shadow transform}\label{Shadow}

A non-trivial step in the explicit evaluation of CPWs is to determine the shadow transform of spinning 3pt conformal structures, which enters in the integral representation \eqref{nbscpw} of CPWs. This is particularly relevant for CPWs with spinning external operators because, owing to the degeneracy of 3pt conformal structures with more than one spinning operator, the shadow transform in this case is not just a change in dimension $\Delta \rightarrow d-\Delta$ but also a rotation in the space of conformal structures. Evaluating the shadow transform for such spinning 3pt conformal structures is the main goal of this appendix. In \S \tcb{\ref{shadowbulk}}, we will also discuss the implementation of the shadow transform from a bulk perspective.

Our approach is to first implement the shadow transform at the level of the canonical 3pt conformal structures \eqref{canonicalnbasis}:
\begin{align}\label{canonicalnbasisA}
\langle \langle {\cal O}_{\Delta_1,l_1}(y_1){\cal O}_{\Delta_2,l_2}(y_2) {\cal O}_{\Delta,l}(y_3)  \rangle \rangle^{(\text{{\bf n}})}&=\frac{{\sf Y}_1^{J_1-n_2-n}{\sf Y}_2^{J_2-n-n_1}{\sf Y}_3^{s-n_1-n_2}{\sf H}_1^{n_1}{\sf H}_2^{n_2}{\sf H}_3^{n}}{(y_{12}^2)^{\tfrac{\tau_1+\tau_2-\tau}{2}}(y_{23}^2)^{\tfrac{\tau_2+\tau-\tau_1}{2}}(y_{31}^2)^{\tfrac{\tau+\tau_1-\tau_2}{2}}}\,,
\end{align}
in terms of which 3pt conformal structures in any basis may be easily expanded. To perform the shadow transform of \eqref{canonicalnbasisA}, it is convenient to express the corresponding integrand in the form \cite{Giombi:2017hpr}
\begin{multline}\label{simpshadcan}
   \langle \langle {\cal O}_{\Delta_1,l_1}(y_1){\cal O}_{\Delta_2,l_2}(y_2) {\tilde {\cal O}}_{\Delta,l}(y_3)  \rangle \rangle^{(\text{{\bf n}})}\\ = \frac{\kappa_{d-\Delta,l}}{\pi^{d/2}}\int d^dy_0\,\mathcal{N}(y_0,y_i)\,{\sf H}_3^{n_0}{\sf Y}_1^{J_1-n_2-n_0}{\sf Y}_2^{J_2-n_0-n_1}\frac{\pl_{h_1}^{n_1}\pl_{h_2}^{n_2}}{\tbinom{J_3}{n_1,n_2}}\,(\zeta\cdot\bar{\zeta})^{J_3},
\end{multline}
where we are taking the shadow transform with respect to the operator ${\cal O}_{\Delta,l}$ and have defined the overall conformal factor
\begin{equation}
    \mathcal{N}(y_0,y_i)=\frac1{(y_{03}^2)^{d-\tau-2r}}\frac1{(y_{12}^2)^{\tfrac{\tau_1+\tau_2-\tau}2}(y_{20}^2)^{\tfrac{\tau_2+\tau-\tau_1}2}(y_{01}^2)^{\tfrac{\tau+\tau_1-\tau_2}2}}\,,
\end{equation}
and the auxiliary vectors:
\begin{subequations}
\begin{align}
    \zeta^a&=\pl_{z_0}^a\left({\sf Y}_0+h_1{\sf H}_1+h_2{\sf H}_2\right)\,,\\
    \bar{\zeta}^a&=\pl_{z_0}^a\,{\sf H}\,,\\
    {\sf H}&=z_{0} \cdot I\left(y_0-y_3\right) \cdot z_3.
\end{align}
\end{subequations}
The auxiliary parameters $h_1$ and $h_2$ serve to keep track of the structures ${\sf H}_1$ and ${\sf H}_2$. The tensor $I_{\mu\nu}$ is the inversion tensor \eqref{inversont}.

By evaluating the contraction between $\zeta$ and ${\bar \zeta}$ in \eqref{simpshadcan} explicitly:
\begin{align}
    \zeta\cdot\bar{\zeta}&=- \frac{\left(y_{13}^2\right) \left(z_3\cdot y_{30}\right)}{\left(y_{01}^2\right) \left(y_{03}^2\right)^2} + \frac{\left(y_{23}^2\right) \left(z_3\cdot y_{30}\right)}{\left(y_{02}^2\right) \left(y_{03}^2\right)^2} -  \frac{2\, h_2 \left(y_{13}^2\right) \left(z_1\cdot y_{10}\right) \left(z_3\cdot y_{30}\right)}{\left(y_{01}^2\right)^2 \left(y_{03}^2\right)^2}\nonumber\\
    &+ \frac{2\, h_2 \left(z_1\cdot y_{13}\right) \left(z_3\cdot y_{30}\right)}{\left(y_{01}^2\right) \left(y_{03}^2\right)^2} -  \frac{2\, h_1 \left(y_{23}^2\right) \left(z_2\cdot y_{20}\right) \left(z_3\cdot y_{30}\right)}{\left(y_{02}^2\right)^2 \left(y_{03}^2\right)^2} + \frac{2\, h_1 \left(z_2\cdot y_{23}\right) \left(z_3\cdot y_{30}\right)}{\left(y_{02}^2\right) \left(y_{03}^2\right)^2}\nonumber\\
    &+ \frac{\left(z_3\cdot y_{31}\right)}{\left(y_{01}^2\right) \left(y_{03}^2\right)} + \frac{2 h_2 \left(z_1\cdot y_{10}\right) \left(z_3\cdot y_{31}\right)}{\left(y_{01}^2\right)^2 \left(y_{03}^2\right)} -  \frac{\left(z_3\cdot y_{32}\right)}{\left(y_{02}^2\right) \left(y_{03}^2\right)} + \frac{2 h_1 \left(z_2\cdot y_{20}\right) \left(z_3\cdot y_{32}\right)}{\left(y_{02}^2\right)^2 \left(y_{03}^2\right)},
\end{align}
the conformal integral \eqref{simpshadcan} in the shadow transform can be expressed as a series, in which each term is expressible as derivatives of the standard scalar conformal integral:
\begin{equation}\label{Ir1r2r3}
    I_{\alpha_1,\alpha_2,\alpha_3}^{r_1,r_2,r_3}\equiv (z_1\cdot\pl_{y_1})^{r_1}(z_2\cdot\pl_{y_2})^{r_2}(z_3\cdot\pl_{y_3})^{r_3}\int d^dy_0\,\frac1{(y_{01}^2)^{\alpha_1}(y_{02}^2)^{\alpha_2}(y_{03}^2)^{\alpha_3}}\,,\qquad \alpha_1+\alpha_2+\alpha_3=d\,.
\end{equation}
Using that 
\begin{subequations}
\begin{align}
   I_{\alpha_1,\alpha_2,\alpha_3}^{0,0,0} &= \int d^dy_0\,\frac1{(y_{01}^2)^{\alpha_1}(y_{02}^2)^{\alpha_2}(y_{03}^2)^{\alpha_3}}\\
    &= \frac{\pi^{d/2}}{(y_{12}^2)^{\tfrac{\alpha_1+\alpha_2-\alpha_3}2}(y_{23}^2)^{\tfrac{\alpha_2+\alpha_3-\alpha_1}2}(y_{31}^2)^{\tfrac{\alpha_3+\alpha_1-\alpha_2}2}},
\end{align}
\end{subequations}
the calculation then reduces to evaluating the action of the derivatives in \eqref{Ir1r2r3} on the above scalar 3pt conformal structure for each term in the series. To this end it is convenient to drop all terms proportional to $z_i\cdot z_j$, whch gives following simple representation of the differential operators:
\begin{subequations}
\begin{align}
    z_1\cdot\pl_{y_1}&=2 \sum_{j\neq 1}z_1\cdot y_{1j}\,\pl_{y_{1j}^2},\\
    z_2\cdot\pl_{y_2}&=2 \sum_{j\neq 2}z_2\cdot y_{2j}\,\pl_{y_{2j}^2},\\
    z_3\cdot\pl_{y_3}&=2 \sum_{j\neq 3}z_3\cdot y_{3j}\,\pl_{y_{3j}^2}.
\end{align}
\end{subequations}

In the basis \eqref{nicebasis}, the series expansion of the conformal integral \eqref{simpshadcan} can be re-summed into a relatively simple form. For simplicity we first present the case of conformal structures with $n_1=n_2=n_0=0$, which gives:
\begin{equation}
    \mathcal{S}_3[\mathfrak{S}_{\bf J,\tau}^{0}]=\alpha_{\bf J,\tau}\widehat{\mathcal{S}}_3[\mathfrak{S}_{\bf J,\tau}^{0}]\,,
\end{equation}
where $\mathcal{S}_3[\mathfrak{S}_{\bf J,\tau}^{0}]$ denotes the shadow transform of the basis element \eqref{nicebasis} with respect to the operator located at $y_3$ and

\begin{multline}\label{Shadow0}
   \widehat{\mathcal{S}}_3[\mathfrak{S}_{\bf J,\tau}^{0}]=\sum_{k=0}^{\text{Min}(J_1,J_2)}\frac{(-1)^k}{2^k\,k!}\\\times\,\tfrac{(\tau_1+J_1-1)_k(\tau_2+J_2-1)_k(J_1-k+1)_k(J_2-k+1)_k\left(-\tfrac{d-2\tau_3}2+J_3\right)_k}{\left(-\tfrac{2(J_1-J_3)-2+d+\tau_1-\tau_2-\tau_3}2\right)_k\left(-\tfrac{2(J_2-J_3)-2+d-\tau_1+\tau_2-\tau_3}2\right)_k\left(\tfrac{\tau_1+\tau_2-\tau_3}2\right)_k\left(J_3+\tfrac{\tau_1+\tau_2+\tau_3-d}2+l-1\right)_k}\\\times\,{\sf H}_3^k \,\mathfrak{S}_{J_1-k,J_2-k,J_3;\tau_1+2k,\tau_2+2k,d-\tau_3-2J_3}\,.
\end{multline}
The overall normalisation $\alpha_{\bf J}$ is given by:
\begin{equation}\label{ShadowNorm}
    \alpha_{\bf J,\tau}=\frac{\Gamma \left(\tfrac{d+\tau_1-\tau_2-\tau_3}{2}\right) \Gamma \left(\tfrac{d-\tau_1+\tau_2-\tau_3}{2}\right)}{\Gamma \left(\tfrac{2 J_3+\tau_1-\tau_2+\tau_3}{2}\right) \Gamma \left(\tfrac{2 J_3-\tau_1+\tau_2+\tau_3}{2}\right)}\,\frac{ \left(\tfrac{d-\tau_1+\tau_2-\tau_3}{2}-J_3\right)_{J_2} \left(\tfrac{d+\tau_1-\tau_2-\tau_3}{2}-J_3\right)_{J_1}}{\left(\tfrac{-\tau_1+\tau_2+\tau_3}{2}\right)_{J_2} \left(\tfrac{\tau_1-\tau_2+\tau_3}{2}\right)_{J_1}}\,,
\end{equation}
such that
\begin{equation}\label{alphacond}
    \alpha_{{\bf J},\tau_i,d-\tau_3-2J_3}\alpha_{\bf J,\tau}=1\,.
\end{equation}

The result for arbitrary $n_0$ and $n_1=n_2=0$ can be obtained in a similar manner and reads:
\begin{multline}\label{shadowgeneralN3}
     \mathcal{S}_3[\mathfrak{S}_{\bf J,\tau}^{n_0}]=\alpha_{\bf J,\tau}\,\frac{\left(J_1-n_0+\tfrac{\tau_1-\tau_2+\tau_3}{2}\right)_{n_0} \left(J_2-n_0+\tfrac{-\tau_1+\tau_2+\tau_3}{2}\right)_{n_0}}{\left(J_1-n_0-J_3+\tfrac{d+\tau_1-\tau_2-\tau_3}{2} \right)_{n_0} \left(J_2-n_0-J_3+\tfrac{d-\tau_1+\tau_2-\tau_3}{2}\right)_{n_0}}\\\times\,{\sf H}_3^{n_0}\widehat{\mathcal{S}}_3[\mathfrak{S}_{J_1-n_0,J_2-n_0,J_3;\tau_1+2n_0,\tau_2+2n_0,\tau_3}^{0}]\,,
\end{multline}
which also allows to explicitly check involutivity:
\begin{equation}
    \mathcal{S}_3\circ\mathcal{S}_3[\mathfrak{S}_{\bf J,\tau}^{n_0}]=\mathfrak{S}_{\bf J,\tau}^{n_0}\,.
\end{equation}

Given the shadow tensor structures $\mathcal{S}_3\left[\mathfrak{S}_{\bf J,\tau}^{n_0}\right]$, the $\alpha_{\bf J,\tau}$ coefficient allows to define the corresponding shadow OPE coefficient ${\tilde c}^{n_0}_{\bf J,\tau}$ as:
\begin{subequations}
\begin{align}
    &c^{n_0}_{\bf J,\tau}\mathfrak{S}_{\bf J,\tau}^{n_0}\rightarrow \tilde{c}^{n_0}_{\bf J,\tau}\widehat{\mathcal{S}}_3\left[\mathfrak{S}_{\bf J,\tau}^{n_0}\right]\,,\\
    \tilde{c}^{n_0}_{\bf J,\tau}&=\alpha_{\bf J,\tau}\,\tfrac{\left(J_1-n_0+\tfrac{\tau_1-\tau_2+\tau_3}{2}\right)_{n_0} \left(J_2-n_0+\tfrac{-\tau_1+\tau_2+\tau_3}{2}\right)_{n_0}}{\left(J_1-n_0-J_3+\tfrac{d+\tau_1-\tau_2-\tau_3}{2} \right)_{n_0} \left(J_2-n_0-J_3+\tfrac{d-\tau_1+\tau_2-\tau_3}{2}\right)_{n_0}}\,c^{n_0}_{\bf J,\tau},
\end{align}
\end{subequations}
where $c^{n_0}_{\bf J,\tau}$ is the original OPE coefficient associated to the tensor structure $\mathfrak{S}_{\bf J,\tau}^{n_0}$.

\subsection{Shadow transform via the bulk and large $N$ CFTs}\label{shadowbulk}

\begin{figure}[h]
\centering
\includegraphics[width=0.9\textwidth]{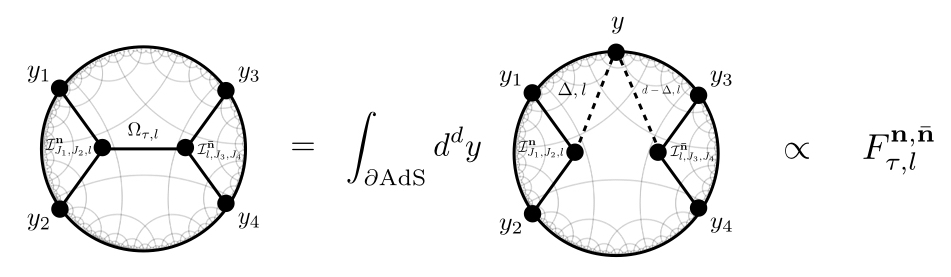}
\caption{CPWs from the bulk}
\label{fig::harmonic}
\end{figure}

 It is well-known that 4pt Witten diagrams with an exchanged bi-tensorial harmonic function ($\tau=\Delta-l$)
\begin{subequations}
 \begin{align}
     \left(\Box_1-\Delta\left(\Delta-d\right)+l\right) \Omega_{\tau,l}(x_1,x_2;u_1,u_2)&=0,\\
     \left(\nabla_1 \cdot \partial_{u_1}\right) \Omega_{\tau,l}(x_1,x_2;u_1,u_2)&=0,\\
     \left(\partial_{u_1} \cdot \partial_{u_1}\right) \Omega_{\tau,l}(x_1,x_2;u_1,u_2)&=0,
 \end{align}
\end{subequations}
correspond to individual CPWs on the boundary upon performing the bulk integration  \cite{Penedones:2007ns,Costa:2014kfa,Bekaert:2014cea,Sleight:2016hyl,Sleight:2017fpc}. See fig \ref{fig::harmonic}. This can be seen via the split representation of harmonic functions \cite{Leonhardt:2003qu}
\begin{align}\label{difference}
     \Omega_{\tau,l}(x_1,x_2;u_1,u_2)=\frac{\left(\tfrac{d}2-\tau-l\right)^2}{\pi\,l!}\int d^dy\,K_{\tau+l,l}(x_1,u_1;y,\hat{\pl}_z)K_{d-\tau-l,l}(x_2,u_2;y,z)\,,
\end{align}
in terms of the bulk-to-boundary propagator $K_{\Delta,l}$, which causes the exchange to factorise into two tree-level 3pt Witten diagrams integrated over their common boundary point. Upon evaluating the 3pt bulk integrals, this gives the integral form \eqref{nbscpw} of a CPW. That the harmonic functions are also given by a difference of bulk-to-bulk propagators $\Pi$ as \cite{Costa:2014kfa}
\begin{align}
    \Omega_{\tau,l}(x_1,x_2;u_1,u_2)&=\frac{\left(\tfrac{d}2-\tau-l\right)}{2\pi}\Big[\Pi_{\tau+l,l}(x_1,x_2;u_1,u_2)-\Pi_{d-\tau-l,l}(x_1,x_2;u_1,u_2)\Big],
\end{align}
trivially confirms the interpretation of the difference \eqref{cpt} of 4pt correlators as a difference of boundary conditions in the dual 4pt bulk amplitude, and that it is given by individual CPWs in the ${\sf s}$-, ${\sf t}$- and ${\sf u}$-channels.

In practice, to evaluate the CPWs via the bulk as in fig \ref{fig::harmonic} we use the ambient space formalism to compute the constituent spinning 3pt Witten diagrams (see e.g. \cite{Sleight:2017fpc}, also for notation). The main subtlety is that we need to properly account for the ambient space uplift of the bulk harmonic functions $\Omega_{\tau,l}(x_1,x_2;u_1,u_2)$. 
The above expressions are intrinsic AdS expressions and when one considers their uplift to ambient space it is necessary to fix their degree of homogeneity by multiplying the ambient space representatives for the above harmonic functions by appropriate powers of $X^2$. Restriction to the AdS manifold recovers $X^2=-1$. Such powers of $X^2$ will be crucial whenever flat ambient space derivatives act on the harmonic functions at both points. The trick used in \cite{Sleight:2017fpc} was to perform integration by parts on the cubic vertex with the effect that no derivative acts on $x_2$ in the harmonic function but only on $x_1$ which has the correct homogeneity degree by construction. This trick allows to conveniently evaluate the shadow transform of any 3pt conformal structure including the corresponding overall normalisation. Luckily for us, indeed, integration by parts in ambient space is completely under control in the ambient formalism \cite{Joung:2011ww,Joung:2012fv,Taronna:2012gb,Joung:2013nma}, upon explicitly introducing a $\delta$-function measure:
\begin{equation}
    \int_{\text{AdS}}I_\Delta=\int d^{d+1}X\,\delta\left(\sqrt{-X^2}-L\right)\,I_{\Delta}
\end{equation}
with $\Delta$ the fixed degree of homogeneity of the ambient space integrand. Furthermore, it turns out convenient to introduce the notation ($\Lambda=1$):
\begin{equation}
    \int_{\text{AdS}}I_\Delta=\int d^{d+1}X\,\delta\left(\sqrt{-X^2}-1\right)\,\lambda^n \,I_\Delta\equiv \int d^{d+1}X\,\delta^{[n]}\left(\sqrt{-X^2}-L\right)\,I_\Delta\,,
\end{equation}
where $\delta^{[n]}$ is the $n$-th derivative of the $\delta$-function distribution. One can then prove that
\begin{equation}
    \lambda^{n}=(-1)^n(d+\Delta)\cdots(d+\Delta-2n+2)\,,
\end{equation}
where $\Delta$ is the degree of homogeneity of $I_{\Delta}$. In this way, given a point-splitted cubic vertex (see eq. \eqref{massivebasis})
\begin{equation}
    \mathcal{V}=f({\cal Y}_i,{\cal H}_i)\,\phi_1\,\phi_2\,\phi_3
\end{equation}
the integration by part operation which integrates by parts all derivatives acting on $\phi_3$ away is realised by the following operator:
\begin{equation}
    \bar{\mathcal{V}}=e^{\lambda {\cal H}_3\pl_{{\cal Y}_1}\pl_{\bar{{\cal Y}}_2}}\,f({\cal Y}_1,-\bar{{\cal Y}}_2,{\cal Y}_3,{\cal H}_i)\,,
\end{equation}
where $\bar{\cal Y}_i=
\pl_{U_i}\cdot \pl_{X_{i-1}}$ are the anti-cyclic contractions of derivatives. One can also define the transformation which brings back a generic vertex function of the type:
\begin{equation}
    \bar{\mathcal{V}}=f({\cal Y}_i,\bar{{\cal Y}}_i,{\cal H}_i)\,,
\end{equation}
to the cyclic basis as:
\begin{equation}
    \mathcal{V}=e^{-\lambda\left[{\cal H}_1\pl_{{\cal Y}_2}\pl_{\bar{{\cal Y}}_3}+{\cal H}_2\pl_{{\cal Y}_3}\pl_{\bar{{\cal Y}}_1}+{\cal H}_3\pl_{{\cal Y}_1}\pl_{\bar{{\cal Y}}_2}\right]}\,f({\cal Y}_i,\bar{{\cal Y}}_i-{\cal Y}_i,{\cal H}_i)\Big|_{\bar{{\cal Y}}_i=0}\,.
\end{equation}
Therefore, starting from a generic cubic vertex
\begin{equation}
    \mathcal{V}=f({\cal Y}_i,{\cal H}_i)\,\phi_1\,\phi_2\,\phi_3\,,
\end{equation}
acting on a field $\phi_3$ of twist $\tau_3$ one thus gets the following modified cubic vertex acting on the shadow field with twist $d-\tau_3-2J_3$:
\begin{equation}
    \widetilde{\mathcal{V}}=e^{(\lambda_{d-\tau_3-2J_3}-\lambda_{\tau_3}){\cal H}_3\pl_{{\cal Y}_1}\pl_{{{\cal Y}}_2}}\,f({\cal Y}_1,{{\cal Y}}_2,{\cal Y}_3,{\cal H}_i)
\end{equation}
where the label on $\lambda_{\tau}$ specifies the degree of homogeneity which should be used for the field $\phi_3$ when evaluating $\lambda_\tau^n$. The above amounts in more detail the the following steps:
\begin{itemize}
    \item Remove derivatives from $\phi_3$ on-shell up to integrations by parts.
    \item Evaluate all insertion of $\lambda$ using:
    \begin{equation}
        \Delta={\cal Y}_i\pl_{{\cal Y}_i}+\bar{{\cal Y}}_i\pl_{\bar{{\cal Y}_i}}+(\tau_i+J_i)\,
    \end{equation}
    in terms of the twists and spins of the field $\phi_3$.
    \item Insert the harmonic function which amounts to effectively change the twist of $\phi_3$ according to
    \begin{equation}
        \tau_3\rightarrow d-\tau_3-2J_3\,,
    \end{equation}
    this can be done now with no problem because no ambient space derivative acts on $\phi_3$.
    \item Go back to the cyclic basis using now the modified degree of homogeneity and evaluate the corresponding bulk integral.
\end{itemize}
Focusing for simplicity on the case $n_1=n_2=n_0=0$, upon resummation one then obtains (see \S\tcb{\ref{NewBasis}} for notation):
\begin{subequations}
\begin{align}
    \mathfrak{W}_{\tau_1,\tau_2,\tau_3}\left[\mathcal{I}_{J_1,J_2,J_3}^{0,0,0}(\tau_i)\right]&=\mathsf{B}(J_i;\tau_i)\mathfrak{S}_{{\bf J};\tau}^{0}\,,\\
     \mathfrak{W}_{\tau_1,\tau_2,d-\tau_3-2J_3}\left[\widetilde{\mathcal{I}}_{J_1,J_2,J_3}^{0,0,0}(\tau_i)\right]&=\mathsf{B}(J_i;\tau_i)\,\mathcal{S}_3[\mathfrak{S}_{\bf J,\tau}^{0}]\,,
\end{align}
\end{subequations}
where for convenience we strip off the same function $\mathsf{B}(J_i;\tau_i)$ which encodes the OPE coefficient of the CFT up to the 2pt function normalisation $\left\langle\mathcal{O}_{J,\tau}(y_1)\mathcal{O}_{J,\tau}(y_2)\right\rangle=\mathcal{C}_{\tau+J,J}\,\tfrac{{\sf H}^J}{(y_{12}^2)^\tau}$. All in all, applying the differential operators, implementing the integrations by parts and evaluating the Witten diagram we get:

{\footnotesize
\begin{multline}\label{Shadow0b}
  \mathcal{S}_3[\mathfrak{S}_{\bf J,\tau}^{0}]=\beta_{\bf s,\tau}\,\sum_{k=0}^{\text{Min}(J_1,J_2)}\frac{(-1)^k}{2^k\,k!}\\\times\,\frac{(\tau_1+J_1-1)_k(\tau_2+J_2-1)_k(J_1-k+1)_k(J_2-k+1)_k\left(-\tfrac{d-2\tau_3}2+J_3\right)_k}{\left(-\tfrac{2(J_1-J_3)-2+d+\tau_1-\tau_2-\tau_3}2\right)_k\left(-\tfrac{2(J_2-J_3)-2+d-\tau_1+\tau_2-\tau_3}2\right)_k\left(\tfrac{\tau_1+\tau_2-\tau_3}2\right)_k\left(J_3+\tfrac{\tau_1+\tau_2+\tau_3-d}2+l-1\right)_k}\\\times\,{\sf H}_3^k \,\mathfrak{S}_{J_1-k,J_2-k,J_3;\tau_1+2k,\tau_2+2k,d-\tau_3-2J_3}\,,
\end{multline}}
with an overall normalisation $\beta_{\bf J,\tau}$ given by:
\begin{multline}\label{ShadowNorm2}
    \beta_{\bf J,\tau}=\underbrace{\frac{\kappa_{d-\tau_3-J_3,J_3}}{\pi ^{d/2} (d-2 \tau_3-2 J_3) \mathcal{C}_{\tau_3+J_3,J_3}}}_{\mathcal{N}_{\tau_3,J_3}}\frac{\Gamma\left(\tfrac{d+\tau_1-\tau_2-\tau_3}2\right)\Gamma\left(\tfrac{d-\tau_1+\tau_2-\tau_3}2\right)}{\Gamma\left(\tfrac{2J_3+\tau_1-\tau_2+\tau_3}2\right)\Gamma\left(\tfrac{2J_3-\tau_1+\tau_2+\tau_3}2\right)}\\\times\,(-1)^{J_1+J_2}\frac{\left(-\frac{d+2 (J_1- J_3)-2+\tau_1-\tau_2-\tau_3}{2}\right)_{J_1} \left(-\frac{d+2 (J_2- J_3)-2-\tau_1+\tau_2-\tau_3}{2}\right)_{J_2}}{\left(\frac{\tau_1-\tau_2+\tau_3}{2}\right)_{J_1} \left(\frac{-\tau_1+\tau_2+\tau_3}{2}\right)_{J_2}}\,.
\end{multline}
As expected the coefficient $\mathcal{N}_{\tau_3,J_3}$, combined with the definition \eqref{difference}, precisely reproduces the normalisation $\kappa_{d-\tau_3-J_3,J_3}/\pi^{d/2}$ we have in the CPW up to the normalisation $1/\mathcal{C}_{\tau_3+J_3,J_3}$. The latter factor is exactly the normalisation needed to get a unit normalisation for the two point function of the exchanged operator. The rest reproduces, as expected, the shadow conformal structure that was obtained in the previous section by evaluating explicitly the shadow conformal integral together with the precise overall normalisation.\footnote{That this must be the case follows from the identity among bulk to boundary propagators:
\begin{equation}
    K_{\Delta_+,J}(x;y)=-(\Delta_+-\Delta_-)\int_{\pl \textbf{AdS}}d^d\bar{y}\,K_{\Delta_-}(x;\bar{y})K_{\Delta_+}(y;\bar{y})\,,
\end{equation}
where the normalisation of the limit of the boundary-to-bulk propagator reproduces the propagator of the $\sigma$-field in large-$N$ CFTs (see e.g. \cite{Leonhardt:2003sn,Giombi:2017hpr}).
} We have also checked the case $n_0>0$ reproducing the result obtained via the conformal integral \eqref{shadowgeneralN3} and we have tested also the cases $n_1\neq 0$ or $n_2\neq 0$ providing a non-trivial consistency checks of the Witten diagram result in \cite{Sleight:2017fpc}. We conclude this section stressing how the shadow transform via the bulk using the explicit form of the bulk-to-boundary map turns out to be computationally simpler then carrying directly the conformal integral and reduces the computation of a spinning CPW to a single conformal integral associated to the bulk harmonic function.

\section{4pt Conformal Integrals}\label{4pt}

Our approach is based on a direct evaluation of the simplest CPW conformal integral of the type:
\begin{multline}\label{canonicbasisciA}
F^{{\bf n},{\bf \bar{n}}}_{\tau,l}\left(y_i\right) =\frac{\kappa_{d-\Delta,l}}{\pi^{d/2}} \int d^dy_0\, \langle \langle {\cal O}_{\Delta_1,J_1}(y_1){\cal O}_{\Delta_2,J_2}(y_2) {\cal O}_{\Delta,l}(y_0)  \rangle \rangle^{({\bf n})} \\\times\,\langle \langle {\tilde {\cal O}}_{d-\Delta,l}(y_0)  {\cal O}_{\Delta_3,J_3}(y_3){\cal O}_{\Delta_4,J_4}(y_4)\rangle \rangle^{({\bf \bar{n}})},
\end{multline}
in terms of which a CPW in any basis can be easily expanded. Recall that
\begin{align}
\langle \langle {\cal O}_{\Delta_1,J_1}(y_1){\cal O}_{\Delta_2,J_2}(y_2) {\cal O}_{\Delta_3,J_3}(y_3)  \rangle \rangle^{({\bf n})}&=\frac{{\sf Y}_1^{J_1-n_2-n}{\sf Y}_2^{J_2-n-n_1}{\sf Y}_3^{s-n_1-n_2}{\sf H}_1^{n_1}{\sf H}_2^{n_2}{\sf H}_3^{n}}{(y_{12}^2)^{\tfrac{\tau_1+\tau_2-\tau}{2}}(y_{23}^2)^{\tfrac{\tau_2+\tau-\tau_1}{2}}(y_{31}^2)^{\tfrac{\tau+\tau_1-\tau_2}{2}}}\,,
\end{align}
where at 4pt level one introduces the following explicit building blocks:
{\small \begin{subequations}\label{4ptstructures}
\begin{align}
& {\sf Y}_{1}= \frac{z_1\cdot y_{12}}{y_{12}^2}-\frac{z_1\cdot y_{10}}{y_{10}^2}, 
&& {\sf Y}_{3}= \frac{z_3\cdot y_{34}}{y_{34}^2}-\frac{z_3\cdot y_{30}}{y_{30}^2},\\
&{\sf Y}_{2}= \frac{z_2\cdot y_{20}}{y_{20}^2}-\frac{z_2\cdot y_{21}}{y_{21}^2},
&& {\sf Y}_{4}= \frac{z_4\cdot y_{40}}{y_{40}^2}-\frac{z_4\cdot y_{43}}{y_{43}^2},\\
& {\sf Y}_{0}= \frac{{z_0}\cdot y_{01}}{y_{01}^2}-\frac{{z_0}\cdot y_{02}}{y_{02}^2},
&& {\sf \bar{Y}}_{0}= \frac{z_0\cdot y_{03}}{y_{03}^2}-\frac{z_0\cdot y_{04}}{y_{04}^2},\\
&{\sf H}_1=\frac1{y_{02}^2}\left(z_0\cdot z_2+2\,\frac{z_0\cdot y_{02}\,z_2\cdot y_{20}}{y_{02}^2}\right)
&&{\sf H}_3=\frac1{y_{04}^2}\left(z_0\cdot z_4+2\,\frac{z_0\cdot y_{04}\,z_4\cdot y_{40}}{y_{04}^2}\right)\\
&{\sf H}_2=\frac1{y_{01}^2}\left(z_0\cdot z_1+2\,\frac{z_0\cdot y_{01}\,z_1\cdot y_{10}}{y_{01}^2}\right)
&&{\sf H}_4=\frac1{y_{03}^2}\left(z_0\cdot z_3+2\,\frac{z_0\cdot y_{03}\,z_3\cdot y_{30}}{y_{03}^2}\right)\\
&{\sf H}_0=\frac1{y_{12}^2}\left(z_1\cdot z_2+2\,\frac{z_1\cdot y_{12}\,z_2\cdot y_{21}}{y_{12}^2}\right)
&&{\sf \bar{H}}_0=\frac1{y_{34}^2}\left(z_3\cdot z_4+2\,\frac{z_3\cdot y_{34}\,z_4\cdot y_{43}}{y_{34}^2}\right),
\end{align}
\end{subequations}}
which can be used to construct 4pt CPWs with the point $y_0$ in common. 

There are two technical steps in the evaluation of the conformal integral in \eqref{canonicbasisciA} -- see appendix A of \cite{Giombi:2017hpr}. The first is to express the general contraction of conformal structures in terms of the relevant Gegenbauer polynomial. The conformal integral is then evaluated and expressed in Mellin form using the Symanzik star formula \cite{Symanzik:1972wj}.

For the first step we evaluate the traceless contraction in the symmetric representation. This can be done by rewriting the CPW in terms of auxiliary vectors $\Xi$ and $\bar{\Xi}$ defined as
\begin{subequations}\label{XiXibar}
\begin{align}
    \Xi^a&=\pl_{z_0}^a\left[{\sf Y}_0+h_1 {\sf H}_1+h_2{\sf H}_2\right]\,,\\
    \bar{\Xi}^a&=\pl_{z_0}^a\left[{\sf \bar{Y}}_0+h_3 {\sf H}_3+h_4{\sf H}_4\right]\,.
\end{align}
\end{subequations}
One can then express the CPW as\footnote{For concision, $\mathcal{N}(y_i,y_0)$ in \eqref{Pseries} represents the prefactor \begin{multline}
    \mathcal{N}(y_i,y_0)=\tfrac{1}{\left(y_{12}^2\right)^{\tfrac{\tau_1+\tau_2-\tau}2} \left(y_{02}^2\right)^{\tfrac{\tau_2+\tau-\tau_1}2} \left(y_{01}^2\right)^{\tfrac{\tau+\tau_1-\tau_2}2}}\\\times\tfrac1{\left(y_{34}^2\right)^{\tfrac{\tau_3+\tau_4-(d-\tau-2r)}2} \left(y_{03}^2\right)^{\tfrac{\tau_3+(d-\tau-2r)-\tau_4}2} \left(y_{04}^2\right)^{\tfrac{(d-\tau-2r)+\tau_4-\tau_3}2}}\,.
\end{multline}}
\begin{multline}\label{Pseries}
    F^{{\bf n},{\bf \bar{n}}}_{\tau,l}=\mathcal{N}(y_i,y_0){\sf Y}_1^{J_1-n_2-n}{\sf Y}_2^{J_2-n-n_1}{\sf Y}_3^{J_3-m_4-m}{\sf Y}_4^{J_4-m_3-m}{\sf H}_0^{n}{\sf \bar{H}}_0^{m}\\\times\frac{\pl_{h_1}^{n_1}}{n_1!}\frac{\pl_{h_2}^{n_2}}{n_2!}\frac{\pl_{h_3}^{m_3}}{m_3!}\frac{\pl_{h_4}^{m_4}}{m_4!}\sum_{k=0}^{[l/2]}\frac{c_{l,k}}{\binom{s}{n_1,n_2}\binom{s}{m_3,m_4}}(\Xi\cdot\bar{\Xi})^{l-2k}[\Xi^2\,\bar{\Xi}^2]^k\,,
\end{multline}
where $c_{s,k}$ are the Gegenbauer polynomial coefficients:
\begin{equation}\label{Gcoeff}
    c_{n,k}=\frac{(-4)^{-k} n! \Gamma \left(\frac{d}{2}-k+n-1\right)}{k! (n-2 k)! \Gamma \left(\frac{d}{2}+n-1\right)}\,,
\end{equation}
and the contractions $\Xi^2$, $\bar{\Xi}^2$ and $\Xi\cdot\bar{\Xi}$ are given explicitly in \S \tcb{\ref{Contractions}}.

Keeping this in mind, equation \eqref{Pseries} can be expanded in terms of standard 4pt conformal integrals of the following type:
\begin{subequations}
\begin{align}\label{stci}
    I^{r_1,r_2,r_3,r_4}_{\alpha_1,\alpha_2,\alpha_3,\alpha_4}&=(z_1\cdot\pl_{y_1})^{r_1}(z_2\cdot\pl_{y_2})^{r_2}(z_3\cdot\pl_{y_3})^{r_3}(z_4\cdot\pl_{y_4})^{r_4}I^{0,0,0,0}_{\alpha_1,\alpha_2,\alpha_3,\alpha_4}\,,\\
   I^{0,0,0,0}_{\alpha_1,\alpha_2,\alpha_3,\alpha_4}&=\int d^d y_0\frac{1}{(y_{01}^2)^{\alpha_1}(y_{02}^2)^{\alpha_2}(y_{03}^2)^{\alpha_3}(y_{04}^2)^{\alpha_4}},
\end{align}
\end{subequations}
with $\alpha_1+\alpha_2+\alpha_3+\alpha_4=d$.
At this point it is useful to switch to Mellin space using the Symanzik star formula \cite{Symanzik:1972wj}, where the above integrals for $r_i=0$ have a simple representation:
\begin{multline}\label{0000conf}
    I^{0,0,0,0}_{\alpha_1,\alpha_2,\alpha_3,\alpha_4}=\tfrac{1}{\left(y_{12}^2\right)^{\frac{1}{2}(\alpha_1 + \alpha_2)} \left(y_{34}^2\right)^{\frac{1}{2}(\alpha_3 + \alpha_4)}}\left(\frac{y_{24}^2}{y_{14}^2}\right)^{\tfrac{\alpha_1-\alpha_2}2}\left(\frac{y_{14}^2}{y_{13}^2}\right)^{\tfrac{\alpha_3-\alpha_4}2}\\\times\,\pi^{d/2}\int\frac{ds\, dt}{(4\pi i)^2}\,\rho_{\left\{\alpha_i\right\}}\left(s,t\right) u^{t/2}v^{-(s+t)/2}\,,
\end{multline}
with Mellin measure
\begin{align}
\rho_{\left\{\alpha_i\right\}}\left(s,t\right)&=\Gamma \left(\tfrac{-t+\tau_1+\tau_2}2\right) \Gamma \left(\tfrac{-t+\tau_3+\tau_4}2\right)\nonumber\\&\hspace{100pt}\times\Gamma \left(\tfrac{s+t}{2}\right) \Gamma \left(\tfrac{-s-\tau_1+\tau_2}2\right) \Gamma \left(\tfrac{-s+\tau_3-\tau_4}2\right) \Gamma \left(\tfrac{ s+t+\tau_1-\tau_2-\tau_3+\tau_4}2\right).
\end{align}

The Mellin representation of the integrals \eqref{stci} with $r_i>0$ is obtained from the result \eqref{0000conf} for $r_i=0$ with the differential operators as in \eqref{stci}. For simplicity one can set to zero $z_i\cdot z_j$ with the full result reconstructable at the end via conformal symmetry. Setting $z_i\cdot z_j = 0$ is moreover preserved by differentiation with respect to $y_i$ and gives a simple expression for the action of the differential operators $z_i \cdot \partial_{y_i}$:
\begin{subequations}
\begin{align}
    z_1\cdot\pl_{y_1}&=2 \sum_{j\neq 1}z_1\cdot y_{1j}\,\pl_{y_{1j}^2}+2\left(\frac{z_1\cdot y_{12}}{y_{12}^2}-\frac{z_1\cdot y_{13}}{y_{13}^2}\right)u\pl_u+2\left(\frac{z_1\cdot y_{14}}{y_{14}^2}-\frac{z_1\cdot y_{13}}{y_{13}^2}\right)v\pl_v\\
    z_2\cdot\pl_{y_2}&=2 \sum_{j\neq 2}z_2\cdot y_{2j}\,\pl_{y_{2j}^2}+2\left(\frac{z_2\cdot y_{21}}{y_{12}^2}-\frac{z_2\cdot y_{24}}{y_{24}^2}\right)u\pl_u+2\left(\frac{z_2\cdot y_{23}}{y_{23}^2}-\frac{z_2\cdot y_{24}}{y_{24}^2}\right)v\pl_v\\
    z_3\cdot\pl_{y_3}&=2 \sum_{j\neq 3}z_3\cdot y_{3j}\,\pl_{y_{3j}^2}+2\left(\frac{z_3\cdot y_{34}}{y_{34}^2}-\frac{z_3\cdot y_{31}}{y_{31}^2}\right)u\pl_u+2\left(\frac{z_3\cdot y_{32}}{y_{23}^2}-\frac{z_3\cdot y_{31}}{y_{13}^2}\right)v\pl_v\\
    z_4\cdot\pl_{y_4}&=2 \sum_{j\neq 4}z_4\cdot y_{4j}\,\pl_{y_{4j}^2}+2\left(\frac{z_4\cdot y_{43}}{y_{34}^2}-\frac{z_4\cdot y_{42}}{y_{24}^2}\right)u\pl_u+2\left(\frac{z_4\cdot y_{41}}{y_{14}^2}-\frac{z_4\cdot y_{42}}{y_{24}^2}\right)v\pl_v
\end{align}
\end{subequations}
Given a spinning CPW in any basis, with the above formalism one can extract the corresponding Mack polynomial.

\section{Contractions}\label{Contractions}

The contractions among the $\Xi$ and ${\bar \Xi}$ defined in \eqref{XiXibar} are given explicitly by
{\allowdisplaybreaks\small
\begin{subequations}
\begin{align}
    \Xi\cdot\bar{\Xi}&=
    - \frac{\left(y_{13}^2\right)}{2 \left(y_{01}^2\right) \left(y_{03}^2\right)}
    + \frac{\left(y_{14}^2\right)}{2 \left(y_{01}^2\right) \left(y_{04}^2\right)} 
    + \frac{\left(y_{23}^2\right)}{2 \left(y_{02}^2\right) \left(y_{03}^2\right)} 
    -  \frac{\left(y_{24}^2\right)}{2 \left(y_{02}^2\right) \left(y_{04}^2\right)} 
    -  \frac{h_2 \left(y_{13}^2\right) \left(z_1\cdot y_{10}\right)}{\left(y_{01}^2\right)^2 \left(y_{03}^2\right)} \\\nonumber
    &+ \frac{h_2 \left(y_{14}^2\right) \left(z_1\cdot y_{10}\right)}{\left(y_{01}^2\right)^2 \left(y_{04}^2\right)} 
    + \frac{h_2 \left(z_1\cdot y_{13}\right)}{\left(y_{01}^2\right) \left(y_{03}^2\right)} 
    -  \frac{h_2 \left(z_1\cdot y_{14}\right)}{\left(y_{01}^2\right) \left(y_{04}^2\right)} 
    -  \frac{h_1 \left(y_{23}^2\right) \left(z_2\cdot y_{20}\right)}{\left(y_{02}^2\right)^2 \left(y_{03}^2\right)}
    + \frac{h_1 \left(y_{24}^2\right) \left(z_2\cdot y_{20}\right)}{\left(y_{02}^2\right)^2 \left(y_{04}^2\right)} \\\nonumber
    &+ \frac{h_1 \left(z_2\cdot y_{23}\right)}{\left(y_{02}^2\right) \left(y_{03}^2\right)} 
    -  \frac{h_1 \left(z_2\cdot y_{24}\right)}{\left(y_{02}^2\right) \left(y_{04}^2\right)} 
    -  \frac{h_4 \left(y_{13}^2\right) \left(z_3\cdot y_{30}\right)}{\left(y_{01}^2\right) \left(y_{03}^2\right)^2} 
    + \frac{h_4 \left(y_{23}^2\right) \left(z_3\cdot y_{30}\right)}{\left(y_{02}^2\right) \left(y_{03}^2\right)^2} 
    \\\nonumber
    &-  \frac{2 h_2 h_4 \left(y_{13}^2\right) \left(z_1\cdot y_{10}\right) \left(z_3\cdot y_{30}\right)}{\left(y_{01}^2\right)^2 \left(y_{03}^2\right)^2} + \frac{2 h_2 h_4 \left(z_1\cdot y_{13}\right) \left(z_3\cdot y_{30}\right)}{\left(y_{01}^2\right) \left(y_{03}^2\right)^2} -  \frac{2 h_1 h_4 \left(y_{23}^2\right) \left(z_2\cdot y_{20}\right) \left(z_3\cdot y_{30}\right)}{\left(y_{02}^2\right)^2 \left(y_{03}^2\right)^2} \\\nonumber
    &+ \frac{2 h_1 h_4 \left(z_2\cdot y_{23}\right) \left(z_3\cdot y_{30}\right)}{\left(y_{02}^2\right) \left(y_{03}^2\right)^2} + \frac{h_4 \left(z_3\cdot y_{31}\right)}{\left(y_{01}^2\right) \left(y_{03}^2\right)} + \frac{2 h_2 h_4 \left(z_1\cdot y_{10}\right) \left(z_3\cdot y_{31}\right)}{\left(y_{01}^2\right)^2 \left(y_{03}^2\right)} -  \frac{h_4 \left(z_3\cdot y_{32}\right)}{\left(y_{02}^2\right) \left(y_{03}^2\right)}\\\nonumber
    & + \frac{2 h_1 h_4 \left(z_2\cdot y_{20}\right) \left(z_3\cdot y_{32}\right)}{\left(y_{02}^2\right)^2 \left(y_{03}^2\right)} -  \frac{h_3 \left(y_{14}^2\right) \left(z_4\cdot y_{40}\right)}{\left(y_{01}^2\right) \left(y_{04}^2\right)^2} + \frac{h_3 \left(y_{24}^2\right) \left(z_4\cdot y_{40}\right)}{\left(y_{02}^2\right) \left(y_{04}^2\right)^2} -  \frac{2 h_2 h_3 \left(y_{14}^2\right) \left(z_1\cdot y_{10}\right) \left(z_4\cdot y_{40}\right)}{\left(y_{01}^2\right)^2 \left(y_{04}^2\right)^2}\\\nonumber
    & + \frac{2 h_2 h_3 \left(z_1\cdot y_{14}\right) \left(z_4\cdot y_{40}\right)}{\left(y_{01}^2\right) \left(y_{04}^2\right)^2} -  \frac{2 h_1 h_3 \left(y_{24}^2\right) \left(z_2\cdot y_{20}\right) \left(z_4\cdot y_{40}\right)}{\left(y_{02}^2\right)^2 \left(y_{04}^2\right)^2} + \frac{2 h_1 h_3 \left(z_2\cdot y_{24}\right) \left(z_4\cdot y_{40}\right)}{\left(y_{02}^2\right) \left(y_{04}^2\right)^2}\\\nonumber
    & + \frac{h_3 \left(z_4\cdot y_{41}\right)}{\left(y_{01}^2\right) \left(y_{04}^2\right)} + \frac{2 h_2 h_3 \left(z_1\cdot y_{10}\right) \left(z_4\cdot y_{41}\right)}{\left(y_{01}^2\right)^2 \left(y_{04}^2\right)} -  \frac{h_3 \left(z_4\cdot y_{42}\right)}{\left(y_{02}^2\right) \left(y_{04}^2\right)} + \frac{2 h_1 h_3 \left(z_2\cdot y_{20}\right) \left(z_4\cdot y_{42}\right)}{\left(y_{02}^2\right)^2 \left(y_{04}^2\right)}\\
\Xi\cdot{\Xi}&=\frac{\left(y_{12}^2\right)}{\left(y_{01}^2\right) \left(y_{02}^2\right)} + \frac{2 h_2 \left(y_{12}^2\right) \left(z_1\cdot y_{10}\right)}{\left(y_{01}^2\right)^2 \left(y_{02}^2\right)} -  \frac{2 h_2 \left(z_1\cdot y_{12}\right)}{\left(y_{01}^2\right) \left(y_{02}^2\right)} -  \frac{2 h_1 \left(y_{12}^2\right) \left(z_2\cdot y_{20}\right)}{\left(y_{01}^2\right) \left(y_{02}^2\right)^2}\\\nonumber& -  \frac{4 h_1 h_2 \left(y_{12}^2\right) \left(z_1\cdot y_{10}\right) \left(z_2\cdot y_{20}\right)}{\left(y_{01}^2\right)^2 \left(y_{02}^2\right)^2} + \frac{4 h_1 h_2 \left(z_1\cdot y_{12}\right) \left(z_2\cdot y_{20}\right)}{\left(y_{01}^2\right) \left(y_{02}^2\right)^2} + \frac{2 h_1 \left(z_2\cdot y_{21}\right)}{\left(y_{01}^2\right) \left(y_{02}^2\right)} + \frac{4 h_1 h_2 \left(z_1\cdot y_{10}\right) \left(z_2\cdot y_{21}\right)}{\left(y_{01}^2\right)^2 \left(y_{02}^2\right)}\\
    \bar{\Xi}\cdot\bar{\Xi}&=\frac{\left(y_{34}^2\right)}{\left(y_{03}^2\right) \left(y_{04}^2\right)} + \frac{2 h_4 \left(y_{34}^2\right) \left(z_3\cdot y_{30}\right)}{\left(y_{03}^2\right)^2 \left(y_{04}^2\right)} -  \frac{2 h_4 \left(z_3\cdot y_{34}\right)}{\left(y_{03}^2\right) \left(y_{04}^2\right)} -  \frac{2 h_3 \left(y_{34}^2\right) \left(z_4\cdot y_{40}\right)}{\left(y_{03}^2\right) \left(y_{04}^2\right)^2}\\\nonumber& -  \frac{4 h_3 h_4 \left(y_{34}^2\right) \left(z_3\cdot y_{30}\right) \left(z_4\cdot y_{40}\right)}{\left(y_{03}^2\right)^2 \left(y_{04}^2\right)^2} + \frac{4 h_3 h_4 \left(z_3\cdot y_{34}\right) \left(z_4\cdot y_{40}\right)}{\left(y_{03}^2\right) \left(y_{04}^2\right)^2} + \frac{2 h_3 \left(z_4\cdot y_{43}\right)}{\left(y_{03}^2\right) \left(y_{04}^2\right)} + \frac{4 h_3 h_4 \left(z_3\cdot y_{30}\right) \left(z_4\cdot y_{43}\right)}{\left(y_{03}^2\right)^2 \left(y_{04}^2\right)}\,,
\end{align}
\end{subequations}
}
where for simplicity we set to zero without loss of generality all structures involving $z_i\cdot z_j$.\footnote{Note that we cannot drop all $z_i\cdot z_j$ before evaluating the contractions $\Xi\cdot\bar{\Xi}$. We first compute the scalar contraction and then drop $z_i\cdot z_j$.}

\section{Continuous Hahn polynomials}\label{Hahn}

In this appendix we review various properties of continuous Hahn polynomials which are relevant for this work. 

Continuous Hahn polynomials \cite{andrews_askey_roy_1999} $Q_l^{(a,b,c,d)}(s)$ are orthogonal with respect to the Mellin-Barnes bilinear product:
\begin{equation}
    \left\langle f(s)g(s)\right\rangle_{a,b,c,d}=\int_{-i\infty}^{i\infty}\frac{ds}{4\pi i}\,\Gamma(\tfrac{s+a}2)\Gamma(\tfrac{s+b}2)\Gamma(\tfrac{c-s}2)\Gamma(\tfrac{d-s}2)\,f(s)\,g(s)\,,
\end{equation}
with normalisation
\begin{multline}\label{Qnorm}
    \left\langle Q_{l}^{(a,b,c,d)}(s)Q_{n}^{(a,b,c,d)}(s)\right\rangle=\delta_{l,n}\\\times\, \underbrace{\frac{(-1)^n 4^n n! \Gamma \left(\frac{a+c}{2}+n\right) \Gamma \left(\frac{a+d}{2}+n\right) \Gamma \left(\frac{b+c}{2}+n\right) \Gamma \left(\frac{b+d}{2}+n\right)}{\left(\tfrac{a+b+c+d}{2}+n-1\right)_n \Gamma \left(\frac{a+b+c+d}{2}+2 n\right)}}_{\mathfrak{N}^{(a,b,c,d)}_n}\,.
\end{multline}
They can be expressed explicitly in terms of a hypergeometric function ${}_3F_2$ in two equivalent forms:
\begin{subequations}
\begin{align}
    Q_{l}^{(a,b,c,d)}(s)&=\frac{(-2)^l \left(\frac{a+c}{2}\right)_l \left(\frac{a+d}{2}\right)_l}{\left(\frac{a+b+c+d}{2}+l-1\right)_l}\, _3F_2\left(\begin{matrix}-l,\frac{a+b+c+d}{2}+l-1,\frac{a+s}{2}\\\frac{a+c}{2},\frac{a+d}{2}\end{matrix};1\right)\,,\\
    Q_{l}^{(a,b,c,d)}(s)&=\frac{2^l \left(\frac{a+d}{2}\right)_l \left(\frac{b+d}{2}\right)_l}{\left(\frac{a+b+c+d}{2}+l-1\right)_l}\, _3F_2\left(\begin{matrix}-l,\frac{a+b+c+d}{2}+l-1,\frac{d-s}{2}\\\frac{a+d}{2},\frac{b+d}{2}\end{matrix};1\right)\,,
\end{align}
\end{subequations}
with unit normalisation for $s^l$ monomial $Q_{l}^{(a,b,c,d)}(s)\sim s^l+\ldots$\,. The two representations above admit the following series expansion in terms of $\left(\tfrac{-s+d}2\right)_n$ and $\left(\tfrac{s+a}2\right)_n$:
\begin{subequations}
\begin{align}
     Q_{l}^{(a,b,c,d)}(s)&=\sum_n\,\frac{2^l (-1)^n \binom{l}{n} \left(\frac{a+d}{2}+n\right)_{l-n} \left(\frac{b+d}{2}+n\right)_{l-n}}{\left(\tfrac{a+b+c+d}{2}+l+n-1\right)_{l-n}}\,\left(\tfrac{-s+d}2\right)_n\,,\label{Qtchannel}\\
     Q_{l}^{(a,b,c,d)}(s)&=\sum_n\,\frac{(-2)^l (-1)^n \binom{l}{n} \left(\frac{a+c}{2}+n\right)_{l-n} \left(\frac{a+d}{2}+n\right)_{l-n}}{\left(\tfrac{a+b+c+d}{2}+l+n-1\right)_{l-n}}\,\left(\tfrac{s+a}2\right)_n\,.\label{Quchannel}
\end{align}
\end{subequations}
The first expansion \eqref{Qtchannel} above is useful when dealing with $\sf{t}$-channel CPWs expanded in the $\sf{s}$-channel. The second expansion \eqref{Quchannel} is instead useful when studying the $\sf{u}$-channel CPWs. 

The following integral identities play a central role in the present work. Defining
\begin{equation}
    \mathfrak{I}_{\alpha_1,\alpha_2,\alpha_3,\alpha_4}^{\beta_1,\beta_2,\beta_3,\beta_4}(l)\equiv \int_{-i\infty}^{+i\infty}\frac{ds}{4\pi i}\,\Gamma\left(\tfrac{s+\alpha_1}2\right)\Gamma\left(\tfrac{s+\alpha_2}2\right)\Gamma\left(\tfrac{-s+\alpha_3}2\right)\Gamma\left(\tfrac{-s+\alpha_4}2\right)Q_l^{(\beta_1,\beta_2,\beta_3,\beta_4)}(s)\,,
\end{equation}
we obtain the following two equivalent expressions:
\begin{subequations}\label{int4F3}
\begin{align}
    \label{t-channel4F3}&\mathfrak{I}_{\alpha_1,\alpha_2,\alpha_3,\alpha_4}^{\beta_1,\beta_2,\beta_3,\beta_4}(l)=2^l\frac{\Gamma \left(\tfrac{\alpha_1+\alpha_3}{2}\right) \Gamma \left(\tfrac{\alpha_1+\alpha_4}{2}\right) \Gamma \left(\tfrac{\alpha_2+\alpha_3}{2}\right) \Gamma \left(\tfrac{\alpha_2+\alpha_4}{2}\right)}{\Gamma \left(\tfrac{\alpha_1+\alpha_2+\alpha_3+\alpha_4}{2}\right)}\sum_{p=0}^l \binom{l}{p} \left(-p+\tfrac{\alpha_4-\beta_4}{2}+1\right)_p\\\nonumber
    &\times\frac{\left(p+\tfrac{\beta_1+\beta_4}{2}\right)_{l-p} \left(p+\tfrac{\beta_2+\beta_4}{2}\right)_{l-p}}{\left(p+\tfrac{2 l+\beta_1+\beta_2+\beta_3+\beta_4-2}{2}\right)_{l-p}}\, _4F_3\left(\begin{matrix}p-l,\tfrac{\alpha_1+\alpha_4}{2},\tfrac{\alpha_2+\alpha_4}{2},\tfrac{\beta_1+\beta_2+\beta_3+\beta_4}{2}+l+p-1\\\frac{\alpha_1+\alpha_2+\alpha_3+\alpha_4}{2},\frac{\beta_1+\beta_4}{2}+p,\frac{\beta_2+\beta_4}{2}+p\end{matrix};1\right)\\
    \label{u-channel4F3}&\mathfrak{I}_{\alpha_1,\alpha_2,\alpha_3,\alpha_4}^{\beta_1,\beta_2,\beta_3,\beta_4}(l)=(-2)^l \frac{\Gamma \left(\tfrac{\alpha_1+\alpha_3}{2}\right) \Gamma \left(\tfrac{\alpha_1+\alpha_4}{2}\right) \Gamma \left(\tfrac{\alpha_2+\alpha_3}{2}\right) \Gamma \left(\tfrac{\alpha_2+\alpha_4}{2}\right)}{\Gamma \left(\tfrac{\alpha_1+\alpha_2+\alpha_3+\alpha_4}{2}\right)}\sum_{p=0}^l\,\binom{l}{p} \left(-p+\tfrac{\alpha_1-\beta_1}{2}+1\right)_p\\\nonumber
    &\times\frac{\left(p+\tfrac{\beta_1+\beta_3}{2}\right)_{l-p} \left(p+\tfrac{\beta_1+\beta_4}{2}\right)_{l-p}}{\left(p+\tfrac{2 l+\beta_1+\beta_2+\beta_3+\beta_4-2}{2}\right)_{l-p}}\, _4F_3\left(\begin{matrix}p-l,\frac{\alpha_1+\alpha_4}{2},\frac{\alpha_2+\alpha_4}{2},\frac{\beta_1+\beta_2+\beta_3+\beta_4}{2}+l+p-1\\\frac{\alpha_1+\alpha_2+\alpha_3+\alpha_4}{2},\frac{\beta_1+\beta_3}{2}+p,\frac{\beta_1+\beta_4}{2}+p\end{matrix};1\right)\,.
\end{align}
\end{subequations}
The first of these sums truncates for some $p<l$ to a finite sum when considering $\sf{t}$-channel CPWs, for in that case $\frac{\alpha_4-\beta_4}2$ is a non-negative integer. The second truncates for some $p<l$ to a finite sum when considering $\sf{u}$-channel CPWs, for which $\frac{\alpha_1-\beta_1}2$ is a non-negative integer. These two relations are obtained using the two representations \eqref{Qtchannel} and \eqref{Quchannel} introduced for the continuous Hahn polynomials and play a ubiquitous role in the evaluation of crossing kernels.

The above integral identities also allow to recover further useful identities for the overlap of two continuous Hahn polynomials with different arguments and spin:
\begin{multline}\label{QQintegral}
 \mathfrak{Z}_{\alpha_1,\alpha_2,\alpha_3,\alpha_4}^{\beta_1,\beta_2,\beta_3,\beta_4}(l,l^\prime)\equiv\left(\mathfrak{N}_{l^\prime}^{(\alpha_1,\alpha_2,\alpha_3,\alpha_4)}\right)^{-1}\int_{-i\infty}^{i\infty}\frac{ds}{4\pi i}\,\Gamma\left(\tfrac{s+\alpha_1}2\right)\Gamma\left(\tfrac{s+\alpha_2}2\right)\Gamma\left(\tfrac{-s+\alpha_3}2\right)\Gamma\left(\tfrac{-s+\alpha_4}2\right)\\\times\,Q_{l^\prime}^{(\alpha_1,\alpha_2,\alpha_3,\alpha_4)}(s)\,Q_l^{(\beta_1,\beta_2,\beta_3,\beta_4)}(s)\,,
\end{multline}
\begin{align}
   \mathfrak{Z}&_{\alpha_1,\alpha_2,\alpha_3,\alpha_4}^{\beta_1,\beta_2,\beta_3,\beta_4}(l,l^\prime)=\frac{(-1)^{l^\prime} 2^{l-{l^\prime}}} {{l^\prime}!} \sum_{p=0}^l\sum_{f=0}^{l^\prime}(-1)^f\binom{{l^\prime}}{f} \binom{l}{p}\left(\tfrac{\alpha_4-\beta_4}{2}+f-p+1\right)_p\\\nonumber&\times\frac{\left(p+\frac{\beta_1+\beta_4}{2}\right)_{l-p} \left(p+\tfrac{\beta_2+\beta_4}{2}\right)_{l-p}}{\left(\tfrac{\alpha_1+\alpha_3}{2}\right)_{{l^\prime}} \left(\tfrac{\alpha_2+\alpha_3}{2}\right)_{{l^\prime}}}\frac{\left(\tfrac{2 {l^\prime}+\alpha_1+\alpha_2+\alpha_3+\alpha_4-2}{2}\right)_{{l^\prime}+1} \left(\tfrac{2 f+\alpha_1+\alpha_2+\alpha_3+\alpha_4}{2}\right)_{{l^\prime}-1}}{\left(\tfrac{2 l+2 p+\beta_1+\beta_2+\beta_3+\beta_4-2}{2}\right)_{l-p}}\\\nonumber&\times\, _4F_3\left(\begin{matrix}p-l,\tfrac{\alpha_1+\alpha_4}{2}+f,\tfrac{\alpha_2+\alpha_4}{2}+f,\tfrac{\beta_1+\beta_2+\beta_3+\beta_4}{2}+l+p-1\\\tfrac{\alpha_1+\alpha_2+\alpha_3+\alpha_4}{2}+f,\tfrac{\beta_1+\beta_4}{2}+p,\frac{\beta_2+\beta_4}{2}+p\end{matrix};1\right)\,,
\end{align}
which implies the following decomposition:
\begin{equation}
    Q_l^{(\beta_1,\beta_2,\beta_3,\beta_4)}(s)= Q_{l^\prime}^{(\alpha_1,\alpha_2,\alpha_3,\alpha_4)}(s)+\sum_{l^\prime=0}^{l-1} \mathfrak{Z}_{\alpha_1,\alpha_2,\alpha_3,\alpha_4}^{\beta_1,\beta_2,\beta_3,\beta_4}(l,l^\prime)\,Q_{l^\prime}^{(\alpha_1,\alpha_2,\alpha_3,\alpha_4)}(s)\,.
\end{equation}
Note that for particular integer values of $\tfrac{\alpha_4-\beta_4}2$ the sum over $p$ and $f$ truncates independently of $l$ or $l^\prime$. Further simplifications also appear when $\alpha_i-\beta_i=2n_i$ with $n_i$ an integer, in which case the Hypergeometric functions simplify via Gauss-type identities.
A particularly simple case which we employ in \S\tcb{\ref{2020corr}} is when only one among $\alpha_i$ and $\beta_i$ is different. There are four such cases which give the following simple results:
\begin{subequations}
\begin{align}
    \mathfrak{Z}_{\alpha_1,\alpha_2,\alpha_3,\alpha_4}^{\alpha_1,\alpha_2,\alpha_3,\beta_4}(l,i)&=\frac{(-2)^{l-i} \binom{l}{i}\left(i+\frac{\alpha_1+\alpha_3}{2}\right)_{l-i} \left(i+\frac{\alpha_2+\alpha_3}{2}\right)_{l-i} \left(i+\frac{\beta_4-\alpha_4}{2}\right)_{l-i}}{\left(\tfrac{\alpha_1+\alpha_2+\alpha_3+\alpha_4}{2}+2i\right)_{l-i} \left(\tfrac{\alpha_1+\alpha_2+\alpha_3+\beta_4}{2}+i+l-1\right)_{l-i}}\,,\\
    \mathfrak{Z}_{\alpha_1,\alpha_2,\alpha_3,\alpha_4}^{\alpha_1,\alpha_2,\beta_3,\alpha_4}(l,i)&=\frac{(-2)^{l-i} \binom{l}{i} \left(i+\frac{\alpha_1+\alpha_4}{2}\right)_{l-i} \left(i+\frac{\alpha_2+\alpha_4}{2}\right)_{l-i} \left(i+\frac{\beta_3-\alpha_3}{2}\right)_{l-i}}{\left(\tfrac{\alpha_1+\alpha_2+\alpha_3+\alpha_4}{2}+2i\right)_{l-i} \left(\tfrac{\alpha_1+\alpha_2+\beta_3+\alpha_4}{2}+i+l-1\right)_{l-i}}\,,\\
    \mathfrak{Z}_{\alpha_1,\alpha_2,\alpha_3,\alpha_4}^{\alpha_1,\beta_2,\alpha_3,\alpha_4}(l,i)&=\frac{2^{l-i} \binom{l}{i}\left(i+\frac{\alpha_1+\alpha_3}{2}\right)_{l-i} \left(i+\frac{\alpha_1+\alpha_4}{2}\right)_{l-i} \left(i+\frac{\beta_2-\alpha_2}{2}\right)_{l-i}}{\left(\tfrac{\alpha_1+\alpha_2+\alpha_3+\alpha_4}{2}+2i\right)_{l-i} \left(\tfrac{\alpha_1+\beta_2+\alpha_3+\alpha_4}{2}+i+l-1\right)_{l-i}}\,,\\
    \mathfrak{Z}_{\alpha_1,\alpha_2,\alpha_3,\alpha_4}^{\beta_1,\alpha_2,\alpha_3,\alpha_4}(l,i)&=\frac{2^{l-i} \binom{l}{i} \left(i+\frac{\alpha_2+\alpha_3}{2}\right)_{l-i} \left(i+\frac{\alpha_2+\alpha_4}{2}\right)_{l-i} \left(i+\frac{\beta_1-\alpha_1}{2}\right)_{l-i}}{\left(\tfrac{\alpha_1+\alpha_2+\alpha_3+\alpha_4}{2}+2i\right)_{l-i} \left(\tfrac{\beta_1+\alpha_2+\alpha_3+\alpha_4}{2}+i+l-1\right)_{l-i}}\,,
\end{align}
\end{subequations}
Applying the above relation in \S\tcb{\ref{1010corr}} and \S\tcb{\ref{2020corr}} we can prove that the overlap between different tensor structures in the CPWs arises within the triangles in fig. \ref{fig:Jphitableaux}, \ref{fig:Tphi} and \ref{fig:Jphi}.

\section{Crossing kernel details}\label{KernelDetails}

In this appendix we give some details about the evaluation of the Mellin-Barnes integrals in \S \tcb{\ref{Crossing Kernels}} for the crossing kernels. This type of computation was first made in \cite{Gopakumar:2016cpb}, in which the result was expressed as a series. In the following we revisit this computation, where we are able to express the general result in terms of hypergeometric functions ${}_4F_3$.

There are two types of integrals which need to be evaluated. The  first corresponds to the ${\sf t}$- to the ${\sf s}$-channel crossing kernels and the second to the ${\sf u}$- to the ${\sf s}$-channel crossing kernels.

\paragraph{From the $\sf t$- to the $\sf s$-channel}
In this case, the seed integral\footnote{In this context \emph{seed integral} refers to the basic integral which all others reduce to.} we need to evaluate is the following Mellin-Barnes integral at fixed $t$:
\begin{multline}
    \mathfrak{J}^{({\sf t})}_{J,l}=\frac{(-1)^l}{l!}\int \frac{ds}{4\pi i}\,\Gamma \left(\tfrac{-s-\tau_1+\tau_2}{2} \right) \Gamma \left(\tfrac{-s+\tau_3-\tau_4}{2}\right) \Gamma \left(\tfrac{s+t+\tau -\tau_2-\tau_3}{2}\right) \Gamma \left(\tfrac{d-2 r+s+t-\tau -\tau_2-\tau_3}{2}\right)\\\times\,\tilde{\mathcal{C}}^{({\sf t})}_{J,\tau}P^{(\sf{t})}_{J,\tau}(s,t|\tau_i)\,Q_{l,t}^{(t,t+\tau_1-\tau_2-\tau_3+\tau_4,-\tau_1+\tau_2,\tau_3-\tau_4}(s)\,,
\end{multline}
with
\begin{equation}
    \tilde{\mathcal{C}}^{({\sf t})}_{J,\tau}=\frac{2^{-2 r} (J+\tau -1)_J \Gamma (2 J+\tau )}{\Gamma \left(\tfrac{d-2 (J+\tau )}{2}\right) \Gamma \left(\tfrac{2 J+\tau +\tau_1-\tau_4}{2}\right) \Gamma \left(\tfrac{2 J+\tau -\tau_1+\tau_4}{2}\right) \Gamma \left(\tfrac{2 J+\tau +\tau_2-\tau_3}{2}\right) \Gamma \left(\tfrac{2 J+\tau -\tau_2+\tau_3}{2}\right)}.
\end{equation}
This integral can be reduced to finite sums of standard hypergeometric Mellin-Barnes integrals using the definitions of Mack and Continuous Hahn polynomials. For the $\sf{t}$-channel crossing kernel, we use the following expansion for the continuous Hahn polynomials:
\begin{multline}
    Q_{l,t}^{(t,t+\tau_1-\tau_2-\tau_3+\tau_4,-\tau_1+\tau_2,\tau_3-\tau_4}(s)\\=\,\sum_{q=0}^{\infty}\underbrace{\frac{2^l \left(\tfrac{t+\tau_1-\tau_2}{2}\right)_l \left(\tfrac{t+\tau_3-\tau_4}{2}\right)_l}{(l+t-1)_l}\frac{(-1)^q \binom{l}{q} (l+t-1)_q}{\left(\frac{t+\tau_1-\tau_2}{2}\right)_q \left(\frac{t+\tau_3-\tau_4}{2}\right)_q}}_{d^{(1)}_{t,l}(q)}\,\left(\tfrac{-s+\tau_3-\tau_4}{2}\right)_q\,,
\end{multline}
while we focus on the factors in the Mack polynomial which are proportional to $s$.
Combining everything we obtain the following expression:
\begin{equation}\label{FullCrossing}
    \mathfrak{J}^{(\sf t)}_{J,l}=\frac{(-1)^l}{l!}\tilde{\mathcal{C}}^{({\sf t})}_{J,\tau}\sum_{k=0}^{[J/2]}\sum_{\sum_i r_i=J-2k}b^{(\sf t)}_{t,\tau}(r_i|k)\,\sum_{q=0}^{\infty}{d}^{(1)}_{t,l}(q)\underbrace{\int \frac{ds}{4\pi i}\,\rho_{M}^{(\sf t)}(r_i|k)\left(\tfrac{-s+\tau_3-\tau_4}{2}\right)_q}_{I^{(\sf t)}_{r_i|k,q}}\ ,
\end{equation}
where we have also defined the Mellin-Barnes measure:
\begin{multline}
    \rho_M^{(\sf t)}(r_i|k)=\Gamma \left(\tfrac{-s-\tau_1+\tau_2+2r_3}{2}\right) \Gamma \left(\tfrac{-s+\tau_3-\tau_4+2r_2}{2}\right)\\\times\,\Gamma \left(\tfrac{s+t+\tau -\tau_2-\tau_3+2k}{2}\right) \Gamma \left(\tfrac{d-2 r+s+t-\tau -\tau_2-\tau_3+2k}{2}\right),
\end{multline}
and the coefficient:
\begin{multline}
    b^{(\sf t)}_{t,\tau}(r_i|k)=c_{J,k}\,\tfrac{\left(\tfrac{\tau +\tau_1-\tau_4}{2}\right)_J \left(\tfrac{\tau -\tau_1+\tau_4}{2}\right)_J}{(J+\tau -1)_J (d-J-\tau -1)_J}\,\frac{(-1)^{r_1+r_4} (r_1,r_2,r_3,r_4)!}{2^{r_1+r_2+r_3+r_4}}\\\times \frac{\left(\tfrac{-t+\tau_3+\tau_4}{2}\right)_{r_1} \left(\tfrac{-t+\tau_1+\tau_2}{2}\right)_{r_4}\left(\tfrac{-d+\tau -\tau_2+\tau_3+2}{2}\right)_{k+r_1+r_2} \left(\tfrac{-d+\tau +\tau_2-\tau_3+2}{2}\right)_{k+r_3+r_4}}{\left(\tfrac{\tau -\tau_1+\tau_4}{2}\right)_{k+r_1+r_3} \left(\tfrac{\tau +\tau_1-\tau_4}{2}\right)_{k+r_2+r_4}}.
\end{multline}
We can now evaluate the Mellin-Barnes integral explicitly using a generating function trick \cite{Gopakumar:2016cpb}. Indeed a generating function for Pochhammer symbols is simply:
\begin{equation}
    (1-z)^{\tfrac{s-\tau_3+\tau_4}{2}}=\sum_{q=0}^\infty\tfrac1{q!}\,(\tfrac{s-\tau_3+\tau_4}2)_q\,z^q.
\end{equation}
Using the latter generating function and after simple change of variables we arrive to:
\begin{multline}
    I_{r_i|k,q}^{(\sf t)}=\pl_{z}^q\int_{-i\infty}^{i\infty}\frac{ds}{4\pi i}\,\rho_M^{(\sf t)}(r_i|k)(1-z)^{\tfrac{s-\tau_3+\tau_4}{2}}\Big|_{z=0}\\=\pl_{z}^q\left[(1-z)^{r_2}\int_{-i\infty}^{i\infty}\frac{ds}{4\pi i}\,\tilde{\rho}_M^{(\sf t)}(r_i|k)(1-z)^{-\tfrac{s}{2}}\right]_{z=0},
\end{multline}
with a new Mellin-Barnes Measure given by:
\begin{multline}
    \tilde{\rho}_M^{(\sf t)}(r_i|k)=\Gamma \left(\frac{s}{2}\right) \Gamma \left(\tfrac{2 k+2 r_2-s+t+\tau -\tau_2-\tau_4}{2}\right)\\\times\, \Gamma \left(\tfrac{-2 r_2+2 r_3+s-\tau_1+\tau_2-\tau_3+\tau_4}{2}\right) \Gamma \left(\tfrac{d-2 J+2 k+2 r_2-s+t-\tau -\tau_2-\tau_4}{2}\right).
\end{multline}
Rewriting the binomial as $(1-z)^{r_2}=\sum_{p=0}^{r_2} \binom{r_2}{p}\,(-z)^p$ and using the identity:
\begin{multline}
    {}_2F_1(a,b;c;z)=\frac{\Gamma(\tfrac{c}{2})}{\Gamma(\tfrac{a}{2})\Gamma(\tfrac{b}{2})\Gamma(\tfrac{c-a}{2})\Gamma(\tfrac{c-b}{2})}\\\times\,\int_{-i\infty}^{i\infty}\frac{ds}{4\pi i}\Gamma\left(\tfrac{s}{2}\right)\Gamma\left(\tfrac{c-a-b+s}{2}\right)\Gamma\left(\tfrac{a-s}{2}\right)\Gamma\left(\tfrac{b-s}{2}\right)(1-z)^{-\tfrac{s}2}\,,
\end{multline}
after evaluating the derivatives with respect to $z$, we then arrive to
\begin{equation}
    I_{r_i|k,q}^{(\sf t)}=\sum_{p=0}^{r_2}I_{r_i|k,q,p}^{(\sf t)}\,,
\end{equation}
with
\begin{multline}
    I_{r_i|k,q,p}^{(\sf t)}=(-1)^p\,p!\,\binom{r_2}{p}\,\binom{q}{p}\\\times\tfrac{\Gamma \left(k+r_3+\tfrac{a_{13}}{2}\right) \Gamma \left(k+r_2+q-p+\tfrac{a_{24}}{2}\right) \Gamma \left(k+r_3+\tfrac{b_{13}}{2}\right) \Gamma \left(k+r_2+q-p+\tfrac{b_{24}}{2}\right)}{\Gamma \left(2 k+(q-p)+r_2+r_3+\frac{c}{2}\right)}\,.
\end{multline}
where
\begin{subequations}
\begin{align}
    a_{13}&=t+\tau-\tau_1-\tau_3\,,\\
    a_{24}&=t+\tau-\tau_2-\tau_4\,,\\
    b_{13}&=t+d-\tau-2J-\tau_1-\tau_3\,,\\
    b_{24}&=t+d-\tau-2J-\tau_2-\tau_4\,,\\
    c&=d-2J+2t-\tau_1-\tau_2-\tau_3-\tau_4.
\end{align}
\end{subequations}
The above result allows to explicitly perform the infinite sum over $q$ in \eqref{FullCrossing} replacing it with a finite sum in $p$ independent from $l$, as
\begin{multline}
    J^{(\sf t)}(r_i|k,p)\equiv\sum_{q=0}^{\infty}d^{(1)}_{t,l}(q)I^{(\sf t)}_{r_i|k,q,p}=\tfrac{(-2)^l p! \binom{l}{p} \binom{r_2}{p}(l+t-1)_p \left(\tfrac{t+\tau_1-\tau_2}{2}\right)_l \left(\tfrac{t+\tau_3-\tau_4}{2} \right)_l}{l!(l+t-1)_l \left(\frac{t+\tau_1-\tau_2}{2}\right)_p \left(\frac{t+\tau_3-\tau_4}{2}\right)_p}\gamma_{r_i|p}^{(\sf t)}\,\\\times\,{} _4F_3\left(\begin{matrix}-l+p,l+p+t-1,k+r_2+\tfrac{b_{24}}2,k+r_3+\tfrac{a_{24}}2\\\tfrac{2p+t+\tau_1-\tau_2}{2},\tfrac{2p+t+\tau_3-\tau_4}{2},2k+r_2+r_3+\tfrac{c}2\end{matrix};1\right),
\end{multline}
with
\begin{align}
    \gamma_{r_i|p}^{(\sf t)}=\frac{\Gamma \left(k+r_2+\tfrac{a_{24}}{2} \right) \Gamma \left(k+r_3+\tfrac{a_{13}}{2}\right) \Gamma \left(k+r_2+\tfrac{b_{24}}{2}\right) \Gamma \left(k+r_3+\tfrac{b_{13}}{2}\right)}{\Gamma \left(2k+r_2+r_3+\tfrac{c}{2}\right)},
\end{align}
reducing the full crossing kernel to a finite sum of hypergeometric functions ${}_4F_3$:
\begin{multline}\label{FullCrossingT}
    \mathfrak{J}^{(\sf t)}_{J,l}=\tilde{\mathcal{C}}^{({\sf t})}_{J,\tau}\sum_{k=0}^{[J/2]}\sum_{\sum_i r_i=J-2k}\sum_{p=0}^{r_2}c_{t,\tau}^{(\sf t)}(r_i|k,p)\gamma_{r_i|p}^{(\sf t)}\\\times \,{} _4F_3\left(\begin{matrix}-l+p,l+p+t-1,k+r_2+\tfrac{b_{24}}2,k+r_3+\tfrac{a_{24}}2\\p+\tfrac{t+\tau_1-\tau_2}{2},p+\tfrac{t+\tau_3-\tau_4}{2},2k+r_2+r_3+\tfrac{c}2\end{matrix};1\right)\,,
\end{multline}
with an overall coefficient weighting the sum given by
\begin{align}
    c^{(\sf t)}_{t,\tau}(r_i|k,p)=\tfrac{(-2)^l p!\binom{l}{p} \binom{r_2}{p}(l+t-1)_p \left(\tfrac{t+\tau_1-\tau_2}{2}\right)_l \left(\tfrac{t+\tau_3-\tau_4}{2} \right)_l}{l!(l+t-1)_l \left(\frac{t+\tau_1-\tau_2}{2}\right)_p \left(\frac{t+\tau_3-\tau_4}{2}\right)_p}\,b^{(\sf t)}_{t,\tau}(r_i|k).
\end{align}
At this stage it is easy to notice that the sum over $r_1$ factorises with respect to the hypergeometric function. Solving the constraint $r_1+r_2+r_3+r_4=J-2k$ as:
\begin{align}
    r_3=J-2k-r_2-j\,, \qquad r_4=j-r_1\,,
\end{align}
we can resum the $r_1$ dependent terms as:
{\allowdisplaybreaks
\begin{multline}
    \sum_{r_1=0}^{J-2k}{\tiny\begin{pmatrix}
    J-2k\\
    r_1,r_2,j-r_1
    \end{pmatrix}}
    \tfrac{\left(\tfrac{-t+\tau_3+\tau_4}{2} \right)_{r_1} \left(\tfrac{-t+\tau_1+\tau_2}{2}\right)_{j-r_1}  \left(\tfrac{\tau -\tau_2+\tau_3+2-d}{2}\right)_{k+r_1+r_2} \left(\tfrac{\tau +\tau_2-\tau_3+2-d}{2}\right)_{J-k-r_1-r_2}}{\left(\tfrac{\tau +\tau_1-\tau_4}{2}\right)_{j+k-r_1+r_2} \left(\tfrac{\tau -\tau_1+\tau_4}{2}\right)_{-j+J-k+r_1-r_2}}\\=
    {\tiny\begin{pmatrix}
    J-2k\\
    r_2,j
    \end{pmatrix}}\frac{\left(\tfrac{-t+\tau_1+\tau_2}{2}\right)_j \left(\tfrac{\tau -\tau_2+\tau_3+2-d}{2}\right)_{k+r_2} \left(\tfrac{\tau +\tau_2-\tau_3+2-d}{2}\right)_{J-k-r_2}}{\left(\tfrac{\tau +\tau_1-\tau_4}{2}\right)_{j+k+r_2} \left(\tfrac{\tau -\tau_1+\tau_4}{2}\right)_{J-j-k-r_2}}\times\\\times\,{\small{}_4F_3\left(\begin{matrix}
        -j,-\frac{d}{2}+k+r_2+\frac{\tau -\tau_2+\tau_3}{2}+1,-j-k-r_2-\frac{\tau+\tau_1-\tau_4 }{2}+1,-\frac{t-\tau_3-\tau_4}{2}\\-j+\frac{t-\tau_1-\tau_2}{2}+1,\frac{d-\tau-\tau_2+\tau_3 }{2}-J+k+r_2,J-j-k-r_2+\frac{\tau-\tau_1+\tau_4 }{2}
    \end{matrix};1\right)}.
\end{multline}}
Now the summation over $r_2$ and $j$ can be extended up to $J$ owing to the vanishing of the overall coefficients for the additional values of $r_2$ and $j$. This implies that we can perform also the sum over $k$ explicitly. After redefining $r_2=i+p-k$ the sum over $i$ and $j$ run from $0$ to $J-p$ and the sum over $k$ gives:
\begin{multline}
    \sum_{k=0}^{[J/2]}{\footnotesize\begin{pmatrix}
    J-2k\\
    j,p+i-k
    \end{pmatrix}}\frac{(-1)^k\binom{i-k+p}{p} \Gamma \left(\frac{d}{2}+J-k-1\right)}{k! (J-2 k)!}\\={\footnotesize\begin{pmatrix}
    J\\
    i,p
    \end{pmatrix}}\tfrac{\binom{i+p}{p}}{J!} \Gamma \left(\tfrac{d+2 J-2}{2} \right) \, \underbrace{_2F_1\left(-i,i+j-J+p;2-\tfrac{d}{2}-J;1\right)}_{\tfrac{\left(\tfrac{d}{2} +j+p-1\right)_i}{\left(\tfrac{d}{2} + J-i-1\right)_i}}.
\end{multline}
To summarise we arrive to the following general expression:
\begin{empheq}[box=\fbox]{align}
\mathfrak{J}^{(\sf t)}_{J,l}(t)&=\mathcal{Z}_{J,l}(t)\,\sum_{p=0}^{
J}d_{p}^{(J,l)}\sum_{i,j=0}^{J-p}a^{(J,l|p)}_{i,j}\\\nonumber
&\times\, {}_4F_3\left(\begin{matrix}
-j,p+i+\tfrac{\tau-d -\tau_2+\tau_3}{2}+1,1-i-j-p-\tfrac{\tau+\tau_1-\tau_4}{2},-\tfrac{t-\tau_3-\tau_4}{2}\\ 1-j+\tfrac{t-\tau_1-\tau_2}{2},p+i-J+\tfrac{d-\tau-\tau_2+\tau_3\ }{2},J-i-j-p+\tfrac{\tau-\tau_1+\tau_4}{2}\end{matrix};1\right)\\\nonumber
&\times {}_4F_3\left(\begin{matrix}-l+p,l+p+t-1,p+i-J+\tfrac{t-\tau_2 -\tau_4}{2}+\tfrac{d-\tau}{2},p+i+\tfrac{t+\tau-\tau_2-\tau_4}{2}\\
p+\tfrac{t+\tau_1-\tau_2}{2},p+\tfrac{t+\tau_3-\tau_4}{2},\tfrac{d-\tau_1-\tau_2-\tau_3-\tau_4}{2}-j+t\end{matrix};1\right)\,,
\end{empheq}
with
{\allowdisplaybreaks
\begin{subequations}
\begin{align}
    \mathfrak{J}^{(\sf t)}_{0,0}(t)&=\tfrac{\Gamma (\tau )}{\Gamma \left(\tfrac{d-2 \tau}{2}\right) \Gamma \left(\tfrac{\tau +\tau_1-\tau_4}{2}\right) \Gamma \left(\tfrac{\tau -\tau_1+\tau_4}{2}\right) \Gamma \left(\tfrac{\tau +\tau_2-\tau_3}{2}\right) \Gamma \left(\tfrac{\tau -\tau_2+\tau_3}{2}\right)}\times\\\nonumber&\qquad\times\tfrac{\Gamma \left(\tfrac{t+\tau -\tau_1-\tau_3}{2}\right) \Gamma \left(\tfrac{t+\tau -\tau_2-\tau_4}{2}\right) \Gamma \left(\tfrac{d+t-\tau -\tau_1-\tau_3}{2}\right) \Gamma \left(\tfrac{d+t-\tau -\tau_2-\tau_4}{2}\right)}{\Gamma \left(\tfrac{d+2 t-\tau_1-\tau_2-\tau_3-\tau_4}{2}\right)},\\
    \mathcal{Z}_{J,l}(t)&=\frac{(-1)^l \, 2^{l-J} (\tau )_{2 J} \left(\tfrac{d-2 (J+\tau )}{2}\right)_J \left(\tfrac{t+\tau_1-\tau_2}{2}\right)_l \left(\tfrac{t+\tau_3-\tau_4}{2}\right)_l}{l!\, (l+t-1)_l\, (d-J-\tau -1)_J \left(\tfrac{\tau +\tau_2-\tau_3}{2}\right)_J \left(\tfrac{\tau -\tau_2+\tau_3}{2}\right)_J}\ \mathfrak{J}^{(\sf t)}_{0,0}(t)\,,\\
    d_p^{(J,l)}(t)&=\frac{p!\, \binom{l}{p}\, (l+t-1)_p}{\left(\tfrac{t+\tau_1-\tau_2}{2}\right)_p \left(\tfrac{t+\tau_3-\tau_4}{2}\right)_p},\\
    a^{(J,l|p)}_{i,j}(t)&=(-1)^j \binom{i+p}{p}{\small\begin{pmatrix}
    J\\ j,i+p
    \end{pmatrix}}\tfrac{\left(\tfrac{\tau -\tau_2+\tau_3+2-d}{2}\right)_{i+p} \left(\tfrac{\tau +\tau_2-\tau_3+2-d}{2}\right)_{J-p-i}}{\left(\tfrac{\tau +\tau_1-\tau_4}{2}\right)_{i+j+p} \left(\tfrac{\tau -\tau_1+\tau_4}{2}\right)_{J-p-i-j}}\\\nonumber
    &\times\,\tfrac{ \left(\tfrac{t+\tau -\tau_2-\tau_4}{2}\right)_{i+p}  \left(\tfrac{t+\tau -\tau_1-\tau_3}{2}\right)_{J-p-i-j}}{\left(\tfrac{d-2 (p+i+j)+t-\tau -\tau_1-\tau_3}{2}\right)_{i+j+p} \left(\frac{d+t-\tau -\tau_2-\tau_4}{2}+p+i-J\right)_{J-p-i}}\\\nonumber
    &\times\,\left(\tfrac{-t+\tau_1+\tau_2}{2}\right)_j\left(\tfrac{d-2 j+2 t-\tau_1-\tau_2-\tau_3-\tau_4}{2}\right)_j\,\tfrac{\left(\tfrac{d}{2}+ j+p-1\right)_i}{\left(\tfrac{d}{2}+J-i-1\right)_i}.
\end{align}
\end{subequations}}

\paragraph{From the $\sf u$- to the $\sf s$-channel}
In this case, the seed integral we need to evaluate is the following Mellin-Barnes integral at fixed $\sf{t}$:
\begin{multline}
    \mathfrak{J}^{(\sf u)}_{J,l}=\frac{(-1)^l}{2l!}\int \frac{ds}{4\pi i}\,\Gamma \left(\tfrac{-s-\tau_1+\tau_2}{2} \right) \Gamma \left(\tfrac{-s+\tau_3-\tau_4}{2}\right) \Gamma \left(\tfrac{s+t+\tau -\tau_2-\tau_3}{2}\right) \Gamma \left(\tfrac{d-2 r+s+t-\tau -\tau_2-\tau_3}{2}\right)\\\times\,\tilde{\mathcal{C}}_{J,l}^{(\sf u)}\,P^{(\sf{u})}_{J,\tau}(s,t|\tau_i)\,Q_{l,t}^{(t,t+\tau_1-\tau_2-\tau_3+\tau_4,-\tau_1+\tau_2,\tau_3-\tau_4}(s)\,,
\end{multline}
with
\begin{equation}
    \tilde{\mathcal{C}}_{J,l}^{(\sf u)}=\frac{2^{1-2 J}  (J+\tau -1)_J \Gamma (2 J+\tau )}{\Gamma \left(\tfrac{d-2 (r+\tau )}{2} \right) \Gamma \left(\tfrac{2 J+\tau +\tau_1-\tau_3}{2} \right) \Gamma \left(\tfrac{2 J+\tau -\tau_1+\tau_3}{2}\right) \Gamma \left(\tfrac{2 J+\tau +\tau_2-\tau_4}{2} \right) \Gamma \left(\tfrac{2 J+\tau -\tau_2+\tau_4}{2}\right)}
\end{equation}
This integral can be reduced as before to finite sums of standard hypergeometric Mellin-Barnes integrals using the definitions of Mack and Continuous Hahn polynomials. In this case it is convenient to use a different expansion of the continuous Hahn polynomials:
\begin{multline}
    Q_{l,t}^{(t,t+\tau_1-\tau_2-\tau_3+\tau_4,-\tau_1+\tau_2,\tau_3-\tau_4}(s)\\=\,\sum_{q=0}^{\infty}\underbrace{\frac{(-2)^l \left(\tfrac{t-\tau_1+\tau_2}{2}\right)_l \left(\tfrac{t+\tau_3-\tau_4}{2}\right)_l}{(l+t-1)_l}\frac{ (-l)_q (l+t-1)_q}{\left(\frac{t-\tau_1+\tau_2}{2}\right)_q \left(\frac{t+\tau_3-\tau_4}{2}\right)_q}}_{d^{(2)}_{t,l}(q)}\,\left(\tfrac{s+t}{2}\right)_q\,,
\end{multline}
while we focus on the factors in the Mack polynomial which are proportional to $s$.
Combining everything we obtain the following expression:
\begin{equation}\label{FullCrossing2}
    \mathfrak{J}^{(\sf u)}_{J,l}=\frac{(-1)^l}{2l!}\,\tilde{\mathcal{C}}_{J,l}^{(\sf u)}\sum_{k=0}^{[J/2]}\sum_{\sum_i r_i=J-2k}b^{(\sf u)}_{t,\tau}(r_i|k)\,\sum_{q=0}^{\infty}{d}^{(2)}_{t,l}(q)\underbrace{\int \frac{ds}{4\pi i}\,\rho_{M}^{(\sf u)}(r_i|k)\left(\tfrac{s+t}{2}\right)_q}_{I^{(\sf u)}_{r_i|k,q}}\ ,
\end{equation}
where now:
\begin{multline}
    \rho_{M}^{(\sf u)}(r_i|k)=\Gamma \left(\tfrac{2 r_1+s+t}{2}\right) \Gamma \left(\tfrac{2 k-s+\tau -\tau_1-\tau_4}{2}\right) \\\times\,\Gamma \left(\tfrac{d-2 J+2 k-s-\tau -\tau_1-\tau_4}{2}\right) \Gamma \left(\tfrac{2 r_4+s+t+\tau_1-\tau_2-\tau_3+\tau_4}{2}\right),
\end{multline}
and
\begin{multline}
    b^{(\sf u)}_{t,\tau}(r_i|k)=c_{J,k}\,\tfrac{\left(\tfrac{\tau +\tau_1-\tau_3}{2}\right)_J \left(\tfrac{\tau -\tau_1+\tau_3}{2}\right)_J}{(J+\tau -1)_J (d-J-\tau -1)_J}\,\frac{(-1)^{r_1+r_4} (r_1,r_2,r_3,r_4)!}{2^{r_1+r_2+r_3+r_4}}\\\times \frac{\left(\tfrac{-t+\tau_1+\tau_2}{2}\right)_{r_2} \left(\tfrac{-t+\tau_3+\tau_4}{2}\right)_{r_3}\left(\tfrac{-d+\tau +\tau_2-\tau_4+2}{2}\right)_{k+r_1+r_2} \left(\tfrac{-d+\tau -\tau_2+\tau_4+2}{2}\right)_{k+r_3+r_4}}{\left(\tfrac{\tau -\tau_1+\tau_3}{2}\right)_{k+r_1+r_3} \left(\tfrac{\tau +\tau_1-\tau_3}{2}\right)_{k+r_2+r_4}}.
\end{multline}
Following similar steps as in the ${\sf t}$ channel we then arrive to:
\begin{multline}\label{FullCrossingU}
    \mathfrak{J}^{(\sf u)}_{J,l}=\tilde{\mathcal{C}}_{J,l}^{(\sf u)}\sum_{k=0}^{[J/2]}\sum_{\sum_i r_i=J-2k}\sum_{p=0}^{r_1}c^{\sf u}_{t,\tau}(r_i|k,p)\gamma^{\sf u}_{r_i|p}\\\times \,{} _4F_3\left(\begin{matrix}-l+p,l+p+t-1,k+r_1+\tfrac{b_{14}}2,k+r_1+\tfrac{a_{14}}2\\p+\tfrac{t+\tau_1-\tau_2}{2},p+\tfrac{t+\tau_3-\tau_4}{2},2k+r_1+r_4+\tfrac{c}2\end{matrix};1\right)\,,
\end{multline}
with an overall coefficient weighing the sum given by
\begin{subequations}
\begin{align}
    c^{(\sf u)}_{t,\tau}(r_i|k,p)&=\tfrac{2^l (-1)^p \binom{r_1}{p}(-l)_p (l+t-1)_p \left(\tfrac{t-\tau_1+\tau_2}{2}\right)_l \left(\tfrac{t+\tau_3-\tau_4}{2} \right)_l}{2l!(l+t-1)_l \left(\frac{t-\tau_1+\tau_2}{2}\right)_p \left(\frac{t+\tau_3-\tau_4}{2}\right)_p}\,b^{(\sf u)}_{t,\tau}(r_i|k),\\
    \gamma^{\sf u}_{r_i|p}&=\frac{\Gamma \left(k+r_1+\tfrac{a_{14}}{2} \right) \Gamma \left(k+r_4+\tfrac{a_{23}}{2}\right) \Gamma \left(k+r_2+\tfrac{b_{14}}{2}\right) \Gamma \left(k+r_3+\tfrac{b_{23}}{2}\right)}{\Gamma \left(2k+r_1+r_4+\tfrac{c}{2}\right)},
\end{align}
\end{subequations}
where
\begin{subequations}
\begin{align}
    a_{14}&=t+\tau-\tau_1-\tau_4\,,\\
    a_{24}&=t+\tau-\tau_2-\tau_3\,,\\
    b_{14}&=t+d-\tau-2J-\tau_1-\tau_4\,,\\
    b_{23}&=t+d-\tau-2J-\tau_2-\tau_3\,,\\
    c&=d-2J+2t-\tau_1-\tau_2-\tau_3-\tau_4.
\end{align}
\end{subequations}
Following similar steps as for the ${\sf t}$ channel, we can explicitly perform two of the sums and arrive to:
\begin{empheq}[box=\fbox]{align}
\mathfrak{J}^{(\sf u)}_{J,l}(t)&=\mathcal{Z}_{J,l}(t)\,\sum_{p=0}^{
J}d_{p}^{(J,l)}\sum_{i,j=0}^{J-p}a^{(J,l|p)}_{i,j}\\\nonumber
&\times\, {}_4F_3\left(\begin{matrix}
-j,p+i+\tfrac{\tau-d +\tau_2-\tau_4}{2}+1,1-i-j-p-\tfrac{\tau-\tau_1+\tau_3}{2},-\tfrac{t-\tau_1-\tau_2}{2}\\ 1-j+\tfrac{t-\tau_3-\tau_4}{2},p+i-J+\tfrac{d-\tau+\tau_2-\tau_4\ }{2},J-i-j-p+\tfrac{\tau+\tau_1-\tau_3}{2}\end{matrix};1\right)\\\nonumber
&\times {}_4F_3\left(\begin{matrix}-l+p,l+p+t-1,p+i-J+\tfrac{t-\tau_1 -\tau_4}{2}+\tfrac{d-\tau}{2},p+i+\tfrac{t+\tau-\tau_1-\tau_4}{2}\\
p+\tfrac{t-\tau_1+\tau_2}{2},p+\tfrac{t+\tau_3-\tau_4}{2},\tfrac{d-\tau_1-\tau_2-\tau_3-\tau_4}{2}-j+t\end{matrix};1\right)\,,
\end{empheq}
where now:
{\allowdisplaybreaks
\begin{subequations}
\begin{align}
    \mathfrak{J}^{(\sf u)}_{0,0}(t)&=\tfrac{\Gamma (\tau )}{\Gamma \left(\tfrac{d-2 \tau}{2}\right) \Gamma \left(\tfrac{\tau +\tau_1-\tau_3}{2}\right) \Gamma \left(\tfrac{\tau -\tau_1+\tau_3}{2}\right) \Gamma \left(\tfrac{\tau +\tau_2-\tau_4}{2}\right) \Gamma \left(\tfrac{\tau -\tau_2+\tau_4}{2}\right) }\\\nonumber&\qquad\times\,\tfrac{\Gamma \left(\tfrac{t+\tau -\tau_2-\tau_3}{2}\right) \Gamma \left(\tfrac{t+\tau -\tau_1-\tau_4}{2}\right) \Gamma \left(\tfrac{d+t-\tau -\tau_2-\tau_3}{2}\right) \Gamma \left(\tfrac{d+t-\tau -\tau_1-\tau_4}{2}\right)}{\Gamma \left(\tfrac{d+2 t-\tau_1-\tau_2-\tau_3-\tau_4}{2}\right)},\\
    \mathcal{Z}_{J}(t)&=\frac{(-1)^J 2^{l-J} (\tau )_{2 J} \left(\tfrac{d-2 (J+\tau )}{2}\right)_J \left(\tfrac{t-\tau_1+\tau_2}{2}\right)_l \left(\tfrac{t+\tau_3-\tau_4}{2}\right)_l}{l! (l+t-1)_l (d-J-\tau -1)_J \left(\tfrac{\tau +\tau_2-\tau_4}{2}\right)_J \left(\tfrac{\tau -\tau_2+\tau_4}{2}\right)_J}\ \mathfrak{J}^{(\sf u)}_{0,0}(t)\,,\\
    d_p^{(J,l)}(t)&=\frac{p!\binom{l}{p} (l+t-1)_p}{\left(\tfrac{t-\tau_1+\tau_2}{2}\right)_p \left(\tfrac{t+\tau_3-\tau_4}{2}\right)_p},\\
    a^{(J,l|p)}_{i,j}(t)&=(-1)^j \binom{i+p}{p}{\small\begin{pmatrix}
    J\\ j,i+p
    \end{pmatrix}}\tfrac{\left(\tfrac{\tau +\tau_2-\tau_4+2-d}{2}\right)_{i+p} \left(\tfrac{\tau -\tau_2+\tau_4+2-d}{2}\right)_{J-p-i}}{\left(\tfrac{\tau -\tau_1+\tau_3}{2}\right)_{i+j+p} \left(\tfrac{\tau +\tau_1-\tau_3}{2}\right)_{J-p-i-j}}\\\nonumber
    &\times\,\tfrac{\left(\tfrac{t+\tau -\tau_1-\tau_4}{2}\right)_{i+p} \left(\tfrac{t+\tau -\tau_2-\tau_3}{2}\right)_{J-p-i-j}}{\left(\tfrac{d-2 (p+i+j)+t-\tau -\tau_2-\tau_3}{2}\right)_{i+j+p} \left(p+i-J+\frac{d+t+\tau -\tau_1-\tau_4}{2}\right)_{J-i-p}}\\\nonumber
    &\times\,\left(\tfrac{-t+\tau_3+\tau_4}{2}\right)_j\left(\tfrac{d-2 j+2 t-\tau_1-\tau_2-\tau_3-\tau_4}{2}\right)_j\,\tfrac{\left(\tfrac{d}{2}+ j+p-1\right)_i}{\left(\tfrac{d}{2}+J-i-1\right)_i}.
\end{align}
\end{subequations}}

\section{Mixed-symmetry CPWs and projectors}\label{Mixed}

In the most general case where external spins are totally symmetric operators, the operators that can contribute in all channels mixed symmetry operators with at most three rows. In order to work with such mixed-symmetry representations, it is convenient to introduce auxiliary variables $z_i$, where $i=1,2,3$ labelling each of the three rows. On can then encode projectors onto such representations in polynomials of $z_i\cdot \bar{z}_j$, $z_i\cdot z_j$, $\bar{z}_i\cdot\bar{z}_j$,
\begin{equation}\label{amixppoly}
    p(z_i\cdot \bar{z}_j,z_i\cdot z_j,\bar{z}_i\cdot\bar{z}_j),
\end{equation}
where the $\bar{z}_i$ correspond to the second set of legs in the projector and \eqref{amixppoly} is invariant under the exchange of the variable $z_i$ with $\bar{z}_i$. The conditions that these polynomials are irreducible is a simple lowest weight condition:
\begin{align}\label{YT}
    z_i\cdot\pl_{z_j}\,p(z_i\cdot \bar{z}_j,z_i\cdot z_j,\bar{z}_i\cdot\bar{z}_j)=0\,,
\end{align}
while to project out traces one should also impose:
\begin{align}
    \pl_{z_1}\cdot\pl_{z_1}\,p(z_i\cdot \bar{z}_j,z_i\cdot z_j,\bar{z}_i\cdot\bar{z}_j)=0\,,
\end{align}
where it is sufficient to impose only one equation when \eqref{YT} is satisfied. I.e. that \label{mixppoly} is in the kernel of $\pl_{z_2}\cdot\pl_{z_2}$ and $\pl_{z_3}\cdot\pl_{z_3}$ is automatic when \eqref{YT} holds.

Some examples of the above are projectors onto $(l,1)$ Young tableaux:\footnote{This is a particular case of the general formula
\begin{equation}
    p_{(l_1,l_2)}=\frac{l_1-l_2+1}{l_1+1}\left[(z_1\cdot\bar{z}_1)^{l_1}(z_2\cdot\bar{z}_2)^{l_2}+\ldots\right]\,.
\end{equation}}
\begin{equation}
    p_{(l,1)}=\frac{l}{l+1}\left[(z_1\cdot\bar{z}_1)^l(z_2\cdot\bar{z}_2)-(z_1\cdot\bar{z}_1)^{l-1}(z_1\cdot\bar{z}_2)(z_2\cdot\bar{z}_1)\right]+\mathcal{O}(z_i\cdot z_i)
\end{equation}
as well as other projectors which can be easily obtained solving the conditions \eqref{YT}. Above we do not give the explicit form of the trace terms (i.e. the $\mathcal{O}(z_i\cdot z_i)$) since these only give descendant contributions to the CPWs and are straightforward to obtain. The above normalisation ensures that the above is a projector:
\begin{equation}
   \frac{1}{l!}\, p_{(l,1)}(z_i,\pl_{\bar{\xi}_i})\,p_{(l,1)}(\xi_i,\bar{z}_i)=p_{(l,1)}(z_i,\bar{z}_i)\,.
\end{equation}

With the above projectors, mixed symmetry representations can be readily encorporated into the formalism of this work upon projecting the OPE structures onto the corresponding irreducible representations. We focus on the structures relevant for double-trace operators, which are those proportional to the mean-field theory OPE structures:
\begin{equation}
    {\sf H}_{1z}^{J_1}{\sf H}_{2v}^{J_2}{\sf Y}_{w}^{J_3-J_1-J_2}\,,
\end{equation}
where we have introduced the auxiliary variables $z$, $v$ and $w$ in place of the single variable $z_3$ to account for all possible Young projection that can be realised:
\begin{subequations}
\begin{align}
        {\sf H}_{1z}&=\frac1{y_{01}^2}\left(z\cdot z_1+2\,\frac{z\cdot y_{01}\,z_1\cdot y_{10}}{y_{01}^2}\right)\,,\\
    {\sf H}_{2v}&=\frac1{y_{02}^2}\left(v\cdot z_2+2\,\frac{v\cdot y_{02}\,z_2\cdot y_{20}}{y_{02}^2}\right)\,,\\
    {\sf Y}_{w}&= \frac{w\cdot y_{01}}{y_{01}^2}-\frac{w\cdot y_{02}}{y_{02}^2}\,,
\end{align}
\end{subequations}
together with their barred counterparts depending on $y_3$ and $y_4$ along the lines of the list in \eqref{4ptstructures}.

We can thus work out irreducible mixed-symmetry  contributions to OPE structures by simply acting with the projectors  above. In the table below we present the OPE structure relevant for double-trace operators contributing to $\mathcal{A}_{1010}$ and $\mathcal{A}_{2020}$ 4pt correlators.
{\small
\begin{center}
\begin{tabular}{ |c|c|c|c|  }
 \hline
 Correlator & $(l,0)$ & $(l-1,1)$& $(l-2,2)$\\
 \hline
 $\mathcal{A}_{1010}$   & ${\sf H}_{1z}{\sf Y}_z^{l-1}$   & $\frac{l-1}l \left({\sf H}_{v1} {\sf Y}_z^{l-1}-{\sf H}_{z1} {\sf Y}_v {\sf Y}_z^{l-2}\right)$  & 0  \\
 \hline
 $\mathcal{A}_{2020}$   & ${\sf H}_{1z}^2{\sf Y}_z^{l-2}$   & $\frac{l-2}l \left({\sf H}_{z1}{\sf H}_{v1} {\sf Y}_z^{l-2}-{\sf H}_{z1}^2 {\sf Y}_v {\sf Y}_z^{l-3}\right)$   &$\tfrac{l-3}{l-1}\left(\begin{matrix}{\sf H}_{z1}^2{\sf Y}_{v}^2{\sf Y}_z^{l-4}-2{\sf H}_{z1}{\sf Y}_{v1}{\sf Y}_z^{l-3}{\sf Y}_v\\+{\sf H}_{v1}^2{\sf Y}_z^{l-2}\end{matrix}\right) $  \\
 \hline
\end{tabular}
\captionof{table}{A table of the OPE structures contributing to double-trace OPE coefficients.}\label{tab2}
\end{center}}
The result for the ${}_i\mathcal{Q}$ polynomials used in this work was obtained by explicitly evaluating the corresponding conformal integrals with the above OPE structures. With the same techniques we can also evaluate more general CPWs associated to more complicated OPE structures.

\section{Twist block operator on Mellin amplitudes}\label{DiffRel}
In this appendix we present the explicit form of the twist block difference operator acting on Mellin amplitudes with arbitrary twist scalar external legs:
{\allowdisplaybreaks
\begin{align}
    \mathcal{T}_{\tau}\,\mathfrak{M}(s,t)&=
    \tfrac{(t-\tau ) (2-2 d+t+\tau)}2\Bigg[p_1(s)\mathfrak{M}(s,t) +(s+t)(s+t+\tau_1-\tau_2-\tau_3+\tau_4) \mathfrak{M}(s+2,t)\\\nonumber
    &\hspace{140pt}+ (s+\tau_1-\tau_2) (s-\tau_3+\tau_4) \mathfrak{M}(s-2,t)\Bigg]\\\nonumber
    &+\tfrac{(t-\tau_1-\tau_2) (t-\tau_3-\tau_4)}{2} \Bigg[p_2(s,t)\mathfrak{M}(s,t-2) +p_3(s,t) \mathfrak{M}(s+2,t-2)\Bigg]\\\nonumber
    &+\frac{1}{2} (t-\tau_1-\tau_2 )(t-\tau_1-\tau_2-2 )(t-\tau_3-\tau_4 )(t-\tau_3-\tau_4 -2) \mathfrak{M}(s+2,t-4)\,,
\end{align}}
where we have defined the following polynomials:
\begin{subequations}
\begin{align}
    p_1(s,t)&=d^2-2 d (\tau +1)+2 s^2+2 s (t+\tau_1-\tau_2-\tau_3+\tau_4)+\tau ^2\\\nonumber&+t (\tau_1-\tau_2-\tau_3+\tau_4+2)+2 \tau -(\tau_1-\tau_2) (\tau_3-\tau_4)\,,\\
    p_2(s,t)&=d^2-d (2 s+2 \tau +\tau_1-\tau_2-\tau_3+\tau_4+4)\\\nonumber
    &+t (2 s+\tau_1-\tau_2-\tau_3+\tau_4+2)+\tau  (\tau +2)+(\tau_1-\tau_2) (\tau_3-\tau_4)\,,\\
    p_3(s,t)&=d^2+d (2 s+2 t-2 \tau +\tau_1-\tau_2-\tau_3+\tau_4-4)\\\nonumber&+t (-2 s-\tau_1+\tau_2+\tau_3-\tau_4+2)-2 t^2+\tau  (\tau +2)-(\tau_1-\tau_2) (\tau_3-\tau_4)\,.
\end{align}
\end{subequations}
Analogously the quadratic Casimir reads as a difference operator:
\begin{align}
    \mathcal{C}_2\,\mathfrak{M}(s,t)&=\frac{1}{2}\,q(s,t) \mathfrak{M}(s,t) +\frac{1}{2} (s+\tau_1-\tau_2) (s-\tau_3+\tau_4) \mathfrak{M}(s-2,t)\\\nonumber
    &-\frac{1}{2} (t-\tau_1-\tau_2) (t-\tau_3-\tau_4) \mathfrak{M}(s,t-2)-\frac{1}{2} (t-\tau_1-\tau_2) (t-\tau_3-\tau_4) \mathfrak{M}(s+2,t-2)\\\nonumber
    &+\frac{1}{2} (s+t)(s+t+\tau_1-\tau_2-\tau_3+\tau_4) \mathfrak{M}(s+2,t)\,,
\end{align}
in terms of the polynomial:
\begin{multline}
    q(s,t)=t^2+t (-\tau_1+\tau_2+\tau_3-\tau_4)-2 d t-2 s^2\\-2 s (t+\tau_1-\tau_2-\tau_3+\tau_4)+(\tau_1-\tau_2) (\tau_3-\tau_4)\,.
\end{multline}

\end{appendix}

\bibliography{refs}
\bibliographystyle{JHEP}

\end{document}